\documentclass[a4paper, 12pt, twoside,notitlepage,useAMS]{book}
\pdfoutput=1
\usepackage{a4,epsfig,fancyhdr,anysize,natbib,amssymb,amsmath,fleqn,color}
\marginsize{3.0cm}{2.0cm}{3.0cm}{3.0cm}
\newcommand{\be}{\begin{equation}}
\newcommand{\ee}{\end{equation}}
\newcommand{\vc}[1]{\boldsymbol{#1}}
\renewcommand{\t}[1]{\it{\bf{#1}}}

\newcommand{\mt}{\mathrm}

\begin{document}
\renewcommand\baselinestretch{1.2}
\renewcommand\thepage{\roman{page}}
\pagestyle{fancy}
\renewcommand{\chaptermark}[1]{\markboth{{}\
   \emph{\small{#1}}}{}}

\renewcommand{\sectionmark}[1]{\markright{\emph{\small{#1}}}}
\fancyhf{}

\fancyhead[LE,RO]{\thepage} 
\fancyhead[RE]{\nouppercase{\leftmark}} 
\fancyhead[LO]{\rightmark}
\renewcommand{\headrulewidth}{0.0pt}
\renewcommand{\footrulewidth}{0pt}

\thispagestyle{empty}
\newcommand{\HRule}{\rule{\linewidth}{1mm}}
\setlength{\parindent}{1cm}
\setlength{\parskip}{0.8mm}
\newpage

\vspace*{0.01cm}
\noindent
\begin{center}
\Huge {Variability of Black-Hole Accretion Discs \\ \vskip0.3cm} \LARGE{\emph{A theoretical study}} \\[5mm]
\end{center}
\HRule
\vskip 2cm
\begin{center} \LARGE B\'arbara Trov\~ ao Ferreira\end{center}
\begin{center} \Large
 Trinity College \\
 University of Cambridge

\vskip 4.5cm

\begin{figure}[!h]
\begin{center}
\includegraphics[width=2.6cm]{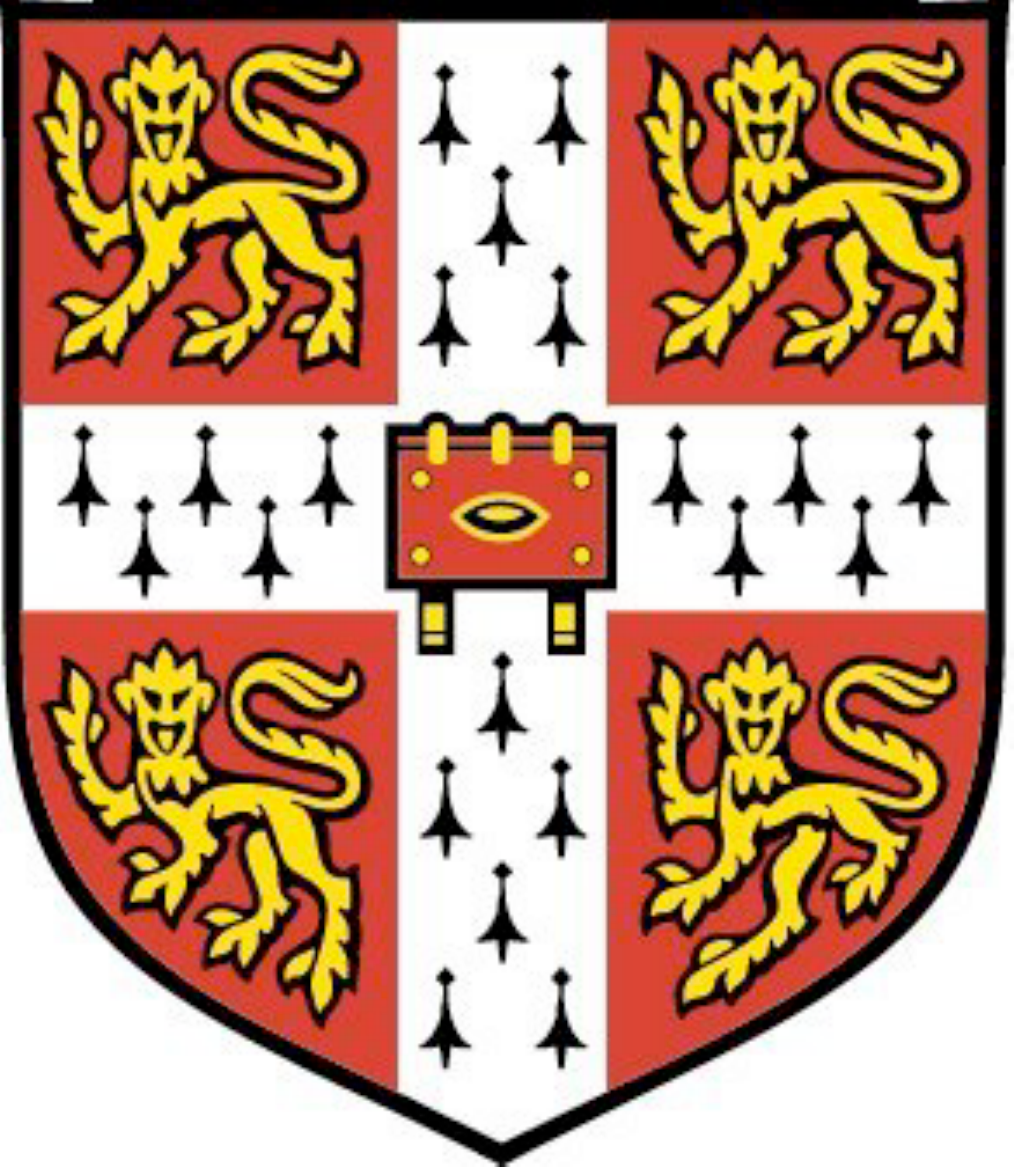}
\end{center}
\end{figure}
\end{center}

\begin{center}\Large {A dissertation submitted for the degree of\\
   \emph{Doctor of Philosophy} \\ May 25, 2010}\end{center}

\newpage
\phantom{thispage}\thispagestyle{empty}

\newenvironment{dedication}   {\newpage \thispagestyle{empty}
\vspace*{\stretch{1}} \begin{center} \em}   {\end{center}
\vspace*{\stretch{3}} \clearpage}
\begin{dedication}\begin{flushright}{\it \Large{Ao meu pai}}\end{flushright}\end{dedication}

\newpage
\phantom{thispage}\thispagestyle{empty}

\newpage
\thispagestyle{empty}
\vspace*{\stretch{0.8}}
\noindent
\section*{Acknowledgments}
\vskip2em
This thesis is a statement of the intellectual debt I owe to many academics
and was made possible through the encouragement of various people. To all
goes my most sincere appreciation.

Above all I acknowledge the support, guidance and patience of my PhD
supervisor, Gordon Ogilvie. My interest in accretion discs was driven by
his stimulating and thorough lectures on Astrophysical Fluid Dynamics
and what I learnt from him on the subject is beyond acknowledgement. I am
deeply thankful for the instructive discussions we had throughout my PhD,
for his always useful remarks and suggestions and for his exhaustive review
of my written work.

I would further like to express my gratitude to Omer Blaes, Chris Fragile
and Ken Henisey for their enthusiasm and for motivating discussions which
greatly contributed to clarify my ideas on various matters. I would also
like to thank John Papaloizou, for comments and suggestions which helped
improve different parts of this thesis, and Michael Proctor and Chris Done
for insightful questions.

I am truly grateful for the many conversations with Toby, Mark and Chris
for either directly assisting my research or taking me away from it
when needed.

On practical matters, I acknowledge the financial support of the Portuguese
Government's \emph{Funda\c c\~ao para a Ci\^encia e a Tecnologia} through a PhD grant and
Trinity College for providing me with a home during my stay in Cambridge
and for being a place of many social distractions.

I am fortunate to be surrounded by many great individuals whose affection,
encouragement and joy carried me through my time in Cambridge. To my close
friends goes my most sincere thanks: the old ones for being a comforting
constant and the new ones for making my graduate years infinitely more
pleasant.

Finally, I would like to thank Lu\'isa and the other Bernardinos for their joyful support and
my father, to whom I dedicate this thesis, for his unconditional love, for
being primarily responsible for who I have become and for supporting me in
my every decision.

\vspace*{\stretch{1}}

\newpage
\phantom{thispage}\thispagestyle{empty}


\newpage
\thispagestyle{plain}
\section*{Declaration}
\vskip2em
This dissertation is based on research done at the Department of
Applied Mathematics and Theoretical Physics from October 2006 to December
2009. The material presented is original unless reference is
made to work of others in the text. The research work described in this thesis was done in collaboration with my supervisor Dr. Gordon Ogilvie. Part of that research is presented in the following papers:

\begin{itemize}
\item \textbf{Chapter \ref{excmech}} $\quad$ B\'arbara T. Ferreira and Gordon I. Ogilvie, {\it On an excitation mechanism for trapped inertial waves in accretion discs around black holes}, Monthly Notices of Royal Astronomical Society 386, 2297 (2008).

\item \textbf{Chapter \ref{warpecc}} $\quad$ B\'arbara T. Ferreira and Gordon I. Ogilvie, {\it Warp and eccentricity propagation in discs around black holes}, Monthly Notices of Royal Astronomical Society 392, 428 (2009).
\end{itemize}
I have also contributed to the following paper, done in collaboration with Ken Henisey, Professor Omer Blaes and Dr. P. Chris Fragile:

\begin{itemize}
\item Ken B. Henisey, Omer M. Blaes, P. Chris Fragile and B\'arbara T. Ferreira, {\it Excitation of trapped waves in simulations of tilted black hole accretion disks with magnetorotational turbulence}, Astrophysical Journal, 706, 705 (2009).
\end{itemize}
Reference to this work, discussed mainly in chapter \ref{conc}, is made occasionally by citing \cite{heniseyetal2009}. The research described in chapters \ref{timeaccretion} and \ref{reflect} remains unpublished.

\vspace{3cm}

\begin{flushright}
B\'arbara Trov\~ao Ferreira\footnote{E-mail: barbara.t.ferreira@gmail.com}

Cambridge, May 25, 2010
\end{flushright}

\newpage
\phantom{thispage}\thispagestyle{empty}
\newpage
\thispagestyle{plain}
\section*{Abstract}
\vskip2em
Accretion discs are fluid-dynamical entities which surround many black holes. Observations reveal that these systems exhibit variability on a range of time scales. This thesis investigates phenomena occurring in black-hole accretion discs which are likely to induce high-frequency quasi-periodic variability. Two classes of pseudo-relativistic theoretical models are investigated. 

The first is based on the stability of transonic accretion flows and its connection to a disc instability that takes the form of propagating waves (viscous overstability). The time-dependent study looks at the conditions under which the transition between subsonic disc-like accretion, which occurs at large radii, and the supersonic flow characteristic of the immediate vicinity of the black hole is stable. In agreement with previous findings, results indicate that the system reaches a steady state for low viscosity. Above that threshold the transonic solutions are unstable to viscous overstability. The overstable inertial-acoustic waves appear to be excited near the maximum of the epicyclic frequency and are global in the sense that their frequency is maintained for a wide range of radii. 

The second class of models looks at accretion-disc oscillations which are trapped due to the non-monotonic variation of the epicyclic frequency in relativistic flows. In particular, it focuses on inertial waves trapped below the maximum of the epicyclic frequency which are excited in deformed, warped or eccentric, discs. The excitation mechanism involves a non-linear coupling between the global deformation, an intermediate wave and the inertial mode and results, under a variety of conditions, in growth of the latter. Excitation is only effective when global deformations are capable of reaching the inner disc with non-negligible amplitude. With that in mind, the conditions favourable to the propagation of warped and eccentric modes from the outer to the inner regions are analysed. Another aspect that is taken into account is the influence of a transonic background, ignored in the coupling calculations, on the propagation of modes in the disc. It is found that, under certain conditions, inertial waves may be severely affected or destroyed in this background. On the other hand, results indicate that the decay rate of inertial waves due to the presence of the radial inflow is small in sufficiently thin discs. In this case, the coupling mechanism can still work to excite trapped inertial modes.

\thispagestyle{plain}

\begin{dedication}\begin{flushright}{\it
      \Large{}}\end{flushright}\end{dedication}
\thispagestyle{plain}

\tableofcontents
\thispagestyle{empty}

\newpage
\thispagestyle{empty}
\renewcommand\thepage{\arabic{page}}
\setcounter{page}{1}



\part{Introduction}
\label{i}

\chapter{Introduction}
\label{intro}

\noindent
Variability is a defining feature of astrophysical objects. Jupiter has a constantly changing atmosphere and its satellite Io sees its surface modified by over 400 active volcanoes. The Sun changes its spots during the 11-year cycle while other stars vary strongly in brightness as they pulsate. In the solar case, variability is intrinsic to the astrophysical object. It can also be extrinsic if, e.g., caused by an eclipsing star in a binary which produces luminosity variations as seen by an observer on Earth. Io's variability is indirectly caused by Jupiter --- tidal effects are responsible for the planet's internal heating and consequent geological activity. Whether intrinsic or extrinsic, variations can be sporadic, irregular, semi-regular or periodic and may occur at a myriad of time scales even in a single object.

One of the most spectacular examples of variability is that of active galactic nuclei (AGN). Observations show that these compact regions, located at the centre of some galaxies, are not only abnormally luminous but also show extremely rapid variations. These phenomena have long \citep{lyndenbell1969} been attributed to the presence of a supermassive black hole which releases energy as it accretes matter in its surroundings.  

The discovery of the first AGN and galactic X-ray sources in the 60s and the theoretical studies that followed established accretion as a major energy production mechanism. Subsequently, accretion around black holes was the target of a series of ground-breaking studies, some of which established the basis of accretion disc theory and remain the standard reference today \citep{pr1972,ss73,nt73}. Other important developments in the theoretical modelling of accretion flows arrived scatteredly in the following 20 years or so with the identification of accretion solutions beyond the standard thin disc \citep[e.g.][]{ichimaru1977,rbbp1982,adaf} and with the re-discovery of an efficient mechanism for angular momentum transport (fundamental for accretion) by \cite{mri}.

The decades of 60 and 70 also saw major improvements in observational astronomy. New techniques were developed and telescopes and satellites able to capture the light emitted by astrophysical objects at the full spectral range began to operate. Of particular importance was the development of X-ray astronomy. This highly energetic emission is expected in accreting compact objects, such as neutron stars or black holes, where the surrounding gas is heated to very high temperatures as it falls in the gravitational field of these bodies.

The first galactic X-ray source, Scorpius X-1, was detected in the early 60s and the studies of \cite{shklovsky1967} connected the X-ray emission from the object to gas accretion onto a neutron star. More X-ray objects were discovered with the launch of various satellites in the 70s. Most sources were believed to be in binaries where a normal star provided the material to be accreted by the neutron star or black hole; the term X-ray binaries was coined to describe such systems. The presence of a normal star permits a lower limit for the mass of the compact object to be determined, providing evidence for the very existence of black holes. Another X-ray source, Cygnus X-1, was the first strong candidate and is now widely believed to be one of these relativistic objects.

Variability in the X-ray emission of galactic sources has been a constant from the very beginning of the observational burst of such systems. For example, Cygnus X-1 has long been known to exhibit variability on time scales down to a millisecond as shown by \cite{rothschildetal1974}. The same authors suggested this could be related to the turbulence in the accretion disc surrounding the source. At longer time scales, variability may be due to changes in the mass accretion rate \citep[e.g.][in the case of Scorpius X-1]{pb1968}.

X-ray astronomy had another major impulse in the last couple of decades with the launch of several X-ray observatories, including NASA's \emph{Rossi X-ray Timing Explorer} (RXTE). It is mainly thanks to these satellites that, so far, about 20 X-ray binaries believed to contain a black hole have been identified and analysed in detail. These new observatories also contributed to important developments in the field of X-ray variability, in particular at very short time scales. One of the most exciting discoveries was that of quasi-periodic oscillations (QPOs) with frequencies up to 450 Hz. The very high frequencies connect these oscillations to the inner regions of the accretion disc, making them fundamental in the study of the relation of relativistic compact objects with their surroundings.

Measurements of X-ray variability and theoretical studies of black-hole disc accretion have therefore been intimately related for the past 30 or 40 years. Combined, they provide a unique window to the physics of strong gravitational fields, allowing for properties of both the accretion disc and the black hole to be inferred. Moreover, the development of X-ray observations and black-hole accretion disc theory is fundamental to test not only stellar evolution theories but also general relativity. 

The launch of new satellites provided a significant boom in the field of observational X-ray astronomy and X-ray variability but black-hole accretion disc theory is, at present, far from being up-to-date with its observational counterpart. With this thesis, I aim to contribute to the development of the theory of accretion discs around black holes with a particular emphasis on the rapid variability of such objects. The research done focuses on two major topics. The first concerns the stability of models that describe the transition between the region where accretion is disc-like and subsonic to the region inside the marginally stable orbit where typical radial velocities are supersonic (Part \ref{ta}). The second focuses on the study of oscillations in black-hole accretion discs with emphasis on an excitation mechanism involving global deformations and on wave-reflection properties at the inner disc boundary (Part \ref{os}). A relation with quasi-periodic variability is explored in both parts. 

In the current chapter, I present the general background of black-hole accretion disc theory and observations and underline the importance of studying the variability of these objects. A more detailed outline of the research described in this thesis is given at the end of Part \ref{i}.

\section{Fundamentals of accretion disc theory}

According to the Oxford English Dictionary accretion can be defined as ``the assimilation of external matter by a growing body". This physical process is of key importance in astrophysics. When the ``growing body" (the central object) is compact, accretion is an extremely efficient energy release mechanism believed to power some of the most fascinating objects in the Universe. The basic idea is simple: a particle in orbit in the gravitational field of a body of mass $M$ and radius $R_*$, located at a distance $R$ from the central object radiates away potential energy as it moves to smaller radii eventually reaching the central body. The radiated energy is proportional to $M/R_*$, making accretion more efficient the heavier and more compact the central object is.

In more realistic situations, accretion of a gas as opposed to accretion of a point mass is considered. Early calculations described such process as being spherically symmetric \citep[e.g.][]{bondi}. However, it is now widely accepted that, in most cases of astrophysical interest, accretion occurs by means of a disc: the accreted material becomes rotationally supported as it approaches the central mass (aside, possibly, from the very inner region). The reason is that the accreted matter will generally possess nonzero angular momentum which has to be removed for accretion to occur.  \enlargethispage{\baselineskip}

This can easily be understood by going back to the point mass accretion case for a moment. It is clear that energy is released when the particle moves inwards. If the point mass moves around the central object in a circular orbit  (the lowest-energy trajectory for a given angular momentum) it possesses a nonzero specific angular momentum equal to $\sqrt{GMR}$, $G$ being the gravitational constant. It is therefore straightforward to see that if the particle is to move inwards, angular momentum needs to be removed. An accretion disc is an example of a flow where energy is released by transferring angular momentum outwards and mass inwards. The outward transport of angular momentum occurs due to the action of viscous torques (friction) within the disc. The exact mechanism in the origin of this transport is still a central issue in accretion-disc theory and is discussed further below.

A disc-like structure is indeed the preferred shape originating from a rotating cloud of gas. Most astrophysical objects form from the gravitational collapse of such gas clouds. If gravity is the main force in the problem, the gas collapses onto a point in a spherically-symmetric fashion. However, if the cloud is initially rotating, the centrifugal force --- which mainly acts in the plane perpendicular to the axis of rotation --- should also be taken into account. The balance between gravity and centrifugal force is such that the result of the collapse is a disc instead of a point. Even if the cloud is initially slowly rotating, by conservation of angular momentum rotation becomes faster and faster as the object collapses.

There are many examples of accretion discs in astrophysics. They are believed to exist around most young stellar objects similar to the primordial Sun, in interacting binary stars and AGNs just to name a few. In addition, direct imaging has been obtained of protoplanetary discs \citep{hh30} and of an obscuring disc surrounding an AGN \citep{agn}. The existence of accretion discs is beyond question.

The focus of this thesis is on black-hole accretion discs typically present in close binary systems. In the case of interest, the more massive (primary) star in the binary evolves more rapidly than the secondary, reaching the end of its life cycle as a black hole. If the normal star, still evolving, and the compact object are close enough, the secondary may fill its Roche lobe and start transferring matter to its companion. Due to the rotation of the binary, the transferred material has too much angular momentum and is unable to fall directly onto the hole. Instead, the matter is spread around the compact object forming an accretion disc which allows for angular momentum to be reduced before the material is accreted. The disc is thought to be thin enough so that packets of gas or fluid elements move in approximately Keplerian orbits (except very close to the black hole or neutron star, where general relativistic effects have to be taken into account). Part of the orbital energy is converted into heat by dissipative processes and eventually some of it is radiated away. Viscosity can provide efficient dissipation and enable angular momentum transport: neighbouring rings orbiting at slightly different velocities rub against each other and the viscous stresses cause the matter in the ring to be spread and to dissipate energy, causing the gas to slowly spiral inwards. Angular momentum is transferred outwards and eventually removed from the outer disc and given to the binary orbit through tidal torques exerted by the secondary.

\subsection{Viscosity}

As seen before, for the gas in a disc to sink further into the gravitational potential, and eventually be accreted by the central object, it needs to lose angular momentum by means of viscous torques. What remains unknown, and is to date not completely understood, is the exact nature of this ``viscosity". The time for the gas to spiral in from the secondary to the object in the centre of the disc can be estimated from observations of dwarf novae and X-ray transients \citep[e.g.][]{kingetal2007}. This gives an indication of how efficient the angular momentum transport mechanism needs to be to explain the measured accretion times. Such estimates ruled out natural viscosity (related to the momentum transport  due to thermal motion of particles), which provides a coefficient of kinematic viscosity $\nu$ several orders of magnitude smaller than that required to explain the observed accretion times. A promising idea to interpret the rate of removal of angular momentum considers an additional form of viscosity: that coming from the turbulent motions in the gas. This ``turbulent viscosity" $\nu_\mathrm{T}$ takes into account the momentum transport due to turbulent eddies which can be thought of as large-scale molecules. This enhanced viscosity provides the main contribution to the total coefficient of kinematic viscosity which, as a result, has the necessary magnitude to agree with observational measurements.

The idea that turbulent motions would enable outward transport of angular momentum was considered by \cite{vonw1948} and later by \cite{ss73}. However, a full discussion of the origin of such turbulence was avoided. Instead of attempting to derive $\nu_\mathrm{T}$ from a turbulence model, Shakura $\&$ Sunyaev adopted a ground-breaking parametrisation for the viscosity based on dimensional considerations: $\nu_\mathrm{T}=\alpha c_\mathrm{s} H$. (Here $c_\mathrm{s}$ is the isothermal sound speed in the disc and $H$ is the vertical semi-thickness or typical scale-height). The magnitude of the viscosity and the efficiency of angular momentum transport are therefore characterised by only one  dimensionless parameter, $\alpha$, which is less than, or possibly comparable to, unity for subsonic turbulence. This simple prescription is valid regardless of the nature of the stress tensor and allowed for accretion disc ($\alpha$-)models to be constructed by bypassing the discussion on the origin of turbulent viscosity.

In years that followed the publication of Shakura and Sunyaev's influential paper, many attempts to isolate instabilities capable of giving rise to turbulent processes were carried out \citep[see review by][]{papaloizoulin1995}. However, only in 1991 a promising mechanism was re-discovered and applied to accretion discs. The idea of \cite{mri} involves the action of a hydromagnetic instability originally discovered by \cite{velikhov1959} and \cite{chandrasekhar1960}. The magnetorotational instability (MRI) is the process by which weak magnetic fields are amplified by differential rotation via axisymmetric disturbances to the circular motion. The principles and action of this process can be understood by analysing the behaviour of two fluid elements of masses $m_1$ and $m_2$ moving in two Keplerian orbits (angular velocity $\Omega=\Omega_\mathrm{K}\propto R^{-3/2}$) around a central body. The element $m_2$ is slightly further away from the centre and therefore has a smaller orbital speed and larger angular momentum than $m_1$. If the system is permeated by a magnetic field, the magnetic forces will act in such a way that the fluid elements can be thought of as being connected by a spring. If the elements are displaced away from each other perpendicularly to a field line, the attractive magnetic tension acts to return the elements to their original position in a fashion similar to a spring under tension. This force acts to pull back the element rotating faster, $m_1$, while $m_2$ is pushed forward. The torque acting on the former/latter is therefore negative/positive implying a loss/gain of angular momentum, that is, the angular momentum of $m_1$ is transferred outwards to $m_2$. This process is unstable because the loss of $m_1$ makes it move closer to the central mass while $m_2$ moves further away since $l\propto \sqrt{R}$. The tension becomes stronger, the torques larger, and the outward transfer of angular momentum unstoppable (until, of course, reconnection and other forms of dissipation enter the picture and control the amplification of the magnetic field). A more detailed explanation of the MRI and its role in the outward transfer of angular momentum can be found in, e.g., \cite{reviewmri}. 

Notwithstanding this important step towards the understanding of accretion, in practice it is common to simply use the traditional Shakura--Sunyaev formula to describe viscous torques. It is widely accepted that discs are indeed turbulent, permeated by magnetic fields and dynamically unstable to axisymmetric disturbances. However, in many problems of interest, it is enough to keep these ideas in the back of one's mind and use a simple viscosity prescription (which, should be noted, is not incompatible with MRI turbulence) and hydrodynamic equations to describe accretion discs. Indeed, magnetic fields and turbulence are not directly taken into account in the calculations made throughout this thesis. (The range of length- and time-scales addressed could not be covered by the 3D numerical simulations required to account for MRI turbulence.) Despite its many simplifying assumptions, including the ones just mentioned, the Shakura--Sunyaev model is still the standard reference for the structure of steady accretion discs today. It is described in detail in the following section.

\subsection{Standard model}

The description of the \cite{ss73} model made here is a short review of the work presented in the original paper, in \cite{accretionpower}, \cite{pringle1981}, \cite{blaesnotes} and in \cite{ogilvienotes}. The basic equations and assumptions of the model are described but the reader should refer to the mentioned articles for a more detailed discussion. In this section (and throughout this thesis), cylindrical coordinates $(R,\phi,z)$ are used to describe discs. The mid-plane is at $z=0$ while the central mass is at $R=z=0$.

\subsubsection{Set of equations and assumptions}

In the famous \cite{ss73} hydrodynamic model, the disc is assumed to be geometrically thin. This is the natural assumption originating from the qualitative picture of a rotating cloud of gas: the gravitational attraction of the central mass and the centrifugal force provide the dominant balance. 
The collective effects of the fluid, such as pressure, viscosity or turbulence, are weak and as a result particles in the gravitational field follow approximately Keplerian orbits, assumed to be circular and coplanar. Standard discs are non-self-gravitating with a total mass negligible when compared to the mass of the central object (which is not significantly altered due to accretion). The mass accretion rate is taken to be constant or slowly varying and the time for accretion is slow compared to other characteristic time scales in the disc. Therefore, Shakura--Sunyaev discs are cool and optically thick since there is enough time available for the gravitational energy released due to accretion to dissipate. 

Many astrophysical objects can be treated as fluids because the typical collisional mean free path of microscopic particles is considerably smaller than the macroscopic length scale associated with such objects. Accretion discs are no exception and are typically described using fluid-dynamical equations. In general, the hydrodynamic equations describing mass and momentum conservation in the flow can be written as
\be
\frac{\partial\rho}{\partial t}+\nabla\cdot\left(\rho\vc{u}\right)=0,
\label{hydro1}
\ee
\be
\rho\left(\frac{\partial\vc{u}}{\partial t}+\vc{u}\cdot\nabla\vc{u}\right)=-\rho\nabla\Phi-\nabla p+\nabla\cdot\t{T},
\label{hydro2}
\ee
where $\rho$, $p$ and $\vc{u}$ are the density, pressure and velocity of the fluid, respectively, $\Phi$ is the gravitational potential and $\t{T}$ is the stress tensor. The only stress component relevant in an axisymmetric system characterised by circular orbital motion is $T_{R\phi}=\mu R d\Omega/dR$, i.e., it is given by the dynamic viscosity, $\mu=\rho\nu$, times the shear rate. The radially-outward transfer of angular momentum is described by the azimuthal component of (\ref{hydro2}). This is the angular momentum conservation equation.

The thin-disc approximation is equivalent to assuming that the disc material lies very close to the mid-plane $z=0$. The hydrodynamic equations are then dramatically simplified. The small $H/R$ ratio of the disc implies that vertically-integrated quantities can be used to describe its radial structure. In an axisymmetric flow, the vertical integration of equations (\ref{hydro1}) and the $\phi$ component of (\ref{hydro2}) results in \citep[e.g.][]{pringle1981}
\be
\frac{\partial \Sigma}{\partial t}+\frac{1}{R}\frac{\partial}{\partial R}\left(R\Sigma \bar{u}_R\right)=0,
\label{masscon}
\ee
\be
R\frac{\partial}{\partial t}\left(\Sigma R^2\Omega\right)+\frac{\partial}{\partial R}\left(\Sigma R^3\Omega\bar{u}_R\right)=\frac{\partial}{\partial R}\left(\bar{\nu}\Sigma R^3\frac{\mt{d}\Omega}{\mt{d}R}\right),
\label{angint}
\ee
where 
\be
\Sigma(R,t)=\int^{\infty}_{-\infty}\rho(R,z,t) \mt{d}z,
\ee
\be
\bar{u}_R(R,t)=\frac{1}{\Sigma}\int^{\infty}_{-\infty}\rho(R,z,t) u_R(R,z,t) \mt{d}z,
\ee 
\be
\bar{\nu}(R,t)=\frac{1}{\Sigma}\int^{\infty}_{-\infty}\mu(R,z,t) \mt{d}z
\ee 
are the vertically-integrated (surface) density, the (density-weighted) mean radial velocity and the (density-weighted) mean kinematic viscosity, respectively\footnote{Note that equation (\ref{angint}) assumes $u_{\phi}=\Omega(R) R$. The consistency of this approximation is verified later on in this section.}. In practice the integrals will extend from the lower disc boundary $-z_0$ to the upper boundary $z_0$. Throughout this section, $z_0=H$ but the reader should be aware that this is not the case in general since $H$ is a pressure or density scaleheight while $z_0$ is the actual disc vertical boundary; the exact relation between $z_0$ and $H$ is dependent on the disc's vertical structure. 

The time for accretion may be taken to be much larger than the dynamical (orbital) time, $t_\mt{dyn}\sim\Omega^{-1}$, so that the disc can be treated as quasi-steady for the purpose of determining its radial structure. (Note that this is just a simplifying assumption and does not hold in every system.) The equations (now in $R$ only) then simplify to
\be
\bar{\nu}\Sigma=\frac{\dot{M}}{3\pi}f, \quad \textrm{with}\quad f=1-\sqrt{\frac{R_\textrm{in}}{R}},
\label{rstructure1}
\ee
where $\dot{M}=2\pi R \Sigma (-\bar{u}_R)>0$ is the constant mass accretion rate. To obtain this equation a couple of assumptions were made. The angular velocity $\Omega$ was taken to be Keplerian, $\Omega_\mathrm{K}=\sqrt{GM/R^3}$, implying
\be
T_{R\phi}=\mu R\frac{\mt{d}\Omega}{\mt{d}R}=-\frac{3}{2}\mu\Omega_\mathrm{K}.
\label{sst}
\ee
In addition, to integrate the angular momentum equation, the location where viscous torques vanish was taken to be the inner disc radius, $R_\mt{in}$.

For a general rotation curve, $\Omega$, the 1D angular momentum conservation equation (\ref{angint}) in the steady state may be integrated to
\be
\dot{M}(l-l_\mt{in})=2\pi R^2(-\tau_{R\phi}),
\label{angconst}
\ee
where $l=\Omega R^2$ is the specific angular momentum and $\tau_{R\phi}=\int T_{R\phi}\mt{d}z$; $l_\mt{in}$ is the angular momentum constant, that is, the value of $l$ at the radius where the viscous torque is zero. Throughout this thesis I assume that this is indeed the location where the disc terminates \citep{notorque,ap2003}.

The remaining components of the momentum equation (\ref{hydro2}) are simple. In the vertical direction the dominant balance is gravity vs. pressure,
\be
0=-\rho\frac{\partial\Phi}{\partial z}-\frac{\partial p}{\partial z}.
\ee 
By expanding the gravitational potential about 
$z=0$, it is simple to verify that the equation of hydrostatic equilibrium can be written as
\be
\frac{\partial p}{\partial z}\approx -\rho\Omega_z^2 z,
\label{hydroeq}
\ee
where $\Omega_z^2(R)=\Phi_{,zz}(R,0)$ is the square of the vertical epicyclic frequency. This is the frequency at which a particle oscillates about its original orbit when it suffers a vertical perturbation; if the perturbation is radial, the characteristic frequency is the radial epicyclic frequency $\kappa$. In a Keplerian disc,
\be
\Phi=-\frac{GM}{\sqrt{R^2+z^2}},
\ee
and
\be
\Omega\equiv\left(\frac{1}{R}\frac{\partial\Phi}{\partial R}\right)^{1/2}_{z=0}=\Omega_\mathrm{K}=\Omega_z=\kappa. 
\ee

An order of magnitude analysis of (\ref{hydroeq}) shows that an appropriate formula for the typical scaleheight of the disc is $H=c_\mathrm{s}/\Omega_z=c_\mathrm{s}/\Omega_\mathrm{K}$. From this definition for $H$ it is easy to see that $\mu=\alpha p/\Omega_\mt{K}$. This means that angular momentum is transferred radially outwards at a rate which is taken to be proportional to the total pressure, with the efficiency of the process being parametrised by a dimensionless quantity $\alpha$. Moreover, (\ref{sst}) implies $T_{R\phi}=-(3/2)\alpha p$. This definition for the stress tensor is a common modification to that presented in the original Shakura--Sunyaev paper where $T_{R\phi}=-\alpha p$. The difference between the two stress prescriptions, $T^{\mathrm{SS}}_{R\phi}=-\alpha p$ and $T^{\mathrm{shear}}_{R\phi}=\mu R\Omega'$, is even more evident when the angular velocity is not Keplerian in which case
\be
T^{\mathrm{shear}}_{R\phi}=\alpha p \frac{R}{\Omega_\mathrm{K}}\frac{\mt{d}\Omega}{\mt{d}R}=\alpha p\left(\frac{\mt{d}\ln \Omega}{\mt{d}\ln R}\right)\left(\frac{\Omega}{\Omega_\mt{K}}\right).
\label{stresstensor}
\ee
The use of this expression for $T_{R\phi}$ may result in significant changes in the mathematical character of the sonic critical point of steady flows \citep{ak1989}. This critical point is present when the radial inflow velocity, neglected in the Shakura--Sunyaev model, is considered. The differences between the two types of stress tensor will be discussed in Part \ref{ta} of the thesis where I'll refer to $T^{\mt{SS}}_{R\phi}$ as the $\alpha p$ type stress tensor and to $T^{\mt{shear}}_{R\phi}$ as the diffusion-type stress \citep{kxlz1993}.

In the radial direction, the thin disc approximation results in the dominant balance being centrifugal,
\be
\frac{u_{\phi}^2}{R}=\frac{\partial\Phi}{\partial R}\approx\left(\frac{\partial\Phi}{\partial R}\right)_{z=0},
\label{radeq}
\ee
since the radial pressure gradient is negligible and $u_R\sim\alpha(H/R)^2u_\phi\ll u_\phi$. This implies that the azimuthal component of the velocity is, to a good approximation, equal to the orbital velocity of a test particle: $u_{\phi}=R\Omega$ as assumed previously. In a Keplerian disc, the thin disc assumption then implies
\be
\frac{H}{R}=\frac{c_\mathrm{s}}{u_{\phi}}\ll1,
\label{se1}
\ee
i.e., the azimuthal flow is supersonic which justifies the neglect of the radial pressure gradient term since
\be
\left.\frac{1}{\rho}\frac{\partial p}{\partial R}\middle/\frac{u_\phi^2}{R}\right.\sim\frac{c_\mt{s}^2}{u_\phi^2}\ll1.
\ee

To fully determine the disc's radial structure, energy considerations have to be taken into account. In traditional models, the gas in the disc is heated by viscous dissipation and cooled effectively by radiative diffusion through the vertical boundaries. The energy balance can be written in the form \citep{accretionpower}
\be
\frac{\partial F}{\partial z}=\mu(R\Omega')^2\Leftrightarrow F(H)-F(0)=\frac{1}{2}\bar{\nu}\Sigma (R\Omega')^2,
\label{eb}
\ee
where $F$ is the radiative flux. Owing to the thin disc approximation, the temperature gradient $\nabla T$ is essentially vertical so $F$ can be written as \citep[e.g.][]{stellar},
\be
F(z)=-\frac{16 \sigma T^3}{3 \kappa_\mathrm{R} \rho}\frac{\partial T}{\partial z},
\label{rdl}
\ee
where $\sigma=5.67\times10^{-5}\,\mathrm{g}\,\mathrm{ s}^{-3}\,\mathrm{ K}^{-4}$ is the Stefan-Boltzmann constant and $\kappa_\mathrm{R}$ is the Rosseland mean opacity. 

Another characteristic feature of the Shakura--Sunyaev analysis is the crude treatment of the disc's vertical structure, which is in fact neglected for the purpose of calculating the radial variation of fluid's quantities. If the disc is isothermal in the vertical direction, integration of (\ref{hydroeq}) shows that the density drops rapidly with height. This justifies a local treatment of the disc's structure which is taken to be governed by the fluid's quantities in the mid-plane. In this case: $\Sigma=2\rho H$, $P=\int p \mt{d}z=2 p H$ (with $P$ and $\Sigma$ related by the isothermal sound speed, $P=c_\mathrm{s}^2 \Sigma$) and 
\be
F\sim \frac{4\sigma T^4}{3\Sigma\kappa_\mathrm{R}},
\ee
where $\rho$ and $T$ can be interpreted as the disc's density and temperature in the mid-plane, respectively. The energy balance equation can then be written as
\be
\frac{4\sigma T^4}{3\Sigma\kappa_\mathrm{R}}=\frac{1}{2}\bar{\nu}\Sigma (R\Omega')^2,
\ee
where it was assumed that the disc's central temperature is significantly larger than its surface temperature.

In summary, in the traditional model, the disc's radial structure is described by the following set of equations:
\begin{equation}
\rho=\Sigma/2H,
\label{struc1}
\end{equation}
\begin{equation}
P=2Hp=c_\mathrm{s}^2 \Sigma,
\label{pres}
\end{equation}
\be
H=c_\mathrm{s}/\Omega_\mathrm{K},
\ee
\begin{equation}
\bar{\nu}\Sigma=\frac{\dot{M}}{3\pi}f,
\label{nusigma}
\end{equation}
\begin{equation}
\frac{4\sigma T^4}{3\Sigma\kappa_\mathrm{R}}=\frac{1}{2}\bar{\nu}\Sigma (R\Omega')^2,
\label{energybal}
\end{equation}
\be
\bar{\nu}\Sigma=\int \mu \mt{d}z=\alpha P/\Omega_\mathrm{K}.
\label{struc2}
\ee

Before going on to solve these equations, a discussion regarding the inner disc boundary is in order since $R_\mt{in}$ is present in $f$ and a mention to its exact location was avoided. In fact, discs around different objects have different inner radii. A disc surrounding a weakly magnetised star may extend to the stellar surface where a thin viscous boundary layer mediates the transition between the disc's angular velocity and the stellar rotation. On the other hand, when the star is strongly magnetised, the disc terminates at the magnetospheric radius beyond which the accretion flow follows the magnetic field lines. In accretion discs around black holes the situation is once more different. One of the fundamental results of the theory of general relativity is the existence of an \emph{innermost stable circular orbit} (ISCO) or \emph{marginally stable orbit}, $R_\mt{ms}$, for test particles around a black hole; e.g. for a non-rotating black hole, $R_\mt{ms}=6 GM/c^2$ (three Schwarzschild radii). This implies that a black-hole accretion disc cannot extend to the event horizon\footnote{The work of \cite{kw1985} showed that in the case of neutron stars with ``soft'' equations of state, the marginally stable orbit may lie outside the stellar surface in which case the disc would terminate there as it happens with black holes. The exact equation of state of real neutron stars is, however, uncertain. Therefore, here I will assume that neutron stars are described by ``stiff'' equations of state in which case the classic picture of the disc extending to the stellar surface holds.\label{footns}}. For radii larger than $R_\textrm{ms}$ the orbits are stable and a viscous torque is required to transport mass inwards. On the other hand, for $R<R_\textrm{ms}$ orbits are unstable and no viscosity is necessary to make the gas spiral rapidly towards the black hole. The inner boundary can therefore be assumed to be $R_\mt{in}\approx R_\mt{ms}$ (the equality is only exact in a pressureless disc where $H\sim 0$). This is the radius where the specific angular momentum and binding energy of a test particle reach a minimum.

\subsubsection{The three disc regions}

To close the system of equations two final ingredients are needed: a relation for the opacity $\kappa_\mathrm{R}=\kappa_\mathrm{R}(\rho,T)$ and an equation of state $p=p(\rho,T)$. The former will depend on the processes that contribute to the opacity: if electron scattering is dominant, $\kappa_\mathrm{R}$ is given by the constant Thomson opacity, $\kappa_\mathrm{T}=0.33\,\mathrm{cm}^2\,\mathrm{g}^{-1}$. On the other hand, if free-free absorption is dominant, opacity is approximately determined by the Kramers formula, $\kappa_\mathrm{K}=5\times10^{24}\rho T^{-7/2}\,\mathrm{cm}^2\,\mathrm{g}^{-1}$ \citep[e.g.][]{accretionpower,stellar}. In general, a combination of the two should be considered. As for the equation of state, discs may be assumed to be composed of both gas and radiation in which case
\be
p=p_{\textrm{g}}+p_{\textrm{r}}=\frac{k_B}{\mu_\textrm{m} m_{\textrm{p}}}\rho T+\frac{4\sigma}{3c}T^4
\label{eqstate},
\ee
were $k_B=1.38\times 10^{-16}\,\mathrm{cm}^2\,\mathrm{g}\,\mathrm{ s}^{-2}\,\mathrm{ K}^{-1}$ is the Boltzmann constant, $\mu_m=0.615$ is the mean molecular weight and $m_\mathrm{p}=1.67\times10^{-24}\,\mathrm{g}$ is the proton mass.

The full consideration of all the processes that contribute to the opacity and to the total pressure gives rise to a complicated disc structure. In the Shakura--Sunyaev approach the analysis is simplified by considering the disc to be composed of three parts: in the innermost regions, (a) and (b), electron scattering is the main contribution to the opacity while free-free absorption is dominant in the outer region (c). In (a) the pressure is determined by the radiation pressure ($p\propto T^4$) while gas pressure ($p\propto \rho T$) dominates in (b) and (c). The radiation-dominated region exists only at very high temperatures, when the accretion rate is larger than a few percent of the Eddington accretion rate (necessary to sustain the Eddington luminosity $L_\mt{Edd}$), being important in accretion discs around black holes.

Assuming that in all three regions the \emph{stress scales with the total pressure}, the disc structure may be described by\newline
\newline
\emph{Region (a):}
\begin{equation}
H=6.1\times10^6\dot{m}\,m\,f,
\label{ss1}
\end{equation}
\begin{equation}
\Sigma=0.03\,\alpha^{-1}\,\dot{m}^{-1}\,r^{3/2}\,f^{-1},
\label{sigmaa}
\end{equation}
\begin{equation}
P=5.0\times10^{22}\,\dot{m}\,\alpha^{-1}\,f\,r^{-3/2},
\end{equation}
\vspace{3pt}
\emph{Region (b):}
\begin{equation}
H=3.0\times10^3\,\alpha^{-1/10}\,\dot{m}^{1/5}\,m^{9/10}\,r^{21/20}\,f^{1/5},
\end{equation}
\begin{equation}
\Sigma=1.6\times10^5\,\alpha^{-4/5}\,\dot{m}^{3/5}\,m^{1/5}\,r^{-3/5}\,f^{3/5},
\end{equation}
\begin{equation}
P=5.0\times10^{22}\,\dot{m}\,\alpha^{-1}f\,r^{-3/2},
\end{equation}
\vspace{3pt}
\emph{Region (c):}
\begin{equation}
H=1.3\times10^3\,\alpha^{-1/10}\,\dot{m}^{3/20}\,m^{9/10}\,r^{9/8}\,f^{3/20},
\end{equation}
\begin{equation}
\Sigma=7.5\times10^5\,\alpha^{-4/5}\,\dot{m}^{7/10}\,m^{1/5}\,r^{-3/4}\,f^{7/10},
\end{equation}
\begin{equation}
P=5.0\times10^{22}\,\dot{m}\,\alpha^{-1}\,f\,r^{-3/2},
\label{ss2}
\end{equation}
where all quantities are in CGS units, $m=M/M_\odot$, $r=R\,c^2/GM=R/R_\mt{g}$ (in units of the gravitational radius) and $\dot{m}=\dot{M}/\dot{M}_\mathrm{Edd}$. The Eddington accretion rate is given by $\dot{M}_\mathrm{Edd}=L_\mt{Edd}/\eta c^2=4\pi G M m_\mt{p}/\eta c\sigma_\mt{T}$ , and is calculated assuming an ``accretion efficiency'' $\eta=0.06$ appropriate for a Schwarzschild black hole; $\sigma_\mt{T}=6.65\times 10^{-25}\,\mt{cm}^2$ is the cross-section for Thomson scattering. The fluid's quantities have exactly the same dependencies in $m$, $\dot{m}$, $\alpha$, $r$ and $f$ as in the original Shakura--Sunyaev model. The differences in the numerical factors are due to the small difference in the values used for $\kappa_T$ and $\mu_\textrm{m}$, to the different normalization of radius, and also because here I consider the $r\phi$ component of the stress tensor, $T_{r\phi}=\mu r\Omega'$, to be $-(3/2)\alpha p$ instead of the usual $-\alpha p$ value. The remaining quantities, $p$, $\rho$ and $T$, may be easily determined from the formulas above.

Despite neglecting to treat the vertical structure of the disc, not considering other contributions to the cooling and heating rates of the gas besides radiation and viscous heating, and despite the simple parametrization for the angular momentum transfer, the Shakura--Sunyaev model is still the standard reference for the radial structure of accretion discs if gas pressure is dominant. Region (a) has always been more problematic. In fact, the assumption that stress scales with the total pressure in the radiation pressure dominated regime leads to viscous and thermal instabilities, analysed further in the following section.

\subsubsection{Instabilities}
\label{instability}

Soon after the publication of the standard thin disc models, some authors \citep{le74,ss76} pointed out that, with the usual viscosity prescription, the very inner (a) region is subject to instabilities acting on the viscous and thermal timescales of the disc.

To study instabilities the time-dependence of the hydrodynamic equations needs to be restored. Equations (\ref{masscon}) and (\ref{angint}) may then be combined into a diffusion equation describing the viscous evolution of a thin disc,
\be
\frac{\partial \Sigma}{\partial t}=\frac{3}{R}\frac{\partial}{\partial R}\left[R^{1/2}\frac{\partial}{\partial R}\left(\bar{\nu}\Sigma R^{1/2}\right)\right],
\ee
where Keplerian rotation was assumed. From equation (\ref{nusigma}) it can be seen that, at fixed radius, the viscosity depends only on the surface density, $\bar{\nu}=\bar{\nu}(\Sigma,R)$. Therefore, if $\Sigma$ (initially independent of $t$ for a disc in equilibrium) is perturbed slightly, $\Sigma(R,t)\rightarrow\Sigma(R)+\delta\Sigma(R,t)$, the viscosity $\bar{\nu}\Sigma(R,t)\rightarrow\bar{\nu}\Sigma(R)+\delta(\bar{\nu}\Sigma)(R,t)$ with $\delta(\bar{\nu}\Sigma)=(\partial(\bar{\nu}\Sigma)/\partial\Sigma)\delta\Sigma$. The viscous evolution equation can then be written as \citep{pringle1981}
\be
\frac{\partial}{\partial t}\delta(\bar{\nu}\Sigma)=\frac{3}{R}\frac{\partial}{\partial R}\left[R^{1/2}\frac{\partial}{\partial R}\left(\delta(\bar{\nu}\Sigma) R^{1/2}\right)\right]\frac{\partial}{\partial\Sigma}(\bar{\nu}\Sigma),
\ee
where the variation of $\partial(\bar{\nu}\Sigma)/\partial\Sigma$ with $R$ was assumed slow. If $\partial(\bar{\nu}\Sigma)/\partial\Sigma<0$, the diffusion coefficient is negative implying that overdense regions are constantly fed with more material while underdense regions become more and more rarified, i.e., the disc breaks up into rings; this is the condition for instability. In region (a) of the Shakura and Sunyaev model, it is simple to see from equations (\ref{struc1})--(\ref{struc2}) that, at fixed $R$,
\be
c_\mt{s}\propto\Sigma^{-1}\Rightarrow\bar{\nu}\Sigma=\frac{\mt{const}}{\Sigma},
\ee
where $\mt{const}>0$ showing that the region is indeed viscously unstable.

The characteristic timescale for this secular instability is the timescale for the radial motion (accretion) in the disc \citep{pringle1981},
\be
t_{\nu}\sim\frac{R}{\bar{u}_R}\sim\frac{R^2}{\bar{\nu}}\sim\alpha^{-1}\left(\frac{H}{R}\right)^{-2} t_\mt{dyn}.
\ee
In a thin disc, a faster timescale corresponds to the time for establishing vertical thermal balance which can be estimated by dividing the thermal energy content per unit area by the dissipation rate,
\be
t_\mt{th}=\frac{c_\mt{s}^2\Sigma}{\bar{\nu}\Sigma\Omega^2}\sim\alpha^{-1} t_\mt{dyn}.
\ee
This is the characteristic timescale of the thermal instability; for $\alpha\ll1$, $t_\mt{th}\gg t_\mt{dyn}$.

During the very short thermal time, the surface density remains practically constant and the disc is maintained in hydrostatic equilibrium. Therefore, it may be considered that in equations (\ref{struc1})--(\ref{struc2}) only $H$ and $T$ (and therefore $p$) vary with time. From the energy balance equation (\ref{energybal}), the heating and cooling rates may be defined to be
\be
Q^+=\frac{1}{2}\bar{\nu}\Sigma(R\Omega')^2
\ee
and
\be
Q^-=\frac{4\sigma T^4}{3\Sigma\kappa_\mt{R}},
\ee
respectively. In region (a) the opacity is constant so that $Q^-\propto T^4$. On the other hand, using (\ref{struc2}) and the first part of (\ref{pres}), the heating rate can be written as
\be
Q^+=\frac{9}{4}\alpha\Omega_\mt{K}Hp.
\ee
The dependence of $H$ on $p$ can be determined from the hydrostatic equilibrium equation and (\ref{struc1}), $H=2p/\Sigma\Omega_\mt{K}^2$, and therefore $Q^+\propto p^2\propto T^8$ when the radiation pressure dominates. From the dependence of $Q^-$ and $Q^+$ it is seen that a small increase in temperature will lead to a thermal runaway since the heating rate increases much faster than the cooling rate. This is the basis of the thermal instability.

According to the Shakura--Sunyaev model, these instabilities would be expected in discs with a luminosity larger than a few percent of the Eddington luminosity. However, and despite attempts to explain the variability of some observational objects using these instabilities \citep[][and references therein]{blaesnotes}, their presence in real systems has never been proven. With the exception of GRS 1915+105, a very unusual and extremely luminous source, observations of X-ray binaries with $0.01\lesssim L/L_\mt{Edd}\lesssim 0.5$ have failed to identify such instabilities \citep{gd2004}. \cite{pringle1981} pointed out that they are not necessarily physical and may rather be regarded as ``self-inconsistencies": some of the approximations made in order to construct the steady disc model may not be consistent with the assumption of steadiness. Indeed, these can be avoided by simply assuming that the stresses scale with the gas pressure \citep{sc81}, or with a combination of both gas and radiation pressure \citep[e.g.][]{tl1984,mf2002}. Still, recent magnetohydrodynamic simulations by \cite{hiroseetal2009} suggest that turbulent stresses (magnetic in nature) may indeed be correlated with the total pressure as predicted in the original model. However, fluctuations in stress are the ones driving pressure fluctuations, contrary to the usual assumption that changes in pressure give rise to changes in stress; this seems to avoid thermal instability (which has a faster growth rate than viscous instability). With these results in mind, throughout this thesis, I consider the $\alpha$ viscosity prescription in the form

\begin{equation}
\mu=\alpha\frac{p}{\Omega_\mt{K}},
\label{avp}
\end{equation}
so that the viscosity $\mu$ and total pressure $p$ correlate, as in the original Shakura \& Sunyaev model.

Another instability of viscous flows, one potentially relevant for explaining quasi-periodic variability, was investigated by Kato (1978). This pulsational instability or viscous overstability (since it takes the form of propagating waves) has a growth rate lower than that of thermal instability by a factor $(H/\lambda)^2$, where $\lambda>H$ is the wavelength of the perturbations. Viscous overstability is further discussed in Part \ref{ta} of this thesis.

\subsection{Beyond the standard model}

\subsubsection{Radial inflow: steady transonic accretion}

In the standard thin disc model, the gravitational attraction of the central object is balanced by rotation and the radial inflow is ignored to lowest order. The disc structure calculated within these assumptions is singular at the inner boundary ($f=0$) where the disc is forced to terminate since the density in region (a) goes to infinity there [cf. equation (\ref{sigmaa})]. The singularity is removed if the radial inflow is taken into account. This is particularly important in discs around black holes: as noted by (e.g.) \cite{nt73}, accretion is always transonic in this case since the flow must be subsonic at large radii and supersonic close to the compact object. As a result, the radial velocity is dynamically important in the inner disc. 

\cite{lt1980} developed a more correct treatment of the innermost region of steady flows and demonstrated that the solution for the radial velocity of thin-disc accretion onto a black hole is transonic and analogous to \cite{bondi} spherical accretion. It should be noted that the transonic accretion model is only relevant very close to the marginally stable orbit and in the plunging supersonic region. The Shakura--Sunyaev model is still applicable a certain distance away from the inner boundary and the transonic solution should be matched with the classical thin-disc solution at large radii. 

The point where the modulus of the radial inflow equals the sound speed is a critical point of the hydrodynamic equations describing 1D steady accretion; a physically acceptable solution must pass through this sonic point regularly. In a thin disc, this critical location is close to the marginally stable orbit and it is approximately where the gas transits from being in stable orbits to freely falling into the black hole. The exact location of this inner boundary is dependent on the flow parameters such as mass accretion rate and viscosity \citep{mp1982}.

As it will be seen in chapter \ref{steadyintro}, transonic accretion is constrained by boundary conditions and regularity demands at the sonic point. For some values of the flow parameters such as sound speed and $\alpha$, steady transonic solutions obeying realistic conditions cannot be found, implying the time variability of the accretion process. Indeed, e.g., \cite{nature} attempted to relate the unsteadiness of the flow to the X-ray quasi-periodic phenomena. This hypothesis is studied in Part \ref{ta} of this thesis where I analyse the transition from subsonic to supersonic accretion in a time-dependent flow.

\subsubsection{Thick discs and ADAFs}

The Eddington luminosity is the limit at which the inward gravitational force is exactly balanced by the outward radiation force. This concept is particularly meaningful in the inner regions of accretion flows around neutron stars or black holes where the gas is heated to high enough temperature for radiation pressure to dominate over gas pressure. In this case, the equations derived to describe region (a) are applicable and (\ref{ss1}) appropriately conveys the dependency of the disc thickness on the mass accretion rate, independently of the viscosity prescription. It is then clear [since $H/R=41.5\dot{m}f/r$ in region (a)] that if the accretion rate is close to Eddington, the thin disc approximation must break down. 

The interest in thick discs arose in an attempt to explain these ``puffed up'' structures or tori relevant for high luminosity accretion flows or that may result from thermal and viscous instabilities in thin discs \citep{pw1980,abramowiczetal1980}. Accretion tori were used to model the very inner accretion flow in AGNs and seemed to provide the necessary energy by accretion for jets and a mechanism for their collimation without the need for magnetic fields \citep{lyndenbell1978}. These thick structures may also be relevant in inefficiently cooling flows where the gas is heated to temperatures such that the local sound speed becomes comparable to the azimuthal velocity [cf. equation (\ref{se1})].

Since radiation pressure forces are important in defining the radial equilibrium of the disc, there is no reason to assume that the rotation is Keplerian [cf. equation (\ref{radeq})]. In principle, there is freedom to chose the distribution of specific angular momentum, provided it does not decrease outwards so that Rayleigh's criterion for stability to axisymmetric perturbations is satisfied. The typical assumption was that of constant angular momentum; this is not only the simplest distribution but is also the one for which the total luminosity of the flow is maximal \citep{abramowiczetal1980}. Models constructed with this assumption are, however, dynamically unstable to global non-axisymmetric modes as discovered by \cite{ppi1}. 

The key location for the development of the Papaloizou-Pringle instability (PPI) is the corotation radius, located within the torus for non-axisymmetric modes. If a wave of frequency $\omega$ and azimuthal number $m$ (cf. chapter \ref{oscilintro}) propagates in a disc rotating with angular velocity $\Omega$, the wave pattern will appear, to an inertial observer, with a Doppler-shifted wave frequency $\hat{\omega}=\omega-m\Omega$; the location where $\hat{\omega}=0$ is the corotation radius, $R_\mt{c}$. Waves propagating outside corotation have $\hat{\omega}>0$ while $\hat{\omega}<0$ for waves propagating inside $R_\mt{c}$. Because the flow is rotating and there is kinetic energy stored in the disc as a result of this rotation, it is possible for a wave propagating in such a medium to have negative energy. This simply means that the total energy of wave+disc is smaller than the energy of the disc on its own; the wave passage decreases the mechanical energy of the gas. This is the case for waves propagating inside their corotation radius as opposed to positive-energy waves that propagate outside (cf. chapter \ref{excmech}). At the corotation resonance, energy exchanges may occur. If negative-energy waves lose energy there their amplitude is increased since they become more energetically negative. Positive-energy waves may collect this energy left at corotation and grow in amplitude as a result. The process leads to instability if negative-energy waves are reflected at the inner boundary of the torus so that the energy exchange at corotation is maintained \citep{reviewmri}.

The instability acts on the very fast dynamical timescale and persists even in flows with non-constant distributions of specific angular momentum \citep{ppi2}. The growth is maximal for constant $l$ ($\Omega\propto r^{-2}$) and decreases rapidly as the Keplerian distribution ($\Omega\propto r^{-3/2}$) is approached \citep{goldreichetal1986}. The discovery of the PPI was striking and threw into doubt the very existence of thick discs. However, it was later shown by \cite{blaes1987} and confirmed by numerical simulations \citep[][and references therein]{reviewmri} that the growth of the instability may be reduced or even halted in accreting flows where an inward radial velocity is taken into account. Thick accretion discs may indeed be astrophysically relevant but there are still theoretical uncertainties regarding their structure and stability.  \enlargethispage{-\baselineskip}

Another interesting accretion solution is that of advective-dominated accretion flows  or ADAFs. They are important when the accreting gas is unable to cool efficiently and most of the thermal energy is advected inwards by accretion of entropy instead of being radiated away. They are expected to have astrophysical relevance in situations where the mass accretion rate is either very low or very high \citep[see examples in][]{adaf}. Although these solutions are ``thick'' in the sense that $H/R$ is not necessarily small and the pressure gradient should be included in the radial momentum equation, they may not be unstable to PPI due to the presence of radial advection.

\begin{figure}[t!]
\begin{center}
\includegraphics[width=80mm]{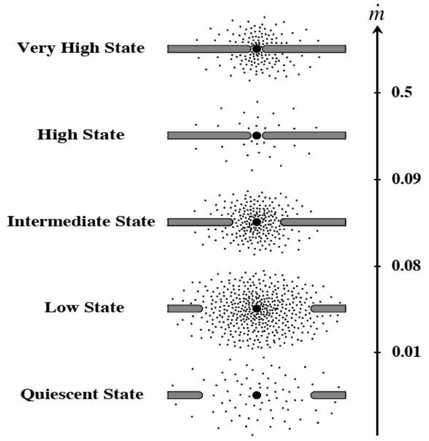}
\caption{Schematic representation of the accretion flow in different spectral states with the ADAF indicated by dots and the thin disc by horizontal bars. The transition between different states (from low to very high) occurs as the mass accretion rate increases and the disc gets closer to the marginally stable orbit \citep[from][]{esinetal1997}.}
\label{esin}
\end{center}
\end{figure}

The most widely accepted model for accretion flows around black holes involves both an ADAF and a thin disc \citep{nmy1996}. Accretion at large radii occurs by means of a thin disc (possibly surrounded by a hot corona) which then switches, at a transition radius $R_\mt{tr}$, to an advection-dominated flow which dominates in the inner regions; the exact location of $R_\mt{tr}$ is dependent on the mass accretion rate. This model has spectral characteristics in agreement with those observed for X-ray binaries \citep[][and references therein]{nmqbook}. Moreover, and despite the complications involved in defining black-hole states and transitions from one to another (cf. section \ref{states}), an elegant and simple theoretical model differentiates the various states by using different transition radii for each of them. The highest state, dominating close to the Eddington limit, has the highest accretion rate and smallest $R_\mt{tr}$ with accretion occurring almost exclusively by means of a thin disc \citep[][and Fig.~\ref{esin}]{esinetal1997}. A more recent version of this state-transition model is presented in \cite{donereview}. 

\subsubsection{Warped and eccentric discs}

In the classical theory of accretion discs the gas is assumed to rotate around a massive central object following coplanar, circular and approximately Keplerian orbits. Although this is the simplest solution for the motion of fluid elements placed in a Newtonian potential well, external forces can give rise to non-planar discs composed of non-circular rings. In fact, the general solution allows for discs to be twisted or tilted, with their orbits presenting a smoothly varying eccentricity and/or inclination. The literature offers strong observational and theoretical evidence \citep[][and references therein]{bardeenpetterson1975,ogilvie2000,ogilvie2001,globalsimulations} to believe that under various conditions discs can become either warped or eccentric. In particular, the precession of deformed discs has been used to explain several phenomena in X-ray binaries \citep[e.g.][]{gerendboynton1976,whitehurst1988,sv1998,lai1999}. 

Warped and eccentric discs are described further in chapter \ref{oscilintro}, where warping and eccentricity are described as global modes, and in chapter \ref{warpecc} where the conditions under which global deformations may appear and propagate to the inner disc region are analysed in more detail.

\section{Astrophysical black holes}

The short discussion on the location of the inner disc boundary of the previous section, emphasised the importance of accretion discs in distinguishing between neutron stars and black holes. More importantly, because black holes themselves cannot emit light (apart from Hawking radiation), observations can only reveal the emission from the accretion discs. Therefore, accretion flows are fundamental tools in the search for black-hole candidates and they also provide means to determine properties of the central object, namely, black-hole mass and spin. In this section I discuss the search for, and characterisation of, astrophysical black holes and the importance of accretion discs in this process. 

\subsection{Black-hole candidates}

To a certain extent stars come in all shapes and sizes or, more precisely and contrary to the popular belief, they are not necessarily exactly spherically or axially symmetric since magnetic fields and rotation may affect their symmetry. However, according to stellar evolution, some stars will end their lives as black holes and the vacuum solutions of Einstein's general relativity equations that describe these objects are highly symmetric. If the star is sufficiently massive and non-rotating it is expected to collapse to a Schwarzschild black hole which is spherically symmetric. On the other hand, a rotating massive star is expected to end its life as a Kerr black hole which is characterised by the electrically-neutral axially-symmetric solution of Einstein field equations. (Charged black holes described by the Reissner-Nordstr\"om solution are not expected to exist in nature as such a body would attract opposite charges and be neutralised as a consequence.) Is there a contradiction between the theories of stellar evolution and general relativity or can a non-symmetric star collapse onto a highly symmetric body? General relativity provides a solution for the apparent contradiction by stating that during the collapse the compact object gets rid of its asymmetry by radiating away gravitational waves. This can be proved mathematically by studying the evolution of small non-axially symmetric perturbations of the gravitational field during the collapse process. While this provides a theoretically satisfying solution for the paradox between stellar evolution and general relativity, both theories are yet to be fully tested observationally. The proof of the existence of Kerr black holes (the Schwarzschid solution can be included as a special case) in nature constitutes one of the fundamental observational tests of these theories. A more detailed description of black holes in general relativity can be found in, e.g., \cite{bh}.

Astrophysical black holes may be divided in four categories depending on their mass \citep[see][and references therein]{bh}. Cosmology theorises mini-black holes with $M < 1 M_\odot$ to have formed in the early Universe when very high densities prevailed. These objects could presumably be detected since they are predicted to emit gamma rays during evaporation due to Hawking radiation, which would contribute to the cosmic gamma-ray background. Measurements so far are limited and inconclusive. Stellar-mass black holes, $M = \mathcal{O} (1-10 M_\odot)$, are the endpoint of stellar evolution of massive stars and the earliest observational evidence of such type of objects comes from X-ray observations of binary star systems. Intermediate-mass black holes, $M = \mathcal{O} (100-1000 M_\odot)$, are probably formed by collision and/or merging of stellar-mass black holes. Highly luminous compact X-ray sources emitting 10-100 times more power in X-rays than stellar-mass black hole candidates have been identified. Such a release of energy can be explained if black holes power these objects. They have mostly been identified in starburst (rapid stellar formation) galaxies. Finally, the most spectacular supermassive black holes power active galactic nuclei or AGN ($M > 10^5 M_\odot$). In the so-called ``active galaxies" an enormous amount of energy is released from a small volume at their centres and their high luminosity can only be reasonably explained by accretion onto a black hole. 

The earliest and probably strongest evidence of the existence of black holes in the Universe comes from observations of binary stars. From the number of binaries known in our Galaxy that consist of a visible (normal) star and an invisible (dark) companion in close orbit, in approximately 20 the dark star is believed to be a black hole \citep{bhbbook}. When fusion of elements in the stellar interior ceases to release energy the star begins to die. According to the theory of stellar evolution, a lower-mass star sheds its envelope into a planetary nebula and the resulting compact degenerate remnant is a white dwarf. Higher-mass stars produce a supernova as a result of collapse. The remnant from this violent explosion is either a neutron star of black hole. In a white dwarf the inward pulling gravity is balanced by electron degeneracy pressure while a neutron star is supported by neutron degeneracy pressure. If the mass of the neutron star is larger than $2-3 M_\odot$ the pressure is insufficient to prevent gravitational collapse, the radius of the star eventually becomes smaller than its Schwarzschild radius and a black hole forms.

The invisible companion in a close binary system can be one of the three endpoints of stellar evolution. If its observationally-determined mass is larger than $3 M_\odot$, it is considered a black-hole candidate (BHC). From orbital theory, the mass-function of a binary may be found to be given by
\begin{equation}
F(M) \equiv \frac{PV_s^3}{2\pi G} = \frac{M\sin^3\phi}{(1+M_s/M)^2},
\end{equation}
where $P$ is the orbital period, $V_s$ is the amplitude of the line-of-sight velocity, $M_s$ and $M$ are the masses of the normal star and its dark companion, respectively and $\phi$ is the angle between the line-of-sight and the normal to the orbital plane. The mass function can be determined observationally and is a lower limit to the mass of the dark companion. Therefore, if $F(M)>3 M_\odot$ an object is considered a black-hole candidate. 

A high X-ray luminosity originating from the inner accretion flow is also a good indication of the existence of a black hole. In the strong-field region (approximately between $1-10 R_g$) where most of the gravitational energy is released, temperatures are of $\mathcal{O}(10^7 \, \mt{K})$ and emission is in the X-ray part of the spectrum. If the high-energy spectra and temporal variability of an object are similar to other BHC, it can also be identified as a candidate. X-ray variability is discussed further in section \ref{observ}.

The absolute proof of the existence of black holes would be the detection of an event horizon. Even if such a region cannot be ``seen", observations of neutron-star binaries and black-hole binaries (BHB) may be compared to look for signatures of a surface in the former. Because the accretion disc extends to the surface of neutron stars, once the gas reaches it, it accumulates there and its kinetic energy is converted into heat. Then it eventually undergoes thermonuclear burning due to the gravitational compression of the gas onto the surface. The burning can be unstable in which case thermonuclear bursts occur. This phenomenon has been observed in neutron star binaries but not in black hole candidates (McClintock et al. 2003), re-inforcing its identification as such (since black holes have no surface thermonuclear burning cannot happen).

\subsection{Measuring black-hole spin}

For general relativity to be tested observationally, the identification of black holes isn't enough. It is necessary to show that the space around them may be described by the Kerr metric which is characterised fully by only two parameters, the mass and the spin of the black hole. Given that most BHCs have determined or constrained masses, the next logical step is to measure spin, which is of more fundamental importance than mass. The latter simply gives a scale while the former changes the geometry of the space-time in the vicinity of the black hole; knowing the spin is essential to model how an accreting black hole interacts with its surroundings. Besides being important in testing general relativity, spin measurements would, e.g., allow us to build more realistic models of black-hole formation and black-hole binary evolution, to make models of relativistic jets which suggest a dependence on spin and to test stellar-collapse models of gamma-ray bursts.

However, measuring the spin of a black hole is much harder than determining its mass. While the latter can be measured in the Newtonian limit (using the stellar companion), the effects of spin are only revealed in the strong gravity regime, i.e., very close to the black hole in the very inner region of the accretion disc. According to general relativity, only certain values are expected for the spin of a black hole. Defining a dimensionless spin parameter $a=cJ/GM^2$ where $J$ is the angular momentum of the hole, the cosmic censorship hypothesis (which says that naked singularities cannot exist in the Universe) limits the values of $a$ to be between $-1$ and $1$. Negative values correspond to a black hole rotating retrogradely with respect to the accretion disc (or binary rotation) and can be excluded for simplicity. If $a=1$, the event horizon of a black hole rotates at the speed of light. In reality, ``extreme-Kerr holes'' are impossible: \cite{thorne1974} proved that the maximum rotation value is limited to $0.998$.

The location of the marginally stable orbit and the accretion efficiency are dependent on spin. In a non-rotating black hole the binding energy released when a particle spirals inwards to the marginally stable orbit is 6 per cent and $R_\mt{ms}=6R_\textrm{g}$. On the other hand, in a maximally-rotating hole the efficiency is 30 per cent and $R_\mt{ms}=R_\textrm{g}$ which is also the radius of the event horizon. As a result, the orbital frequencies associated to $R_\mt{ms}$ are also dependent on spin. For a Schwarzschild black hole $f_{\textrm{ms}}=\Omega_{\textrm{ms}}/2\pi=220(M/10M_\odot)^{-1}$ Hz; for a maximally rotating Kerr black hole this value increases to $1615(M/10M_\odot)^{-1}$ Hz. These differences are crucial for some observational determinations of spin.

\subsubsection{Methods of spin measurement}

There are currently four promising methods for the observational determination of black-hole spin. I summarise each one in what follows.\newline

\emph{X-ray polarimetry:} Polarisation features can be strongly affected by general relativistic effects \citep{csp1980}. The idea is based on the assumption that high energy photons are expected to come from closer to the black hole than low energy ones. Because of the strong gravitational bending of light rays, as the photon energy increases, the plane of linear polarisation swings gradually through an angle which is dependent on spin. Therefore, spin can be determined from the parameters characterising the polarised accretion disc spectra.\newline  \enlargethispage{\baselineskip}

\emph{Iron line profile:} The prominent Fe K$\alpha$ X-ray spectral line has its origin in the gas orbiting close to the hole and is relativistically broadened. Doppler and general relativistic redshifts, frame-dragging and relativistic beaming contribute to create an asymmetric line profile. In principle, mass, angular momentum of the black hole and orientation of the accretion disc can be determined from these line profiles \citep{rn2003}.\newline

\emph{Continuum fitting:} Thin discs around black holes are truncated at the marginally stable orbit, which is dependent on mass and spin. If mass is known in advance, a measure of $R_\mt{ms}$ provides a measure of spin \citep{zcc1997}. Using blackbody radiation theory, radii of stars can be determined from the radiation flux $F$ received from the star and the temperature $T$ of the continuum radiation, provided the distance $D$ is known (luminosity $L=4\pi D^2F=4\pi R^2\sigma T^4 \Leftrightarrow (R/D)^2=F/\sigma T^4$). In a similar way, $R_\mt{ms}$ can be measured from the radius of the ``hole'' of the disc emission. But it should be mentioned that a slightly more complicated formula is used to determine this radius to, e.g., consider the $T(R)$ profile of discs and take into account the disc inclination $i$.\newline

\emph{High-frequency QPOs:} These high-frequency modulations seen in the light curve of some black-hole candidates have spectra characteristic of the inner disc and very stable frequencies, insensitive to luminosity variations. This suggests that their frequencies are primarily dependent on the fundamental properties of the black hole, $M$ and $a$, rather than the properties of the accretion flow. Provided there is a theoretical model for the frequencies of QPOs, only the mass of the hole is necessary to determine its spin \citep[see, e.g.,][and section \ref{hfqposob}]{klisbook}. 

\subsubsection{Spin-measurement critique}
\label{spincritique}

X-ray polarimetry is one of the methods for spin measurement where theoretical methods are developed the most, with computational models that compute the Stokes parameters of a polarised accretion disc spectrum already available \citep{dky2004}. However, very sensitive instruments need to be built to measure polarimetric information and thus far this method has produced no results.

The iron line profile method, on the other hand, has already produced some results. Observations indicate that, e.g., the black-hole candidate GRS 1915+105 is slowly spinning while XTE J1650--500, XTE J1655--40, XTE J1550--564 have nearly maximal spin \citep[see][and references therin]{remimcclin2006}. However, this method has some caveats and it is unclear if the reported spin measurements are reliable. The determination of spin from iron line profiles is dependent on disc models, the contribution to the line profile coming from $R<R_\mt{ms}$ is uncertain, and it is useful to know the inclination of the disc to better fit the continuum. Moreover, non-relativistic effects such as scattered radiation may also contribute to line broadening and observations have limited resolution \citep{rn2003}.

The model which as, so far, produced more spin measurements is that of continuum fitting and, from an observational point of view, is the most promising. The values of $F$ and $T(R)$ can be obtained from X-ray observations; therefore, given precise measures of $M$, $i$ and $D$ (from ground-based observations) as inputs, it is possible to fit X-ray spectral data to a fully relativistic model of the disc emission and get $a$ as a fit parameter. State-of-the-art theoretical models include corrections from gravitational redshift, ray deflections and disc atmosphere, just to name a few. Spins measured by this method include GRO J1655--40: $0.65<a<0.75$, 4U 1543--47: $0.7<a<0.8$ and GRS 1915+105, $0.98<a<1.0$ \citep[][and references therein]{remimcclin2006}.

The contrasting difference between the spins measured for GRS 1915+105 using the iron-line and the continuum-fitting methods should remind the reader that both are far from being reliable. Indeed, the X-ray continuum fitting is not free from caveats. Once again it is a model-dependent method and assumes the disc to be thin and non-warped which may not necessarily be the case, particularly close to the black hole. It requires the inclination of the disc as a fitting parameter, and as pointed out by \cite{maccarone2002}, the orbital inclination may differ significantly from that of the black-hole spin axis, relevant in the inner region. An accurate spin determination also requires the disc to be in the thermal (high-soft) state where the spectra is black-body like and the disc is expected to terminate at $R_\mt{ms}$ \citep[see next section for more details on these observational characteristics or][]{bhbbook}. Indeed, different fits to the data give rise to contrasting spin measurements. For example, \cite{reisetal2009} measured a spin larger than $0.9$ for GRO J1655--40 different from the value mentioned above. In addition, a different estimate for the spin of GRS 1915+105 is reported by \cite{mdgd2006} who found $a\sim0.7$.

The possibility of measuring black-hole spin using high-frequency QPOs provides significant stimulus for research amongst the community of theoretical astronomers. These mysterious oscillations, a prime example of black-hole accretion disc variability, are clearly a general relativistic phenomenon. Unfortunately, their origin and observational properties are yet to be explained. However, once the correct model providing a relation between QPO frequency and black-hole spin and mass is determined, they offer the most reliable method for spin measurement, almost independent of observational parameters. 

One of the motivations behind the research described in this thesis is precisely the search for a high-frequency QPO model. Although these oscillations are quasi-periodic, they are usually explained by periodic models: the damping caused by viscous-turbulent effects in the disc or the superposition of frequencies are factors of aperiodicity, transforming periodic oscillations into quasi-periodic. In Part \ref{os} of this thesis I will discuss a model based on relativistic wave trapping: if QPOs are identified with trapped waves and if the trapping is not perfect, the ``leakage'' out of the region where these waves propagate may also explain the aperiodicity. 

In order to construct realistic theoretical models for high-frequency QPOs, it is important to know about the observational characteristics of these modulations. Therefore, in the next section, I summarise the current understanding of X-ray variability phenomena where quasi-periodic oscillations are included.

\section{Observations}
\label{observ}

Accretion flows are expected to be inhomogeneous due to, e.g., turbulence, magnetic fields and density fluctuations. These phenomena may result in time variations in the emission coming from accretion discs. Although these variations happen on a wide range of timescales, here and throughout this thesis the interest will go to the fastest variability components (in particular quasi-periodic phenomena but broad-band noise is also included in this category) detected in black-hole candidates.

\subsection{Timing properties and black-hole states}
\label{states}

Timing is the study of X-ray variability. Since rapid variations are a stochastic phenomenon (due to the turbulent nature of accretion), statistical techniques are appropriate for its study and Fourier analysis is the tool commonly used for it.
A very detailed explanation of X-ray variability is given by \cite{klisbook}. 

The power density spectrum (PDS)\footnote{The PDS is a measure of variability of energy with time. It shouldn't be confused with energy spectrum [$S(E)=E\times N(E)$, where $E$ is the energy and $N(E)$ the photon number or photon spectrum] taken at each time $t$. If there is variability, $S(E)[t=t_0]\neq S(E)[t=t_1]$ and a ``variability spectrum'' can be constructed by selecting a range of energies and comparing the differences in $S(E)$ at different times. As a result, a variability feature may be stronger at, e.g., a lower range of energies rather than a higher range.} is formed from the group of power-spectral components or X-ray lightcurve [Fourier power density, $P_{f}(f)$] segments: after taking the discrete Fourier transform of each of the components, the squared amplitudes of each transform are averaged together resulting in the PDS. The PDS of BHCs often shows transient peaks, discrete, subtle features that can be modelled using Lorentzian profiles,
\begin{equation}
P_f(f)\propto\frac{\lambda}{(f-f_0)^2+(\lambda/2)^2},
\end{equation}   
where $\lambda$ is the full-width at half maximum (FWHM) of the feature and $f_0$ is its centroid frequency. If the feature has a relatively high coherence\footnote{Note that the coherence of QPOs is only high when compared to the coherence of peaked noise. In fact, the quality factor of QPOs in BHCs is much lower than the quality factor of other modulations such as the dwarf nova oscillations or QPOs in neutron stars.} or, equivalently, a high quality factor $Q\equiv f_0/\lambda$ (typically, $Q\gtrsim2$), it is considered a QPO instead of peaked noise. Therefore, QPOs are sharp, narrow features in the PDS while broad structures are identified as noise. The range of centroid frequencies for QPOs is $0.01-450$ Hz. Another important property is the strength or variance of the signal which is typically expressed in terms of its fractional root-mean-squared amplitude, $rms \propto(\int P_f(f)\mt{d}f)^{1/2}$, often given in percentage.

The energy spectra of BHCs can display thermal and non-thermal components as well as transitions between the two, where either one can dominate. The thermal component is blackbody-like ($kT \sim 1$ keV) with origin in the inner, optically-thick part of the accretion disc. The non-thermal component is usually well-fitted by a power law, characterised by a photon index $\Gamma$ (photon spectrum $\propto\textrm{energy}^{-\Gamma}$) and it is thought to have origin in an accretion disc ``corona'' or an ADAF \citep[][and references therein]{donereview} composed of high-energy photons. Black-hole states can, roughly,  be defined according to the type of component that dominates in the X-ray spectrum. \cite{remimcclin2006} present the following classification:

\begin{itemize}
\item\textbf{Thermal or high/soft\footnote{The terms \emph{soft} and \emph{hard} refer to X-ray colours. A photometric approach to the study of the X-ray broad-band spectrum quantifies it using X-ray colours: a hard colour is related to the photon count in a higher energy band while a soft colour is referent to a lower energy band.} state:} X-ray flux is dominated by the thermal component; QPOs are either not observed or very weak in this state.

\item \textbf{Hard or low/hard state:} energy spectrum is dominated by a power-law component with $\Gamma\sim 1.7$; QPOs may be present.

\item\textbf{Steep power law (SPL) or very high state:} a power-law component with $\Gamma\sim 2.5$ dominates; QPOs are frequent. This is the dominating state close to the Eddington limit.
\end{itemize}
Intermediate states between these three can be considered as well. Of particular importance is the SPL-hard intermediate state where low-frequency QPOs are often detected. QPOs of high frequency prefer the very high state. 

Most BHBs are X-ray novae that remain in a faint, quiescent state most of their lifetime. When the first outburst occurs, they are detected since the luminosity increases by many orders of magnitude. It is in the outburst that the above ``active'' states are used to classify the black hole.

Quasi-periodic oscillations have different properties depending on their centroid frequency. Very low frequency QPOs ($f_0<0.1$ Hz) are extremely rare and poorly understood. Low-frequency QPOs have values of $f_0$ between $0.1$ and $40$ Hz while high-frequency ones have $f_0>40$ Hz. These are the centroid frequency limits defined by \cite{remimcclin2006}; \cite{klisbook} considers that high-frequency phenomena have $f_0\gtrsim100$ Hz and that the low-frequency features have $0.01$ Hz $< f_0<100$ Hz.

\subsection{Low-frequency quasi-periodic oscillations (LFQPOs)}

LFQPOs are very strong and coherent oscillations: their $rms$ amplitudes can be considerable ($rms > 0.15$) and the quality factor is typically of $\mathcal{O}(10)$. The maximum values of $rms$ are observed when the steep power law contributes significantly (more than $20\%$) to the flux at energies between 2--20 keV, with the peak in $rms$ being for 6--10 keV, but LFQPOs are also observed at much higher energies. 

If the power law needs to have a considerable contribution in order for the oscillations to be detected, it is reasonable to assume that they are related to the non-thermal component of the spectrum. However, as argued by \cite{remimcclin2006}, this does not necessarily mean that the LFQPOs don't have origin in the accretion disc and can even be considered important tools to understand the relation between the thermal and power-law components of the spectrum as their properties possibly have origin in this relation. 

Another property of the oscillations that suggests a relation between them and the accretion flow is the fact that they are quasi-stable features: often they can be observed in the PDS over several weeks, although they can have some variability in frequency on short-time scales. However, it is complicated to connect LFQPOs with the inner accretion disc since their frequencies are much lower than the ones expected for the orbits in this region. \cite{bhbbook} also mention that the distinctiveness and strength of the oscillations require the emitting region to have global properties.

\begin{table}
\begin{center}
\begin{tabular}{ccccccc}
\hline
Object & & $f_0$ (Hz) & & Commensurability & & $M/M_\odot$\\
\hline
GRO J1655--40 & & 300,450* & & 2:3 & & 6.0--6.6 \\
XTE J1550--564 & & 184,276* & & 2:3 & & 8.4--10.8\\
GRS 1915+105 & & [41,67], [113,168] & & $\sim$3:5, $\sim$2:3 & & 10.0--18.0\\
H 1743--322 & & 165,241 & & $\sim$2:3 & & N/A\\
XTE J1859+266 & & 190 & & N/A & & 7.6--12* \\
XTE J1650--500 & & 250 & & N/A & & N/A \\
4U 1630--47 & & 184 & & ** & & N/A  \\
\hline
\end{tabular} 
\caption{Confirmed detections of high-frequency QPOs in black hole candidates and mass estimates \citep[][and references therein]{remimcclin2006}. The * denotes uncertainty --- in the case of frequencies the error is of $\mathcal{O}(\pm20\textrm{ Hz})$ \citep{remillardmicroquasar2002}. For the object 4U 1630--47, \cite{kleinwolfetal2004} claim to have found a pair of HFQPOs with frequencies in a 4:1 ratio: 170, 42 Hz. However, these features have a very low quality factor and therefore this commensurability (**) is not included in the table. Spin estimates are not included due to the lack of agreement between measurements made by different methods (cf. section \ref{spincritique}).}
\label{tableqpos}
\end{center}
\end{table} 

\subsection{High-frequency quasi-periodic oscillations (HFQPOs)}
\label{hfqposob}

Unlike LFQPOs, the HFQPOs are considered one of the fundamental tools in the study of the relation of relativistic compact objects and their surroundings. Although they are not as strong as their low-frequency ``siblings'' ($rms\sim0.01$), their frequencies are the ones expected for orbits in the inner region of the accretion disc. They are considerably more stable than LFQPOs since substantial changes in luminosity only result in minor variations in frequency.

With the exception of the X-ray binary XTE J1550--564 where an HFQPO was detected at a photon count rate below the very high state \citep[see][and references therein]{klisbook}, all BHBs where QPOs of this type have been detected were in this high-luminosity state. This may be indicative of a relation between high-frequency QPOs and high mass accretion rate.

An interesting property of HFQPOs is the fact that, in some objects, they are observed (most often not simultaneously) in pairs with frequencies in a ratio of $n_1$:$n_2$, with $n_1$, $n_2$ integers. The first discovery of these ``twin'' high-frequency QPOs in black hole candidates revealed a 3:2 ratio for the frequencies of the oscillations; there are currently 4 objects with QPOs frequencies with this ratio (table \ref{tableqpos}). When pairs of HFQPOs are detected, the one with higher frequency is the more stable to luminosity variations \citep[][and references therin]{remimcclin2006}. 

The stability and range of frequencies led various authors \citep[e.g.][and references therein]{bhbbook} to connect these oscillations to fundamental properties of the black hole and to oscillations in the inner accretion disc. In fact, a plot of the highest-frequency QPOs observed in some objects versus the mass of the black hole \citep{bellonietal2006}, reveals an approximate $1/M$ relation. This agrees with the relativistic prediction that the frequencies of inner disc oscillations vary with $1/M$ if the values of $a$ of the analysed black holes are alike. A model for HFQPOs should, therefore, be intrinsically relativistic so that characteristic frequencies scale with $c^3/GM$.

Provided that they are understood theoretically, high-frequency QPOs are one of the most attractive tools to measure black hole parameters, to understand the behaviour of the accretion flow in strong field regions and to scrutinize the validity of the theory of general relativity in these regions. Theoretical models for this variability phenomena are presented in the following section.

\section{Variability of black-hole accretion discs}

Rapid X-ray variability is a ubiquitous phenomenon in systems thought to be powered by accretion onto a black hole and the accretion flow is the most obvious source of variability. Although similar occurrences are also observed in, e.g., cataclysmic variables or discs around neutron stars, theoretical modelling of rapid variability is particularly attractive in the case of black-hole accretion flows. They are thought to be less complicated, in the sense that there is no surface or stellar magnetic fields to account for, and there is an interesting mass scaling relation between stellar-mass and supermassive systems. Indeed, the $1/M$ dependence for the frequency of quasi-periodic oscillations suggests that they might also be commonly present in AGNs, although with considerably longer periods, and may provide an estimate for the mass of such systems. Despite the difficulties involved in detecting long-period oscillations, \cite{gierlinskietal2008} reported the detection of a strong QPO with a period of $\sim1$ hour in the X-ray emission of active galaxy RE J1034+396. This discovery, together with previous analysis of similarities between broad-band variability in AGNs and galactic X-ray sources \citep{mchardyetal2006}, corroborates the idea that physical processes occurring in accretion discs are the same at all mass scales \citep[see][and references therein]{kingetal2004}. 

In this dissertation the focus goes to physical processes that may result in rapid aperiodic variability, namely high-frequency QPOs. The theoretical work presented here is, in principle, applicable to both AGNs and X-ray binaries (and even weakly-magnetised neutron stars) although observational comparison is made essentially with data from galactic X-ray sources which are more widely available. It should be noted, however, that the main motivation for the research described here is not that of explaining observational phenomena in all their complexity. I acknowledge the importance of building theoretical models that correctly explain all the observational characteristics of X-ray variability phenomena, but the existing data are complex and evolving fast while accretion flow models are few and based on a variety of simplifications. The theoretical modelling of variability in accretion discs is complicated and the progress consequently slow. Many state-of-the-art high-frequency QPO models mainly focus on explaining the typical values for the frequencies of these oscillations using general relativistic effects of some sort \citep[][and references therein]{bhbbook}. The reason for this is that the theoretical community is mostly interested in finding a relation between such frequencies and the fundamental parameters of black holes. Indeed, according to \cite{rebusco2008} ``All models of QPOs are essentially dynamical models, that miss any emission mechanisms and any connections to the spectral states of the sources.''. Moreover, many QPO models are too crude to include a proper hydrodynamical treatment. There is yet to be an agreement on the basic dynamics involved in explaining high-frequency oscillations and my contribution is that of providing a strong backup on the dynamical level to some of the models by, e.g., showing how a particular oscillation frequency may be excited in the accretion flow.

The dynamical frequencies characteristic of accretion discs around black holes are presented next, and a review of current high-frequency QPO models follows.

\subsection{Characteristic frequencies}

The bulk of the X-ray emission from BHCs originates in the very inner region of accretion flows and HFQPOs have frequencies proportional to $1/M$ and approximately independent of luminosity (or equivalently accretion rate). Therefore, relativistic effects are expected to be important in modelling rapid variability phenomena and the relevant timescales are likely to be dynamical. In this section I introduce the most relevant frequencies of thin, pseudo-Newtonian or relativistic discs. It is assumed that characteristic frequencies in the accretion flow can be approximated by particle-orbit expressions.

In previous sections, some frequencies relevant for test particles orbiting an object of mass $M$ were discussed. In Newtonian dynamics, the azimuthal frequency of orbiting particles is Keplerian, $\Omega_\mt{K}=\sqrt{GM/R^3}$. In this case, other relevant frequencies such as the vertical and radial epicyclic frequencies have the same variation in radius as $\Omega_\mt{K}$. However, using the \cite{pw1980} pseudo-Newtonian potential,
\begin{equation}
\Phi_\mt{PW}(R,z)=-\frac{GM}{(R^2+z^2)^{1/2}-R_\mt{S}},
\end{equation}
where $R_\mt{S}=2GM/c^2=2R_\mt{g}$ is the Schwarzschild radius, the angular and epicyclic frequencies are different and the latter is no longer a monotonically decreasing function of radius. Indeed,
\begin{equation}
\Omega_\mt{PW}^2\equiv\left(\frac{1}{R}\frac{\partial \Phi_\mt{PW}}{\partial R}\right)_{z=0}=\frac{GM}{R(R-R_\mt{S})^2},
\label{omegapw}
\end{equation}
\begin{equation}
\kappa_\mt{PW}^2\equiv\frac{2 \Omega}{R}\frac{\mt{d}(R^2\Omega)}{\mt{d}R}=\frac{GM(R-3R_\mt{S})}{R(R-R_\mt{S})^3},
\label{kappapw}
\end{equation}
and is easily seen that $\kappa$ has a maximum at $R=(2+\sqrt{3})R_\mt{S}$. 

\begin{figure}[!t]
\begin{center}
\includegraphics[width=110mm]{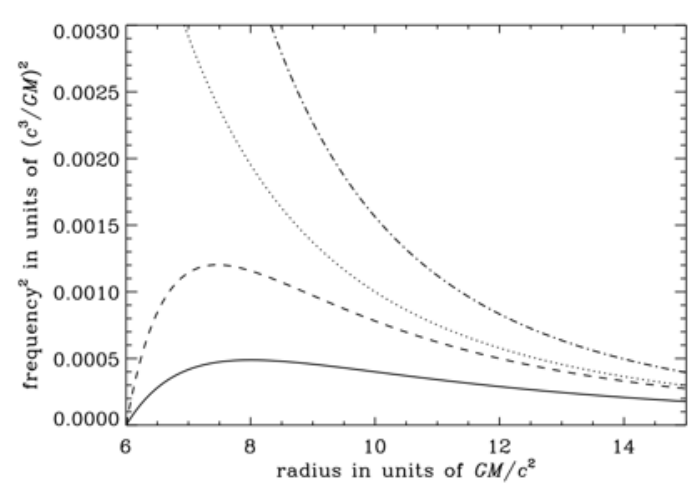}
\caption{Profiles of some characteristic frequencies squared. In a Keplerian disc, $\Omega^2=\kappa^2$ and the variation with radius is monotonic (dotted line). In a Paczy\'nski-Wiita disc, the profile of the angular velocity squared is that represented by the dash-dotted line while the epicyclic frequency varies non-monotonically with $R$ as showed by the dashed line. When a relativistic expression is used for this frequency the variation with $R$ is similar but with a maximum at a different radius (full line).}
\label{freqintro}
\end{center}
\end{figure}

The difference between the pseudo-Newtonian and Keplerian cases is due to the fact that the Paczy\'nski--Wiita potential simulates relativistic effects. In fact, for a Schwarzschild black hole, the relativistic expression for the epicyclic frequency is \citep{okazakietal1987}:
\begin{equation}
\kappa_\mt{S}^2=\frac{GM(R-3R_\mt{S})}{R^4},
\end{equation}
which also has a maximum but at $R=4R_\mt{S}$. The presence of a maximum in the epicyclic frequency is a characteristic of relativistic discs and its value, only dependent on the mass and spin of the black hole, may be of great importance to the phenomena of high-frequency QPOs as will be seen later on in this thesis. As $R$ decreases, the epicyclic frequency increases, reaches a maximum and then goes to zero at the radius of the marginally stable orbit. For $R>R_\mt{ms}$, $\kappa^2>0$ and a particle will oscillate at the epicyclic frequency when it suffers a radial perturbation. For $R<R_\mt{ms}$, $\kappa^2<0$ and the orbits are unstable since a small radial perturbation will cause the particle to depart exponentially from its original trajectory. Fig.~\ref{freqintro} shows the variation of $\Omega_\mt{K}^2$, $\Omega_\mt{PW}^2$, $\kappa_\mt{PW}^2$ and $\kappa_\mt{S}^2$ with radius.

When a rotating black hole is considered, the location of the marginally stable orbit moves inwards. In this case the angular velocity and relativistic epicyclic frequency may be written as \citep{okazakietal1987,nowaklehrchapter}

\begin{equation}
\Omega=(r^{3/2}+a)^{-1},
\label{relomega}
\end{equation}
\begin{equation}
\kappa=\Omega\sqrt{1-\frac{6}{r}+\frac{8a}{r^{3/2}}-\frac{3a^2}{r^2}},
\label{relkappa}
\end{equation}
where the lengths have been normalised to $R_\mt{g}=GM/c^2$ and the frequencies to $c^3/GM$. In this case, the vertical epicyclic frequency, equal to the angular velocity for a non-rotating black hole, is given by \citep{kato1990}

\begin{equation}
\Omega_z=\Omega\sqrt{1-\frac{4a}{r^{3/2}}+\frac{3a^2}{r^2}}.
\label{omegazfreq}
\end{equation}

Another important frequency is the Lense--Thirring precession frequency which is different from zero only for a rotating black hole. In this case, gas falling towards the compact object is dragged due to the rotation of the latter. The frequency at which this frame dragging process occurs is the Lense--Thirring precession frequency given by
\begin{equation}
\Omega_\mt{LT}=\Omega-\Omega_z\approx\frac{2a}{r^3},
\label{lt}
\end{equation}
for small $a$.

Even though relativistic effects are likely to be important in the theoretical modelling of rapid variability phenomena, they are only partially included in the research described in this thesis. A fully relativistic model is not necessary since most important effects can be included by supplementing a Newtonian treatment with the correct relativistic expressions for the characteristic frequencies in the disc \citep{rkato2001}. In Part \ref{ta} relativistic effects are mimicked by using a pseudo-Newtonian Paczy\'nski--Wiita potential while in Part \ref{os}, where the influence of spin is taken into account, relativistic expressions for $\Omega$, $\kappa$ and $\Omega_z$ are used. These expressions correctly reproduce the position of the marginally stable orbit, appropriately describe apsidal and nodal relativistic precession rates, related to $\Omega_z^2-\Omega^2$ and $\kappa^2-\Omega^2$, and rightly represent the frequency and stability of orbits in the Kerr metric. Although some information related to the metric coefficients is lost in the pseudo-relativistic approach, many important physical processes may be correctly described in this approximation. Reference to work done using purely relativistic calculations to describe similar phenomena to those I attempt to model in this thesis will be made when appropriate.

\subsection{Theoretical models for high-frequency QPOs}
\label{hfqpomodel}

In this section I present the existing theoretical models for rapid X-ray variability in black-hole accretion discs. Although kHz QPOs in neutron stars were discovered a few years earlier, high-frequency QPOs in black-hole candidates only started being detected in the late 90s and early 2000s \citep[][and references therein]{klisbook}. Most of the current models applicable to high-frequency aperiodic variability in black-hole accretion discs have therefore been developed in the past decade or so. Despite hints that a common physical mechanism could operate in both neutron stars and black-hole candidates and explain both kHz QPOs and HFQPOs \citep{pbk1999}, current observations indicate that the rapid aperiodic oscillations of neutron stars are more complex, e.g., their frequencies vary systematically. Since, at present, a correlation between the two phenomena is unclear \citep{klisbook,rezzollaetal2003}, I will focus mainly on models which are only applicable to HFQPOs detected in accretion flows surrounding black holes.

\subsubsection{Instability of transonic flows}

There are, however, some models developed in the 80s and 90s to explain QPOs in neutron stars that may be of interest for HFQPOs in BHCs that have been left out of modern reviews on the subject. Of interest is the flow model of \cite{kaaretetal1997} that involves instabilities at the marginally stable orbit. [The reader should recall that \cite{kw1985} showed that in some situations $R_\mt{ms}$ may lie outside the stellar surface (cf. footnote \ref{footns}).] Although it was initially developed to explain those kHz QPOs in neutron stars where the frequency was not correlated with the source count rate, the same mechanism can operate in black-hole accretion discs and offers a possible explanation for HFQPOs which have a frequency almost independent of luminosity variations.

The QPO mechanism of the model by \cite{kaaretetal1997} was originally proposed by \cite{nature} and is based on the finding of Muchotrzeb-Czerny \citep{m1983,m1986} that steady flows are not supported near the inner boundary, where the flow changes from subsonic to supersonic, for high values of viscosity. The unsteadiness of the flow could give rise to observable oscillations (``via Doppler beaming or eclipses'') at the Keplerian orbital frequency at the sonic point \citep[see also][]{kmw1990,mlp1996}.

The idea is interesting and has, to my knowledge, not been developed further in the literature in connection to HFQPOs in BHCs. There are some aspects of this theory which are not clear: the exact nature of the luminous structure giving rise to observable oscillations is not specified and the physical process in the origin of the instability is not studied. The problem is of considerable theoretical interest and of possible observational relevance and therefore merits further study. This is carried out in Part \ref{ta} of this thesis.

\subsubsection{Diskoseismology}

The most developed set of theoretical models aiming at explaining quasi-periodic variability in BHCs relies on hydrodynamic disc oscillations. Part \ref{os} of this dissertation is dedicated to this class of models and will provide further insight; here I only summarise the key results and aspects of this theory. 

\cite{katofukue1980} first realised that accretion discs could support discrete modes of oscillation and also shown that some of these waves, the inertial-acoustic (f or inner-p) modes, can be trapped in a region which has the marginally stable orbit as its inner boundary due to general relativistic effects. The trapping occurs because the radial epicyclic frequency has a maximum at a particular radius and is of fundamental importance for real disc oscillations as the lack of reflective boundaries forbids their growth \citep{rkato2001}. The field was developed further by \cite{okazakietal1987} who found that another class of oscillations, the inertial (r or g) modes, could also be trapped but in the region below the maximum of the epicyclic frequency (if the modes are axisymmetric). The term ``diskoseismology'' was coined by \cite{nowakwagoner1991,nowakwagoner1992} who noticed that both classes of modes could be described by the same dispersion relation. 

The frequencies of all classes of trapped modes scale with the inverse black-hole mass for similar spins, are only weakly affected by changes in density or pressure when the disc is thin, and are in the range expected for high-frequency QPOs \cite[e.g.][]{nowaklehrchapter}. The inertial trapped modes are particularly attractive since they don't rely on a reflective inner disc boundary and may give rise to large luminosity modulations \citep{perezetal1997}. These facts led \cite{nowaketal1997} to associate the 67 Hz feature detected in GRS 1915+105 with a trapped inertial mode which has a frequency close to the maximum of $\kappa$. The idea can also be applied to other stable high-frequency oscillations which were detected in subsequent years \citep{nowaklehrchapter}. The authors suggested that the modes could be excited by the ``negative radiation damping mechanism'' but they noted that large mode amplitudes were still required to make them observable.

\subsubsection{3:2 resonance models}

A potential drawback of diskoseismic models came with the discovery of commensurate frequencies QPOs since models to date were only taking into account the trapping of axisymmetric r~modes (all of similar frequency). Following these detections, Abramowicz, Klu\'zniak and their collaborators \citep[e.g.][]{kluzniakabramowicz2001,abramowiczkluzniak2001} dedicated their attention to models involving resonances between two oscillation modes of disc fluid elements. A possible explanation for the commensurate frequencies is that $n_1$:$n_2$ resonances are excited near radii in the disc where the epicyclic and the orbital frequencies are in a ratio of $n_1$:$n_2$ (1:3, 1:2, 2:3). The coupling between the two oscillations is suggested to occur due to gravity and pressure effects but the exact location and mechanism for excitation is rather arbitrary \citep{aetal2004}. 

The ``resonance model'' suffers from several caveats. To start there is no special feature in both the epicyclic and orbital frequencies at the resonant radii, apart from the integer ratio coincidence, to justify having them as key locations. The exact excitation and coupling mechanisms are never described in the theory. Indeed, the models are simplistic, only analyse the motion of a single particle, rely on an ad-hoc parameter which is added to the equations to artificially increase the modes' amplitudes and the coupling between the resonant modes provided by gravity and pressure is too weak to transfer energy between the oscillations \citep{rebusco2008}. The too localised description of the models doesn't allow for wave propagation and, as noted by \cite{rezzollaetal2003}, the radial extent where the resonance occurs may be too small to result in the observed modulation in emissivity. 

An interesting model combining features of both the diskoseismic and resonance theories was proposed by \cite{rezzollaetal2003}. They theorise the existence of an oscillating torus surrounding the black hole and identify QPOs with inertial-acoustic modes of this structure. The torus is considered to have a finite extent $L$ and provides the resonant cavity necessary for mode growth. The fundamental and first overtone modes of frequency are approximately in a 3:2 ratio if $L$ is small, are within the expected range for HFQPOs if the spin is large but depend on the size of the torus. Moreover, it is not clear that such a structure dominates the accretion flow in the very high accretion states \citep[][and cf. Fig.~\ref{esin}]{donereview}. Observational constraints were, however, not what most hindered this theory. The crucial drawback came from numerical simulations of fluid tori which found the inertial-acoustic oscillations to be significantly damped by the PPI (despite the radial inflow) and particularly by MRI turbulence \citep{fragile2005}.

Attempts at models involving orbital motions of density clumps or ``hot spots'' were made by \cite{sb2004} and \cite{s2005} \citep[see also][]{svm1999}. These models use relativistic ray-tracing to explain the spectra of X-ray lines produced by oscillating test particles and do not treat the disc as a fluid-dynamical entity. The frequencies depend on the size and luminosity, and radial positions of the hot spots and it is unclear if these elements can survive the disc's differential rotation and turbulent motions.

\subsubsection{Excitation mechanisms}
\label{disko}

The last few years saw the return of diskoseismic models. However, state-of-the-art magnetohydrodynamic (MHD) simulations, now available, indicate that MRI turbulence does not excite, and may indeed damp, the inertial modes while inertial-acoustic modes are more likely to exist in discs threaded by magnetic fields \citep{arrasetal2006,rm2009}. This may be explained by the fact that the power spectrum of MRI turbulence peaks at a frequency close to the maximum of $\kappa$, making axisymmetric trapped inertial modes likely to be affected by this instability. Moreover, \cite{fulai2009} showed that the trapping region of $r$ modes is modified or destroyed when strong poloidal magnetic fields are taken into account. It is however unclear if the magnitude of poloidal magnetic fields in real discs is large enough to modify the trapping region significantly. 

These facts led some authors to set the inertial modes aside and look for excitation mechanisms for inertial-acoustic modes instead. A particularly interesting hydrodynamic instability is that studied by \cite{laitsang2009} which relies on wave absorption at the corotation resonance. Their linear analysis shows that non-axisymmetric f modes with frequencies $\omega\approx(0.5-0.7)m\Omega_\mt{ms}$, where $m$ is the azimuthal wave-number and $\Omega_\mt{ms}$ is the angular velocity at the marginally stable orbit may be excited by means of this instability being likely candidates to explain the HFQPOs. 

Varni\`ere, Tagger and their collaborators \citep{vrtd2002,rvt2002,tv2006} developed a complicated theory based on unstable versions of the diskoseismic modes to explain both low-frequency and high-frequency QPOs. In the case of HFQPOs, the instability responsible for the excitation of inertial-acoustic modes combines features of the accretion-ejection and MHD Rossby wave instabilities and may result in significant growth of these modes in magnetised discs. The predicted QPO frequencies are of the same order of magnitude as those of \cite{laitsang2009}.

Two important caveats accompany these models. First, if the spin of the black hole is significant, the predicted frequencies are too high to explain the range of frequencies of rapid quasi-periodic oscillations particularly if $m>1$. Second, and most importantly, both instabilities require the inertial-acoustic modes to be reflected at the inner boundary. As shown by \cite{blaes1987}, the rapid radial inflow at the marginally stable orbit damps these modes, making them unlikely relevant in realistic discs. Moreover, in the very high state where HFQPOs are almost exclusively detected, the disc is expected to extend to the marginally stable orbit and accretion rate and radial inflow are likely to be significant. Wave reflection at the sonic point will be analysed further in chapter \ref{reflect}.

On the other hand, and following the studies of Abramowicz and Klu\'zniak which emphasised the importance of resonances in the physical processes giving rise to HFQPOs, \cite{katowarp2004,kato2008} proposed that oscillations could be resonantly excited in deformed (warped or eccentric) discs. The first study of local wave excitation in a tidally distorted disc via the
parametric instability was made by \cite{goodman1993}. The idea of using a warp as an excitation mechanism for disc oscillations goes back to \cite{papterquem1995}, who mentioned the possibility of parametric generation of inertial waves. Detailed calculations are reported by \cite{gammieetal2000}. Kato, using a different approach, studied this problem analytically for thin, relativistic discs
with non-rotating central objects. He made simple estimates for the growth rates of inertial-acoustic modes and inertial modes, which are of considerable interest. As he pointed out, the only oscillations with observational relevance are those which are trapped and are resonantly excited in their propagation region. According to Kato, these would be the axisymmetric inertial mode with frequency close to $\mt{max}(\kappa)$, and two non-axisymmetric inertial-acoustic modes trapped between the inner boundary of the disc and the resonant radius located at $R\approx 4R_\mt{S}$ for $a=0$ \citep[see, e.g.,][]{kato2007}. The $m=1$ f mode would have a frequency $\omega_1=(\Omega-\kappa)_\mt{r}$ while the $m=2$ would have $\omega_2=(2\Omega-\kappa)_\mt{r}$, where the subscript indicates that the frequencies are measured at the resonance radius. Owing to the location of the resonance, $(\Omega-\kappa)_\mt{r}\approx\mt{max}(\kappa)$ so that only two frequencies remain significant. Kato then proceed to identify the 3:2 HFQPOs with the two non-axisymmetric inertial-acoustic modes. However, to account for the fact that the frequencies had to be in a 3:2 ratio, he argued that the one-armed oscillation would be detected with a frequency $2\omega_1$ due to observational effects so that $\omega_2:2\omega_1\approx3:2$ \citep[see, e.g.,][]{katofukue2006}.

Although interesting and worth further attention, many details of Kato's model are dubious. To start, there are many uncertainties in his calculations of growth rates and he did not discuss the origin or nature of the global deformations. Despite the fact that calculations were done only in the zero-spin case, Kato attempted to use its predictions to calculate spins of BHCs. Also, the oscillation frequencies involved, $2\omega_1$ and $\omega_2$, may be too high in a black hole with small mass and significant rotation. Indeed, an estimate of the spin of GRS 1915+105 using these frequencies to explain its HFQPOs results in a possibly (depending on the mass used) negative $a$ \citep{katofukue2006}. On the other hand, as in the models of \cite{tv2006} and \cite{laitsang2009}, identification of QPOs with inertial-acoustic modes requires a reflective boundary at $R_\mt{in}$.

Taking these factors into account, I go back to the suggestion that inertial modes, trapped away from the inner boundary, are more significant to explain HFQPOs within this model. In chapter \ref{excmech} I develop and generalize Kato's ideas on this excitation mechanism and make detailed numerical calculations of trapped inertial modes and their growth rates for rotating black holes. A simple dynamical treatment of the warp and eccentricity is presented in chapter \ref{oscilintro} and used in \ref{excmech}, while a broader discussion of the origin and global propagation of these deformations is given in chapter \ref{warpecc}.

A preliminary confirmation of the possible excitation of modes with frequencies at or below the maximum of the epicyclic frequency, which are partially inertial in nature, came with the fully relativistic MHD simulations of \cite{heniseyetal2009} of warped discs. In agreement with previous simulations, the authors find no signs of inertial waves in the case where the disc is untilted. However, when a deformation is present, the presence of inertial variability is clear. It is complicated to identify this variability with an inertial mode because the disc is highly deformed, the background flow is complex and the mode structure may be partially modified through the coupling mechanism. The structure analysed in the paper seems to be a composite oscillation with signatures of both inertial and inertial-acoustic modes of frequency close to $\mt{max}(\kappa)$. These results show that inertial modes may indeed exist in flows with MRI turbulence provided an excitation medium, such as a deformation in  the disc, is present. 

\subsubsection{Viscous overstability and the return of the unsteady transonic flow}

The attentive reader will have noticed that the simulations of \cite{heniseyetal2009} show both inertial and inertial-acoustic signatures at a frequency close to the maximum of the epicyclic frequency. Although the modes' frequencies are not in a 3:2 ratio and the inertial-acoustic oscillations are likely to be absorbed at the inner boundary, this, together with the results of \cite{laitsang2009} and \cite{tv2006}, turns my attention back to the f modes. However, I'm interested in a different excitation mechanism, not investigated by these authors and yet to be mentioned, as it may be related to the unsteadiness of the transonic flow mentioned previously. \cite{chentaam1995} suggested that inertial-acoustic waves may be relevant to explain HFQPOs in a viscously overstable flow. These waves are globally viscously overstable if the viscosity parameter is sufficiently high. The investigations of \cite{khm1988} and \cite{ap2003} suggested that the global character of this instability and the presence of viscously overstable inertial-acoustic waves in realistic discs may be connected with the unsteadiness of the transonic flow, which happens at high viscosity. A more thorough study of both physical processes and the possibility of a connection between them is the topic of Part \ref{ta} of this thesis.  \enlargethispage{-\baselineskip}

\subsection{Thesis outline}

The core of this dissertation is divided in two parts. In Part \ref{ta} I investigate the problem of isothermal transonic accretion in a time-dependent framework with the aim of studying the stability of such solutions beyond local analysis. Part \ref{os} focuses on different types of oscillations in black-hole accretion discs and different problems related to this topic. Part \ref{ta} is composed of two chapters while four constitute Part \ref{os}; a detailed outline of each of them follows.

In chapter \ref{steadyintro} I extensively review the work done previously on the structure and stability of transonic accretion flows. I start by considering the literature on steady transonic accretion, with particular emphasis on the work of \cite{ap2003}, and go on to the articles where the stability of these accretion solutions is investigated, both locally and globally with the aid of numerical simulations. The topic of viscous overstability naturally arises in these stability considerations and is also assessed. 

In chapter \ref{timeaccretion} I solve the time-dependent version of the equations solved in \cite{ap2003} which describe a thin, 1D transonic flow with isothermal equation of state and essentially an $\alpha P$-type stress tensor. I thoroughly analyse the solutions obtained for different values of sound speed and a range of viscosities with particular importance given to the region of the parameter space where global inertial-acoustic oscillations are visible in the simulations. These are most likely due to the onset of viscous overstability. Since stable solutions are obtained for low values of $\alpha$ (in agreement with the findings of the steady problem), I discuss the possibility of a relation between the unsteadiness of transonic accretion and the presence of viscously overstable waves in the disc. A relation between these phenomena and high-frequency QPOs is briefly alluded to.

In chapter \ref{oscilintro} I introduce the fundamentals of disc oscillations with a particular focus on the trapping regions of inertial and inertial-acoustic waves characteristic of black-hole accretion discs. The dependence of perturbed quantities in all three coordinates ($R,\phi,z$) is considered and, for simplicity, viscosity and the background radial inflow are neglected although the latter is considered in chapter \ref{reflect}. I also examine disc deformations such as warping and eccentricity in the context of global modes. 

Chapter \ref{excmech} is dedicated to an excitation mechanism for trapped inertial waves. In the basis of this mechanism is the coupling between the mentioned oscillations, a global deformation and an intermediate wave. The mathematical analysis involves solving non-linear equations for the inertial and intermediate oscillations coupled by the warp or eccentric mode. Excitation occurs because the intermediate waves resulting from the coupling have negative energy; if they are damped as they approach the inner boundary of the disc or their corotation resonance, positive energy becomes available for the growth of the inertial mode. The dependence of the growth rate on the spin of the black hole, the sound speed and deformation amplitude are determined, showing that this excitation mechanism can be effective under a wide variety of conditions. Provided the global modes reach the inner region with a non-negligible amplitude, inertial modes can be excited, being likely candidates to explain HFQPOs. 

Chapter \ref{warpecc} focuses precisely on the conditions under which global modes may reach the inner region of black-hole accretion discs. The radial dependence of eccentricity and warp tilt is considered within a more realistic viscous disc model and taken to be described by less simplistic equations than those used in chapter \ref{oscilintro}. Propagation from the outer regions is facilitated for high accretion rate and low viscous damping. Results show that the high and very high states of black-hole accretion where the mass accretion rate is probably close to the Eddington limit are likely to be the ones where the global deformations reach the inner disc more easily.

In chapter \ref{reflect} I connect the work done in both Part \ref{ta} and Part \ref{os}. In the former, I emphasised the importance of considering a non-negligible radial inflow in the inner regions of black-hole accretion discs and determined stable transonic solutions for low values of viscosity. However, in most of Part \ref{os} this transonic background was ignored, despite being important close to the inner boundary of the disc (defined by the sonic point when the radial inflow is considered). This final core chapter is therefore dedicated to the influence of the transonic background on the propagation of waves. Results relating to trapped inertial modes are particularly interesting as they show that, under some conditions, their structure may be severely modified or destroyed by the background inflow. 

Finally, conclusions are presented in the only chapter of Part \ref{c} where I summarise the work described in this dissertation and refer to its observational and theoretical relevance.

\newpage
\thispagestyle{empty}

\thispagestyle{empty}
\part{Transonic Accretion and Viscous Overstability}
\label{ta}
\chapter{Introduction}
\label{steadyintro}


This part of the thesis is dedicated to the structure and stability of transonic accretion flows and to the onset of viscous overstability in these models. In this chapter I provide a thorough introduction to both topics and extensively review the model of \cite{ap2003} for steady transonic accretion in a thin, isothermal disc onto a black hole. Their work reveals the existence of an upper value for the viscosity parameter $\alpha$ above which no physically acceptable steady state solutions are found. Motivated by this finding, in chapter \ref{timeaccretion} I study the equivalent time-dependent flow to understand what happens when $\alpha$ is above that threshold value. The results obtained show viscously overstable waves propagating in the disc when the viscosity is above that limit, hinting a possible relation between the two phenomena.

In this part of the thesis different symbols will be used to represent different limits for the viscosity parameter $\alpha$. The reader should refer to Table \ref{gloss} for a glossary.


\section{Steady transonic accretion}
\label{steady}

\subsection{Introduction}

\begin{table}
\begin{center}
\begin{tabular}{cl}
\hline
Symbol & Meaning: value of $\alpha$ above which: \\
\hline
$\alpha^*$ & physically acceptable steady state, transonic accretion solutions \vspace{-1mm} \\ & cannot be found \vspace{2mm} \\ 
$\alpha_1$ & the time-dependent system no longer settles into a steady state \vspace{2mm} \\
$\alpha_\mt{sn}$ & the flow goes through a critical point of nodal type (for $\alpha<\alpha_\mt{sn}$  \vspace{-1mm}  \\ & the point is a saddle)  \vspace{2mm} \\
$\alpha_\mt{nn}$ & the passage through a nodal point is no longer made in the dire-  \vspace{-1mm} \\ & ction of the eigenvector corresponding to the largest eigenvalue \vspace{2mm}  \\ 
$\alpha_\mt{sp}$ \vspace{2mm} & the sonic point is unstable according to local analysis \\
$\alpha_\mt{vo}$ & \emph{global} viscous overstability sets in \\
\hline
\end{tabular} 
\end{center}
\caption{Glossary of the different symbols used to represent different (but possibily equivalent) limits for the viscosity parameter $\alpha$ used throughout this part of the thesis.}
\label{gloss}
\end{table} 

The \cite{ss73} steady disc model introduced in the previous chapter is singular at the inner disc boundary. If the disc surrounds a black hole, it is assumed to terminate at the marginally stable orbit of a test particle where the Keplerian angular momentum is minimum and a ``no torque boundary condition'' (which defines $R_\mt{in})$ is taken to be valid at that radius. Therefore, $R_\mt{in}=R_\mt{ms}$ and accretion is disc-like for $R>R_\mt{in}$ while the matter is free-falling onto the black hole for $R<R_\mt{in}$ \citep[e.g.][]{pbk1981}.

The singularity at the inner disc boundary is artificial and is removed if the radial velocity, dynamically important in the inner region of black-hole accretion discs, and the radial pressure gradient are not neglected in the radial momentum equation. This was first done in the thin disc case by \cite{lt1980} who found that, when $u_R$ is included, the solution has one or more critical points where $u_R$ changes from subsonic to supersonic. Indeed, the radial momentum equation, combined with the mass conservation equation, can be written as \citep{mp1982}
\be
\frac{\mt{d}u_R}{\mt{d}R}=\frac{f}{u_R^2-c_\mt{s}^2},
\label{cpradial}
\ee
where $f$ is a function of $R$, $u_R$ and of the parameters of the flow (e.g., viscosity, accretion rate, etc.). This equation has critical points $R_0$ where $u_R^2=c_\mt{s}^2$. Although more than one point is mathematically possible, a physically acceptable accretion solution is subsonic in the outer disc, passes through only one of these critical points, and becomes supersonic close to the compact object \citep[e.g.][]{nt73}.


When the radial drift and pressure gradient terms are important, the distribution of angular momentum deviates from the Keplerian one and the disc can no longer be taken to terminate exactly at $R_\mt{ms}$. The disc \emph{per se} can be thought of as the region where accretion is driven by angular momentum transfer. Since turbulent processes cannot transport angular momentum at supersonic speeds (i.e. $\alpha<1$), it makes sense to think of the disc as existing only in the subsonic region and terminating at $R_0$. Accretion is a free-fall process occurring at constant angular momentum for $R<R_0$. Indeed, the ``no torque boundary condition'' should be valid there since no information can propagate from the supersonic region to the subsonic \citep{pbk1981,notorque} except possibly by magnetic stresses \citep{krolik99,gammie99}. In summary, in the transonic thin-disc accretion scenario $R_\mt{in}\approx R_0$ and this location is close to, but not necessarily at, $R_\mt{ms}$ \citep{mp1982}.

The case of isothermal thin-disc accretion, i.e., $c_\mt{s}=\mt{constant}$, $c_\mt{s}^2/c^2\ll 1$, $u_\mt{R}<0$, with an $\alpha P$ stress tensor is simple yet illustrative. In this model the vertically-integrated, steady state hydrodynamic equations can be reduced to a single ordinary differential equation of the form (\ref{cpradial}) which has, in general, two critical points. These sonic points are localised near the marginally stable orbit: one has $R_\mathrm{0}\le R_\mathrm{ms}$ and another $R_\mathrm{0}\ge R_\mathrm{ms}$ \citep{mkfo1984}. By linearising the equations around the critical points and determining the eigenvalues of the Jacobian matrix one finds that the points can be either of nodal, saddle or spiral type. The topology of the flow is dependent on the parameters of the problem, namely the angular momentum constant [$l_\mt{in}$, cf. equation (\ref{angconst}), or equivalently the location $R_0$], the sound speed and the viscosity parameter $\alpha$ \citep{m1986}.

For a transonic solution to be globally acceptable it should extend all the way from the outer disc to the centre going through the sonic point regularly. Realistic boundary conditions at large radii are Keplerian, i.e., the transonic solution should match to the subsonic Shakura--Sunyaev model with the disc rotating with angular velocity $\Omega_\mt{K}$ (or, more appropriately for a pseudo-Newtonian disc, $\Omega_\mt{PW}$) for $R\gg R_0$. For $R\ll R_0$ the flow is supersonic. Furthermore, the critical point should be either of nodal or saddle type since spirals are unphysical. For each point of the parameter space ($l_\mt{in}$ or $R_0$, $c_\mt{s}$ or $\dot{M}$, $\alpha$), these boundary conditions and regularity demands select the relevant solution. According to \cite{lt1980}, although without a rigorous proof, ``the \emph{critical transonic solution} whenever it exists is unique in relevant situations''. The uniqueness of the transonic solution is also verified by \cite{ap2003} for low viscosity.  \enlargethispage{\baselineskip}

\subsubsection{Constraints}

However, there may be points in the parameter space for which no unique solution regular at the sonic point satisfies the boundary conditions. Steady transonic black hole accretion is, therefore, constrained \citep[][and references therin]{ak1989}, i.e., in principle, it may only exist as a physically acceptable solution in certain regions of the parameter space. Of particular interest is the result of \cite{m1983} who found steady transonic solutions (going through a saddle-type critical point) only for $\alpha<\alpha^*$, and suggested that accretion would probably be unsteady for larger viscosities. (She worked with a non-isothermal disc model --- with a rather arbitrary vertical structure --- and found $\alpha^*$ to be approximately $0.02$ and weakly dependent on the mass accretion rate.)

On the contrary, \cite{mkfo1984} stated that accretion is possible for values of $\alpha$ above the limit of \cite{m1983} but with differences in the disc structure and in the nature of the sonic point relative to the low $\alpha$ case. In the small viscosity regime, the gas infall in the transonic region is caused by the pressure gradient force and the critical point, $R_0\lesssim R_\mathrm{ms}$, is of saddle type. On the other hand, in the high viscosity case, accretion is due to viscous effects and the sonic point is a node and is located outside $R_\mathrm{ms}$. The limiting value of $\alpha$ at which the type of critical point changes, $\alpha_\mt{sn}$, is roughly 0.05 \citep[according to the authors' estimates, made within a model different to that used by][]{m1983}.

Analytical considerations by \cite{m1986} confirmed this result that both types of solutions with ($\alpha<\alpha_\mt{sn}$ and $\alpha>\alpha_\mt{sn}$) are physically acceptable. Furthermore, she noticed another fundamental difference in the low- and high-viscosity regimes. In the case of saddle-type points, only one curve going through the sonic point can match with the Keplerian solution at large radii, while an infinite number is possible in nodal-type points. In the saddle case, the boundary conditions uniquely define a value for the angular momentum constant $l_\mt{in}$ for fixed $(\alpha,c_\mt{s})$. On the other hand, for fixed viscosity and sound speed at high $\alpha$, a range $l_\mt{in,min}<l_\mt{in}<l_\mt{in,max}$ may be obtained such that boundary conditions are satisfied. In this case where the sonic point is nodal, the physical constraints don't seem to uniquely determine an accretion solution. 

The paper by \cite{m1986} goes further and is, to my knowledge, the first dealing with the stability of transonic solutions \emph{at} the sonic point. Within an isothermal model, she finds that perturbations at the sonic point are damped if the point is a saddle, while they can either decay or grow if the point is nodal. This, together with the nonuniqueness discussed in the previous paragraph, might indicate that accretion doesn't proceed in a stationary way for $\alpha>\alpha_\mt{sn}$ because ``the solution may travel in a regular or irregular way between the external values of $l_\mt{in,min}$ and $l_\mt{in,max}$''. As a consequence, the sonic point would oscillate between $R_\mt{ms}$ and the outermost nodal critical point possible for a given $\alpha$ and $c_\mt{s}$.  

The isothermal transonic accretion problem was revisited by \cite{ak1989} who, in opposition to \cite{m1983}, were able to find solutions satisfying boundary and regularity conditions for any value of $\alpha$ (even though such conditions forbid some regions of the parameter space of the problem). It is however unclear if their solutions are \emph{unique} for all viscosity parameters. Another important novelty of this work was the classification of critical points in the case where the stress tensor has a diffusion form, i.e., when it is proportional to the gradient of angular velocity rather than to the pressure only. Interestingly, the authors find that the only physical critical points are of saddle type; nodal points do not exist in an isothermal flow characterised by $T_{R\phi}\propto \mt{d}\Omega/\mt{d}R$ [this is, however, not true of non-isothermal flows \citep{chentaam1993}]. This is to show that different expressions for the stress tensor may indeed result in different mathematical characters of the sonic point for the same value of $\alpha$, as previously indicated in Part \ref{i}.  \enlargethispage{\baselineskip}

The more recent investigation of \cite{ap2003} saw the return of the $\alpha^*$ limit. Within an isothermal, thin disc model, the authors find physically acceptable \emph{unique} steady state solutions for $\alpha<\alpha^*=0.14(100 c_\mathrm{s}/c)^{1/3}$. Moreover, the authors find that steady accretion is possible whether the sonic point is saddle or nodal\footnote{Note that a slightly different form of the stress tensor was used in this work. The results of \cite{ap2003} and \cite{mkfo1984,m1986} can be compared, in the isothermal case, by noting that $\alpha_\mathrm{SS}=2\frac{\Omega}{\Omega_\mathrm{PW}}\alpha_\mathrm{AP}\sim2\alpha_\mt{AP}$, where $\alpha_\mathrm{SS}$ refers to the viscosity parameter used in the 80s works while $\alpha_\mathrm{AP}$ refers to that used by \cite{ap2003}.}. The authors point out that steady solutions no longer exist or are unstable when the viscosity is such that the flow passes through a nodal critical point in the \emph{slow direction}. This is the direction of the eigenvector associated with the smaller absolute-valued eigenvalue of the Jacobian, as opposed to the (fast) direction related to the largest eigenvalue in modulus. Therefore, it seems that $\alpha_\mathrm{sn}$ and $\alpha^*$ are not necessarily related as believed by \cite{m1986} since in this more recent work $\alpha^*=\alpha_\mt{nn}$. As it will be seen in section \ref{apsection}, the differences between the two works relate to the fact that \cite{m1986} did not take into account the differences between the passage through a nodal point in the fast and slow directions.

\subsubsection{Observations}

From an observational point of view, \cite{nature} suggested that the unsteady flow at high viscosity would be responsible for variations in the luminosity of the boundary layer between a neutron star and the accretion disc. Moreover, and as mentioned in Part \ref{i}, he proposed that these variations could give rise to X-ray quasi-periodic oscillations (QPOs). On the other hand, \cite{ak1989} suggested that the constraints on steady disc accretion could be relevant in explaining the transient behaviour of X-ray sources, which switch from low to high states. Since steady accretion is only possible in a certain region of parameter space, if the situation is such that the flow is located in one of the forbidden regions, it cannot be stationary and could presumably keep on switching between different accretion states.

\subsection{The Afshordi--Paczy\'nski model}
\label{apsection}

In this section I analyse in more detail the most recent model for steady, thin-disc transonic accretion, that of \cite{ap2003} since a time-dependent version of their equations is solved in the next chapter. 

\subsubsection{Equations and assumptions}

In the Afshordi--Paczy\'nski model, the flow is isothermal and vertically-integrated quantities are used to represent fluid variables. The sound speed in the disc is, therefore, constant and it remains much smaller than the speed of light so as to maintain the thinness of the flow. The model is simple but useful to understand the key features of transonic accretion flows, namely the nature of the critical point, the uniqueness of the passage through this point and the region of the parameter space where physically acceptable (stable) steady solutions are possible.

An important point to notice regards the form of the viscous stress used by \cite{ap2003}. The inattentive reader may believe that the viscous stress is assumed proportional to both the total pressure and the shear rate,
\be
\tau_{R\phi}=-\alpha P \left(-\frac{\mt{d}\ln\Omega}{\mt{d}\ln r}\right)\left(\frac{\Omega}{\Omega_\mt{PW}}\right),
\label{torque}
\ee
where $\alpha$ is constant, as indicated by equation (6) in the paper. (Note that they use the symbol $\Omega_\mt{K}$ to represent the particle-orbit pseudo-Newtonian angular velocity.) However, further down, the authors make an approximation that effectively changes the character of the viscous stress. They assume, as in the classical theory, that the angular momentum is conserved near and inward of $R_\mt{ms}$ (which, the reader should recall from the introduction, is close to the sonic point and inner boundary). Since the interest is in studying the flow passage through $R_0$, and for simplicity, the authors assume
\be
l=\Omega R^2\approx\mt{constant}\Rightarrow \left(-\frac{\mt{d}\ln\Omega}{\mt{d}\ln r}\right)\approx 2
\label{torqueapprox}
\ee
in the expression for the stress tensor. As a result, 
\be
\tau_{R\phi}\approx-2\alpha P\left(\frac{\Omega}{\Omega_\mt{PW}}\right).
\label{torquefinal1}
\ee
In a thin disc where $\Omega\approx\Omega_\mt{PW}$ (this will be shown to be valid in chapter \ref{timeaccretion}), the stress used is equivalent to that of the original \cite{ss73} paper with $\alpha\approx\alpha_\mt{SS}/2$. Since the stress tensor is essentially $\alpha P$ and not of diffusion type, both saddle and nodal points are expected.

When the pressure gradient and radial velocity are taken into account, the vertically-integrated steady-state radial momentum equation can be written as \citep{mkfo1984}
\be
u_R\frac{\mt{d}u_R}{\mt{d}R}-\frac{l^2-l_\mt{PW}^2}{R^3}+\frac{1}{\Sigma}\frac{\mt{d}P}{\mt{d}R}+\frac{P}{\Sigma}\frac{\mt{d}\ln\Omega_\mt{PW}}{\mt{d}R}=0.
\label{radmom}
\ee
Here $u_R$ is the vertically-integrated radial velocity (comparing with the notation used in the previous chapter, $\bar{u}_R$, the bar has been dropped for convenience) and $l_\mt{PW}=\Omega_\mt{PW}R^2$. As before, $P=c_\mt{s}^2\Sigma$ and the mass and angular momentum conservation equations are in the form $\dot{M}=2\pi R\Sigma(-u_R)$ and $\dot{M}(l-l_\mt{in})=2\pi R^2(-\tau_{R\phi})$, respectively. Note that the last term on the LHS of (\ref{radmom}) is frequently not included in the vertically-integrated radial momentum equation \citep[e.g.][]{m1983,chentaam1993,mt1996}. However, according to \cite{mkfo1984}, it is a correction for the decrease of the radial component of $\nabla\Phi$ away from the disc mid-plane and should be included as it is of the same order in $H/R$ as the radial pressure gradient. In any case, it is a constant term in isothermal discs that doesn't significantly change the system of vertically integrated equations describing an accretion flow.

These equations can all be combined into the form (\ref{cpradial}) which can then be written as a two-dimensional autonomous dynamical system,
\be
\frac{\mt{d}u_r}{\mt{d}\vartheta}=f(r,u_r,l_\mt{in},\alpha,c_\mt{s})\quad\mt{and}\quad\frac{\mt{d}r}{\mt{d}\vartheta}=u_r^2-c_\mt{s}^2,
\label{system}
\ee
where the velocities are now in units of $c$, the lengths are scaled to $R_\mt{g}$ and frequencies are in units of $c^3/GM$ and
\be
f=\frac{l_\mt{in}^2 u_r^3/r^2}{\left[-u_rr^{1/2}+2(2-r)\alpha c_\mt{s}^2\right]^2}-\frac{u_r}{(r-2)^2}+u_rc_\mt{s}^2\left(\frac{3}{2r}+\frac{1}{r-2}\right).
\ee 
Here $\vartheta$ is a dummy variable and the components of (\ref{system}) are equivalent to equations (17a) and (17b) of \cite{ap2003}. A solution of this system representing a realistic accretion flow should satisfy the following boundary conditions: $|u_r|\ll c_\mt{s}$ and $u_r$ Keplerian at large radii, $u_r=-c_\mt{s}$ at the critical point and $|u_r|\gg c_\mt{s}$ at small radii. \cite{ap2003} further assume the physical solution to be analytic, i.e., with $u_r(r)$ single-valued at each location and infinitely differentiable; in particular, regularity at the sonic point demands $f=0$ there.

\subsubsection{Solution topology}

\begin{figure}
\begin{center}
\includegraphics[width=75mm]{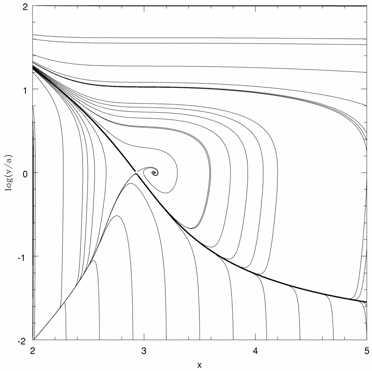} \\
\vspace{2mm}
\includegraphics[width=75mm]{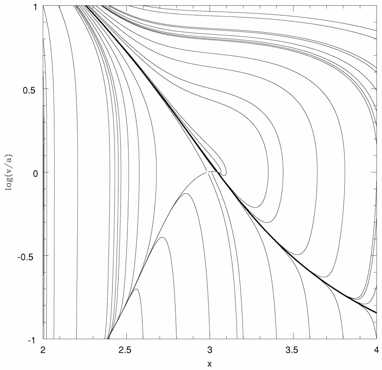}
\includegraphics[width=75mm]{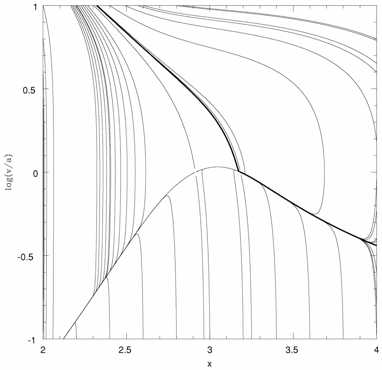} \\
\caption{Phase portraits corresponding to system (\ref{system}) with $v=-u_r$, $a=c_\mt{s}$ and $x=r/2$ for $c_\mt{s}=0.01$ and various values of $\alpha$. The curve that all solutions merge with at $x>x_0$ is the Keplerian trajectory. The thick curve represents the solution of the system that obeys boundary and regularity conditions. In the top panel, $\alpha=0.04$, $x_0=2.930$ and the solution goes through a saddle critical point. In the bottom left, $\alpha=0.1$, $x_0=3.047$ and the flow crosses the nodal sonic point in the fast direction. In the bottom right panel, $\alpha=0.3$, $x_0=3.183$ and the solution goes through the critical point in the slow direction. The thick curve in this last figure doesn't represent a realistic steady solution: a sharp change in the slope of $u_r(r)$ is evident \citep[from][]{ap2003}.}
\label{apdiagrams}
\end{center}
\end{figure}

At the critical points of (\ref{radmom}), the right-hand sides of equations (\ref{system}) vanish. The nature of these critical points can be determined by linearising the system around each of them and calculating the eigenvalues $\lambda$ of the Jacobian matrix. Phase portraits in the plane $(r,u_r)$ can then be plotted. This procedure is standard [although not simple in the particular case of system (\ref{system})] and more details about it can be found in any good textbook on differential equations. A thorough discussion on the nature of critical points is presented in \cite{ferrarietal1985}.

Depending on the values of $\alpha$, $c_\mt{s}$ and $l_\mt{in}$ adopted, \cite{ap2003} find that the equations have, in general, two fixed points of saddle and spiral or saddle and nodal type. The physical solution can only go through one of these critical points, of saddle or nodal type. The authors solve equations (\ref{system}) numerically everywhere except near the fixed points where an analytic expansion is used. Examples of phase portraits corresponding to system (\ref{system}) where the parameters of the problem are such that a physical solution is possible are presented in Fig.~\ref{apdiagrams}. The physical solution is found by looking for a combination of parameters such that the (outer) Keplerian solution passes through a sonic point regularly. It is, however, unclear that the thick curve in the lower right panel may represent a realistic stable solution due to the sharp change in slope immediately after the sonic point. This will be discussed in more detail in a few paragraphs.

\begin{figure}[t!]
\begin{center}
\includegraphics[width=70mm]{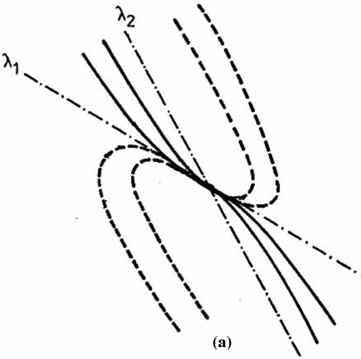}
\includegraphics[width=45mm]{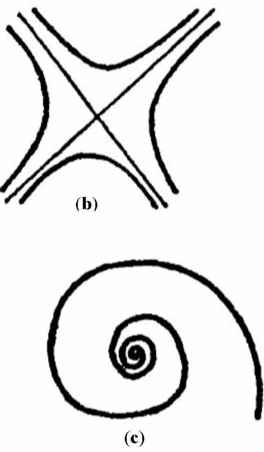}
\caption{Behaviour of the flow near different types of critical points represented in the $(r,-u_r)$ phase plane or, equivalently, in the $(x, \log[v/a])$ phase plane of the panels of Fig. \ref{apdiagrams}. (a) Vicinity of a nodal-type critical point: the eigenvalues of the Jacobian matrix, $\lambda_1$ and $\lambda_2$, are both positive or both negative; $\lambda_2$ is the largest eigenvalue in modulus while $\lambda_1$ is associated with the slow direction. The S-shaped, multiple-valued curves are represented by dashed lines while the single-valued curves are represented by solid lines. (b) Saddle-type critical point: $\lambda_2>0>\lambda_1$ and only the curves in the direction of the eigenvectors go through the sonic point. (c) Spiral-type critical point: $\lambda_1$, $\lambda_2$ complex; unphysical case \citep[adapted from][]{m1986}.}
\label{node}
\end{center}
\end{figure}

For $c_\mt{s}=0.01$ and $\alpha<0.08=\alpha_\mt{sn}$ the Keplerian solution matches onto a saddle-type sonic point and a physical solution is found for all values of the viscosity parameter. In this case, the analyticity of the solution is directly implied by the passage of the flow through the sonic point because only the two solutions in the direction of the eigenvectors of the Jacobian matrix can pass through this point [Fig.~\ref{node} (b)]. Furthermore, only one of these two curves satisfies $|u_r|<c_\mt{s}$ at large radius. Therefore, and in agreement with the qualitative analysis of \cite{m1986}, the boundary conditions pick an unique solution, i.e., for fixed $c_\mt{s}$, $\alpha$, the value of $l_\mt{in}$ such that the flow goes through the sonic point regularly is determined uniquely.

For the same sound speed but larger values of $\alpha$, the Keplerian curve connects to a nodal-type point. In this case, an infinite number of solutions obeying $|u_r|<c_\mt{s}$ at large $r$ can go through the critical point and satisfy $|u_r|>c_\mt{s}$ at small radii. There is, however, a possibility for a unique passage through the sonic point. As it can be seen in Fig.~\ref{node} (a), all solutions merge with the slow direction near the critical point, except for the solution in the fast direction. Therefore, if the flow is to cross the sonic point in this direction the passage is unique and similar to the saddle point case. According to \cite{ap2003}, this happens for $0.08<\alpha<0.14=\alpha_\mt{nn}$ when $c_\mt{s}=0.01$, i.e., for these values of $\alpha$ the Keplerian solution matches onto the fast direction. In this sense, a solution with a nodal-type critical point can be unique provided it goes through the sonic point in the fast direction. The differences between the two types of passage weren't taken into account by \cite{m1986}.

What happens for $\alpha>0.14$, when the passage is no longer made in the fast direction, is unclear. In the interpretation of \cite{ap2003}, the analyticity of the solutions restricts the passages through the sonic point to the slow and fast directions. Therefore, a physical solution could still be determined by requiring, for $\alpha>\alpha_\mt{nn}$, the passage to be made in the direction of the eigenvector associated with the smaller eigenvalue. The problem with this type of analytic passage would then be the sharp change in slope just after the sonic point, as seen in the lower right panel of Fig.~\ref{apdiagrams}, which seems to be unphysical. However, it is not straightforward that the requirement of analyticity is enough to choose only one direction. Indeed, as argued by \cite{m1986}, there may be other curves besides the fast and the slow directions which are single-valued near the critical point [full lines in Fig.~\ref{node} (a)]. The requirement of infinite differentiability at the critical point would restrict the possibilities even more. Even though, while one can't actually ensure that the full lines of Fig.~\ref{node} (a) are analytic at the sonic point, it is also not possible to exclude every single one of them for every possible combination of parameters in the system (\ref{system}) \citep[see also discussion in][]{mkfo1984}. 

In any case, it is clear that the passage through a nodal point in the slow direction is problematic. To verify the uniqueness of the physical solutions of (\ref{system}) picked in the phase diagrams, \cite{ap2003} used two methods. In one of them they started the integration with a Keplerian model at large radii (for fixed $c_\mt{s}$, $\alpha$) and fine-tuned $l_\mt{in}$ until the solution passed through the sonic point. Alternatively, a guess for the location of the critical point is made and then adjusted until the Keplerian solution at large radii is found, providing an unique value of $l_\mt{in}$ for fixed $\alpha$ and $c_\mt{s}=0.01$. It should be noted that the integration method used is the same whether the critical point is a saddle or a node. For $\alpha<0.14$, an unique value of $l_\mt{in}$ and $R_0$ was determined for given $\alpha$ and $c_\mt{s}$ with both methods providing similar results. For larger viscosity the authors state that they were unable to find steady state solutions. Unfortunately, it is not mentioned if this was because more than one value of $l_\mt{in}$ was possible \citep[as in the analysis of][]{m1986} or no value at all was found. 

\subsubsection{Limiting viscosity}

The calculations of \cite{ap2003} show that $\alpha_\mt{nn}=\alpha^*=0.14(100c_\mt{s})^{1/3}$ is a limiting value for steady state accretion. This may be because physical solutions cannot pass a nodal point in the slow direction because of the sharp change in the slope of the curve $u_r(r)$ at the sonic point \citep[due to the instability of the Keplerian curve,][and see bottom right of Fig.~\ref{apdiagrams}]{ap2003}. Alternatively, the limiting viscosity may be due to the possible nonuniqueness of the passage through the critical point when $\alpha>\alpha_\mt{nn}$ \citep{m1986}.

Notwithstanding these mathematical details, what is relevant is what actually happens to the time-dependent accretion solutions when $\alpha$ is such that a steady state flow cannot be found. \cite{m1983} mentions that the most probable scenario is that accretion cannot proceed in a stationary fashion for $\alpha>\alpha^*$ possibly due to some ``new kind of instability''. In fact, the most likely possibility is that the instability that sets in for $\alpha>\alpha^*$ is not completely new even for the 80s. This is the topic of the following sections and a more thorough investigation of this problem, in a time-dependent framework, will be presented in the next chapter.

\section{Viscous overstability and the stability of transonic accretion}

\subsection{Introduction}
\label{vointro}

Although often regarded as ``self-inconsistencies'' the classical thermal and viscous instabilities mentioned in Part \ref{i} can be used to explain dwarf nova outburst variability and the light-curves of soft X-ray transients. However, more regular, periodic and quasi-periodic variability demand for different theoretical models where an oscillation of some sort is considered. In an attempt to connect disc instabilities to periodic phenomena, and to find possible explanations for hydrodynamic turbulence leading to viscosity, \cite{kato1978} studied the stellar pulsational instability in the context of accretion discs. 

\subsubsection{Local analysis}

Local analysis reveals that radial axisymmetric oscillations (inertial-acoustic modes) cause density changes and consequently variations in the viscous stress, which can potentially draw out rotational kinetic energy from the disc. Some of this energy is then available for the inertial-acoustic modes to grow, i.e., the pulsational instability sets in as oscillations of increasing amplitude being more appropriately designated as \emph{viscous overstability}. 

More precisely, both thermal and dynamical processes due to shear motion of the disc are important in the generation of this instability. Most (if not all) of the thermal energy in Shakura--Sunyaev discs is supplied via viscous dissipation with the efficiency of the process taken to be proportional to the local pressure. As a result, in the compressed phase of radial oscillations the viscous energy generation rate increases and oscillations are amplified. From a dynamical point of view, the instability occurs when the oscillations of the azimuthal (longitudinal) component of the viscous force are in phase with the longitudinal motion of the wave, in which case the work done on the oscillations by the viscous force is positive \citep{kato1978}.  \enlargethispage{\baselineskip}

The study of overstability of axisymmetric oscillations by \cite{blumenthal1984} generalised Kato's analysis to include different ratios of gas and radiation pressure and different opacities. Their results show that geometrically thin discs are overstable under a variety of conditions and that regions that are both secularly and thermally stable, can be pulsationally unstable. In this sense, viscous overstability is more robust than the classical instabilities and is widely agreed to be a physical instability that cannot be avoided by, e.g., changing the viscosity prescription.

\subsubsection{Global instability}

Nonetheless, and despite the attempts to attribute quasi-periodic variability of real systems to viscously overstable axisymmetric modes \citep[e.g.][]{blumenthal1984}, caution must be taken when generalising results obtained through a local analysis to a global disc. A local model doesn't account for stabilising effects such as the possibility of the escape rate for perturbations being larger than the growth rate due to viscous overstability, or damping due to phase mixing between modes \citep{khm1988,wallinder1990}. Moreover, since this instability is intimately related to travelling waves, it is unlikely to be relevant in systems with open boundaries where the waves can only grow by some finite factor before leaving the disc. 

As noted by \cite{papaloizoustanley1986}, who used a Navier-Stokes viscosity,
\be
\t{T}=\mu \left[\nabla \vc{u}+(\nabla\vc{u})^\mt{T}\right],
\ee
(in which case the viscous force in the radial direction is non-zero) in their calculations, viscous instabilities disappear for large wavenumber $k$. Since the dispersion relation for inertial-acoustic waves in a Keplerian disc [cf. equation (\ref{drf}) below with $\kappa=\Omega$] dictates that $k$ increases as the angular velocity $\Omega$ decreases, the waves dissipate as they propagate outwards, being unlikely to grow in real discs. Even if the viscous stress is such that viscous overstability survives for small wavelengths, a confined propagation region able to work as a resonant cavity is still required for the instability to be effective in exciting the waves before they escape through the system's boundaries. Alternatively, a region where inertial-acoustic oscillations slow down (e.g. near a turning point where the direction of wave speed changes) would in principle be susceptible to pulsational instability.

\begin{figure}[!t]
\begin{center}
\includegraphics[width=110mm]{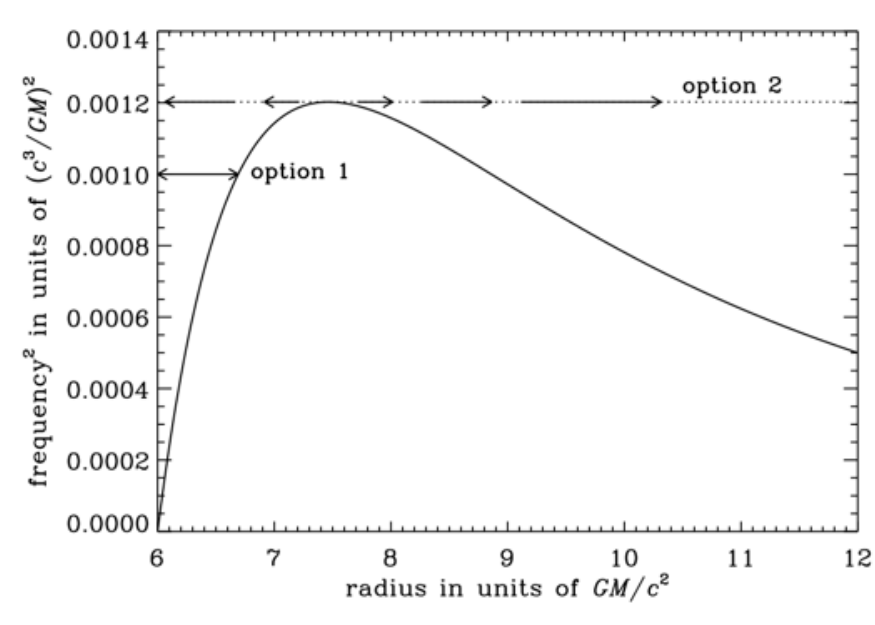}
\caption{Waves susceptible to global viscous overstability and how their frequencies are related to the square of the epicyclic frequency in a Paczy\'nski--Wiita disc (full line). The oscillation with frequency below the maximum of $\kappa$ (option 1) is likely to attain a global character since it is trapped in a small region. On the other hand, the wave with frequency just above $\max(\kappa)$ (option 2) may also be globally overstable since it slows down near the maximum of the epicyclic frequency (single-headed arrows represent the wave's group velocity).}
\label{vowaves}
\end{center}
\end{figure}

While real Keplerian discs are unlikely to reveal pulsationally unstable phenomena \citep[due to propagation characteristics and confinement limits of inertial-acoustic waves,][]{papaloizoustanley1986,kleyetal1993}, certain regions of black-hole accretion discs are potential hotspots for viscous overstability. In relativistic discs, the angular velocity and epicyclic frequency have distinct radial variation. While the former is monotonic, the latter increases inwards up to a maximum value and then decreases in the very inner region, being zero at the marginally stable orbit (see Fig.~\ref{freqintro}). The potential of this non-monotonic variation of $\kappa(r)$ and its possible relation to periodic and quasi-periodic events was realised by \cite{katofukue1980}. These authors showed that inertial-acoustic waves with frequencies smaller than the maximum of $\kappa$ can be trapped in the region between the marginally stable orbit and the epicyclic curve, or more precisely, their Lindblad resonance where the wave frequency $\omega$ equals $\kappa$. This is easily understood by analysing the dispersion relation for such modes. In a geometrically thin, inviscid and isothermal disc with sound speed $c_\mathrm{s}$, this relation has the familiar form (cf. Part \ref{os})\footnote{This dispersion relation is obtained from imposing linear perturbations with frequency $\omega$ and local wavenumber $k$ to a steady disc. The perturbed quantities are, therefore, assumed proportional to $\textrm{e}^{-\textrm{i}\omega t+\textrm{i}\int k(r)\textrm{d}r}$. Details of this process can be found in chapters \ref{timeaccretion} and \ref{oscilintro}.},

\begin{equation}
\omega^2=\kappa^2+c_\mathrm{s}^2k^2.
\label{drf}
\end{equation}
It is evident that a wave can only propagate where $\omega^2-\kappa^2>0$ (see chapter \ref{oscilintro} and Fig.~\ref{freqfig}). The mode is evanescent in other regions where $k^2<0$, with $k^2=0$ at the Lindblad resonance (turning point). Therefore, an inertial-acoustic mode can be trapped in the very inner region of a relativistic disc and be potentially excited by viscous overstability (Fig.~\ref{vowaves}, option 1). However, due to the uncertain behaviour near the marginally stable orbit where the waves might be transmitted or absorbed, it is not clear if this trapping region can be pulsationally unstable. 

An alternative is to consider a wave with a frequency just above $\max(\kappa)$, which is not trapped anywhere in the disc. However, since the group velocity of inertial-acoustic waves is proportional to $k$, there is a region close to the radius where the epicyclic frequency peaks where this oscillation slows down. In principle, viscous overstability could be effective there and such wave could acquire a global character (Fig.~\ref{vowaves}, option 2). 

Moreover, since the inner region is where most gravitational energy is released due to accretion, variations in the energy output due to overstable waves propagating there would, in principle, be observable. Indeed, these waves, which have a characteristic time-scale much shorter than the classic thermal-viscous instabilities, may be related to high-frequency quasi-periodic oscillations detected in X-ray binaries \citep{chentaam1995}.

\subsection{Sonic-point instability}

It should be noted that the first few analytical calculations of viscous overstability \citep{kato1978,blumenthal1984} ignored radial velocity and radial pressure gradients in the equilibrium state of the disc. They are therefore not directly applicable in the case of transonic black-hole accretion. The axisymmetric stability of these solutions was firstly addressed by \cite{m1986} and later on by \cite{khm1988,khm1988b}, \cite{ak1989} and \cite{chentaam1993}. In the context of the work of Kato and his collaborators on the stability of transonic flows, a new \emph{form} of viscous overstability was discovered. It was designated \emph{sonic-point instability}.

\subsubsection{Perturbations at the sonic point}

When steady accretion solutions are perturbed and the time dependence of perturbed quantities is assumed $\propto \mt{e}^{-\mt{i}\omega t}$, the ordinary differential equations describing the radial variation of perturbations are singular at the sonic point \citep{m1986}. Far from this location, the local stability analysis can be done in the usual fashion by simply assuming the dependence in $R$ to be of the form $\mt{e}^{\mt{i}\int k(R)\mt{d}R}$ and analysing the solutions of the dispersion relation. By perturbing the transonic solutions of \cite{mkfo1984} (isothermal disc, standard $\alpha P$ viscosity prescription), the following dispersion relation is obtained \citep[adapted from][]{khm1988}:
\be
\mt{i}(\omega-ku_R)\left[(\omega-ku_R)^2-c_\mt{s}^2k^2-\kappa^2\right]=-2\alpha\Omega c_\mt{s}^2k^2,
\label{vodd}
\ee
where $u_R$ is the background radial velocity. (\ref{vodd}) describes the propagation of two, outward- and inward-propagating, viscously overstable waves [the third solution is the classical viscous or secular instability of \cite{le74} which acts on a much slower timescale]. This equation shows that, \emph{far from the sonic point}, inertial-acoustic waves are locally unstable (in the sense that $\mt{Im}(\omega)>0$ for real $k$) to viscous overstability for all values of viscosity, as in Shakura--Sunyaev discs. [This conclusion is also valid, although the dispersion relation is changed slightly, when the stress tensor is of diffusion type \citep{ak1989}.] However, (\ref{vodd}) is not valid at the sonic point since certain terms neglected in deriving this formula are important close to that location.

The stability of transonic solutions at, and in the immediate vicinity of, the critical point needs to be analysed by expanding the perturbed quantities around $R_0$ \citep{m1986,khm1988}. This process reveals the presence of two different types of modes: a propagating one which is the analytical continuation of the waves obtained from local analysis far from $R_0$, and a standing wave localised at the sonic point. The presence of the standing mode can be understood from the fact that the outward propagation velocity of perturbations relative to the flow equals the radial inflow at the sonic point. That is, the two modes of perturbation are essentially the usual viscously overstable inertial-acoustic modes, although one of them is standing at the critical point. This is the mode responsible for the sonic-point instability which is revealed when \citep{khm1988}
\begin{equation}
\alpha\Omega(R_0)>\left|\frac{\mathrm{d}u_R}{\mathrm{d}R}\right|_{R_0}.
\label{crit}
\end{equation}

Numerical calculations by \cite{mkh1988} confirmed that both standing and propagating modes are revealed around $R_0$ when (\ref{crit}) is satisfied. They also mentioned that, in the non-linear stage of the instability, the amplitude of the sonic-point viscously overstable modes is modulated quasi-periodically with a period comparable to that of QPOs observed in low-mass X-ray binaries.

\subsubsection{Relation to the topology of the flow}

For the viscosity law used in \cite{khm1988}, the criterion (\ref{crit}) indicates that a flow which goes through a saddle critical point is stable while one with a nodal critical point is unstable. [Although the correspondence between the stability-instability transition and the saddle-nodal transition isn't exact, owing to a discontinuity in $\mt{d}u_R/\mt{d}R$ across the critical point, (\ref{crit}) effectively states that nodal-type flows are unstable while those with a saddle point are stable \citep{ak1989}]. Calculations in non-isothermal discs show that the criterion for the sonic-point instability is modified slightly but still corresponds to the transition between saddle and nodal-type points \citep{khm1988b}. These results seem to indicate that the sonic-point instability is related to the topology of the flow around the critical point.

In 1988, the situation seemed clear: there is a value of $\alpha^*$, which corresponds to $\alpha_\mt{sn}$, above which no unique steady solution can be found because nodal-type points are unstable to the sonic-point instability. Indeed, the analysis of \cite{khm1988} reveals that the sonic point has no essential effects on the character of the propagating mode, while the presence of a standing growing perturbation at $R_0$ is intimately related to the critical point. In this sense, it is understood that the stationary mode should be the one connected to the regular passage of the flow through the critical point, and be the main cause of the instability of transonic accretion.

Unfortunately, the work of \cite{ap2003} shows that $\alpha^*\neq\alpha_\mt{sn}$, which either means that the criterion (\ref{crit}) doesn't correspond to the saddle-nodal transition in their problem or, alternatively, that the existence of $\alpha^*$ is not related to the sonic-point instability. This problem will be explored further in the next chapter.

In addition, shortly after 1988, the very existence of the sonic-point instability in realistic discs was challenged by \cite{ak1989}, \cite{kxlz1993} and \cite{kato1994}. They found that the presence or absence of a nodal-type critical point in the accretion flow and of the sonic-point instability is intimately related to the viscosity prescription adopted, as discussed in the following section.
 




\subsection{Different types of stress tensor and viscosity laws}

The work of \cite{ak1989} hinted that the sonic point was probably stable in isothermal discs with a diffusion-type stress tensor, as the nodal-type sonic point isn't present in that case. This possibility was confirmed by \cite{kxlz1993} who analysed the stability of the sonic point for different types of the stress tensor and concluded that the critical point is stable when the diffusion-type stress tensor is adopted. On the basis of their results they speculated that, even in more general non-isothermal cases, a saddle point is always stable to local perturbations while a nodal point is always unstable.

Calculations with a modified form of viscosity for which $\nu$ vanishes at the sonic point and for $R<R_0$ \citep[to avoid viscous transport at supersonic speed and causality violation, see][]{nara1992}, are also of particular interest to the stability of the sonic point. \cite{kato1994} analysed the stability of transonic accretion solutions using a causality-limited viscosity prescription and a stress tensor proportional to $\mt{d}\Omega/\mt{d}R$ and confirmed that the sonic point, of saddle type, is stable. However, \cite{chentaam1993} found nodal-type points in their non-isothermal transonic flows with diffusion-type stress and causal viscosity but \emph{no} sonic-point instability, putting in doubt a relation between the topology and instability of the critical point in a general case.

Before performing a stability analysis, \cite{chentaam1993} numerically calculated the structure of their transonic flows, similarly to what had been done before for $\alpha P$ flows \citep[see also][]{ps1994}. A limiting value $\alpha^*$ above which such solutions can't be found is never mentioned, either because the viscosity was never high enough or because $\alpha^*$ doesn't exist when the stress has a diffusion form. If the latter option is valid, a relation between $\alpha^*$ and the sonic-point instability is likely.   \enlargethispage{\baselineskip}



While the existence of the sonic-point instability is deeply model dependent, the viscous overstability of modes satisfying the local dispersion relation (\ref{vodd}) --- or equivalent for a non-isothermal disc --- isn't. In fact, even when the stress tensor is of diffusion type and a causality-limited viscosity is employed, the disc is still locally unstable to viscous overstability \citep{ak1989,kxlz1993,kato1994}. The question is whether or not this instability can attain a global character as discussed in the following chapter \citep[see also][]{wallinder1990}.



In the next section, I'll consider the topic of \emph{global} viscous overstability which is most easily probed within a numerical framework. I will also comment on the influence of radial viscosity on the stability of the disc.

\subsection{Time-dependent numerical calculations}

As mentioned in section \ref{vointro}, care must be taken when generalising the results of local analysis to real discs. Numerical simulations are required to test if inertial-acoustic oscillations can be pulsationally unstable beyond local analysis. Here, the focus goes to numerical calculations concerning axisymmetric instabilities of black-hole accretion discs.

\subsubsection{Global oscillations}

The first paper on this matter is perhaps that of \cite{mkh1988} which revealed the presence of viscously overstable modes in the inner region of transonic isothermal discs for $\alpha>0.05-0.1$ (using an $\alpha P$ stress). The authors claim this result to be a confirmation of the existence of the sonic-point instability beyond local analysis. However, the initial conditions may have influenced this outcome: since their work was geared towards finding instabilities at the sonic point, the disc was initiated \emph{with a small perturbation at $R_0$}. In any case, the basic result seems to be that overstable oscillations are visible in the time-dependent simulations when $\alpha$ is large enough.

In rough agreement with the previous paper, the global non-linear investigation of \cite{chentaam1995} revealed nonsteady behaviour of the inner disc for high enough $\alpha$ and low enough mass accretion rate. To avoid classic instabilities, the authors used a viscously and thermally stable viscosity prescription with the kinematic viscosity given by

\begin{equation}
\nu=\frac{2}{3}\alpha\beta_\mt{gt} \frac{c_\mathrm{s}^2}{\Omega_\mt{PW}},
\label{vp} 
\end{equation}
where $\beta_\mt{gt}$ is the ratio of gas to total pressure. According to their equation (3) (but not their abstract) the stress tensor used is proportional to the gradient of angular velocity. The disc is initiated with the transonic solution of \cite{chentaam1993}. Using a piecewise parabolic method, they solved the 1D hydrodynamic time-dependent equations to find variability at a range of timescales for $\alpha\gtrsim\alpha_\mathrm{vo}\approx0.2$ and $\dot{M}\lesssim0.3\dot{M}_\mathrm{Edd}$. Since the local analysis predicts instability for all values of viscosity, independently of the accretion rate \citep[e.g.][]{blumenthal1984,chentaam1993}, \cite{chentaam1995} attribute the stability of the disc at small $\alpha$ and high $\dot{M}$ to global effects. Namely, they claim that in those conditions the growth rate of perturbations cannot overcome the escape rate.

A particularly interesting result of \cite{chentaam1995} is the oscillation with global characteristics found at a frequency close to the maximum of the epicyclic frequency ($115$ and $135$ Hz) in two of their test runs. A similar oscillation with frequency $\sim\mt{max}(\kappa)$ was also found by \cite{hmk1992} \citep[although they initiated their calculations with a small perturbation at $R_0$ as in][]{mkh1988}.

\cite{mt1996} extended the previous study to a larger region of the parameter space $(\dot{M},\alpha)$, and used two different viscosity prescriptions: $\nu$ constant or given by (\ref{vp}). Their results indicate that the feature at $\mt{max}(\kappa)$ is expected for $0.01\dot{M}_\mathrm{Edd}\lesssim\dot{M}\lesssim0.25\dot{M}_\mathrm{Edd}$, $0.2\lesssim\alpha\lesssim1$ and that the amplitude of the oscillations increases for large $\alpha$ and small $\dot{M}$. The authors point out that the global mode is produced at $\sim7.5 R_\mathrm{g}$, i.e., the radius where $\kappa$ is maximum in the Paczy\'nski--Wiita potential used in their calculations, and affects the disc for $R\lesssim32R_\mathrm{g}$. The discs are stable for higher accretion rates because, according to the authors, ``the sound speeds are high and perturbations quickly diffuse away before they have time to grow''. When $\dot{M}$ is reduced substantially for fixed $\alpha$, the global feature disappears, the disc exhibits local inertial-acoustic oscillations at all radii at frequency equal to $\kappa(r)$ and the spectra is rather noisy. Finally, the paper concludes that the stability is dependent only on the magnitude of the viscosity and not on its functional form. Moreover, they predict oscillations to be present in a finite region of the parameter space in any disc where the dynamic viscosity increases when the density $\rho$ increases.

More recently, \cite{oneill2009} solved hydrodynamic equations in cylindrical coordinates with $\alpha$-viscosity ($\nu\sim\alpha c_\mathrm{s}^2/\Omega$) using a 2.5 dimensional model, i.e., which enforces azimuthal symmetry but allows the azimuthal velocity to be non-zero. Their simulations are representative of a pseudo-Newtonian accretion disc with $p\propto\rho^{5/3}$. They find axisymmetric waves generated near the maximum of the epicyclic frequency that can propagate outwards up to $R\sim18R_\mathrm{g}$ for $\alpha\ge\alpha_\mathrm{vo}=0.05$. These oscillations have frequencies at or just below $\mt{max}(\kappa)$ and can either be inertial or inertial-acoustic waves or ``continuous extensions in frequency and power'' of the previous in the cases where they propagate to forbidden regions where, according to the dispersion relation, the modes should be evanescent \citep[see, e.g.,][]{rkato2001}. As the authors point out, the inertial (\emph{r} or \emph{g}) and inertial-acoustic (\emph{f} or inner \emph{p}) modes are apparently indistinguishable in their simulations due to the possibility of exchange of power between different velocity components.

Reanalysing the results table of \cite{mt1996}, it is easy to see that an oscillation (although very weak) is detected in a simulation with $\alpha$ as low as $0.05$ in agreement with \cite{oneill2009}, as also pointed out in the latter paper. It should be noted, however, that different calculations, with distinct viscosity prescriptions and equations of state will naturally result in slightly different limiting values of $\alpha$ above viscously overstable waves are detected.

\subsubsection{Stress tensors}

Although the results presented indicate that inertial-acoustic waves with a frequency $\sim\mt{max}(\kappa)$ can be viscously overstable even when global effect are considered, it is important to note that, in all cases mentioned, the stress tensor was assumed to have a rather simple form. In the works of \cite{mkh1988} and \cite{hmk1992}, the only non-zero stress component is $T_{R\phi}$ which is assumed to have the standard $\alpha P$ form. In \cite{chentaam1995} and \cite{mt1996}, the stress tensor is of diffusion-type and, similarly, $T_{R\phi}$ is the only non-zero component. The radial component of the viscous force wasn't included in any of the simulations mentioned.

Previous studies focusing on Keplerian discs show that the radial viscous force may have a stabilising effect on the global outcome of pulsational instability. The paper by \cite{chentaam1992} on radial oscillations in discs around neutron stars where the authors compare the results of \cite{papaloizoustanley1986} and \cite{okudamineshige1991} is particularly interesting. While the latter work, which did not include the effect of viscous forces in the radial direction, claims that Keplerian discs with $\alpha$-viscosity generally exhibit overstable radial oscillations, the former shows that propagation characteristics of such modes prevent viscous overstability. Indeed, \cite{chentaam1992} found overstable oscillations to depend sensitively on the magnitude of both azimuthal and radial viscosities, with the modes being damped when the two are comparable \citep[in accordance with the results of][]{papaloizoustanley1986}. Despite the prediction from linear analysis of instability even in presence of viscosity in the radial direction \citep{blumenthal1984}, the simulations show that non-linear effects can have a stabilising outcome. This is an indication that viscous overstability may not be able to operate in a real Keplerian disc.

The results from black-hole accretion disc simulations are more encouraging. The 2D $(R,z)$ time-dependent equations solved by \cite{mt1997} ignore radiative and bulk viscosity but include \emph{all} components of the stress tensor defined by,
\be
T_{ik}=2\rho\nu\left(D_{ik}-\frac{\delta_{ik}}{3}D_{jj}\right) \quad \mt{where} \quad D_{ik}=\frac{1}{2}\left(\frac{\partial u_i}{\partial x_k}+\frac{\partial u_k}{\partial x_i}\right).
\ee
Their results confirm the instability of discs to global modes with frequency $\mt{max}(\kappa)$ for low accretion rate and high enough viscosity (although the threshold value of $\alpha$ is significantly higher), strengthening the possibility of inertial-acoustic oscillations introducing variability in real discs. \newline

The numerical results collated in this section show that viscously overstable modes can definitely occur in discs around black holes when an $\alpha$ viscosity is used. The oscillations are excited when the mass accretion rate is low and the viscosity is high and are axisymmetric and most likely of inertial acoustic type. The simulations of \cite{oneill2009} show that it might also be possible for inertial modes (which have a non-zero vertical component of the velocity) to be viscously overstable, as previously shown analytically by \cite{orw2000}. It is, however, unlikely for these modes to be excited by viscous overstability as the action of shear on their vertical structure would act to damp them. The result of \cite{orw2000} is indeed believed to be incorrect as it is in direct contradiction with the findings of \cite{latterogilvie2006}. These authors made a full treatment of a 3D shearing-sheet disc model and analysed the possibility of different modes growing due to viscous overstability. Their numerical results show that only inertial-acoustic modes have the potential to grow.  

\section{Summary and open questions}

The body of literature on the structure and stability of transonic accretion flows is overwhelming and, understandably, a reader of this dissertation will probably have skipped some (if not all) parts of my extensive review on the subject. Fortunately, a bullet-point summary is presented below: \enlargethispage{\baselineskip}

\begin{itemize}
\item{\textbullet $\;$}  Flows around black holes are transonic being subsonic at large radii and supersonic at small radii. In a thin disc, the sonic point $R_0$ (which is a critical point of the hydrodynamic equations) is located close to the radius of the marginally stable orbit.

\item{\textbullet $\;$}  In the case of isothermal discs with standard $\alpha P$ viscosity, there is a limiting value of $\alpha$, $\alpha^*\propto (c_\mt{s}/c)^{1/3}$, above which no unique, stable, physically acceptable steady state transonic accretion solutions can be found. The existence of $\alpha^*$ seems to be related to the unphysicalness or nonuniqueness of the regular passage of the flow in the slow direction of a nodal-type point. It is unclear if $\alpha^*$ exists in discs with a diffusion-type stress tensor as studies so far have not mentioned such a limit.

\item{\textbullet $\;$}  Local analysis around the sonic point of isothermal discs reveals that saddle-type points are stable while nodes are unstable. More precisely, this \emph{sonic-point instability} kicks in when $\alpha_\mt{sp}\Omega(R_0)=\left|\mathrm{d}u_R/\mathrm{d}R\right|_{R_0}$, if the disc is isothermal and the stress is $\alpha P$. This form of viscous overstability is weak in the sense that it doesn't exist in isothermal discs with diffusion-type stress tensors (which have only saddle critical points). Furthermore, it is unlikely to exist when a causality-limited viscosity, null at the sonic point, is used. When it exists according to local analysis, it is also revealed in time-dependent numerical simulations although possibly only when the disc is initialised with a perturbation at the sonic point.

\item{\textbullet $\;$}  In general, the instability of transonic solutions seems to be related, not to this sonic-point instability, but to propagating overstable modes. According to local analysis these modes may be present under a wide variety of conditions. When global effects are considered, inertial-acoustic waves are only excited when the viscosity is sufficiently high ($\alpha>\alpha_\mt{vo}$) and the sound speed or mass accretion rate is sufficiently low. In other words, $\alpha_\mt{vo}$ is dependent on $c_\mt{s}$, being lower for lower sound speed. (Presumably this is because perturbations propagate away faster in thick discs and are unable to grow before escaping through the system's boundaries.) Overstable modes are present in simulations modelling discs with different equations of state and stress tensors, i.e., $\alpha_\mt{vo}$ exists even in the cases where $\alpha_\mt{sp}$ and $\alpha^*$ probably don't. 

\item{\textbullet $\;$}  In $\alpha P$, isothermal discs, $\alpha^*=\alpha_\mt{nn}$ while $\alpha_\mt{sp}=\alpha_\mt{sn}$ \citep[but see section 6.2 of][for a different view on this last correspondence]{ap2003}. To my knowledge, the global behaviour of viscous overstability and its limiting value $\alpha_\mt{vo}$ have never been analysed in an $\alpha P$ disc with the flow being initialised \emph{without} a perturbation at $R_0$.

\end{itemize}

Unfortunately, the summary is rather inconclusive and many questions remain unanswered. What happens to the solutions of the time-dependent equations describing a transonic flow when $\alpha>\alpha^*$? If nodal-type points are unstable, how did \cite{ap2003} manage to find steady state solutions for $\alpha>\alpha_\mt{sn}$? Can the sonic-point instability be revealed in time-dependent $\alpha P$ calculations if a perturbation isn't imposed at the sonic point to start with? How does viscous overstability develop in that case? What is the relation $\alpha_\mt{vo}(c_\mt{s})$ in an isothermal $\alpha P$ disc? In the case where $\alpha^*$ exists, is there any relation between it and the onset of global viscous overstability? If so, why?

One of the reasons why some of these questions remain unanswered relates to the difficulty in comparing the different works on these subjects. More precisely, the most recent paper where a value of $\alpha^*$ is found \citep{ap2003} applies a rather unusual form of the viscosity prescription (although the stress tensor is essentially $\alpha P$) and considers an isothermal, thin disc. On the other hand, most of the work on global viscous overstability models non-isothermal discs with a diffusion-type stress tensor.

In conclusion, a direct comparison between both problems where the same assumptions and approximations are used in the steady and time-dependent hydrodynamic equations is lacking. This is the topic of the next chapter where I solve the time-dependent version of the \cite{ap2003} problem for various values of the viscosity parameter $\alpha$. In my calculations, I won't impose any perturbation at the sonic point to initialise the disc so that I can analyse the behaviour of global viscous overstability in this case. I'll also look for signs of the sonic-point instability in the case where $\alpha>\alpha_\mt{sp}$. Most importantly, I'll see what happens to the solution of the time-dependent equations when $\alpha>\alpha^*$ and I'll be able to tell if any relation between $\alpha^*$ and $\alpha_\mt{vo}$ exists. My attempt at answering the questions presented above is described in the next chapter.

\thispagestyle{empty}
\chapter{Time-dependent transonic accretion} 
\label{timeaccretion}

The aim of this chapter is to revisit the problem described in the introductory section of this part of the thesis and attempt to answer the questions posed at the end of the previous chapter. The main goal is to look for a relation between the unsteadiness of transonic flows and the presence of instabilities in the disc: either viscously overstable propagating modes or the sonic-point instability (see chapter \ref{steadyintro} and references therein). 

As mentioned before, to facilitate the comparison between the two phenomena I solve the time-dependent equations of the \cite{ap2003} steady state model. These authors were unable to find unique solutions obeying both realistic boundary conditions and regularity constraints at the sonic point for $\alpha>\alpha^*=0.14(100c_\mt{s}/c)^{1/3}$. The model considered in the present chapter represents an approximately isothermal, axisymmetric, 1D hydrodynamic flow with essentially a standard $\alpha P$ viscosity. The only input parameters of the problem are the sound speed, $c_\mathrm{s}$, and the magnitude of the viscosity, $\alpha$. 

As it will be seen in the remainder of the chapter, numerical simulations show that physically acceptable steady state thin disc solutions exist for $\alpha \lesssim0.14 (100 c_\mathrm{s}/c)^{1/3}=\alpha^*$, therefore confirming the results of the above-mentioned paper. Moreover, it is found that for higher viscosity (up to a certain limit) the disc settles into an oscillatory state where global oscillations with frequencies close to the maximum of the epicyclic frequency are present. The waves excited at high viscosity have all characteristics of viscously overstable inertial-acoustic waves. These results seem to indicate that, at least in the simple model considered, $\alpha^*=\alpha_\mt{vo}$. 

This chapter is organised as follows. In section \ref{enm} I present the equations to be solved and the numerical method used in the process. The results obtained are described and interpreted in section \ref{transresults} and discussed in \ref{transdiscussion}. Conclusions and possible observational relevance of the results obtained are presented in section \ref{transconclusions}.

\section{Equations and numerical method}
\label{enm}

\subsection{Assumptions}

In this study of the stability of accretion flows surrounding black holes, the approximations made and assumptions used are basically the same as those of \cite{ap2003}. The vertical thickness of the flow is assumed to be much less than $R$ so that vertically-averaged quantities can be used in the hydrodynamic equations. Furthermore, relativistic effects are mimicked by the Paczy\'nski--Wiita potential and the flow is assumed axisymmetric, reducing the problem to one spatial dimension. Particles in this gravitational field move around the black hole in circular orbits characterised by a pseudo-Newtonian angular velocity $\Omega_\mt{PW}$ and epicyclic frequency $\kappa_\mt{PW}$. The square of these frequencies is given by [cf. equations (\ref{omegapw},\ref{kappapw})],
\begin{equation}
\Omega_\mathrm{PW}^2=\frac{1}{r(r-2)^2},
\label{kepler}
\end{equation}
\begin{equation}
\kappa_\mt{PW}^2=\frac{(r-6)}{r(r-2)^3},
\label{kappa}
\end{equation}
where $r=R/R_\mt{g}$ and the frequencies are in units of $c^3/GM$ (typical velocities are given in units of $c$). Because the flow has a finite thickness, $\Omega\neq\Omega_\mathrm{PW}$ in general. However, since the disc is assumed thin, these particle orbit expressions should approximately represent the characteristic frequencies for $r>r_\mt{ms}$. 

The epicyclic frequency is maximum at $r=2(2+\sqrt{3})\approx7.5$ where it reaches a value of $112 (10 M_\odot/M)$ Hz. At the marginally stable orbit $r_\mathrm{ms}=6$, $\kappa_\mt{PW}=0$ and the particle-orbit angular momentum (per unit mass),
\begin{equation}
l_\textrm{PW}=\frac{r^{3/2}}{r-2},
\end{equation}
and binding energy,
\begin{equation}
e_\textrm{PW}=\Phi+\frac{r^2\Omega_\textrm{PW}^2}{2}=-\frac{r-4}{2(r-2)^2},
\end{equation}
reach a minimum. For radii smaller than $r_\textrm{ms}$ matter can no longer be kept in stable orbits around the black hole and spirals inwards eventually reaching the event horizon. As discussed in chapter 2, the inner boundary of the disc is expected to be close to this location. In the numerical problem described here, the innermost radius is taken to be located at $r=3<r_\mathrm{ms}$ to better study the transition from the subsonic disc to the supersonic region. (Readers are reminded that the critical sonic point of the flow may be located slightly inside the marginally stable orbit.) As in \cite{ap2003} it is assumed, although only to initialise the flow, that as the matter spirals inwards from $r_\mathrm{ms}$ it approximately conserves its angular momentum and binding energy, i.e.,
\begin{equation}
l\approx l_\mathrm{ms}=3\sqrt{1.5},\quad e\approx e_\mathrm{ms}=-1/16,\quad\textrm{for $r\lesssim r_\mathrm{ms}$.}
\label{ini}
\end{equation}

Similarly to \cite{ap2003}, the viscous stress is assumed to be given by equation (\ref{torquefinal1}).
Despite the breakdown of assumption (\ref{torqueapprox}) outside the marginally stable orbit, it is taken to be valid throughout the disc as this greatly simplifies the structure of the equations describing the accretion flow. It is also the assumption made by \cite{ap2003} so it should be regarded as valid to better compare the results obtained here with those of the paper. The important aspect to retain about the form of the stress tensor is that it is \emph{essentially an $\alpha P$ stress} with $\alpha\approx\alpha_\mt{SS}/2$, where $\alpha_\mt{SS}$ is the usual Shakura--Sunyaev $\alpha$ parameter.

As in \cite{ap2003}, the vertically-integrated pressure $P$ is assumed to be equal to $c_\mathrm{s}^2\Sigma$ but here the disc is taken to be only approximately isothermal in the radial direction. If the disc was regarded as strictly isothermal, the thickness $H$ would become too large in the outer disc where $\Omega_\mathrm{PW}$ is small. To avoid the numerical complications associated with this problem I take

\begin{equation}
c_\mathrm{s}(r)=c_\mathrm{s,AP}\sqrt{\frac{106}{r+100}},
\label{csap}
\end{equation}
which decays with radius but is slowly varying in the inner region. This function is practically equal to the constant Afshordi--Paczy\'nski sound speed $c_\mathrm{s,AP}$ near the marginally stable orbit $r=6$ and gives constant $H/r$ at large $r$. The thin-disc approximation is assured if $c_\mathrm{s,AP}$, an input parameter, is small.

\subsection{Hydrodynamic equations}

Within the previous assumptions, the partial differential equations describing the conservation of mass, momentum and angular momentum can be written in the form \citep{mkfo1984}

\begin{equation}
\frac{\partial \Sigma}{\partial t}+\frac{1}{r}\frac{\partial}{\partial r}(r\Sigma u_r)=0,
\label{eqap1}
\end{equation}
\begin{equation}
\frac{\partial u_r}{\partial t}+u_r\frac{\partial u_r}{\partial r}+\frac{c_\mathrm{s}^2}{\Sigma}\frac{\partial \Sigma}{\partial r}=\frac{u_\phi^2}{r}-\frac{\mathrm{d}\Phi}{\mathrm{d} r}-\frac{\mathrm{d}c_\mathrm{s}^2}{\mathrm{d} r}-c_\mathrm{s}^2\frac{\mathrm{d}\ln\Omega_\mathrm{PW}}{\mathrm{d}r},
\end{equation}
\begin{equation}
\frac{\partial u_\phi}{\partial t}+u_r\frac{\partial u_\phi}{\partial r}=-\frac{u_ru_\phi}{r}+\frac{1}{2\pi r^2\Sigma}\frac{\partial}{\partial r}\left(2\pi r^2 \tau_{r\phi}\right),
\label{eqap2}
\end{equation}
where $u_r<0$ is the radial drift while $u_\phi=l/r=\Omega r$ is the azimuthal velocity. Note that if $\Sigma$ is multiplied by an arbitrary constant the remaining quantities are unchanged. Apart from the $\mt{d}c_\mt{s}^2/\mt{d}r$ term, these equations are the time-dependent version of the ones solved by \cite{ap2003}.

Defining $\zeta=\ln\Sigma$, (\ref{eqap1})--(\ref{eqap2}) can be written in the non-conservative form:

\begin{equation}
\partial_t\vc{U}+\mathbf{A}\partial_r\vc{U}=\vc{B},
\label{pde}
\end{equation}
where $\vc{U}(r,t)$ is the column vector whose components are the vertical-integrated quantities describing the flow $(\zeta,u_r,u_\phi)$. The matrix $\mathbf{A}(r,t)$ is given by
\begin{equation}
\mathbf{A}=
\begin{bmatrix}
u_r & 1 & 0 \\
c_\mathrm{s}^2 & u_r & 0 \\
\frac{2\alpha c_\mathrm{s}^2}{r\Omega_\mathrm{PW}} u_\phi & 0 & u_r+\frac{2\alpha c_\mathrm{s}^2}{r\Omega_\mathrm{PW}}
\end{bmatrix},
\end{equation}
and the ``source-terms'' vector $\vc{B}(r,t)$ can be written as

\begin{equation}
\vc{B}=
\begin{bmatrix}
-\frac{u_r}{r} \\
\frac{u_\phi^2}{r}-\frac{\mathrm{d}\Phi}{\mathrm{d} r}-c_\mathrm{s}^2\frac{\mathrm{d}\ln\left(\Omega_\mathrm{PW}c_\mathrm{s}^2\right)}{\mathrm{d}r} \\
-\frac{u_r u_\phi}{r}-\frac{u_\phi}{r^2}\frac{\mathrm{d}}{\mathrm{d}r}\left(\frac{2\alpha c_\mathrm{s}^2r^2}{r\Omega_\mathrm{PW}}\right)
\end{bmatrix}.
\end{equation}
Here $\Phi=-1/(r-2)$ is the \cite{pw1980} potential and $\Omega_\mathrm{PW}$ and $c_\mathrm{s}$ are given by equations (\ref{kepler}) and (\ref{csap}), respectively while $\alpha$ and $c_\mathrm{s,AP}$ are input parameters of the problem.

\subsection{Numerical method}

The system (\ref{pde}) of quasilinear partial differential equations in $r$ and $t$ is hyperbolic since the square matrix $\mathbf{A}$ is diagonalizable and has real eigenvalues. To solve for $\vc{U}$, I use a numerical code originally written by Dr. Gordon Ogilvie and modified by me that employs a wave-speed splitting method with second-order upwind differencing. The scheme is explicit and uses a first-order forward-time differencing with a variable time step.

\subsubsection{Upwind wave-speed splitting}

In a one-dimensional system with waves propagating to the right/left, a point in the space-time diagram is only influenced by the points to its left/right. In other words, if the information in a flow field is propagating to the right/left, the physical domain of dependence lies to the left/right; this is the so-called upwind direction. Numerical methods that correctly model this physical behaviour are designated upwind schemes and tend to enhance stability, at least for simple forward- or backward-time stepping methods, and are generally stable if the CFL condition is satisfied \citep{gasdynamics}. Hyperbolic partial differential equations are discretised using more upwind than downwind points, i.e., the differencing is biased in the direction opposing wave propagation. For example, if the wave speed is positive, the information propagates to the right and therefore a backward finite difference should be used to represent spatial derivatives. Upwind schemes which use second-order finite differences are less diffusive and have better spatial accuracy than first-order methods.

The concept of upwind direction is complicated in subsonic flows with waves propagating both to the left and to the right. In problems of this nature, upwinding is achieved by using flux or wave-speed splitting techniques. The former uses the governing equations in conservative form while the latter uses a non-conservative form. Flux splitting is preferable in most cases as it yields conservative numerical equations but for the Navier-Stokes equations the two methods are essentially equivalent \citep{gasdynamics}. The code uses the wave-speed splitting scheme since (\ref{pde}) is in non-conservative form.

The basic idea of the method is simple. The system under consideration has three equations and therefore three families of waves [cf. dispersion relation (\ref{vodd})] with different characteristic speeds given by the eigenvalues of the matrix $\mathbf{A}$. The numerical code uses the \textsc{dgeev} routine of \textsc{lapack} to calculate the eigenvalues and the left- and right-eigenvectors at each radius and time step. If the wave speed is positive/negative, the upwind direction lies on the left/right and a backward-/forward-biased finite difference formula is used to calculate spatial derivatives. The time step is variable and depends on the wave speeds in order to ensure that the CFL condition is satisfied.

Since this study focuses on the inner region of the disc, where the transition from subsonic to supersonic occurs, a logarithmic spatial grid is used. Within this non-uniform grid, a non-standard finite-difference formula is required to represent derivatives. For example, when the eigenvalue is positive, the derivative $(\mathrm{d}y/\mathrm{d}r)_i$ is approximated by a backward finite difference formula of the form:

\begin{equation}
\frac{(r_i-r_{i-2})^2(y_i-y_{i-1})-(r_i-r_{i-1})^2(y_i-y_{i-2})}{(r_i-r_{i-1})(r_i-r_{i-2})(r_{i-1}-r_{i-2})},
\end{equation}
which reduces to the standard formula if the grid is uniform. 1601 grid points logarithmically distributed between the inner boundary at $r=3$ and the outer boundary placed at $r=800$ are used in the calculations.

\subsubsection{Initial and boundary conditions}

In order to set up an approximate analytical initial profile for the disc, the following simple assumptions are made. In the initial state, the mass conservation is expressed in the form $\zeta=\ln(-1/ru_r)$. The disc ($r\ge r_\mathrm{ms}$) is assumed to be Keplerian, $u_\phi=r\Omega_\mathrm{PW}$, with the angular momentum conservation expressed as $u_r=-\alpha c_\mathrm{s}^2/u_\phi$ (this is based on the solution for steady Keplerian discs). In the plunging region ($r<r_\mathrm{ms}$), the angular momentum is taken to be approximately constant so that $u_\phi=l_\mathrm{ms}/r$ while 
\be
u_r=-\max{\left[\alpha c_\mathrm{s}^2/u_\phi,\sqrt{2(e_\mathrm{ms}-\Phi)-u_\phi^2}\right]}.
\ee
This formula is based on the solution for a particle spiralling inwards from the marginally stable orbit. The square-root term comes from the definition of Keplerian binding energy and $l_\mathrm{ms}$ and $e_\mathrm{ms}$ are given by (\ref{ini}). The final state of the disc is \emph{not} expected to depend on these initial conditions since the flow is evolved for $t\sim5\times10^4t_\mathrm{dyn}$, where $t_\mathrm{dyn}$ is the orbital timescale at the marginally stable orbit. Characteristic modulations in the disc would be captured on a much shorter timescale, but such a long integration time is necessary since the time required for the flow to reach steady-state equilibrium (expected for low $\alpha$) is much larger than $t_\mathrm{dyn}$.

Since a second-order finite-difference expression for the spatial derivatives is used, it is necessary to specify the value of all three quantities at the two innermost and two outermost grid points. At the inner boundary, the fluid quantities are specified by assuming that the second derivative vanishes at the two innermost grid points. Because the sonic point is located near $r=6$, the inner boundary of the problem is placed in the supersonic region and therefore these conditions don't affect the structure of the subsonic disc or of any part of the domain away from the boundary. At the two outermost points, $u_\phi$ is taken to be Keplerian while the vertically-integrated density is assumed constant and $\mathrm{d}^2u_r/\mathrm{d}r^2=0$. Once again, the structure of the inner disc is not particularly sensitive to these conditions since a wave-damping region is placed near the outer boundary. More specifically, for $r>100$, a friction term is added to the radial equation which damps the waves at a rate $\gamma(r)$ given by

\begin{equation}
\gamma(r)=\frac{\Omega_\mathrm{PW,o}}{2}\left[1+\tanh\left(\frac{r-500}{100}\right)\right],
\end{equation}
where $\Omega_\mathrm{PW,o}\approx0.0001$ is the angular velocity in the wave-damping region. \\

The initial and boundary conditions presented here ensure that the quasilinear hyperbolic initial-value problem to be solved is well-posed. Also, as mentioned in the previous section, the CFL condition is satisfied in the numerical scheme used ensuring the stability of the method. These two conditions are necessary for the convergence of the method.

\section{Results}
\label{transresults}

The system (\ref{pde}) was solved for four different values of $c_\mathrm{s,AP}$, $0.003$, $0.005$, $0.01$ and $0.02$ and a range of $\alpha$. Note that the vertical averaging done to obtain the equations solved in the code is appropriate only when the disc is thin, hence the choices of small sound speed values. In the analysis of the results I only considered the last half of the integration time to assure the system had reached its final state.  Moreover, and because I am interested in studying the inner region of the disc, I focused only on the inner half of the spatial grid (801 points) which extends up to about $r=50$ and where the resolution is considerably higher than in the outer half. The contour plots shown in this section are zoomed in to a time interval of about $5\times 10^3$ in units of $GM/c^3$. This interval is chosen in order to capture variations with frequencies larger than about $25 (10M_\odot/M)$ Hz (but note that time averages are calculated using the entire last half of the integration time). 

For all four values of sound speed used in the problem the system reaches a steady state if the viscosity parameter is sufficiently small, in accordance with \cite{ap2003}. As $\alpha$ is increased, accretion stops being steady and oscillations with a frequency close to the maximum of the epicyclic frequency are present in the disc. This phenomenon is global in the sense that the waves approximately conserve their frequency for a wide range of radii. If the viscosity is increased further, the system reaches a fairly chaotic state where no clear global oscillations are visible. In summary, the final state of the system is strongly dependent on the viscosity:

\begin{itemize}
\item {$\vc{\alpha<\alpha_1}\quad$} steady;
\vspace{5pt}
\item {$\vc{\alpha_1<\alpha<\alpha_2}\quad$} oscillatory;
\vspace{5pt}
\item {$\vc{\alpha>\alpha_2}\quad$} chaotic.
\end{itemize}

\begin{figure}[t!]
\begin{center}
\includegraphics[width=0.49\linewidth]{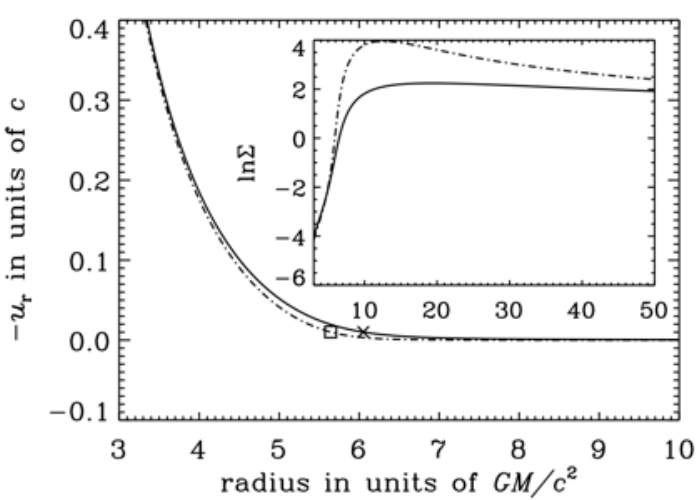}
\includegraphics[width=0.49\linewidth]{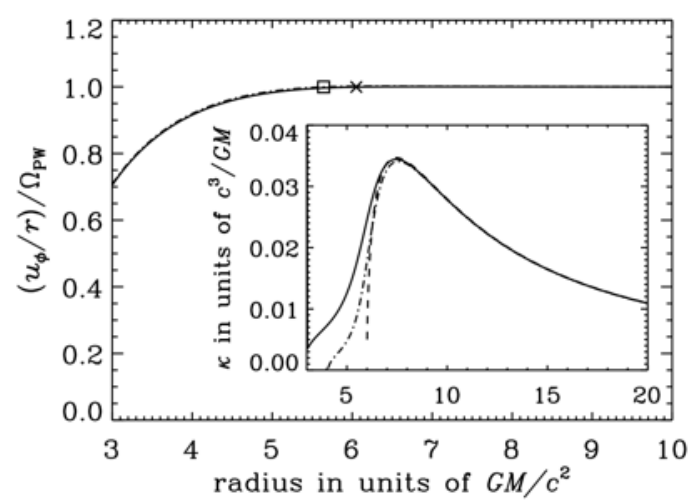}
\caption{Example of the variation of disc characteristics with radii in the case where the system reaches a steady state and the transonic solution is stable. The full lines represent a solution with $c_\mt{s}=0.01$ and $\alpha=0.1$ while the dash-dotted lines represent a flow with the same value of sound speed and $\alpha=0.01$. The square indicates the location of the sonic point, $r_0$, in the lower $\alpha$ case $(r_0=5.64)$ while the cross marks the critical point when $\alpha=0.1$ $(r_0=6.05)$. The dashed line corresponds to the particle-orbit expression for the epicyclic frequency $\kappa$; the fluid's epicyclic frequency was determined from the numerical $\Omega$ using the usual Newtonian relation.}
\label{stable}
\end{center}
\end{figure}

When $\alpha<\alpha_1$ the transonic solutions are stable and the sonic point is well defined since the radial inflow is independent of time. In addition, the numerical  profile obtained for the radial inflow is in agreement with that expected from the time-independent calculations of \cite{ap2003}, serving as a test of the numerical scheme used when $\alpha<\alpha_1$. Fig.~\ref{stable} shows how the flow parameters vary with radii in this stable configuration for $c_\mt{s}=0.01$ and two different values of $\alpha$. As it can be seen, the subsonic disc remains thin in both cases as $\Omega/\Omega_\mt{PW}\approx1$ everywhere. The disc structure is not changed significantly when the sound speed is changed between the values used in the simulations.

The differences in viscosity only result in minor changes in the disc structure: small changes are visible in the epicyclic frequency in the supersonic region and the density is higher in the inner disc when the viscosity is higher. Most importantly, the location of the sonic point changes, being located at $r_0<r_\mt{ms}$ for small $\alpha$ and at $r_0>r_\mt{ms}$ for large viscosity. This is expected from previous steady state studies in isothermal discs with $\alpha P$ viscosity. Although I don't directly investigate the topology of the flow around the sonic point, the results of \cite{ap2003} indicate that the low $\alpha$ flow has a point of saddle type while the high $\alpha$ has a nodal-type critical point.
 
Interestingly, the simulations show that $\alpha_1\approx\alpha^*=0.14(100 c_\mathrm{s})^{1/3}$ as seen in Table \ref{alpha1}. This is a numerical confirmation of the result of \cite{ap2003} that physically acceptable steady state solutions only exist for $\alpha<\alpha^*$. In the next section I focus on the cases where the input parameters are such that the final state is oscillatory to understand what happens when the viscosity is increased above $\alpha^*$.

\begin{table}[b!]
\begin{center}
\begin{tabular}{ccccc}
\hline
$c_\mathrm{s,AP}$ & & $\alpha_1$ & & $\alpha^*=0.14(100 c_\mathrm{s})^{1/3}$ \\
\hline
$0.003$ & & $\sim0.095$ & & $0.094$ \\
$0.005$ & & $\sim0.110$ & & $0.111$ \\
$0.01$ & & $\sim0.140$ & & $0.140$ \\
$0.02$ & & $\sim0.175$ & & $0.176$ \\
\hline
\end{tabular} 
\caption{Viscosity parameter $\alpha_1$ above which oscillations are visible in the simulations vs. the \cite{ap2003} $\alpha^*$ above which the transonic solution is unsteady.}
\label{alpha1}
\end{center}
\end{table} 

\begin{figure}[!t]
\begin{center}
\includegraphics[width=134mm]{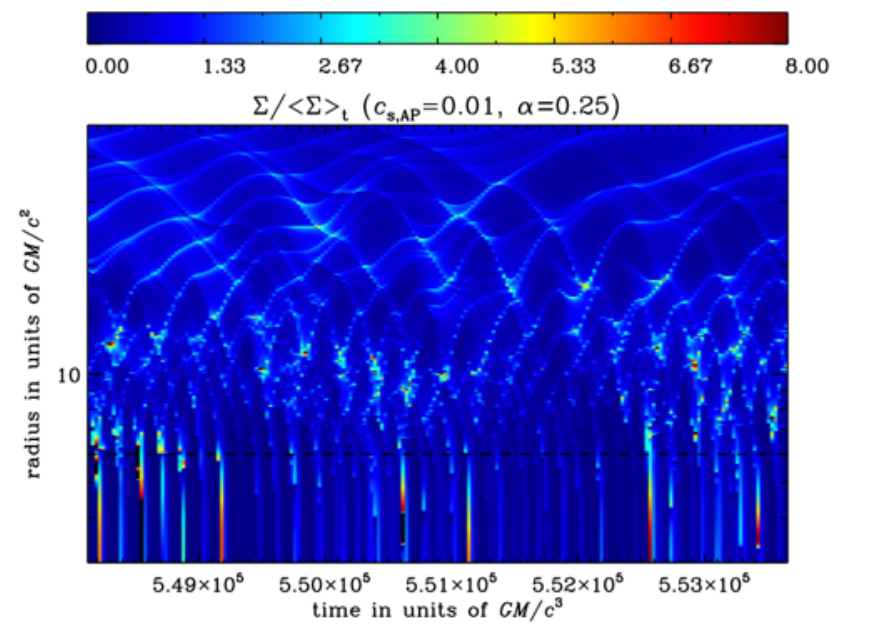}
\caption{Surface density divided by its time-averaged value for $c_\mt{s,AP}=0.01$ and $\alpha=0.25>\alpha_2$. The dashed line indicates the location of the marginally stable orbit. Localised black regions appear in the figure because $\Sigma/<\Sigma>_t$ was limited to a maximum value for clearer visualisation of the contours.}
\label{chaos}
\end{center}
\end{figure}

The limit $\alpha_2$ is difficult to define in a quantitative way since the transition between oscillatory and chaotic states happens gradually. As $\alpha$ is increased above $\alpha_1$ the wave amplitude becomes larger and larger until the oscillations become highly non-linear and the results show signs of wave interactions. Individual oscillations are increasingly difficult to perceive as the system transitions to the chaotic state. For example, for $c_\mt{s}=0.01$ the passage from oscillatory to chaotic occurs at $\alpha_2$ with $0.18<\alpha_2\lesssim0.2$. Fig.~\ref{chaos} shows contours of surface density divided by its time-averaged value for $\alpha>\alpha_2$. It is clear that the response of the flow to the instability becomes chaotic in this state and global oscillations are no longer visible. Possibly, the disc breaks into segments which oscillate at a local average of the epicyclic frequency.

Given the difficulty in restricting $\alpha_2$ to a single value and in characterising the chaotic state, throughout the rest of this chapter I'll focus on the limit $\alpha_1$ only and on the oscillatory state present for $\alpha_1<\alpha<\alpha_2$. It should be noted that in the parameter-space region of interest discontinuities or shocks are not present in the solutions.

\subsection{Oscillatory state}
\label{osre} 

\begin{figure}[!p]
\begin{center}
\includegraphics[width=134mm]{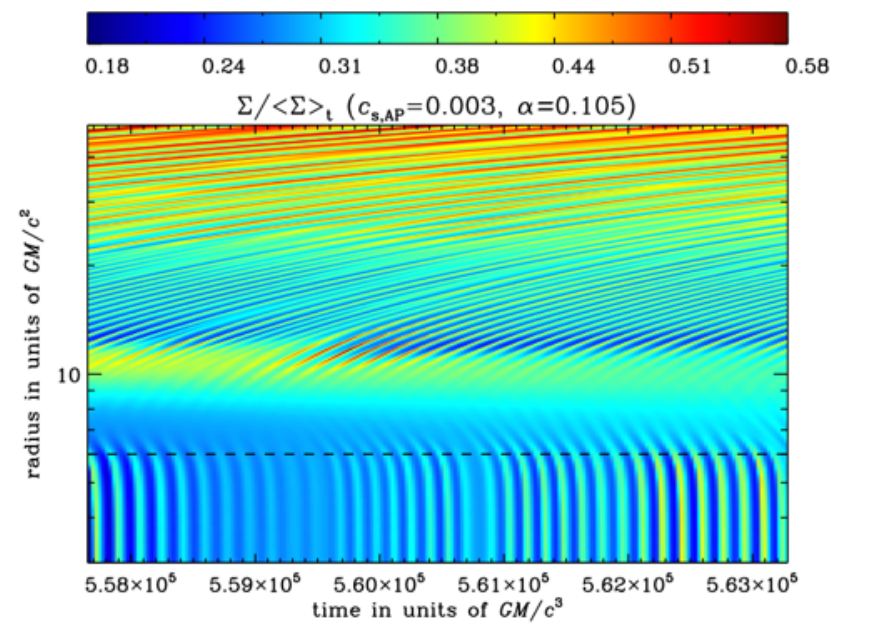} \\
\vspace{0.5cm}
\includegraphics[width=134mm]{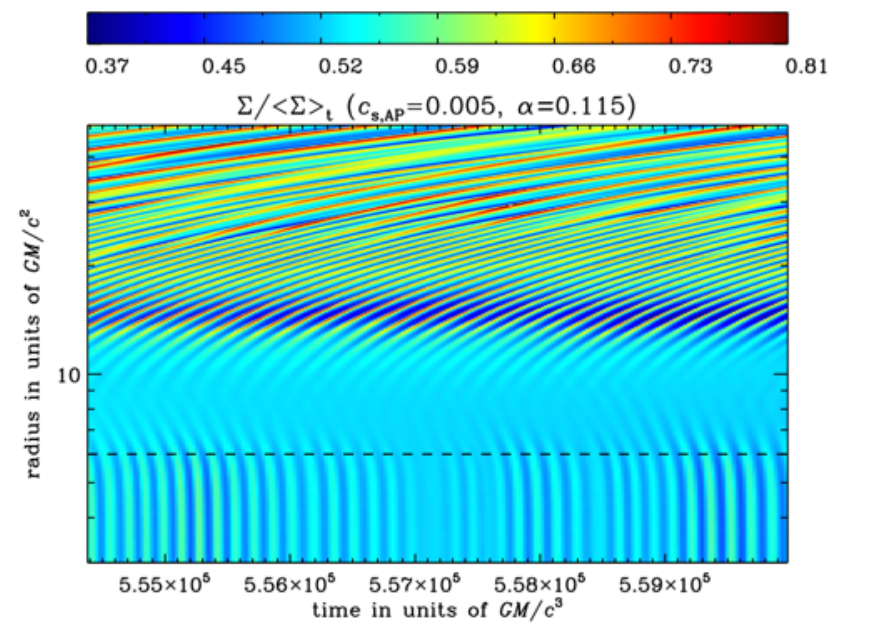}
\caption{Surface density divided by its time-averaged value for different values of sound speed and $\alpha$. The dashed line indicates the location of the marginally stable orbit.}
\label{oscilstate1}
\end{center}
\end{figure}

\begin{figure}[!p]
\begin{center}
\includegraphics[width=134mm]{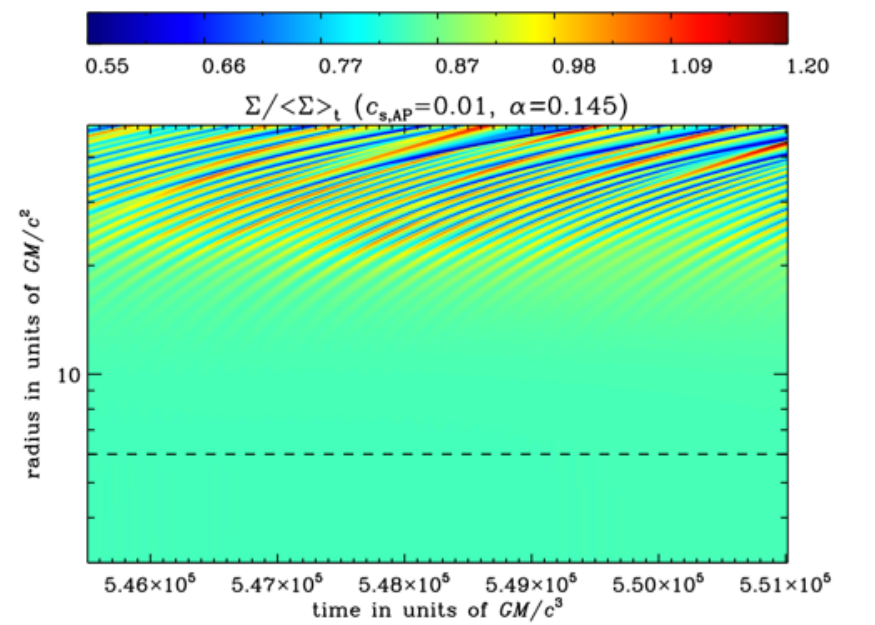} \\
\vspace{0.5cm}
\includegraphics[width=134mm]{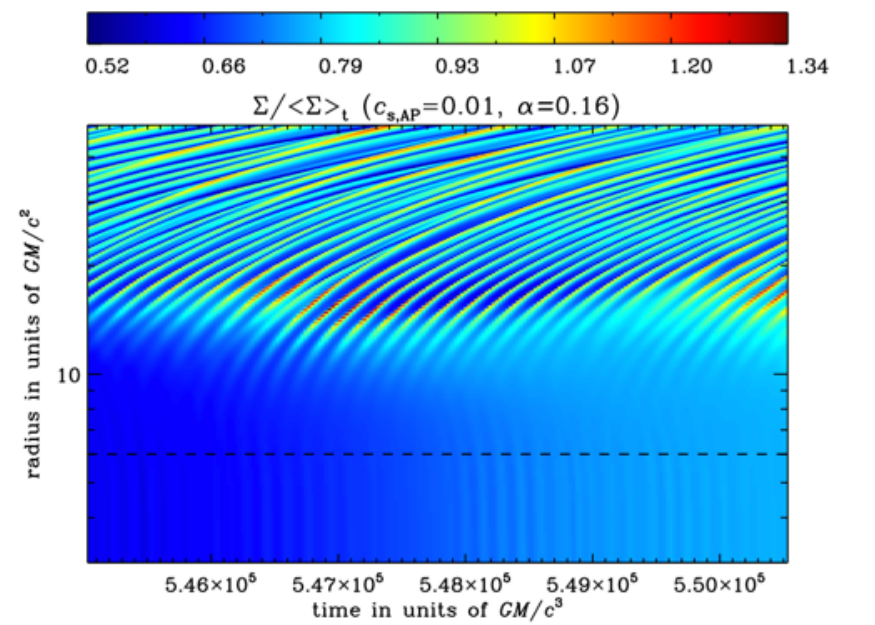}
\caption{Surface density divided by its time-averaged value for $c_\mt{s,AP}=0.01$ and two different values of $\alpha$. The dashed line indicates the location of the marginally stable orbit.}
\label{oscilstate2}
\end{center}
\end{figure}

As mentioned in the previous section, when the viscosity parameter $\alpha$ is increased above the threshold value $\alpha_1=\alpha^*=0.14(100c_\mt{s})^{1/3}$, oscillations are visible in the simulations. \enlargethispage{\baselineskip} These can be seen by, e.g., looking at the contours of surface density $\Sigma(r,t)$ divided by its time-averaged value; typical results are shown in Figs. \ref{oscilstate1} and \ref{oscilstate2}.

It is particularly clear in the panels of Fig.~\ref{oscilstate1} that density fluctuations are practically inexistent in a region RE between $r\approx7$ and $r\approx9$. (For reference, the reader should recall that $r_\mt{ms}=6$ and the maximum of the epicyclic frequency is at $r\approx 7.5$ in the units used.) As the waves propagate away from this region --- inwards for $r<7$ and outwards for $r>9$ --- their amplitude increases. The inward-travelling waves quickly disappear onto the plunging region while the outward-propagating oscillations are visible at an approximately constant frequency for a wide range of radii.  

\begin{figure}[t!]
\begin{center}
\includegraphics[width=0.49\linewidth]{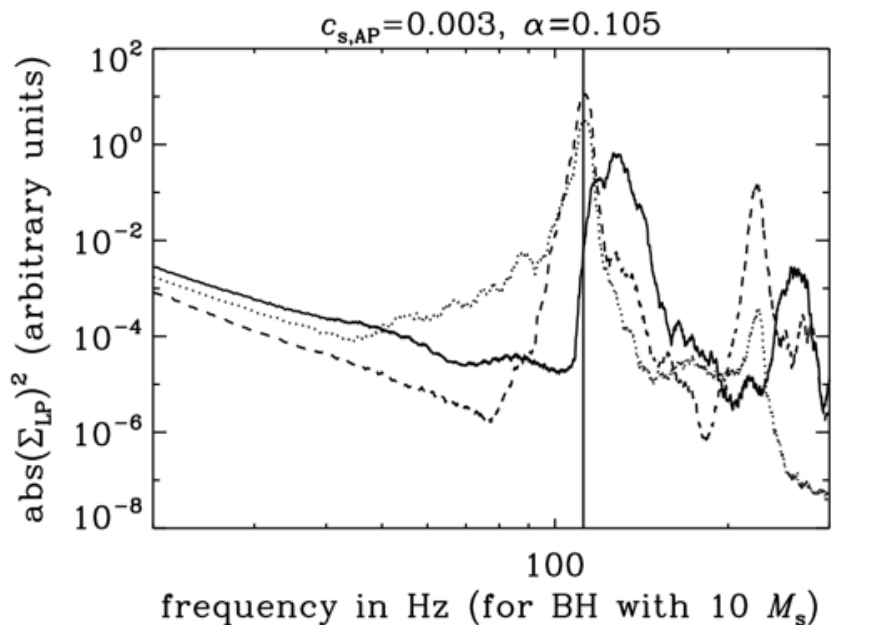}
\includegraphics[width=0.49\linewidth]{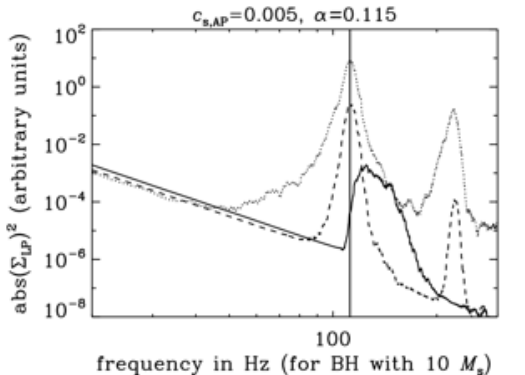} \\
\vspace{5mm}
\includegraphics[width=0.49\linewidth]{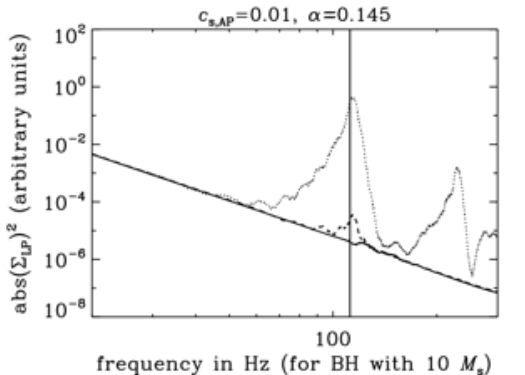}
\includegraphics[width=0.49\linewidth]{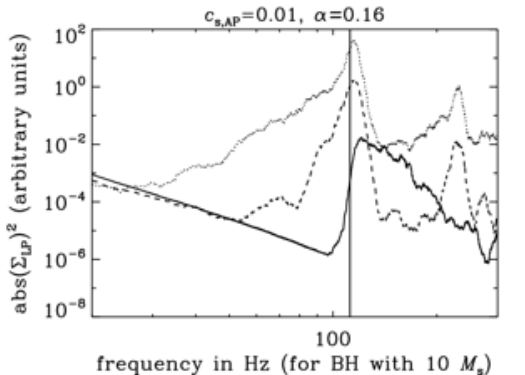}
\caption{Smoothed power spectra of $\Sigma(r_\mt{p},t)$ at $r_\mt{p}=r_\mt{ms}$ (full line), $r_\mt{p}=12$ (dashed line) and $r_\mt{p}=18$ (dotted line). The vertical line indicates the maximum value of $\kappa$. The top left and top right panels correspond to the top and bottom panels of Fig.~\ref{oscilstate1}, respectively while the bottom left and bottom right panels correspond to the top and bottom plots of Fig.~\ref{oscilstate2}.}
\label{powerspectra}
\end{center}
\end{figure}

To determine the frequency of the waves visible in the contour plots it is useful to calculate the Fourier Transform. However, since the simulations have a variable time step and the data are, consequently, unevenly spaced, the Lomb normalised Periodogram (LP) is determined instead using the \textsc{fasper} routine. The LP of the surface density is calculated at three different radii; the power spectra, given by $|\Sigma(r_\mt{p},t)_\mt{LP}|^2=|\Sigma(r_\mt{p},f)|^2$, where the frequency $f=\omega/2\pi$ is in Hz, are shown in Fig.~\ref{powerspectra}. The dominant frequency for radii outwards from RE is very close to the maximum of the epicyclic frequency for all values of $c_\mt{s,AP}$ and $\alpha$ shown. Peaks at harmonic values of $\mt{max}(\kappa)$ are also visible but this is an artefact of the Lomb normalised Periodogram since the waves with frequency close to $\mt{max}(\kappa)$ are not perfectly sinusoidal. The spectra at the marginally stable orbit, possibly influenced by the transonic flow, are more complicated and have a maximum at a frequency slightly higher than $\mt{max}(\kappa)$ when a broad peak is visible. The oscillations visible in Figs. \ref{oscilstate1} and \ref{oscilstate2} are clearly quasi-periodic since the spectral peaks have a finite width.

The isolated influence of viscosity in the oscillations is obvious in both Fig.~\ref{oscilstate2} and the corresponding power spectra at the bottom of Fig.~\ref{powerspectra}, where the sound speed is kept constant at $c_\mt{s,AP}=0.01$ and $\alpha=0.145,\,0.16$. When the viscosity parameter is just slightly above the threshold value at $0.145$, the oscillations are only visible in the outer disc, $r\gtrsim20$. For higher viscosity, $\alpha=0.16$, fluctuations are evident with higher amplitude for $r\gtrsim10$. It is also clear from the power spectra in Fig.~\ref{powerspectra} (and Figs. \ref{oscilstate1} and \ref{oscilstate2}) that the amplitude of the waves increases as they propagate away from region RE. The amplitude of the waves seems to decrease in the outer disc but this may be a numerical artefact: because the grid is logarithmic it becomes increasingly difficult to resolve oscillations as $r$ increases. On the other hand, physically, these waves are expected to dissipate through shocks which would also explain the decrease in amplitude.

The effect of the sound speed can be most easily analysed by comparing the top right ($c_\mt{s,AP}=0.005$) and the bottom left ($c_\mt{s,AP}=0.01$) panels of Fig.~\ref{powerspectra}. Although the viscosity parameter is different in both plots, it is just slightly above the corresponding threshold value so that the influence of viscosity can be understood as equivalent in both cases. Based on these considerations, a comparison of the dotted and dashed curves shows that, at the same radius, waves have larger amplitude and are evident closer to the RE region when the sound speed is smaller.

Before going on to interpret the results described here, it should be noted that there are strong arguments favouring the physical, as opposed to numerical, nature of the oscillations visible in the simulations. They are a robust feature of the calculations: waves are present and have similar characteristics when the resolution or the courant value are changed. Also, similar results are obtained when first-order finite differences are used in the numerical scheme.  Moreover, as seen in the introduction, similar oscillations have been seen in simulations which rely on numerical methods very distinct from the one used here. For example, \cite{chentaam1995} found oscillations at similar frequencies using a code based on an explicit piecewise parabolic method to solve the hydrodynamic equations. This serves as a test of the numerical method employed when $\alpha_1<\alpha<\alpha_2$.

\section{Discussion}
\label{transdiscussion}

\subsection{Local stability analysis}

In this section I'll focus on interpreting the results shown previously for $\alpha_1<\alpha<\alpha_2$. The oscillatory state plots show clear signatures of viscous overstability. To better understand this phenomenon, and for completeness, I'll start by doing a local stability analysis of the \cite{ap2003} problem. The idea is to study the evolution of small perturbations to a transonic \emph{steady} background.

The local dispersion relation can be determined by imposing linear perturbations to a steady disc and see, using equations (\ref{eqap1})--(\ref{eqap2}), how they evolve. In practice this means that quantities such as surface density, radial velocity and azimuthal velocity may be written as $q\rightarrow q+q'$, where terms of $\mathcal{O}(q'^2)$ or higher are assumed much smaller than terms of $\mathcal{O}(q')$. 

In the steady background state, $\Sigma=\Sigma_\mt{b}$, $u_\phi=\Omega r$ and $u_r<0$; these quantities satisfy the time-independent version of equations (\ref{eqap1})--(\ref{eqap2}). Contrary to the standard theory of oscillations in accretion discs \citep[][and Part \ref{os} of this thesis]{lubowpringle1993,rkato2001} here I assume $u_r\neq0$ in the basic state given the focus on the transonic nature of the flow. The perturbations acting on this steady state may be written as
\begin{equation}
q'(r,t)=\textrm{Re}\left[\widetilde{q'}(r)\exp(-\mt{i}\omega t)\right].
\end{equation}

Assuming that the radial variation of perturbed quantities is much faster than that of equilibrium quantities, the equations may then be written as
\be
\left(-\mt{i}\omega+u_r\frac{\mt{d}}{\mt{d}r}\right)\Sigma'=-\Sigma_\mt{b}\frac{\mt{d}u_r'}{\mt{d}r},
\ee
\be
\left(-\mt{i}\omega+u_r\frac{\mt{d}}{\mt{d}r}\right)u_r'-2\Omega u_\phi'=-\frac{c_\mt{s}^2}{\Sigma_\mt{b}}\frac{\mt{d}\Sigma'}{\mt{d}r},
\ee
\be
\left(-\mt{i}\omega+u_r\frac{\mt{d}}{\mt{d}r}\right)u_\phi'+\frac{\kappa^2}{2\Omega}u_r'=-\frac{2\alpha c_\mt{s}^2}{r\Omega_\mt{PW}}\frac{\mt{d}u_\phi'}{\mt{d}r}-\frac{2\alpha c_\mt{s}^2}{\Sigma_\mt{b}}\frac{\Omega}{\Omega_\mt{PW}}\frac{\mt{d}\Sigma'}{\mt{d}r},
\label{wave3}
\ee
where the tildes were dropped for simplification\footnote{Note that I've assumed $T_{r\phi}=-2\alpha c_\mt{s}^2\Sigma \Omega/\Omega_\mt{PW}=-2\alpha c_\mt{s}^2\Sigma u_\phi/r\Omega_\mt{PW}$ and perturbed $u_\phi$. Hence the presence of a term involving $u_\phi'$ on the RHS of equation (\ref{wave3}).}. Note that these equations are not valid very near the sonic point where the radial variation of $u_r$ is important. They are, however, approximately valid in the region where $\kappa$ is maximum and beyond, i.e., where the waves are most prominent.

From the equations above it is easy to verify that perturbations with local wavenumber $k$ ($q'\propto\mt{e}^{\mt{i}\int k(r)\mt{d}r}$) satisfy
\be
\left(k^2 c_\mt{s}^2-\widetilde{\omega}^2+\kappa^2\right)\widetilde{\omega}=-\frac{2\alpha c_\mt{s}^2k}{r\Omega_\mt{PW}}\left(2\mt{i}\Omega^2 kr-k^2c_\mt{s}^2+\widetilde{\omega}^2\right),
\label{apdr}
\ee
where $\widetilde{\omega}=\omega-ku_r$. Recalling that $\alpha_\mathrm{SS}\approx2\frac{\Omega}{\Omega_\mathrm{k}}\alpha_\mathrm{AP}=2\frac{\Omega}{\Omega_\mathrm{k}}\alpha$, this formula is almost equivalent to that obtained by \cite{khm1988} [cf. (\ref{vodd})]. The extra terms showing up on the RHS of (\ref{apdr}) are due to the slightly different stress tensor used here.

In the limit of no viscosity, equation (\ref{apdr}) has a trivial solution corresponding to the classical viscous mode (which will be ignored in the remainder of this analysis) and two solutions corresponding to two inertial-acoustic waves propagating in opposite directions in the disc. When viscosity is taken into account, equation (\ref{apdr}) can be solved in two fashions: one can either consider the wave number $k$ to be real and solve for the complex frequency, or take $\omega$ to be real and determine the complex wave number. In the former case, $\mt{Im}[\omega(r)]$ gives the (positive) growth rate of the inertial modes due to viscous overstability while in the latter case, $\mt{e}^{-\int\mt{Im}[k(r)]\mt{d}r}$ gives the variation of the mode amplitude due to the instability.

\begin{figure}[t!]
\begin{center}
\includegraphics[width=0.78\linewidth]{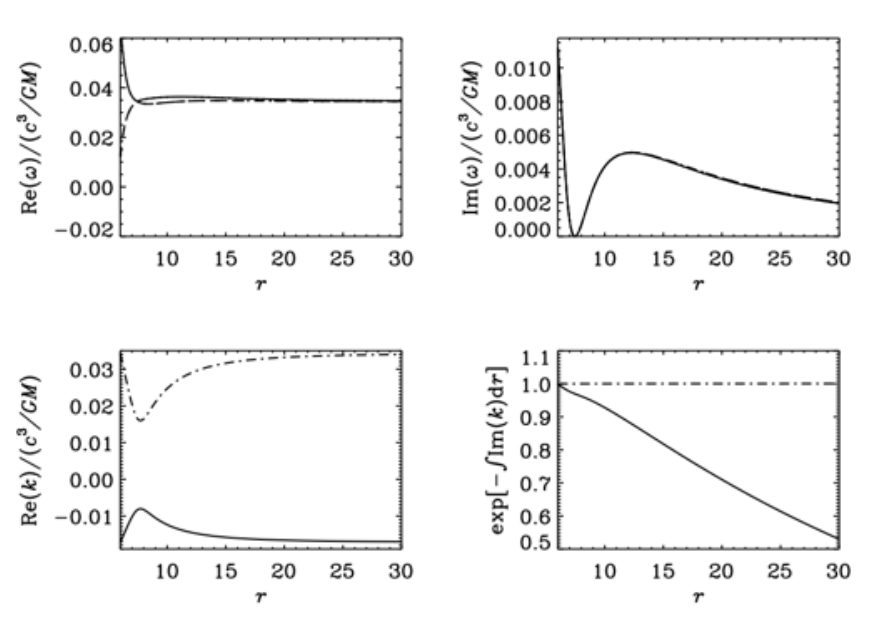}
\caption{Variation of the complex frequency and complex wave number with radius for $c_\mt{s}=0.01$ and $\alpha=0.16$. The lower right panel shows the variation of the wave's amplitude with $r$. The dash-dotted line corresponds to the outward-travelling waves (positive frequency, positive wave number) while the full line corresponds to the inward-propagating waves (positive frequency, negative wave number). The curves coincide in the upper right panel.}
\label{localdr}
\end{center}
\end{figure}

I solve (\ref{apdr}) numerically by both methods to understand the radial dependence of local viscous overstability. When $k$ is assumed real it is taken to be $\pm (\mt{max}(\kappa)^2-\kappa^2)/c_s^2$ [cf. equation (\ref{drf}) with $\omega=\mt{max}(\kappa)$]; when $\omega$ is real it is equal to the maximum of the epicyclic frequency. These choices are related to the fact that the oscillations present in the simulations have a frequency close to $\mt{max}(\kappa)$. For the purpose of the calculations, I take $u_r$ and $\Omega$ to be given by the solutions of equations (\ref{eqap1})--(\ref{eqap2}) for parameters such that a steady state is obtained. Solutions are presented in Fig.~\ref{localdr}. 

As it can be seen, both inward- and outward-propagating waves grow in time by the same amount in practically all regions in the disc. This result differs from the findings of \cite{chentaam1993} and \cite{wallinder1995} who predicted the outgoing wave to be more unstable in certain regions and the ingoing in others. The difference is due to the fact that here the disc is isothermal while heating and cooling rates are considered in the works mentioned. The difference between the outward- and inward-propagating waves is revealed in the lower right panel: at a fixed point in time, the amplitude of the latter decreases with radius while the former suffers no amplitude changes.


In conclusion, the local analysis reveals that the inertial-acoustic modes are unstable practically everywhere in the disc and for all values of $\alpha$, as already stated in previous studies on the stability of transonic discs.
 
\subsection{Oscillatory state and viscous overstability}

It is clear from the analysis of the previous section that the modes that arise in the accretion flow modelled in the simulations are, aside from a small modification due to viscosity, inertial-acoustic in nature. Moreover, according to the \emph{local} dispersion relation, they grow due to viscous overstability for all values of $\alpha$ with a growth rate \citep[adapted from][]{khm1988},
\be
\omega_\mt{i}\approx\frac{2\alpha\Omega^2}{\Omega_\mt{PW}}\frac{c_\mt{s}^2k^2}{c_\mt{s}^2k^2+\kappa^2}.
\label{omegagr}
\ee


\subsubsection{Global outcome}

As it was seen in the introduction, viscous overstability is a propagating instability that relies on mode confinement to effectively excite the inertial-acoustic waves before they escape through the boundaries of the system. Waves with a frequency close to the maximum of $\kappa$ spend a long time near the region where the epicyclic frequency peaks as the wave speed is minimum there. Potentially, viscous overstability may be effective only where these waves slow down.  \enlargethispage{\baselineskip} As the newly excited oscillations propagate away from the excitation region RE (to both the innermost and outer parts of the disc) their amplitude is expected to grow due to the very nature of the propagating instability. 

Viscous overstability may be qualitatively understood as a convective instability \citep[e.g.][]{absconvec}. If an amplifying impulse acts on the flow at a certain location ($x_\mt{sr}$, the source), the perturbations may respond in two fashions. The localised disturbances may be kept close to the source, in which case they grow in amplitude at a fixed point in space and the system is said to be \emph{locally absolutely unstable}. Alternatively, they may be swept away from $x_\mt{sr}$ in which case the flow if \emph{locally convectively unstable}. In the latter case, the part of the response that stays at the source dies away while the part that travels away from $x_\mt{sr}$ grows in amplitude.

In the problem under study, the source\footnote{Here source simply refers to the location in the disc where the waves are amplified. The ultimate source (cause) of the disturbances will be discussed later on.} of the convective instability may be considered to be the region around the $\max(\kappa)$ where the group velocity $v_\mt{g}\propto k/\omega$ is minimum. Although  the local dispersion relation predicts a zero (temporal) growth rate at the maximum of the epicyclic frequency for waves with $\omega=\mt{max}(\kappa)$ (see upper right panel of Fig.~\ref{localdr}), it quickly increases just on either side of this point. The global outcome of the process then depends on a competition between the local growth rate (\ref{omegagr}), proportional to $\alpha$, and the speed at which waves are driven away from it. For frequencies close to $\mt{max}(\kappa)$, inertial-acoustic waves travel with a group velocity
\be
v_\mt{g}\sim c_\mt{s}\sqrt{\frac{\mt{max}(\kappa)^2-\kappa^2}{\mt{max}(\kappa)^2}},
\label{gv}
\ee
that is, the global instability should be favoured for high $\alpha$ (large growth at the source) and low sound speed (waves remain closer to the source for longer periods).

\subsubsection{Discussion}

The results presented in the section \ref{osre} are consistent with this scenario. The excited waves have indeed a frequency close to $\mt{max}(\kappa)$, as seen in Fig.~\ref{powerspectra} and fluctuations are weak in the RE region around the maximum of the epicyclic frequency at $r\approx7.5$. Their amplitude increases and the frequency is kept constant as the waves propagate to the outer disc. Moreover, the growth is more effective for large $\alpha$ and small $c_\mt{s,AP}$. Even though both an inward- and an outward-propagating wave are excited where the group velocity is minimum, the wave going outwards reaches higher amplitudes because it can travel further away from its origin. The wave going inwards is transmitted \citep[partially or completely, depending on the reflection properties at the sonic point,][see also chapter \ref{reflect}]{laitsang2009}  to the plunging region. Wave power is manifested at a range of frequencies at and beyond the disc boundary --- presumably due to the supersonic speeds --- as indicated by the complicated power spectra at $r_\mt{ms}$.

A different idea is that the excitation source is not at the maximum of the epicyclic frequency but at the sonic point \citep{mkh1988}. Presumably, for high enough $\alpha$, the flow coming for large radii has difficulty in passing through the sonic point regularly and undergoes small perturbations that spread throughout the disc. Although this would be a more convenient way of relating $\alpha^*$ and $\alpha_\mt{vo}$, this explanation doesn't seem to be borne out in the results as these perturbations would propagate outwards \emph{everywhere} in the subsonic region. For frequencies lower than $\mt{max}(\kappa)$ they could reflect at the epicyclic barrier and return to the sonic point effectively creating inward-propagating waves in the inner region. Unfortunately, the modes seen there have a frequency slightly higher than $\mt{max}(\kappa)$ (see full curve of Fig.~\ref{powerspectra}), making reflection at the epicyclic barrier impossible. The most likely explanation for the results obtained is that mentioned before: the source of the instability is located close to the maximum of $\kappa$ and waves propagate inwards and outwards from there.

Preliminary results of a simulation where a modified gravitational potential was used, making the epicyclic frequency peak at a radius further away from $r_\mt{ms}=6$, confirm that excited waves propagate inwards and outwards from the region around $\mt{max}(\kappa)$.
\color{black}
Oscillations with frequency close to the maximum of the epicyclic frequency propagating inwards for radii smaller than $\mt{max}(\kappa)$ and outwards for large radii were also found by \color{black}\cite{hmk1992} and by Taam and collaborators, as mentioned in the chapter \ref{steadyintro} \citep[see also][]{oneill2009}.

\color{black}
\subsection{Sonic-point instability}

In the simulations of isothermal discs with a standard $\alpha P$ stress of \cite{mkh1988}, the authors claim the sonic-point instability to be the cause of the disturbances seen in their flow. Hence the idea that the critical point is the excitation source. However, the results described in the previous section clearly show that oscillations are excited in the inner region but not at the sonic point hinting that the sonic-point instability isn't in action.
\color{black}

To test this, I directly verify from the simulations for $c_\mt{s}=0.01$ and various values of $\alpha$, when the criterion
\begin{equation}
\alpha_\mt{SS}\Omega(R_0)>\left|\frac{\mathrm{d}u_R}{\mathrm{d}R}\right|_{R_0},
\label{crit1}
\end{equation}
is satisfied. The stress tensor used in this problem is slightly different from that used by \cite{khm1988} to derive this instability condition. However, (\ref{crit1}) should still be valid if one recalls that $\alpha_\mt{SS}\approx2\alpha\Omega/\Omega_\mt{PW}$ which is still a constant in the thin discs modelled in the simulations (see right panel of Fig.~\ref{stable}). 

When $\alpha<\alpha_1$, the sonic point is well defined and the values of $\Omega(R_0)$ and $\left|\mathrm{d}u_R/\mathrm{d}R\right|_{R_0}$ can be easily obtained. The simulations show very clearly that (\ref{crit1}) is satisfied when $\alpha>0.08=\alpha_\mt{sp}$, that is, the sonic point should be unstable for values of $\alpha$ well below $\alpha_1$. However, no signs of instability are visible for $\alpha_\mt{sp}<\alpha<\alpha_1$.

It should be mentioned at this stage that \cite{ap2003} say that (\ref{crit}) is satisfied for $\alpha\gtrsim 0.18$, according to Fig.~6 in their paper. Unfortunately, it is unclear where this value comes from as they give no details of their calculations of $\left|\mathrm{d}u_R/\mathrm{d}R\right|_{R_0}$. They only point out that, ``contrary to the conventional picture'', $\left|\mathrm{d}u_R/\mathrm{d}R\right|_{R_0}$ is small only when the passage is made in the slow direction of a nodal-type point and not for every node. But the results from the current simulations seem to be in favour of the ``conventional picture'': $0.08$ is precisely the value of $\alpha$ where the point changes from saddle to nodal, i.e., $\alpha_\mt{sp}=\alpha_\mt{sn}$. This result is in agreement with the idea of \cite{kato1994} that saddle points are stable to the sonic-point instability while nodes are unstable.

Notwithstanding these differences, it is relevant to recall that none of the values ($0.08$ and $0.18$) corresponds to the value of $\alpha$ above which the numerical solutions are unstable. I believe that the sonic-point instability is not present in the current simulations, perhaps because it cannot exist beyond local analysis. Although \cite{mkh1988} claim to see the sonic-point instability in their simulations, the reader should recall that they initialised the disc with a perturbation placed at the sonic point. Most importantly, they analysed the very inner region only and no other simulations show any signs of this mode.

\subsection{State transitions and unanswered questions}

When the viscosity is such that the system reaches a physically acceptable steady state, ``the flow is subsonic and the disc is Keplerian at large radii, the flow passes through a critical point, and it becomes supersonic at small radii'' \citep{ap2003}. The fluid is in a stable equilibrium and, as seen before, axisymmetric oscillations are supported. Moreover, the system is locally viscously overstable to these type of perturbations, i.e., according to the local dispersion relation waves can grow in the disc for any given value of $\alpha>0$. However, previous works and the current simulations indicate that local instabilities may not result in global variations of the fluid parameters, as local perturbations may propagate away before attaining reasonable amplitudes. Therefore, it seems qualitatively plausible that a higher value of viscosity is required to see viscously overstable modes in the system, i.e., stable solutions are possible for small values of $\alpha$.

The studies of \cite{ap2003}, indicate that a steady state transonic flow that satisfies physically acceptable boundary conditions can only exist for $\alpha<\alpha^*=0.14(100 c_\mathrm{s})^{1/3}$. In addition, the results reported here show that viscously overstable global oscillations exist for $\alpha_1<\alpha<\alpha_2$ with $\alpha_1=\alpha^*$. This numerical coincidence seems to indicate that the non-steadiness of a physically acceptable steady state flow is indeed related to the onset of viscous overstability, that is, $\alpha^*=\alpha_\mt{vo}$. But how? Why would viscous overstability be triggered when the parameters in the disc are such that the passage through the nodal sonic point is made in the slow direction? The onset of viscous overstability cannot simply be seen as a competition between viscosity ($\propto\alpha$) and the escape rate for perturbations ($\propto c_\mt{s}$) since numerical results indicate that $\alpha_\mt{vo}\propto c_\mt{s}^{1/3}$ and not simply $\alpha_\mt{vo}\propto c_\mt{s}$. 

\cite{ap2003} suggest a relation to the instability of the Keplerian curve in the supersonic region (although the cause of this instability is never addressed). When the flow goes through a nodal critical point in the slow direction the slope of $u_r$ changes sharply after passing this point and no steady solutions can be found. According to \cite{ap2003}, the explanation for the sharp change is related to the Keplerian curve, which is connected to the slow direction for $\alpha>\alpha^*$. The flow follows this curve at large radius where it is stable. However, the Keplerian curve becomes unstable in the supersonic regime which results in the physical solution departing from it and changing its slope sharply, as seen in the lower right panel of Fig.~\ref{apdiagrams}. The numerical results shown here indicate that viscous overstability is triggered at this stage. However, the simulations also show that the instability visible for $\alpha>\alpha^*$ takes the form of propagating waves which are excited near the maximum of the epicyclic frequency. It is therefore unlikely for such an instability to be directly triggered by what's happening at the sonic point. 

An attractive way of thinking about this problem is as follows. Transonic solutions are \emph{always} \emph{locally} pulsationally unstable. However, they only become visibly unstable, i.e., only attain a \emph{global} character, when $\alpha$ and $c_\mt{s}$ are such that steady solutions are no longer possible due to the unphysicalness or nonuniqueness of the passage in the slow direction of a nodal-type sonic point. The system is then forced to chose a different state, that where global viscous overstability exists. Transonic solutions are unstable to global overstable modes when accretion cannot proceed in a physically acceptable way.

If no steady solution is possible, some may argue that instability (or overstability) is meaningless given that the system can never adjust itself to equilibrium. Nonetheless, this situation can be analysed from a different viewpoint: the absence of a steady solution enforces perpetual disturbances on the flow resulting in overstable growth even in a system with open boundaries.


More calculations are needed to understand whether or not the explanation given before to justify the fact that $\alpha^*=\alpha_\mt{vo}$ in isothermal discs with $\alpha P$ viscosity can be applied to discs with different stress tensors. In an isothermal disc with a diffusion-type stress the sonic point is always saddle \citep{ak1989}. However, this does not necessarily mean that physically acceptable steady solutions are always possible. Even in this case there should be differences in the accretion process at low and high $\alpha$, with viscous effects (as opposed to the pressure gradient) being the main cause of infall of matter to the black hole in the latter situation, in which case the critical point is located outside $r_\mt{ms}$. Presumably, a limit $\alpha^*$ may exist in the diffusion-type stress tensor case, e.g., in association with the different accretion scenarios and different locations of the sonic point in the low- and high-viscosity regimes. Therefore, a connection between the existence of $\alpha^*$ and the onset of global viscous overstability could still be possible.

\section{Conclusions}
\label{transconclusions}

In this chapter, I solved the time-dependent version of the \cite{ap2003} transonic accretion problem where a thin, isothermal disc with essentially a standard $\alpha P$ viscosity is considered. The aim was to answer some, if not all, of the questions that originated from an extensive review of the existing literature on steady transonic accretion and viscous overstability.

In particular, I wished to understand what happens to the transonic time-dependent solutions in the limit above which \cite{ap2003} were unable to find physically acceptable, unique steady state solutions. The answer to the first of those questions is clear: steady solutions cannot be found for $\alpha>\alpha^*$ because accretion is unstable in those cases. For $\alpha$ above $\alpha^*$ (up to a certain limit), oscillations of frequency close to $\mt{max}(\kappa)$ propagate throughout the disc, inwards and outwards from the radius where the epicyclic frequency peaks.

The discussion of the previous section indicates that the oscillations visible in the simulations for $\alpha>\alpha^*$ are, in all likelihood, overstable inertial-acoustic modes. Therefore, it seems that in isothermal discs with standard $\alpha P$ viscosity, $\alpha_\mt{vo}=\alpha^*=0.14(100 c_\mt{s})^{1/3}$ where $c_\mt{s}$ in units of $c$. This result indicates that, at least in these flows, there is a relation between the lack of a steady solution and the onset of global viscous overstability. It is however unclear if this is the case for discs characterised by non-isothermal equations of state and/or different stress tensors where the very existence of $\alpha^*$ is disputed.

Other questions concerned the sonic-point instability which relates to an inertial-acoustic viscously overstable mode which stands at the critical point where its outward escape velocity equals the radial inflow. The simulations done show no signs of such instability casting doubts on its existence beyond local analysis. This would explain why \cite{ap2003} were able to find steady solutions for values of viscosity larger than $\alpha_\mt{sp}$ which is determined to be 0.08 from the simulations with $c_\mt{s}=0.01$.

The results obtained here are possibly relevant to explain observations of high-frequency QPOs in black-hole systems. According to \cite{kingetal2004}, observations indicate that the viscosity parameter $\alpha_\mt{SS}$ in real systems should be between 0.1 and 0.4, that is, well within the range where steady accretion is no longer possible\footnote{The values 0.1 to 0.4 correspond to the usual Shakura--Sunyaev $\alpha$. For this viscosity, the \cite{ap2003} limit translates to $\approx0.28(100 c_\mt{s}/c)^{1/3}$.}. In addition, there is numerical evidence coming from MRI simulations that $\alpha$ increases as $r_\mt{ms}$ is approached because of the increase in the shear/vorticity ratio due to the different angular velocity profile in the inner disc \citep{abl1996}. Therefore, in some systems, viscously overstable oscillations should be detectable. In addition, and similarly to the models involving trapped inertial oscillations, the frequency of such modes is approximately $\mt{max}(\kappa)$ and is therefore within the range of high-frequency QPOs. In this model, QPOs would only be detected in states where the cool disc reaches the marginally stable orbit since viscous overstability is triggered only when the sound speed in the region close to the maximum of the epicyclic frequency is low. Indeed, high-frequency QPOs are almost exclusively detected when the black hole is in the very high state where, according to the model of \cite{esinetal1997}, the thin disc reaches $r_\mt{ms}$. 

A caveat of the model presented in this chapter is related to the fact that viscously overstable oscillations prefer low mass accretion rates. As shown by \cite{mt1996}, inertial-acoustic modes are visible in the simulations when $\dot{M}\lesssim0.25\dot{M}_\mathrm{Edd}$. In the very high state where HFQPOs are detected the luminosity is expected to be close to its Eddington value, suggesting that viscous overstability is probably not visible. In any case, the typical values of $\alpha$ and $\dot{M}$ in this state are unknown and, therefore, the presence of overstable inertial-acoustic modes can't be ruled out.

Regardless of the observational applications of the unsteadiness of transonic accretion and viscously overstable waves, the problems discussed in this section are of considerable theoretical interest. 
A connection between both phenomena and the understanding of the conditions under which a disc may become viscously overstable merits further investigation.

\thispagestyle{empty}
\part{Oscillations in Black-Hole Accretion Discs}
\label{os}
\thispagestyle{empty}
\chapter{Introduction}
\label{oscilintro}

\section{Motivation}

One of the most important examples of astrophysical variability is that concerning periodic or quasi-periodic variations intrinsic to astrophysical objects, namely stars and accretion discs. The development of the theory of stellar pulsation or of the theory of disc oscillations is fundamental to understand the origin of such variations and, together with observational studies of these objects, allows for the properties of stars or black holes in the centre of discs to be inferred.


Classic variable stars such as the Cepheids and RR Lyrae have long been observed to have periodic variations in luminosity. Current theoretical models state that these objects vary in brightness because the stars pulsate radially, that is, they expand and contract while keeping their spherical shape. On the other hand, stars such as our Sun have considerably more complicated (and with much lower amplitude) modes of oscillation. Not only more than one mode may be excited at the same time but oscillations that don't preserve the spherical symmetry of the object may also be present. Current observations indicate that several thousands of individual radial and nonradial modes exist in our star \citep{cd2002}. 

Observational studies of classic Cepheids allowed for the discovery of a precise period-luminosity relation of high importance to measure distances in astronomy. On the other hand, theoretical analysis of the solar nonradial oscillations provides information about the Sun's interior. Asteroseismology is the study of stellar oscillations: interior properties of stars can be inferred by analysing the frequency spectra of these oscillations \citep{dalsgaard}. This is one of the main stimuli for the research done on periodic variability in stars. \enlargethispage{-\baselineskip}

From a theoretical point of view, oscillations are often studied using a fluid-dynamical treatment. Mathematically, oscillations are viewed as small perturbations to the fluid quantities such as density and velocity. This analysis can in principle be applied to any fluid and, therefore, to any astrophysical object that can be treated as such.

Oscillations are expected to exist not only in planets and stars but also in galaxies and accretion discs. In the case of spiral galaxies, which can be modelled as collisionless self-gravitating discs, the spiral structure can be explained in terms of propagating density waves \citep[see review by][]{toomre1977}. On the other hand, the study of oscillations in collisional and non-self-gravitating accretion discs is expected to provide information not only about the discs themselves but also their central objects.

However, it should be taken into account that a parallel with asteroseismology is not straightforward since whether or not oscillations in accretion discs can be observed is much more controversial than for stars. Not only are accretion discs far away in comparison to the Sun making small amplitude perturbations hard to detect, but friction in the disc is likely to damp these modes. In both the case of spiral density waves in galaxies and that of pulsations in collisional discs, excitation mechanisms should be taken into account. In the former case, forcing by an external potential or a central bar may be required to maintain the spiral structure \citep{gt1979}. Alternatively, the spiral structure may be self-excited by gravitational instability \citep{linshu}. As for non-self-gravitating discs, oscillations can only be detected from Earth if some mechanism or instability capable of exciting waves to observable amplitudes is present.

Nevertheless, observations indicate that oscillations likely to originate in discs are indeed detected. As pointed out by Kato in the first paper where pulsations of viscous accretion discs were studied \citep{kato1978}, some X-ray sources and quasars are observed to have periodic or quasi-periodic variability. Since these objects are modelled as black holes surrounded by accretion discs, the variability must originate in the disc (or possibly in a jet), which, contrary to the hole, is visible. One of the theoretical motivations behind the study of viscous overstability was precisely the need to corroborate such models of X-ray sources and active galaxies. Within the viscous overstability scenario, oscillations in the disc can be detected because they are excited by viscosity. \enlargethispage{-\baselineskip}

The motivation behind some of the first papers focused on pulsations of viscous discs \citep{hww1980,coxeverson1983}, was the discovery of quasi-periodic oscillations in dwarf novae. This type of variability is found in addition to the short-period coherent oscillations observed previously. The latter type results in a narrow peak in the spectra and is thought to have origin in nonradial pulsations of the white dwarf. The quasi-periodic oscillations show up as broad peaks in the spectra of these objects and have properties coherent with a stochastic origin. The extensive study of the observational characteristics of these oscillations by \cite{robinsonnather1979} concluded that they must originate in the accretion disc. Even though the field has developed since then, current non-magnetic models for QPOs in cataclysmic variables still invoke oscillations of the accretion disc; different explanations are possible in magnetic stars \citep[][and references therein]{warner2004}. Given the possibility of quasi-periodic variability components originating in the disc, their decay times may contain information about the magnitude of the accretion flow viscosity \citep{coxeverson1983}. 

In the past decade or so, quasi-periodic oscillations were detected in some black hole candidates. As mentioned in the introduction of this thesis, some authors \citep{nowaketal1997,reldisko,nowaklehrchapter,rkato2001} argued that these oscillations can be explained in terms of modes arising in the accretion disc that surrounds the black hole. Turbulent viscosity, characteristic of accretion flows, generally prevents waves from propagating across the disc to form coherent global modes. As mentioned previously, a possible way for such modes to exist is for them to be trapped in a small region in the inner part of the disc which would work as a resonant cavity. Although trapped waves may reveal relatively little about the properties of the accretion disc itself, they are a promising tool in the study of the central object.

As summarised by \cite{rkato2001}, there are important distinctions between the theory of stellar oscillations and that of discs. One of the main differences concerns the force balance in both classes of objects. While in stars rotation is typically regarded as a small correction to their structure, accretion discs are centrifugally supported in the radial direction. As a consequence, disc rotation is the major restoring force of disc oscillations and the epicyclic frequency provides the characteristic frequency of inertial and long-wavelength inertial-acoustic modes. The radial distribution of $\kappa$ influences not only the properties of the waves but also their trapping regions and, possibly, excitation mechanisms. Viscous processes are  also relevant when it comes to exciting or damping oscillations, as hinted by the studies of viscous overstability. Viscosity and rotation are, therefore and contrary to the stellar case, two fundamental ingredients in the analysis of pulsations in accretion discs.

The importance of viscosity in exciting disc oscillations was emphasised in Part \ref{ta} of this thesis. The study described in that section and the introduction provided here, motivate a more in-depth study of oscillations in black-hole accretion discs. Important background on the subject is given by \cite{lubowpringle1993,nowaklehrchapter,rkato2001}. In this part of the thesis I consider the simple case of isothermal motions of non-self-gravitating thin inviscid discs. These approximations are justified in discs with mass negligible when compared to the central object and where the timescales relevant for oscillations are much faster than the viscous timescale. 

The emphasis of Part \ref{os} goes to the influence of rotation and relativistic effects on disc oscillations. The effects of radial inflow are considered only in chapter \ref{reflect} where the reflection of waves at the inner disc boundary is studied. The excitation of trapped waves and the propagation of global deformations in relativistic discs is analysed in chapters \ref{excmech} and \ref{warpecc} respectively. A more detailed outline is given below. Throughout this part of the thesis, relativistic effects are included in a pseudo-relativistic fashion by applying the relativistic expressions (\ref{relomega}), (\ref{relkappa}) and (\ref{omegazfreq}) to an otherwise Newtonian model.

\subsection{Outline}

In the remainder of this chapter, I provide a summary of the basic theory of oscillations in black-hole accretion discs. The different modes of oscillation are described and their trapping regions are analysed. I also look into deformations of discs such as warping and eccentricity in the context of global modes.

In chapter \ref{excmech} I work out in detail a non-linear coupling mechanism suggested by Kato in which a global warping or eccentricity of the disc has a fundamental role. These large-scale deformations combine with trapped modes to generate ``intermediate'' waves of negative energy that are damped as they approach either their corotation resonance or the inner edge of the disc, resulting in amplification of the trapped waves. The results obtained indicate that this coupling mechanism can provide an efficient excitation of trapped inertial waves, provided the global deformations reach the inner part of the disc with non-negligible amplitude.

Since chapter \ref{excmech} revealed the importance of warped and eccentric discs, chapter \ref{warpecc} is dedicated to these global deformations. I'll consider the inward propagation of warping and eccentric disturbances in discs around black holes under a wide variety of conditions. The calculations done use secular theories of warped and eccentric discs and assume the deformations to be stationary and occurring in a disc model similar to regions (a) and (b) of Shakura and Sunyaev discs. Results show that the propagation of deformations to the innermost regions of the disc is facilitated for low viscous damping and high accretion rate.

In the final chapter, I dedicate my attention to the influence of a non-negligible basic-state radial inflow on the propagation of waves. I start by analysing perturbations of a one-dimensional flow of an isothermal gas under the action of a potential $\Psi$ which has a maximum where the flow passes from subsonic to supersonic. I determine the reflection of waves at that point where the equations describing the behaviour of linear perturbations proportional to $\exp(-\mathrm{i}\omega t)$ are singular. The implications of the results obtained within this toy model are analysed taking into account the more general case of inertial-acoustic waves propagating in a steady transonic flow. I further investigate how a radial inflow and a sonic point modify the structure of trapped inertial modes and their ``leakage'' from the trapping region to the inner boundary. 

\section{Mathematical analysis of disc oscillations}

As seen previously, the theoretical analysis of oscillations starts with the equations of fluid dynamics. For simplicity and clarity I consider a strictly isothermal disc with a ratio of specific heats $\gamma=1$. Ignoring viscosity and magnetic fields and considering isothermal oscillations, equations describing the flow read,

\be
\frac{\partial\vc{u}}{\partial t}+\vc{u}\cdot\nabla\vc{u}=-\nabla\Phi-\nabla h,
\label{motion1}
\ee
\begin{equation}
\frac{\partial h}{\partial t} +\vc{u}\cdot\nabla h =-c_\mt{s}^2\nabla\cdot\vc{u},
\label{energy1}
\end{equation}
where $h=c_\mt{s}^2\ln\rho$ is the enthalpy. Throughout this part of the thesis, self-gravitation is neglected and a fixed axisymmetric gravitational potential $\Phi(r,z)$ is considered where $(r,\phi,z)$ are cylindrical polar coordinates. (Since, as in previous sections, frequencies are normalized to $c^3/GM$ and radii are normalized to $GM/c^2$, I use the symbol $r$ to represent the radial coordinate.)

As in the case of stars, to study oscillations one needs to analyse what happens to this system of equations when enthalpy and velocity are perturbed, $q=q_0+q'$. Similarly to the analysis of section \ref{transdiscussion}, non-linear terms in the perturbed quantities are ignored. In the equilibrium state, the radial inflow is neglected, $\vc{u_0}=\vc{\Omega}\times\vc{r}=r \Omega(r, z)\vc{e}_\phi$ and $\mathbf{\nabla}h_0=r\Omega^2\mathbf{e}_r-\mathbf{\nabla}\Phi$. Since the basic state is independent of time and azimuth, the perturbations acting on it can be written as

\begin{equation}
q'(r,\phi,z,t)=\textrm{Re}\left[\widetilde{q'}(r,z)\exp(\mt{i}m\phi-\mt{i}\omega t)\right],
\label{perturbations}
\end{equation}
where $m$ is the azimuthal mode number and $\omega$ is the oscillation frequency. Neglecting self-gravitation and dropping, for simplicity, the tildes and zeros, the linearised equations for the perturbed quantities read

\begin{equation}
-\mt{i}\hat{\omega}u'_r - 2\Omega u'_\phi=-\frac{\partial h'}{\partial r},
\label{motion1.1}
\end{equation}
\begin{equation}
-\mt{i}\hat{\omega}u'_\phi +\frac{1}{r}\left(u'_r\frac{\partial}{\partial r} + u'_z\frac{\partial}{\partial z}\right)(r^2\Omega)=-\frac{imh'}{r},
\label{motion1.2}
\end{equation}
\begin{equation}
-\mt{i}\hat{\omega}u'_z =-\frac{\partial h'}{\partial z},
\label{motion1.3}
\end{equation}
\begin{equation}
-\mt{i}\hat{\omega}h'+u'_r\frac{\partial h}{\partial r}+u'_z\frac{\partial h}{\partial z}=-c_\mt{s}^2\left[\frac{1}{r}\frac{\partial (ru'_r)}{\partial r}+\frac{imu'_\phi}{r}+\frac{\partial u'_z}{\partial z}\right],
\label{energy2}
\end{equation}
where $\hat{\omega}=\omega-m\Omega$ is the Doppler-shifted wave frequency.
 
In thin accretion discs, the angular velocity $\Omega(r,z)$ can be regarded as independent of height since the variation of $\Omega$ with $z$ is of order $(H/r)^2$; within this approximation, $\partial h/\partial z=-\Omega_z^2z$. Further neglecting the term $u_r' \partial h /\partial r$ in the last equation, (\ref{motion1.1})--(\ref{energy2}) can be written as

\begin{equation}
-\mathrm{i}\hat{\omega}u'_r - 2\Omega u'_\phi=-\frac{\partial h'}{\partial r},
\label{eq1}
\end{equation}
\begin{equation}
-\mathrm{i}\hat{\omega}u'_\phi +\frac{\kappa^2}{2\Omega}u'_r =-\frac{\textrm{i}mh'}{r},
\end{equation}
\begin{equation}
-\mathrm{i}\hat{\omega}u'_z =-\frac{\partial h'}{\partial z},
\end{equation}
\begin{equation}
-\mathrm{i}\hat{\omega}h'-\Omega_z^2zu'_z=-c_\mathrm{s}^2\left[\frac{1}{r}\frac{\partial (ru'_r)}{\partial r}+\frac{\textrm{i}mu'_\phi}{r}+\frac{\partial u'_z}{\partial z}\right],
\label{eq2}
\end{equation}
where the square of the epicyclic frequency was defined as usual, $\kappa^2=(2\Omega/r)\mt{d}(\Omega r^2)/\mt{d}r$.

Furthermore, it can be assumed that the azimuthal variation of the perturbed quantities is slow compared with the radial and vertical variations so that the terms $imu'_\phi/r$ and $imh'/r$ can be neglected. Another approximation frequently used in the literature is the WKB one: the radial wavelength of perturbed quantities is assumed to be much smaller than the characteristic scale for radial variations of the equilibrium quantities,

\begin{equation}
\lambda\sim\left|\frac{q'}{\partial q'/\partial r}\right|\ll\left|\frac{q}{\partial q/\partial r}\right|\sim r.
\end{equation}
Using this simplification the term $u_r'/r$ can be neglected when compared with $\partial u'_r/\partial r$ and equations can be combined as

\begin{equation}
-\frac{\partial}{\partial r}\left(\frac{1}{\hat{\omega}^2-\kappa^2}\frac{\partial h'}{\partial r}\right)=\mathcal{L}h',
\label{eqcc}
\end{equation}
where
\be
\mathcal{L}=\frac{1}{\hat{\omega}^2}\frac{\partial^2}{\partial z^2}-\frac{\Omega_z^2z}{c_\mt{s}^2\hat{\omega}^2}\frac{\partial}{\partial z}+\frac{1}{c_\mt{s}^2},
\label{liso}
\ee
is a second-order (partial) differential operator in $z$ acting on $h'$ that depends on the azimuthal wave number $m$ only through the dependence in $\hat{\omega}$.

Equation (\ref{eqcc}) is also valid in discs with more general equations of state provided the operator $\mathcal{L}$ takes a different form. In a polytropic disc with general $\gamma$, 

\begin{equation}
\mathcal{L}=\frac{\partial}{\partial z}\left[\frac{1}{\hat{\omega}^2-N_z^2}\left(\frac{\partial}{\partial z}-\frac{N_z^2}{g}\right)\right] -\frac{\rho}{\gamma p}\frac{g}{\hat{\omega}^2-N_z^2}\left[\frac{\partial}{\partial z}-\frac{\hat{\omega}^2}{g}\right].
\label{Lpoly}
\end{equation}
The quantities 

\begin{equation}
N_z^2=g\left(\frac{1}{\gamma p}\frac{dp}{dz}-\frac{1}{\rho}\frac{d\rho}{dz}\right) \;\textrm{ and } \;g=z\Omega_z^2=-\frac{1}{\rho}\frac{dp}{dz},
\end{equation}
are the vertical buoyancy frequency (squared) and gravitational acceleration, respectively. 

Note that (\ref{eqcc}) is (locally) separable in $r$ and $z$ as the $z$-dependent quantities in $\mathcal{L}$ are regarded as independent of the radial coordinate, due to the approximations made before. (The issue of separability is discussed further in the next section.) In what follows, I analyse equation (\ref{eqcc}) in different locations in the disc. Although I focus on isothermal $\gamma=1$ discs, the discussion is kept general and the results obtained are valid also for a polytropic flow [provided $\mathcal{L}$ is given by (\ref{Lpoly})] except where otherwise indicated.

\subsection{Far from a Lindblad resonance --- dispersion relation}
\label{farfrom}

The WKB approximation is valid everywhere except close to the ``turning points'' or singularities of the problem. In the case under analysis, the approximation doesn't work close to the so-called Lindblad and corotation resonances where the equilibrium quantities $\hat{\omega}^2-\kappa^2$ and $\hat{\omega}$ are small, and their radial variation cannot be neglected. The Lindblad resonance is the point where the waves change from a region where they are able to propagate to one where they are evanescent. At this location, the group velocity and wave number are zero. Far from this turning point, the WKB approximation is appropriate and the ansatz

\begin{equation}
h'(r,z)=\widetilde{h'}(r,z)\exp{\left[i\int k(r)dr\right]},
\end{equation}
where the variation of $\widetilde{h'}$ with $r$ is slow, can be used. Equation (\ref{eqcc}) can then be written as

\begin{equation}
k^2\widetilde{h'}=(\hat{\omega}^2-\kappa^2)\mathcal{L}\widetilde{h'},
\label{eqkl}
\end{equation}
which can be effectively considered an ordinary differential equation in $z$ at each $r$, separately. Designating $\Lambda_{n,\hat{\omega}}$ the eigenvalue of the operator $\mathcal{L}$ associated with the eigenfunction $\widetilde{h'}$, 
\begin{equation}
\mathcal{L}\widetilde{h'}=\Lambda_{n,\hat{\omega}}\widetilde{h'},
\label{ev}
\end{equation}
the dispersion relation can be written as $k^2=(\hat{\omega}^2-\kappa^2)\Lambda_{n,\hat{\omega}}$. $\Lambda_{n,\hat{\omega}}$ is a constant or possibly slowly varying function of $r$ and is, in general, unknown and the dispersion relation partially undetermined. The eigenvalue equation was solved numerically by \cite{korycanskypringle1995} in the case of a vertically polytropic disc.

Equation (\ref{eqkl}) has a simple analytic solution, resulting in a dispersion relation independent of $z$, if the disc is locally isothermal in the vertical direction. In the particular case of $\gamma =1$ \citep[the case of general $\gamma$ is treated in][]{lubowpringle1993}, $\mathcal{L}$ takes the form (\ref{liso}) and the equation for $\widetilde{h'}$ reads

\begin{equation}
\frac{d^2\widetilde{h'}}{dz^2}-\mathcal{A}z\frac{d\widetilde{h'}}{dz}+\mathcal{B}\widetilde{h'}=0,
\label{ssw}
\end{equation} 
with $\mathcal{A}=\Omega_z^2/c_\mt{s}^2=1/H^2$ and $\mathcal{B}=\hat{\omega}^2/c_\mt{s}^2-\hat{\omega}^2k^2/(\hat{\omega}^2-\kappa^2)$ constants in $z$. Looking for a solution in terms of a power series and requiring it to be finite, the following dispersion relation is found \citep{okazakietal1987},

\begin{equation}
k^2=\frac{(\hat{\omega}^2-\kappa^2)(\hat{\omega}^2-n\Omega_z^2)}{\hat{\omega}^2c_\mt{s}^2}, \quad n\in\mathbb{N}_0.
\label{ssdr}
\end{equation}
The dispersion relation is local, i.e., it is different at each radius since the characteristic frequencies (\ref{relomega}), (\ref{relkappa}) and (\ref{omegazfreq}) depend on $r$, and is valid only where the wavelength $\lambda=2\pi/k\ll r$.

Requiring the series to be finite is equivalent to requiring\footnote{Note that equation (\ref{ssw}) is similar to the one obtained when solving the Schr\"{o}dinger equation for the 1D harmonic oscillator, after assuming the wave function to be proportional to $f(z)\textrm{e}^{-\textrm{const}\times z^2}$. The requirement used here is equivalent to the one used in the quantum mechanics case for the wave function to be finite.}

\begin{equation}
\left|\exp{\left(-\frac{\mathcal{A}}{4}z^2\right)}\widetilde{h'}(z)\right|\to0 \quad \textrm{as} \quad |z|\to\infty,
\end{equation}
which implies \citep[as noted by][]{okazakietal1987} that the energy density for perturbations is bounded at large distances. The solution of (\ref{ssw}) that obeys this condition is

\begin{equation}
\widetilde{h'}(z)\propto\textrm{He}_n\left(\sqrt{\mathcal{A}}z\right)=\textrm{He}_n\left(\frac{z}{H}\right),
\end{equation} 
where $\textrm{He}_n$ is the modified Hermite polynomial of order $n$ \citep{abramowitzstegun}, with $n=0,1,2,3,\dots$ being the vertical mode number. For odd/even $n$ the mode will have odd/even parity, i.e., $\widetilde{h'}(z)=0$ or $\partial_z\widetilde{h'}(z)=0$ at $z=0$. 

It should be noted that the separation of variables in $r$ and $z$ is not exact since $H$ depends on $r$. [As described by \cite{rkato2001}, this separation is valid to lowest WKB order; \cite{nowakwagoner1992} use a slowly varying function of $r$ to separate variables.] The variation of $H$ with $r$ couples different vertical modes \citep{tanakaetal2002} but this effect is weak when the radial wavelength is short, and is neglected here and in chapter \ref{excmech}.

\subsubsection{Diskoseismic modes and trapping regions}

The simple dispersion relation (\ref{ssdr}) obtained in the isothermal disc $\gamma=1$ approximation already contains very relevant information. Its solutions for different values of $n$ can be analysed as follows:

\begin{itemize}
\item {$\bf{n=0 \quad}$} In this case, the dispersion relation for the non-trivial mode becomes $\hat{\omega}^2=k^2c_\mt{s}^2+\kappa^2$ and waves can propagate where $\hat{\omega}^2>\kappa^2$; this is the inertial-acoustic mode. In non-isothermal discs the closest equivalent of this mode behaves like a surface gravity wave or stellar f~mode \citep{ogilvie1998,lubowogilvie1998} but here it is referred as $n=0$ mode or 2D mode, since it involves a purely horizontal motion independent of $z$.

\item {$\bf{n\neq0 \quad}$} The dispersion relation admits two types of solutions: a high-frequency one with $\hat{\omega}^2>\textrm{max}(\kappa^2,n\Omega_z^2)=n\Omega_z^2$ (p~modes) and a low-frequency one with $\hat{\omega}^2<\textrm{min}(\kappa^2,n\Omega_z^2)=\kappa^2$ (r~modes). Therefore, both classes of waves propagate in different regions in the disc.
\end{itemize}

The inertial or r~modes \citep[so called by][]{korycanskypringle1995} are nearly
incompressible since they are restored by inertial forces, avoiding
acoustic effects.  (They are often called g~modes in the literature,
but they are not related to internal gravity waves or stellar
g~modes.)  Acoustic or p~modes have pressure as their main restoring
force and are essentially compressible.  In an isothermal disc, the 2D
mode is a purely horizontal compressible mode with rotation modifying its otherwise acoustic nature.

The propagation regions of the p and r~modes allow us to define important radii in the disc, the resonant radii \citep[e.g.][]{lubowogilvie1998}. Non-axisymmetric waves can have three types of resonances: corotation where $\hat{\omega}=0$, Lindblad resonances where $\hat{\omega}^2-\kappa^2=0$, and vertical resonances where $\hat{\omega}^2-n\Omega_z^2=0$ (for $n\neq 0$).  Lindblad resonances are turning points for the r and 2D modes, while vertical resonances are turning points for p~modes.

\begin{figure}[!t]
\begin{center}
\includegraphics[width=110mm]{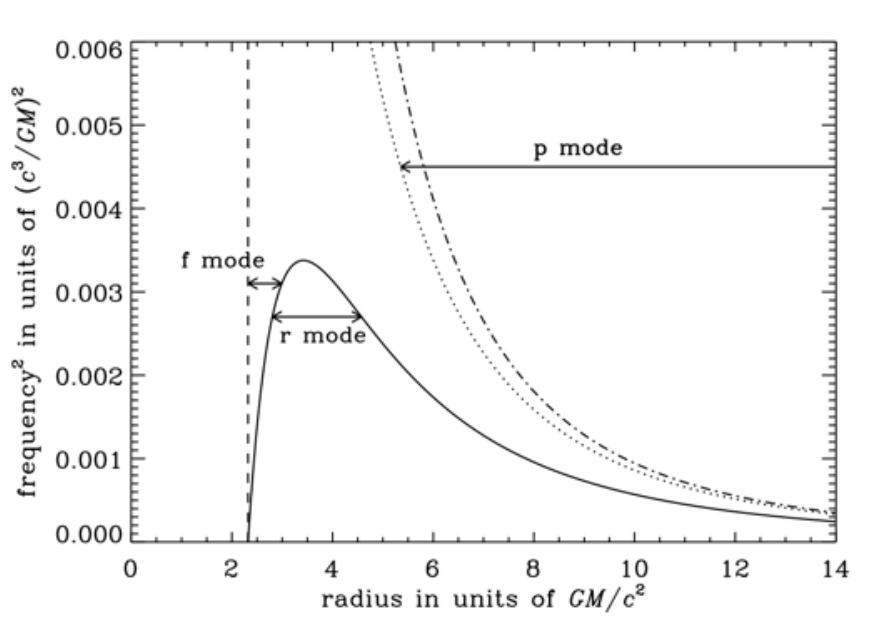}
\caption{Square of the orbital (dash-dotted line), vertical (dotted line) and epicyclic (full line) frequencies for particles orbiting a black hole with spin $a=0.9$. If the accretion disc is thin, the particle orbit expressions for the characteristic frequencies are appropriate, and the three different types of axisymmetric disc modes propagate in the regions indicated by the arrows: f~modes have $n=0$ and $\omega^2<\kappa^2$, while r, p~modes have $n>0$ and $\omega^2<\kappa^2$, $\omega^2>n\Omega_z^2$, respectively. The vertical dashed line represents the marginally stable orbit.}
\label{freqfig}
\end{center}
\end{figure}

The three different types of modes propagate in different regions in the disc (Fig.~\ref{freqfig}). If the epicyclic frequency has a maximum at some particular radius (as it happens in black-hole discs), r and 2D modes can be trapped in the inner part of the disc, while p~modes always propagate to the outer boundary beyond the outer vertical resonance. As mentioned in chapter \ref{intro}, \cite{katofukue1980} originally considered the trapping of 2D modes in the very inner region of the disc. These modes can be trapped between the radius of the marginally stable orbit and the inner Lindblad resonance. However, the conditions at $r_\mathrm{ms}$ are not well understood and it is not clear if this trapping region can work as a resonant cavity. Therefore, in section \ref{close_rs} and chapter \ref{excmech} I focus on the trapping of r~modes, which happens below the maximum of the epicyclic frequency between two Lindblad resonances, a resonant cavity naturally created by the non-monotonic variation of $\kappa$ with radius \citep{okazakietal1987}.

Of particular importance is the axisymmetric trapped wave with frequency $\omega\approx\textrm{max}(\kappa)$, and with the simplest possible radial structure. This mode is trapped in a small region
close to the maximum of the epicyclic frequency and it is of relevance for several reasons. It is naturally confined and therefore more probable to occur in the presence of turbulent viscosity. In addition, its simple structure makes this mode likely to be observed as it may produce a net luminosity variation of the disc without cancellations. Finally, its frequency can be identified with $\max(\kappa)$ which depends only on the properties of the black hole: its mass $M$ and angular momentum $a$. Therefore, by measuring the frequency of this mode and determining the mass of the central object (e.g. through binary orbital dynamics) its spin can be determined.

\subsection{Close to a Lindblad resonance, $\vc{r_\mt{L}}$}
\label{closeLR}

When $r$ is close to $r_\mt{L}$, $\hat{\omega}$ is close to $\kappa$ and therefore the variation of $\hat{\omega}^2-\kappa^2$ cannot be ignored. Therefore, near a Lindblad resonance the WKB approximation is not appropriate. To study this case, it is useful to define a new radial coordinate

\begin{equation}
x=r-r_\mt{L} \ll 1.
\label{xl}
\end{equation}
The quantity $\hat{\omega}^2-\kappa^2$ can be expanded in a Taylor series around $r=r_\mt{L}\Leftrightarrow x=0$. Keeping only linear terms in $x$,

\begin{equation}
\hat{\omega}^2-\kappa^2\approx\mathcal{D}_{1\mt{L}}x,
\label{wkl}
\end{equation}
where $\mathcal{D}_{1\mt{L}} =[d(\hat{\omega}^2-\kappa^2)/dr]_{r=r_\mt{L}}$. Using (\ref{xl}) and (\ref{wkl}) on equation (\ref{eqcc}) and separating variables,

\begin{equation}
-\frac{d}{dx}\left(\frac{1}{x}\frac{dA}{dx}\right)=\Lambda_{n,\hat{\omega}}\mathcal{D}_{1\mt{L}}A,
\end{equation}
where $h'(x,z)=\widetilde{h'}(z)A(x)$ and $\widetilde{h'}$ satisfies (\ref{ev}) or (\ref{ssw}) in the $\gamma=1$ simplified case. Locally, $\Lambda_{n,\hat{\omega}}$ does not depend on $x$ ($\Lambda_{n,\hat{\omega}}\approx\Lambda_{n,\hat{\omega}_\mt{L}}$) and therefore Airy (prime) functions solve this equation. Since $A(x)$ should tend to zero at large $x$ so that the solutions close to the Lindblad resonance and far from it can be asymptotically matched, the solution is

\begin{equation}
A(x)\propto\textrm{Ai}'\left(X\right), \qquad X=\left(-\Lambda_{n,\hat{\omega}_\mt{L}}\mathcal{D}_{1\mt{L}}\right)^{1/3}x.
\end{equation}
Note that the wave is evanescent for $X>0$, and propagates where $X<0$.

\subsection{Close to a stationary point of $\vc{\hat{\omega}^2-\kappa^2}$, $\vc{r_\mt{s}}$}
\label{close_rs}

As mentioned previously, an inertial mode of interest is the one with frequency $\omega\approx\textrm{max}(\kappa+m\Omega)$. This mode is trapped near the maximum of $\kappa+m\Omega$, or equivalently near the stationary point of $\hat{\omega}^2-\kappa^2$, and can avoid being absorbed at the corotation resonance if $m$ is relatively small. In this case, the two Lindblad resonances are very close to each other and therefore a different treatment from the one used before is required. In this case, it is more convenient to combine equations (\ref{eq1})--(\ref{eq2}) into a single equation for $u_r'$ instead of $h'$, 

\begin{equation}
-\frac{\partial^2u_r'}{\partial r^2}=(\hat{\omega}^2-\kappa^2)\mathcal{L}u_r'.
\label{eqccu}
\end{equation}
The two Lindblad resonances are very close to the stationary point of $\hat{\omega}^2-\kappa^2$, $r_\mt{s}$. To study this region it is useful to define a coordinate

\begin{equation}
x=r-r_\mt{s} \ll 1.
\label{xs}
\end{equation}
As before, $\hat{\omega}^2-\kappa^2$ can be expanded in a Taylor series around $r=r_\mt{s}\Leftrightarrow x=0$. However, the linear term in $x$ is now zero by definition of stationary point and, therefore, it is necessary to keep the $\mathcal{O}(x^2)$ term to take some relevant information out of (\ref{eqccu}). The expansion then reads

\begin{equation}
\hat{\omega}^2-\kappa^2\approx\mathcal{C}+\mathcal{D}_{2\mt{s}}x^2,
\label{wks}
\end{equation}
where $\mathcal{D}_{2\mt{s}} =(1/2)[d^2(\hat{\omega}^2-\kappa^2)/dr^2]_{r=r_\mt{s}}$ and $\mathcal{C}=\hat{\omega}_\mt{s}^2-\kappa_\mt{s}^2$ is the value of $\hat{\omega}^2-\kappa^2$ at its stationary point.

Using (\ref{xs}) and (\ref{wks}) on equation (\ref{eqccu}) and separating variables,

\begin{equation}
-\frac{d^2B}{dx^2}=\left(\mathcal{C}+\mathcal{D}_{2\mt{s}}x^2\right)\Lambda_{n,\hat{\omega}}B,
\label{pce}
\end{equation}
where $u_r'(x,z)=\widetilde{h'}(z)B(x)$ and $\widetilde{h'}$ satisfies (\ref{ev}) or (\ref{ssw}) in the simplified case. The solutions to this equation are parabolic cylinder functions \citep{abramowitzstegun} since, locally, $\Lambda_{n,\hat{\omega}}$ does not depend on $x$:

\begin{equation}
\Lambda_{n,\hat{\omega}}\approx\Lambda_{n,\hat{\omega}}(r=r_\mt{s})=\Lambda_{n,\hat{\omega}_\mt{s}}.
\end{equation}
If $\mathcal{C}\Lambda_{n,\hat{\omega}_\mt{s}}>0$ and $\mathcal{D}_{2\mt{s}}\Lambda_{n,\hat{\omega}_\mt{s}}<0$, equation (\ref{pce}) has trapped non-diverging solutions. In this case, the equation can be written in the standard parabolic cylinder form \citep{abramowitzstegun}

\begin{equation}
\frac{d^2B}{dy^2}-\left(\frac{y^2}{4}+b\right)B=0,
\label{pceq}
\end{equation}
with
\begin{equation}
y=x\left(2\sqrt{-\mathcal{D}_{2\mt{s}}\Lambda_{n,\hat{\omega}_\mt{s}}}\right)^{1/2} \qquad \textrm{and} \qquad b=-\frac{\mathcal{C}\Lambda_{n,\hat{\omega}_\mt{s}}}{2\sqrt{-\mathcal{D}_{2\mt{s}}\Lambda_{n,\hat{\omega}_\mt{s}}}}<0.
\end{equation}
Using a boundary condition similar to the one used before, i.e., assuming that the energy density for perturbations is bounded at large distances, the solution of (\ref{pceq}) is

\begin{equation}
B(y)\propto\exp{\left(-\frac{y^2}{4}\right)}\textrm{He}_l(y),
\label{B}
\end{equation} 
where $l=-b-1/2$ is a nonnegative integer.



For every mode trapped near the stationary point of $\hat{\omega}^2-\kappa^2$, (\ref{B}) is valid. Therefore, these waves can be labelled by their three wave numbers $(l,m,n)$, where $l$ is the radial wave number, $m$ the azimuthal number and $n$ the vertical number. An implicit expression for the frequency $\omega=\hat{\omega}(r)+m\Omega(r)=\hat{\omega}_\mt{s}+m\Omega_\mt{s}$ of these modes can be found using the definitions of $l$ and $b$:

\begin{equation}
l=-b-\frac{1}{2}=\frac{\mathcal{C}\Lambda_{n,\hat{\omega}_\mt{s}}}{2\sqrt{-\mathcal{D}_{2\mt{s}}\Lambda_{n,\hat{\omega}_\mt{s}}}}-\frac{1}{2}=\frac{(\hat{\omega}_\mt{s}^2-\kappa_\mt{s}^2)\Lambda_{n,\hat{\omega}_\mt{s}}}{2\sqrt{-\frac{1}{2}[d^2(\hat{\omega}^2-\kappa^2)/dr^2]_{r=r_\mt{s}}\Lambda_{n,\hat{\omega}_\mt{s}}}}-\frac{1}{2}
\end{equation}
In the simple case of an isothermal disc with $\gamma=1$, $\Lambda_{n,\hat{\omega}_\mt{s}}=(\hat{\omega}_\mt{s}^2-n\Omega_{zs}^2)/c_\mt{s}^2\hat{\omega}_\mt{s}^2$, and the frequency of the trapped r~modes with $\hat{\omega}_\mt{s}$ close to $\kappa_\mt{s}$ is given by

\begin{equation}
\left(\frac{\omega-m\Omega_\mt{s}}{\kappa_\mt{s}}\right)^2\approx1-\delta, \qquad \delta=\frac{-\mathcal{C}}{\kappa_\mt{s}^2}=(2l+1)\frac{c_\mt{s}}{\kappa_\mt{s}}\sqrt{-\frac{\mathcal{D}_{2\mt{s}}}{\kappa_\mt{s}^2-n\Omega_{zs}^2}}\ll1.
\label{frequency}
\end{equation}

In chapter \ref{excmech}, I describe an excitation mechanism for inertial trapped waves in the disc by focusing on the simplest mode with $(l,m,n)=(0,0,1)$. The analysis presented here provides insight on the properties of this axisymmetric oscillation. According to equation (\ref{frequency}), this mode has a frequency $\omega^2\approx(1-\delta)\kappa_\mt{s}^2$, where $\delta\propto c_\mt{s}$. Therefore, the larger the sound speed in the disc, the wider is the trapped region and the frequency shift below the maximum of $\kappa$. Furthermore, the simplest possible trapped inertial wave has a Gaussian form centred on $r_\mt{s}$ and horizontal velocities proportional to $z$. 

\subsection{Potential analogy}
\label{potan}

\begin{figure}[!t]
\begin{center}
\includegraphics[width=110mm]{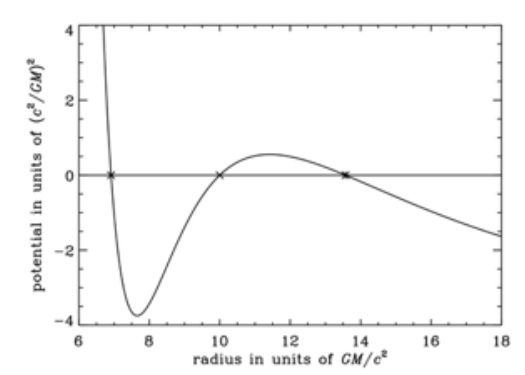}
\caption{Potential $U(r)$ in a disc with $c_\mt{s}=0.01$ and $a=0$ for an axisymmetric, $n=1$ wave of frequency close to $\max{(\kappa)}$. The r~mode is trapped between the two Lindblad resonances (crosses) where the potential is similar to that of a 1D harmonic oscillator. The wave is evanescent between the Lindblad and vertical resonances where a potential barrier is present and propagates as a p~mode outside the vertical resonance (asterisk).}
\label{potextra}
\end{center}
\end{figure}

A very interesting analogy between this problem and a quantum particle/wave being acted by a potential was noted by \cite{lietal2003}. In the isothermal $\gamma=1$ case, this potential is

\begin{equation}
U(r)=-k(r)^2=-\frac{(\hat{\omega}^2-\kappa^2)(\hat{\omega}^2-n\Omega_z^2)}{\hat{\omega}^2c_\mt{s}^2}.
\label{potential}
\end{equation}
Using this description it can be seen that oscillations exist where $U(r)<0$. Otherwise, potential barriers are present and waves are evanescent in agreement with both the quantum mechanical description and the analysis of the dispersion relation done before. The potential for a $m=0,n=1$ wave with frequency close to $\max{(\kappa)}$ is shown in Fig.~\ref{potextra}.

\begin{figure}
\begin{center}
\includegraphics[width=0.49\linewidth]{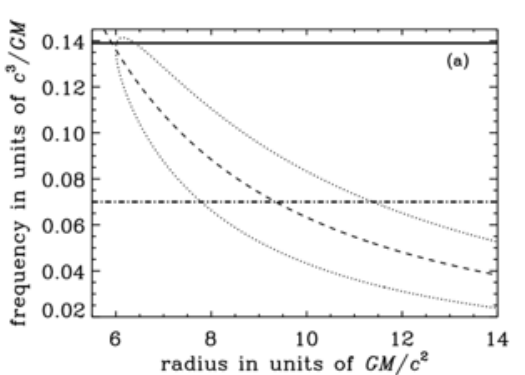}
\includegraphics[width=0.49\linewidth]{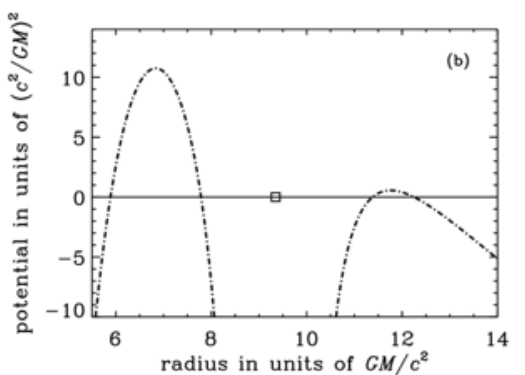}\\
\vspace{5mm}
\includegraphics[width=0.49\linewidth]{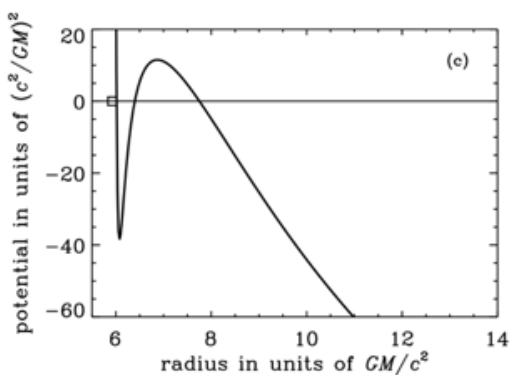}
\includegraphics[width=0.49\linewidth]{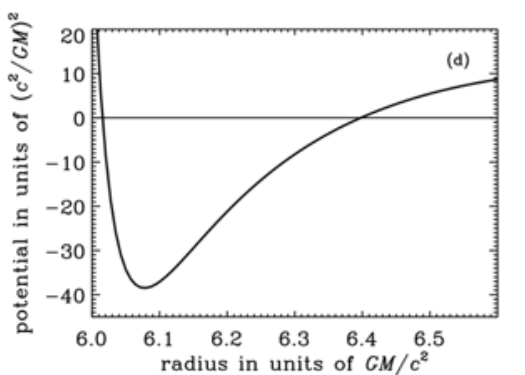}
\caption{ {(a) r~modes propagate between their two Lindblad resonances where $\omega$ intersects the dotted curves, $m\Omega-\kappa$ and $m\Omega+\kappa$, with the exception of the corotation radius $(r_\mt{cr})$, where $\omega$ intersects the dashed curve, $m\Omega$. If the frequency is high enough (full thick line), the corotation resonance is outside the trapping region. (b) If the corotation resonance is not avoided (thick dash-dotted line), the potential is singular ($U(r)\rightarrow-\infty$) at $r_\mt{cr}$ (marked with a square). (c) If the frequency is high enough, the form of the potential is different: there is no singularity for $r>r_\mt{ms}$ ($=6$ for the case $a=0$ represented) since $r_\mt{cr}<r_\mt{ms}$ ($r_\mt{cr}$ again marked by a square). (d) Close-up of the potential well in the case where the corotation resonance is avoided --- the potential in the inner region of the accretion disc is roughly equivalent to the harmonic oscillator one. The modes represented have $m=2$ and $n=1$ and the disc has $c_\mt{s}=0.01$.}}
\label{potentialfig}
\end{center}
\end{figure}

In the above mentioned paper, the authors analyse wave propagation in this potential within the WKB approximation and compare the amplitudes of reflected and incident waves at Lindblad and corotation resonances. Their aim is to see whether or not trapped oscillations can be amplified at these locations. They conclude that $n=0$ modes can be amplified while nonaxisymmetric waves with nodes in the vertical direction (r or p~modes) are strongly damped at corotation. The authors also argue that the growth rates attained by $n=0$ modes are unlikely to be high enough to allow detection.

The conclusions of this paper shouldn't discourage the supporters of diskoseismic modes as candidates for HFQPOs. Even if the waves aren't amplified at corotation, if they aren't strongly damped they can still rely on an external mechanism to be excited. In particular, axisymmetric waves which don't have a corotation resonance in the disc, or nonaxisymmetric inertial modes that are trapped in a region such that corotation is avoided, can still be promising candidates.

In fact, the analysis of the paper does not strictly apply in the case of interest: close to the stationary point of $\hat{\omega}^2-\kappa^2$. In \cite{lietal2003}, the frequency of $n\neq0$ waves is taken to be such that the corotation resonance is in the propagation region [Fig.~\ref{potentialfig} (b)] and the potential is singular at that resonance. However, for $m$ low enough ($m=0,1,2,...$) it is possible to consider a frequency, very close to the stationary point of $\hat{\omega}^2-\kappa^2$ --- or equivalently very close to the maximum of $\kappa+m\Omega$ --- such that corotation is avoided. In other words, the radius $r_\mt{cr}$ where the resonance occurs is not in the accretion disc since $r_\mt{cr}<r_\mt{ms}$ [Fig.~\ref{potentialfig} (c)]. In this case, r~modes can propagate without being absorbed at the corotation resonance; the potential has no singularities in the trapping region.

Interestingly, in the case where the wave frequency is close to the maximum of $\kappa+m\Omega$, the potential in the trapping region is approximately that of a 1D harmonic oscillator potential [Fig.~\ref{potextra} and Fig.~\ref{potentialfig} (d)]. This means that stationary states with wave function

\begin{equation}
\Phi(x)\propto \exp\left({-\frac{\mathcal{K}^2x^2}{4}}\right)\textrm{He}_l(\mathcal{K}x), \quad \mathcal{K}=\textrm{constant}
\label{ho}
\end{equation}
are expected, in agreement with the previous analysis [cf. (\ref{B})].

\section{Global oscillations}
\label{globaloscil}

Low-frequency modes with azimuthal mode number $m=1$ have been widely studied in the context of (quasi-)Keplerian accretion disc theory. As showed by \cite{kato1983,kato1989} these oscillations are global, and have relatively long wavelengths when compared to the other modes described by the local dispersion relation, being the most likely to exist in a turbulent disc. A global $m=1$ mode with one node in the vertical direction ($n=1$) is typically identified with a warp
\citep{papaloizoulin1995}, while $n=0$ modes correspond to eccentric
discs. This is easily seen by focusing on the action of each of these
modes on a ring. The vertical displacement of a $m=1,n=1$ mode is
independent of $z$ and proportional to $\cos(\phi-\textrm{constant})$
at fixed $r$ and $t$, which corresponds to a tilting, as the displacement
\emph{with respect to} the disc plane is different at each azimuthal
angle. In the case of the $n=0$ mode, the radial displacement is the
one that is independent of $z$, and proportional to
$\cos(\phi-\textrm{constant})$ at fixed $r$ and $t$. This means that the
displacement \emph{within} the disc plane varies with $\phi$, creating an
elliptical orbit. 

In this section, I briefly discuss how these global deformation modes can be produced by the presence of a binary companion or by instabilities. In addition, I calculate the variation of disc tilt and disc eccentricity with radius using a simple treatment of warp and eccentric deformations as global modes. Such description is used in chapter \ref{excmech} in the context of mode coupling in the inner disc region.

\subsection{Local wavelength}

Warped or eccentric disturbances of small amplitude can be thought of as linear perturbations of a circular and coplanar disc, as mentioned previously. In the simplest case of a vertically isothermal disc with unit ratio of specific heats, these wave modes obey the dispersion relation calculated before [cf. equation (\ref{ssdr})] with $\omega\approx0$ and $(m,n)=(1,1)$ and $(m,n)=(1,0)$ in the case of the warp and eccentricity, respectively.

\begin{figure}
\begin{center}
\includegraphics[width=0.49\linewidth]{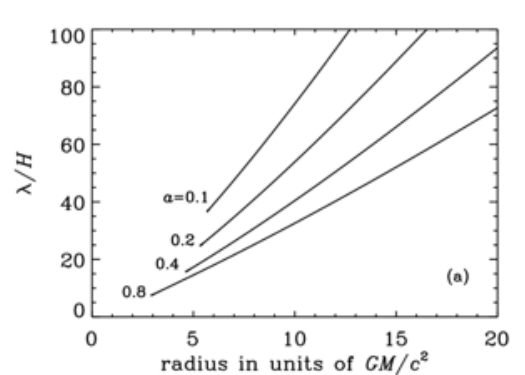}
\includegraphics[width=0.49\linewidth]{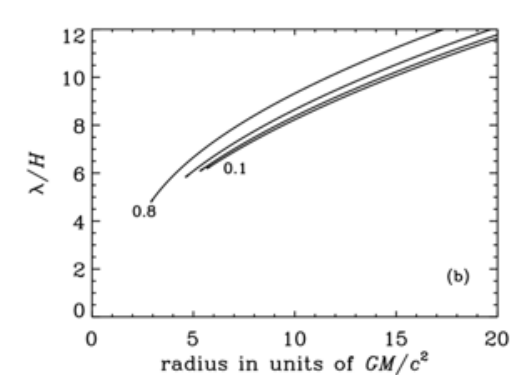}
\caption{Local wavelength of (a) $(\omega,m,n)=(0,1,1)$ and (b) $(0,1,0)$ modes (stationary warp and eccentricity, respectively) in units of $H$ for several values of black-hole spin. The inner edge of the disc is taken to be the marginally stable orbit where $\kappa=0$.}
\label{lambdah}
\end{center}
\end{figure}

The $m=1$ modes are global in the sense that only they can have a wavelength much larger than the semi-thickness of the disc over a wide range of radius. This can be confirmed in Fig.~\ref{lambdah} where I show the variation with $r$ of the local wavelength in units of $H$, for several values of $a$, for both the warp and eccentricity.

\subsection{Warped discs}
\label{warps}

It is believed that warps exist in many astrophysical discs. Precessing warped discs have been used successfully to explain long-term light-curve variations in Her X-1 and some other X-ray binaries \citep{katz1973,gerendboynton1976}. If the central
object is a compact radiation source, warps can be induced by
radiation pressure forces \citep{pringle1996,wijerspringle1999}.  However, this mechanism is less likely to operate in
systems containing a black-hole primary because the disc needs to be
very large \citep{ogilviedubus2001}. The exception is GRS 1915+105 which has a disc extending to large enough radii \citep{rauetal2003}.

If the central object is a rotating black hole, its axis of rotation
might not be perpendicular to the plane of the binary orbit in which
the accretion disc forms. The disc is then said to be
misaligned or tilted. There is both theoretical and observational
evidence for this tilting \citep[see][and references
therein]{globalsimulations}. The misalignment of the orbital angular momentum of the disc
and the spin angular momentum of the black hole results in important
changes in the structure of the inner disc as it will be subject to
Lense-Thirring precession \citep{bardeenpetterson1975}. This
differential precession tends to twist the disc, which may adopt a stationary warped shape. Depending on the ``amount of viscosity'', the induced warps can propagate
either diffusively, roughly speaking if the \cite{ss73} viscosity parameter
$\alpha$ is greater than $H/r$, or in a wave-like manner if $\alpha$
is smaller than $H/r$.  \cite{ivanovillarionov1997} showed that the warp has an oscillatory radial structure in a low-viscosity disc [with wavelength varying with radii as shown in Fig.~\ref{lambdah} (a)], and this was investigated further by \cite{lubowetal2002}.

\subsubsection{Variation of disc tilt with radius}

\begin{figure}
\begin{center}
\includegraphics[width=0.49\linewidth]{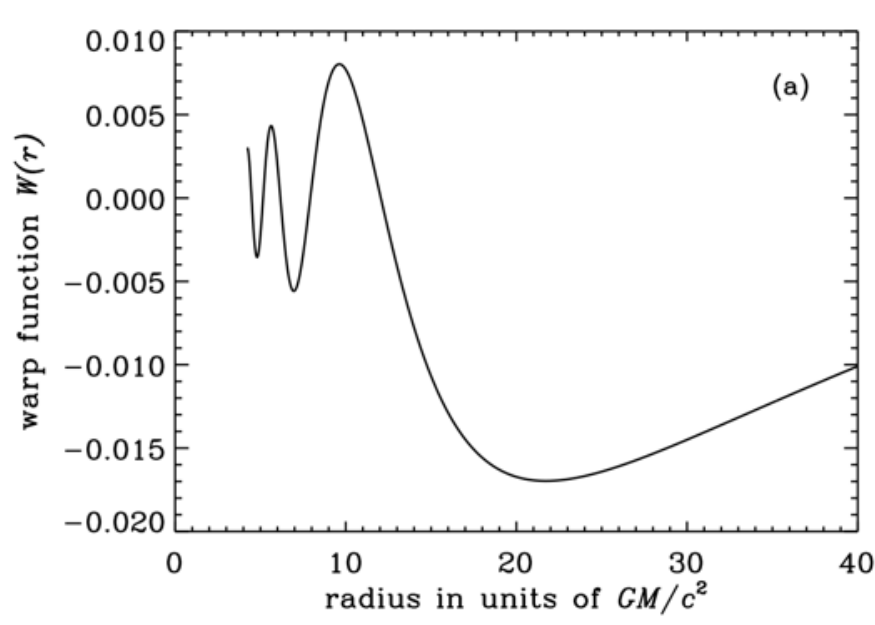}
\includegraphics[width=0.49\linewidth]{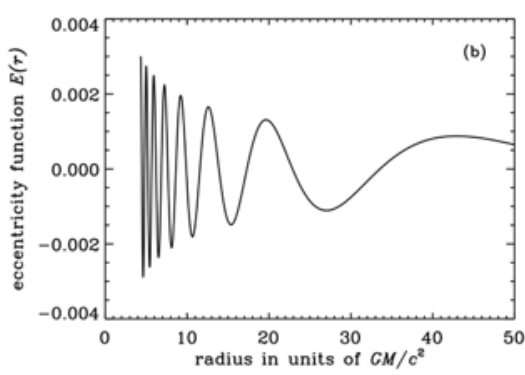}
\caption{(a) Warp function $W(r)$ and (b) eccentricity function $E(r)$ for $a=0.5$. The sound speed is $0.01$ in units of $c$, $W(r_\mathrm{in})=W_0=0.003$, and $E(r_\mathrm{in})=E_0=0.003$.}
\label{eccwarp}
\end{center}
\end{figure}

In a vertically isothermal (pseudo-)relativistic disc with
$\gamma=1$, a zero-frequency mode with $n=1$, $m=1$ can propagate at
all radii if $a>0$ (i.e.~if the disc and black hole rotate in the same
sense).  This can be seen from the dispersion relation
(\ref{ssdr}) and expressions (\ref{relkappa}) and
(\ref{omegazfreq}): in this case $\hat\omega^2=\Omega^2$ is
greater than both $\kappa^2$ and $\Omega_z^2$ and therefore $k^2>0$.
Such a stationary warp can be described by linear perturbations of the
form
\begin{equation}
(u'_{\textrm{W}r},u'_{\textrm{W}\phi},h'_\textrm{W})=\left[u_{\textrm{W}r}(r),u_{\textrm{W}\phi}(r),h_\textrm{W}(r)\right] z,
\label{uprime}
\end{equation}
\begin{equation}
u'_{\textrm{W}z}=u_{\textrm{W}z}(r),
\end{equation}
where the subscript W refers to warp quantities, and the
dependence $\textrm{e}^{\textrm{i}\phi}$ is understood. The simplest
possible warp solution is the rigid tilt, valid for a non-rotating
black hole ($a=0$, $\Omega=\Omega_z$). It is described by $u_{\textrm{W}z}=W\Omega
r$, $u_{\textrm{W}r}=-W\Omega$, $u_{\textrm{W}\phi}=-\mathrm{i}W\textrm{d}(r\Omega)/\textrm{d}r$ and
$h_\textrm{W}=-\mathrm{i}W\Omega^2r$, where $W$ is the constant tilt inclination
[see \cite{papaloizoulin1995} but note that they use $g$ instead of $W$ to represent the disc tilt]. If $W$ varies with $r$, equations (\ref{eq1})--(\ref{eq2}) with $m=1$, $\omega=0$ admit a solution of the
form
\begin{equation}
u'_{\textrm{W}r}=-\Omega Wz+rz\frac{\textrm{d}W}{\textrm{d}r}\frac{\Omega^3}{\Omega^2-\kappa^2},
\end{equation}
\begin{equation}
u'_{\textrm{W}\phi}=-\mathrm{i}Wz\frac{\textrm{d}}{\textrm{d}r}(r\Omega)+\textrm{i}\frac{\Omega\kappa^2}{2(\Omega^2-\kappa^2)}zr\frac{\textrm{d}W}{\textrm{d}r},
\end{equation}
\begin{equation}
u'_{\textrm{W}z}=\Omega r W,
\end{equation}
\begin{equation}
h'_\textrm{W}=-\mathrm{i}\Omega^2Wrz,
\end{equation}
where $W(r)$ is the solution of
\begin{equation}
\frac{\textrm{d}}{\textrm{d}r}\left(\frac{\Omega^2}{\kappa^2-\Omega^2}\frac{\textrm{d}W}{\textrm{d}r}\right)+\frac{1}{r}\frac{\textrm{d}W}{\textrm{d}r}=\frac{\Omega^2-\Omega_z^2}{c_\mathrm{s}^2}W.
\label{eqg}
\end{equation}
This equation\footnote{The above analysis uses the relation
$\kappa^2=4\Omega^2+2r\Omega\,\mathrm{d}\Omega/\mathrm{d}r$, which is
not exactly true of the relativistic expressions because $r^2\Omega$
is not quite the specific angular momentum in relativity. On the other
hand, since the pseudo-relativistic treatment is not fully
self-consistent, if this Newtonian relation is not used here, the
rigid tilt solution is not obtained for $a=0$, contrary to what is
expected physically. Therefore, I choose to use the Newtonian
relation in this simplified treatment of the warp. 
}  is closely related, but
not identical, to equation~(17) of \cite{lubowetal2002}, which was
derived from an analysis of global warps in discs that are not
necessarily isothermal. I solve this equation numerically, using a
4th order Runge--Kutta method with the boundary condition
$\mathrm{d}W/\mathrm{d}r(r_\mathrm{in})=0$, corresponding to zero
torque at the inner edge.  The amplitude of this linear solution may
be fixed by specifying the value $W_0=W(r_\mathrm{in})$ at the inner
boundary, i.e., at the marginally stable orbit; this corresponds to
the (small) inclination of the inner edge of the disc with respect to
the equator of the black hole.  A typical solution is shown in
Fig.~\ref{eccwarp} (a). 

The warp has an oscillatory behaviour, as found by
\cite{ivanovillarionov1997}, with the wavelength increasing with
radius, consistent with the local dispersion relation. This
non-monotonic behaviour of the inclination contrasts with the
Bardeen--Petterson effect \citep{bardeenpetterson1975}, which was
derived using an incorrect equation for the warp. One would normally expect $W(r)$ to tend to a constant value at large
$r$, corresponding to the inclination of the outer part of the disc
with respect to the equator of the black hole.  Unfortunately this is
not true of the approximate equation (\ref{eqg}), which does not hold
accurately at large $r$ because the wavelength becomes comparable to
the radius. However, I'll only use this simplified description of the warp in the next chapter, where the interaction of the warp with waves that propagate in the \emph{inner} disc is considered, so this is not expected to significantly affect the final results.

\subsection{Eccentric discs}
\label{eccdiscs}

Interacting binary stars with mass ratio $q\lesssim0.3$ are believed to
have eccentric accretion discs.  This phenomenon is well documented in
the case of cataclysmic variable stars, where superhumps are observed
during the superoutbursts of the SU UMa class of dwarf novae \citep{pattersonetal2005} and in other
systems of low mass ratio.  In these systems, a resonant interaction
of the orbiting gas with the tidal potential of the companion star
allows a growth of eccentricity \citep{whitehurst1988,lubow1991a,lubow1991b}. Superhumps are also observed in an increasing
number of low-mass X-ray binaries, and systems exhibiting black-hole
HFQPOs are likely to have mass ratios $q\lesssim0.3$ and therefore to have
eccentric discs during at least some phases of their outbursts.

\subsubsection{Variation of eccentricity with radius}
 
As before, consider the set of equations (\ref{eq1})--(\ref{eq2}). A global eccentric mode corresponds to a zero-frequency wave with $m=1$ and $n=0$.  (If the global eccentric mode precesses freely, the frequency is not exactly zero but is completely negligible compared to the characteristic frequencies in the inner part of the disc.) In this case, variables are independent of $z$ and equations (\ref{eq1})--(\ref{eq2}) are reduced to
\begin{equation}
\textrm{i}\Omega u_{\textrm{E}r}-2\Omega u_{\textrm{E}\phi}=-\frac{\textrm{d}h_\textrm{E}}{\textrm{d}r},
\end{equation}
\begin{equation}
\textrm{i}\Omega u_{\textrm{E}\phi}+\frac{\kappa^2}{2\Omega}=-\textrm{i}\frac{h_\textrm{E}}{r},
\end{equation}
\begin{equation}
\textrm{i}\Omega u_{\textrm{E}z}=0,
\end{equation}
\begin{equation}
\textrm{i}\Omega h_\textrm{E}=-c_\mathrm{s}^2\left[\frac{1}{r}\frac{\textrm{d}}{\textrm{d}r}(ru_{\textrm{E}r})+\textrm{i}\frac{u_{\textrm{E}\phi}}{r}\right],
\end{equation}
where the subscript E refers to eccentric mode quantities. This system admits a solution of the form
\begin{equation}
u_{\textrm{E}r}=\textrm{i}E\Omega r,
\end{equation}
\begin{equation}
u_{\textrm{E}\phi}=\frac{c_\mathrm{s}^2 r^2 \Omega}{\Omega^2 r^2 - c_\mathrm{s}^2}\frac{\textrm{d}E}{\textrm{d}r}-\frac{\kappa^2}{2}rE,
\end{equation}
\begin{equation}
u_{\textrm{E}z}=0,
\end{equation}
\begin{equation}
h_\textrm{E}=-\frac{c_\mathrm{s}^2r^2\Omega}{\Omega^2r^2-c_\mathrm{s}^2}\frac{\textrm{d}E}{\textrm{d}r},
\end{equation}
where $E(r)$ is the eccentricity of the disc at radius $r$, and satisfies
\begin{equation}
(\kappa^2-\Omega^2)E=\frac{1}{r^3}\frac{\textrm{d}}{\textrm{d}r}\left(\frac{r^5c_\mathrm{s}^2\Omega^2}{\Omega^2r^2-c_\mathrm{s}^2}\frac{\textrm{d}E}{\textrm{d}r}\right).
\label{eqe}
\end{equation}
Again, this equation is closely related, but not
identical, to equation~(21) of \cite{goodchildogilvie2006},
which was derived from an analysis of global eccentricity in a two-dimensional disc.  Radially propagating solutions are obtained because $\kappa^2<\Omega^2$ in a relativistic disc.
As for the warp tilt $W(r)$, I solve this equation numerically, using a 4th order Runge--Kutta method with boundary conditions $E(r_\mathrm{in})=E_0$, and $\textrm{d}E/\textrm{d}r(r_\mathrm{in})=0$, where $E_0$ is an arbitrary value for the eccentricity at the inner boundary, i.e., at the marginally stable orbit. A typical solution for $E(r)$ is shown in Fig.~\ref{eccwarp} (b). The eccentricity has an oscillatory
behaviour, with the wavelength decreasing with radius, consistent with the local
dispersion relation.
 
 \section{Summary}

In this chapter, I introduced the topic of oscillations in black-hole accretion discs. A dispersion relation for diskoseismic modes in a simple isothermal disc with $\gamma=1$ was derived and the properties and propagation regions of the different types of oscillations represented by this dispersion relation were analysed. This treatment is useful to better understand the trapping of inertial modes; an excitation mechanism for these waves is presented in the next chapter. The mechanism relies on the presence of global deformations in the disc. These were introduced in section \ref{globaloscil} where warping and eccentricity disturbances were treated as global $m=1$ oscillations. In the next chapter, the expressions for $(u_{\mt{W}r},u_{\mt{W}\phi},u_{\mt{W}z},h_\mt{W})$ and $(u_{\mt{E}r},u_{\mt{E}\phi},u_{\mt{E}z},h_\mt{E})$ determined here are used in the equations representing the coupling mechanism between inertial modes and these global deformations. A more realistic treatment of the propagation of warp and eccentricity into the inner part of the disc is deferred to chapter \ref{warpecc}.

\thispagestyle{empty}
\chapter{Excitation Mechanism}
\label{excmech}


\section{Introduction}

The theoretical study of periodic variability in black-hole accretion discs is particularly relevant in the context of high-frequency quasi-periodic oscillations detected in several X-ray binaries. The characteristics of such variability phenomena suggest a connection to the inner accretion flow \citep{remimcclin2006} and these oscillations are thought of as fundamental tools to the study of strong gravitational fields. Their relativistic properties indicate that their frequencies may depend only on the mass and spin of the central object. A model capable of explaining HFQPOs and relating their frequencies to the characteristics of the black hole would, therefore, be a key method for spin measurement. 

As emphasised in the introduction of this thesis, diskoseismic modes are promising candidates to explain HFQPOs. In this chapter, I focus on the theory that these oscillations can be identified with inertial modes trapped below the maximum of the epicyclic frequency \citep[e.g.][]{nowaketal1997,rkato2001}. For trapped oscillations to explain HFQPOs, an excitation mechanism for these modes is required, as their amplitudes need to reach values high enough to allow detection. A possibility can reside in the interaction between waves in the disc and a global deformation (warping or eccentricity), as suggested by \cite{katowarp2004,kato2008}.

Kato's calculations are of interest but, as mentioned in section \ref{disko}, they have many uncertainties. One of the main inconsistencies relates to the prediction of growth of the inertial modes in a warped disc around a non-rotating black hole. In this case, the metric is spherically symmetric and the ``warp'' is only a rigid tilt, which means that nothing changes in the disc except its inclination angle. If there are no relevant changes in the disc, there should be no relevant changes in the oscillations, i.e., a rigid tilt should not work to excite diskoseismic modes. This property is not regarded in Kato's results.

In this chapter, Kato's ideas on this coupling mechanism are developed and generalised for discs around rotating black holes. I use the simple dynamical treatment of the warp and eccentricity presented previously and make numerical calculations of both coupled and uncoupled inertial modes. In the case where inertial modes couple with global deformations and are excited as a result, the variation of the growth rate with various parameters, namely black-hole rotation, is calculated.

In section \ref{trapping} I review the trapping of inertial oscillations in a simple, pseudo-relativistic disc model and make numerical calculations of uncoupled r~modes. In section \ref{coupling} I describe the excitation mechanism for trapped inertial modes, which relies on a coupling between waves in the disc and global deformations. In section \ref{grdisc} I discuss the dependence of the inertial modes' growth rates on disc parameters and black hole spin. Conclusions are presented in section \ref{grconc}.

\section{Trapped inertial oscillations}
\label{trapping}

\subsection{Linearized equations and wave modes}

As seen in the previous chapter, the trapping of oscillations can be easily understood by analysing the
fluid equations in a simple isothermal disc model \citep{lubowpringle1993,rkato2001}. 
In this chapter, I use the same disc model as in section \ref{globaloscil}. As argued before, the most important relativistic effects on wave propagation can be included by using the relativistic expressions for the characteristic frequencies presented in equations (\ref{relomega}), (\ref{relkappa}) and (\ref{omegazfreq}).

Here I start by considering the set of equations (\ref{eq1})--(\ref{eq2}). Variables can be further separated in $r$ and $z$, using (as in section \ref{farfrom})

\begin{equation}
(u'_{r},u'_{\phi},h')=\left[u_{r}(r),u_{\phi}(r),h(r)\right]\textrm{He}_n\left(\frac{z}{H}\right),
\end{equation}
\begin{equation}
u'_{z}=u_z(r)\textrm{He}_{n-1}\left(\frac{z}{H}\right).
\end{equation}
(Since $\textrm{He}_{-1}$ is not defined, $u'_z=0$ for $n=0$.) As noted in section \ref{farfrom},  this separation of variables is not exact since $H$ depends on $r$ but is valid to lowest WKB order. The final set of ordinary differential equations in $r$ for the perturbed quantities reads

\begin{equation}
-\mathrm{i}\hat{\omega}u_r - 2\Omega u_\phi=-\frac{\mathrm{d}h}{\mathrm{d}r},
\label{free1}
\end{equation}
\begin{equation}
-\mathrm{i}\hat{\omega}u_\phi +\frac{\kappa^2}{2\Omega}u_r =-\frac{\mathrm{i}mh}{r},
\end{equation}
\begin{equation}
-\mathrm{i}\hat{\omega}u_z =-n\frac{h}{H},
\end{equation}
\begin{equation}
-\mathrm{i}\hat{\omega}h-\Omega_z^2Hu_z=-c_\mathrm{s}^2\left[\frac{1}{r}\frac{\mathrm{d} (ru_r)}{\mathrm{d} r}+\frac{\mathrm{i}mu_\phi}{r}\right].
\label{free2}
\end{equation}
These equations will be solved numerically for an axisymmetric, $n=1$ inertial mode in the next section.

As seen before, the dispersion relation for wave modes in the disc can be determined by further assuming that the radial wavelength of the perturbed quantities is much smaller than both the azimuthal wavelength and the characteristic scale for radial variations of the equilibrium quantities. It can then be verified that perturbations with local radial wavenumber $k$ obey the dispersion relation (\ref{ssdr}).

\subsection{Numerical calculation of trapped r~modes}

To find the radial structure of trapped modes it is necessary to solve the system of equations (\ref{free1})--(\ref{free2}) subject to appropriate boundary conditions. Numerical calculations of waves
trapped near the maximum of the epicyclic frequency were first performed by \cite{okazakietal1987}. Here I focus on the simplest possible trapped inertial modes, with $m=0$ and $n=1$. The approximate
analysis of equations (\ref{free1})--(\ref{free2}) close to the maximum of the epicyclic frequency performed in the previous chapter shows that, between the two Lindblad resonances, these solutions are described by parabolic cylinder functions. These involve Hermite polynomials of order $l=0,1,2,\dots$, centred at the maximum of $\kappa$, like the solutions of the quantum harmonic oscillator \citep[][see also section \ref{potan}]{perezetal1997}\footnote{It should be noted that \cite{perezetal1997} use $n$ as the radial mode number. Here the three quantum numbers are $(l,m,n)$ corresponding to the three coordinates $(r,\phi,z)$.}. The lowest order mode ($l=0$) has a Gaussian structure in $r$. In this section I solve the same problem using numerical methods and fewer approximations, as a prelude to an analysis of the
non-linear mode couplings that cause these modes to grow.

Since trapped modes are only expected for some discrete values of the oscillation frequency, I solve the set of equations as a generalized eigenvalue problem, of the form
\begin{equation}
\mathbf{A}\mathbf{U}=-\textrm{i}\omega\mathbf{B}\mathbf{U}.
\end{equation}
Here $\mathbf{U}$ is the column vector whose components are the r~mode quantities $(u_r,u_\phi,u_z,h)$ evaluated at a set of discrete points, $\mathbf{A}$ is the matrix representing the system of equations (\ref{free1})--(\ref{free2}), and $\mathbf{B}$ can be different from the identity matrix depending on the boundary conditions used. To solve this problem numerically, I apply a pseudo-spectral method with Chebyshev polynomials. I use a Gauss--Lobatto grid, $x(i)=\textrm{cos}(\pi i /N)$, where $N$ is the number of grid points, and the Chebyshev coordinate $-1 < x < 1$ is related to the radial coordinate by
\begin{equation}
r=-x\frac{r_\mathrm{out}-r_\mathrm{in}}{2}+\frac{r_\mathrm{out}+r_\mathrm{in}}{2}.
\end{equation}
Therefore, $r_\mathrm{in} < r < r_\mathrm{out}$, where $r_\mathrm{in}=r_\mathrm{ms}$ and $r_\mathrm{out}$ is an outer radius chosen to be larger than the outer Lindblad resonance for the r~mode. The representation of the first derivatives in the matrix $\mathbf{A}$ is achieved using the Chebyshev collocation derivative matrix, defined by \citep{boydbook}
\begin{equation}
D_{ij}=\begin{cases}
(1+2N^2)/6 & i=j=0 \\
-(1+2N^2)/6 & i=j=N \\ 
-x_j/[2(1-x_j^2)] & i=j, \, j\neq0, N \\
(-1)^{i+j}p_i/[p_j(x_i-x_j)] & i\neq j 
\end{cases}, 
\end{equation}
where $p_0=p_N=2$, and $p_j=1$ otherwise. The generalized eigenvalues and eigenfunctions of $\mathbf{A}$ are then calculated numerically, using \textsc{idl}'s eigenvalue solver, \textsc{la eigenproblem}, which uses a \textsc{qr} decomposition, and is based on \textsc{lapack} routines. The boundary conditions used are the following:
\begin{itemize}
\item {\textbf{At $\mathbf{r_\mathrm{in}}$}}$\quad$ $u_{r}=0$. According to the dispersion relation, the r~mode is expected to be exponentially decaying for radii smaller than its innermost Lindblad resonance and therefore its velocity is supposed to be approximately zero at the marginally stable orbit, which justifies the choice of this boundary condition. A more realistic analysis of the behaviour of this mode at the inner boundary is done in chapter \ref{reflect}.
\vspace{5pt}
\item {\textbf{At $\mathbf{r_\mathrm{out}}$}}$\quad$ $\textrm{d}u_{r}/\textrm{d}r=\textrm{i}ku_{r}$, where $k$ is given by the dispersion relation (\ref{ssdr}) for $m=0$, $n=1$ at $r_\mathrm{out}$, and using $\omega\approx\textrm{max}(\kappa)$. Since the r~mode can propagate again as a p~mode outside the vertical resonance (Fig.~\ref{potextra}), I choose an outer radius beyond this location, so that the mode is oscillatory there, i.e., $k$ is real. In addition, the outgoing wave solution is selected by choosing $k>0$ so that the group velocity of the waves at $r_\mathrm{out}$ is positive. This condition allows the wave to lose energy through the outer boundary and minimizes artificial wave reflection there.
\end{itemize}

The frequency $\omega$ is possibly complex, in which case its imaginary part is the growth rate of the disturbance. In fact, in the absence of non-linear mode couplings, slowly decaying solutions with $\mathrm{Im}(\omega)<0$ are obtained as a result of the outgoing-wave outer boundary condition.

\begin{figure}
\begin{center}
\includegraphics[width=0.49\linewidth]{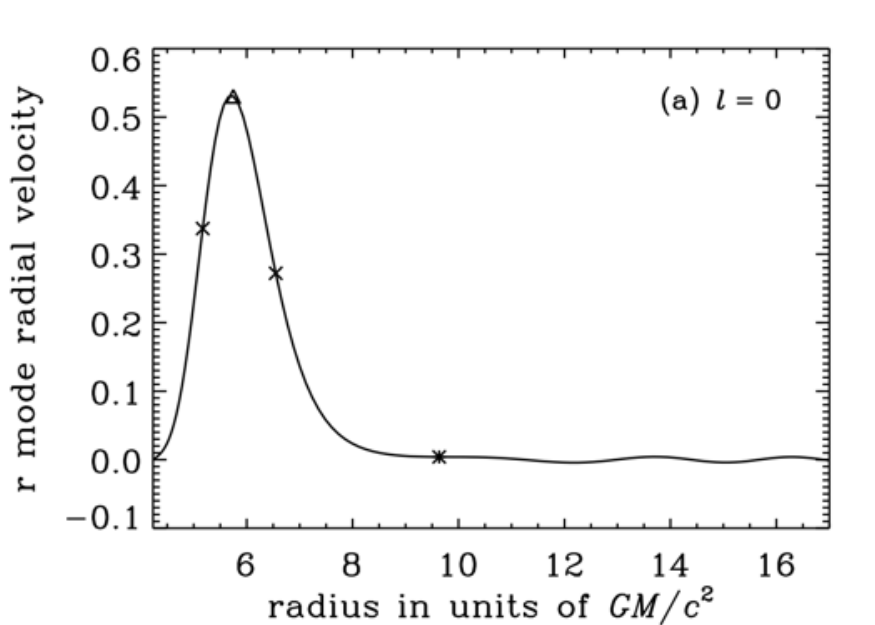}
\includegraphics[width=0.49\linewidth]{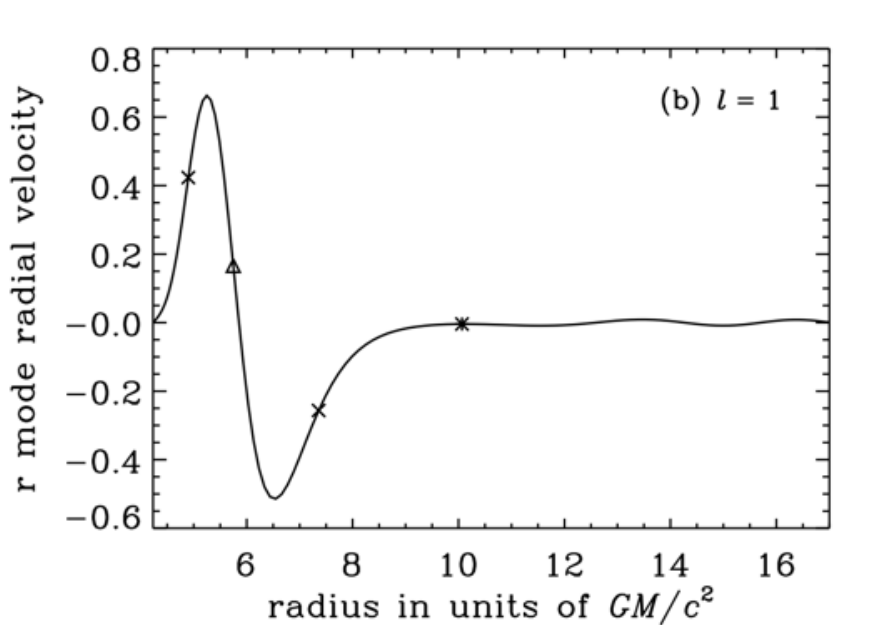}
\caption{Variation of the real part of the radial component of the axisymmetric, $n=1$ r~mode velocity with radius for $c_\mathrm{s}/c=0.01$ and $a=0.5$ for (a) $l=0$ and (b) $l=1$. The triangle indicates the radius where the epicyclic frequency is maximum, and the crosses and asterisks the Lindblad and vertical resonances, respectively.}
\label{freerm}
\end{center}
\end{figure}

Fig.~\ref{freerm} shows the variation of $u_r$ with radius for two typical trapped solutions, corresponding to two different radial mode numbers $l=0$ and $l=1$. (Since I'm solving an eigenvalue
problem, solutions are multiplied by an arbitrary amplitude.) The complex frequencies of the modes represented in Fig.~\ref{freerm} are $0.03196-3.6884\times10^ {-7}\textrm{i}$ and
$0.02989-1.1846\times10^ {-7}\textrm{i}$ (in units of $c^3/GM$), respectively. In these units, and for the value of $a$ used, the maximum of $\kappa$ is 0.03312. These modes are slightly damped because of the boundary condition used at the outer radius, which selects the outgoing wave only. If the sound speed is smaller, the frequency of the modes is closer to the value of the epicyclic frequency at its maximum, and the trapping region and damping rate are smaller (see Table \ref{tablefree}). Very similar results are obtained when the two-point boundary-value problem is solved using a shooting method.

\begin{table}
\begin{center}
\begin{tabular}{ccc}
\hline
$c_\mathrm{s}/c$ & & frequency \\
\hline
$0.002$ & & $0.03289+0.0\textrm{i}$ \\
$0.005$ & & $0.03254-1.276 \times 10^{-10}\textrm{i}$ \\
$0.01$ & & $0.03196-3.688 \times 10^{-7}\textrm{i}$ \\
$0.02$ & & $0.03079-2.417 \times 10^{-5}\textrm{i}$ \\
\hline
\end{tabular} 
\caption{Dependence of the real and imaginary part of the $l=0$ r~mode frequency (in units of $c^3/GM$) on the sound speed $c_\mathrm{s}$, for $a=0.5$. Values obtained with the shooting method are very similar to the ones shown. I use 150 collocation points and a value of 18.2331 for the outer radius, in units of $GM/c^2$ ($r_\mathrm{in}=r_\mathrm{ms}=4.2331$).}
\label{tablefree}
\end{center}
\end{table} 

As seen in section \ref{potan}, these inertial modes can be thought of as waves trapped in a potential, $U(r)=-k(r)^2$ \citep{lietal2003}, which for a frequency close to the maximum of $\kappa$ is similar to the harmonic oscillator potential. If $U(r)<0$ waves can propagate, being evanescent in the regions where potential barriers exist. Also, as in quantum mechanics, these trapped inertial waves can escape through the potential barriers and propagate on the other side, as p~modes (Fig.~\ref{potextra}). The inertial modes are evanescent between the inner radius and the first Lindblad resonance and between the second Lindblad resonance and the vertical one. The ``leakage''  through the potential barrier is also verified, as the results show small-amplitude oscillations after the vertical resonance (Fig.~\ref{freerm}). The larger the sound speed, the larger the width and smaller the height of the barrier, and more ``leakage''  through the barrier is expected, which is verified numerically (Table \ref{tablefree}). This ``leakage''  was first predicted by \cite{okazakietal1987}.

As described above, the inertial mode characterized by $(l,m,n)=(0,0,1)$ is likely to be relevant to the interpretation of observed oscillations. In the next section I describe an excitation mechanism for this mode, based on non-linear wave coupling. If the modes are not strongly coupled, the mechanism is expected to mainly affect the growth rate $\mathrm{Im}(\omega)$, so that the structure of
the wave still resembles a Gaussian centred at the maximum of $\kappa$, as shown in Fig.~\ref{freerm} (a).

\section{Growth of oscillations in deformed discs}
\label{coupling}

\subsection{Warped discs}

In this section I describe a mechanism, first suggested by \cite{katowarp2004}, where a $(m=1,n=1)$ warp wave couples with the simplest trapped inertial mode. The warp is described by the quantities $(u_{\mt{W}r},u_{\mt{W}\phi},u_{\mt{W}z},h_\mt{W})$ determined in section \ref{warps}. 

\subsubsection{Coupling mechanism}

\begin{figure}
\begin{center}
\includegraphics[width=0.35\linewidth]{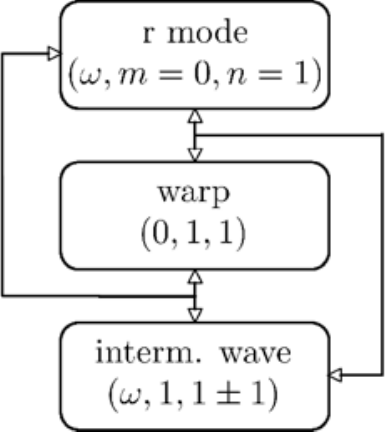}
\caption{Diagram representing the interactions involved in the coupling mechanism involving a warped disc.}
\label{diagram}
\end{center}
\end{figure}

The non-linearities in the basic equations (\ref{motion1}) and (\ref{energy1}) provide couplings between the different linear modes of the system. The interest goes to those couplings that lead to amplification of the trapped modes. The basic idea of the excitation mechanism in warped discs \citep{katowarp2004} is that the warp interacts with an oscillation in the disc (the trapped r~mode) giving rise to an intermediate wave. This intermediate wave can then couple with the warp to feed back on the original oscillations (see Fig.~\ref{diagram}), resulting in growth of the latter.

For the r~mode to be excited, it needs to gain energy in this coupling. Since the warp has null frequency, its energy is essentially zero and so the energy exchanges only happen between the r~mode and the intermediate wave and the disc. It is widely agreed, and certainly true in the short-wavelength limit,
although a general proof is lacking, that a wave that propagates inside its corotation radius has negative energy. (As explained in chapter \ref{intro}, the wave energy is negative in the sense that the total energy of the disc is reduced in the presence of the oscillation; this is possible because the disc is rotating.) On the other hand an axisymmetric wave such as the r~mode, or one
that propagates outside its corotation radius, has positive energy. Suppose that, through coupling with the warp, the r~mode generates an intermediate wave that propagates inside its corotation radius and therefore has negative energy. In the process of generating this wave, the r~mode gains energy and is amplified.  For sustained growth of the r~mode, the intermediate wave must be damped so that its negative energy is continually replenished by the r~mode. (The damping process itself draws positive energy from the rotation of the disc.) Therefore a dissipation term should be included in the equations for
the intermediate wave. Here I choose to damp this wave locally at a rate $\beta\Omega$, where $\beta$ is a dimensionless parameter. The origin of this term is not discussed here but if interpreted as some type of viscous dissipation or friction in the disc, the intermediate wave, which propagates in a larger region in the disc, is expected to be more affected by it than the r~mode, as the latter is trapped in a small region, and has a simpler radial structure. Also, the intermediate wave approaches its corotation radius (or $r_\mt{ms}$), where it is expected to be absorbed, and this effect is implicitly included in the intermediate wave equations when the friction term is included. Therefore, I neglect the dissipation term in the equations for the r~mode. The growth rate obtained for the trapped mode should be compared with estimates of its damping rate due to turbulent viscosity or to a non-negligible radial inflow (chapter \ref{reflect}).

For the coupling to occur the waves need to propagate in the same region in the disc and the parameters $\omega$ and $m$ for the 3 oscillations need to follow some basic coupling rules,
\begin{equation}
\omega_\textrm{R} \pm \omega_\textrm{W}=\omega_\textrm{I}, \quad m_\textrm{R} \pm m_\textrm{W}=m_\textrm{I},
\label{crules}
\end{equation}
where the subscripts R, W and I refer to r~mode, warp and intermediate wave quantities, respectively.  These rules follow from the quadratic nature of the non-linearitites in the basic equations (\ref{motion1}) and (\ref{energy1}).  Also, it is simple to get information about the vertical mode number $n_\textrm{I}$ of the intermediate wave by remembering that the vertical dependence is given by Hermite polynomials. The warp quantities are proportional to $z$ and the simplest possible r~mode has one node in the vertical direction, therefore its quantities are proportional to $\textrm{He}_1\sim z$. When the warp and this r~mode interact, the coupling terms will be proportional to $z^2/H^2=(z^2/H^2-1)+1\sim\textrm{He}_2+\textrm{He}_0$, i.e., they give rise to two intermediate waves, one with 2 nodes in the vertical direction, $n_\textrm{I}=2$, and a 2D wave with $n_\textrm{I}=0$. I consider both possibilities.  Note that the frequency of these intermediate waves is that of the r~mode, and they are present because they are forced through the couplings.  They could not exist as free waves satisfying the boundary conditions at this frequency. \enlargethispage{\baselineskip}

\begin{figure}[!t]
\begin{center}
\includegraphics[width=110mm]{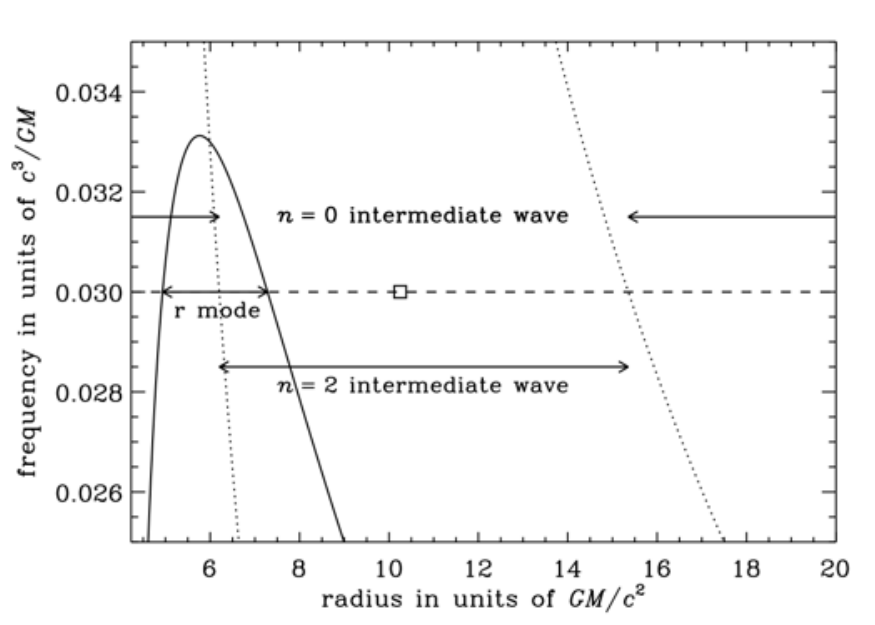}
\caption{Propagation regions of the waves participating in the interaction with a warp when $a=0.5$ and the frequency is chosen to be $\omega=0.03$ (dashed line). The axisymmetric r~mode is trapped in between its two Lindblad resonances where the frequency equals $\kappa$ (full curve). The $m=1,n=0$ intermediate wave propagates where $\omega<\Omega-\kappa$ and $\omega>\Omega+\kappa$ where the dotted lines indicate the curves $\Omega\pm\kappa$. The $m=1,n=2$ intermediate wave propagates where $\Omega-\kappa<\omega<\Omega+\kappa$ with the exception of its corotation resonance indicated by a square. Note that, for the purpose of this diagram, I assumed that the propagation regions are those of uncoupled waves.}
\label{propr}
\end{center}
\end{figure}

According to rules (\ref{crules}), if $\omega_\textrm{R}=\omega$, then $\omega_\textrm{I}=\omega$. In the case of the azimuthal mode numbers, $m_\textrm{I}=m_\textrm{R}\pm1$. By analysing the propagation regions for these waves, and considering $\omega\approx \textrm{max}(\kappa+m\Omega)$, it is possible to conclude that the $n=0$ mode with $m_\textrm{I}=m_\textrm{R}-1$ and frequency $\omega$ does not have any Lindblad resonance, i.e., the point where the wave should be excited, in the disc. On the other hand, the mode with $m_\textrm{R}+1$ and frequency $\omega$ has an inner Lindblad resonance close to the region of propagation of the r~mode (Fig.~\ref{propr}). For these reasons, I choose the azimuthal mode number of the intermediate wave to be $m_\textrm{R}+1=1$, if the r~mode is axisymmetric. This mode propagates inside its corotation resonance, while the r~mode has positive energy. The presence of an inner Lindblad resonance means both that the intermediate wave attains a larger amplitude than it would in the case of non-resonant forcing, and that the flow of energy is such as to amplify the r~mode.

If the vertical mode number of the intermediate wave is chosen to be 2 instead of 0, the r~mode with $(\omega,m,n)=(\omega,0,1)$ interacts with a $(\omega,1,2)$ intermediate wave. The former propagates where $\omega^2<\kappa^2$, while the latter propagates where $(\omega-\Omega)^2<\min(\kappa^2,2\Omega_z^2)$, where $\omega$ is slightly less than $\max{(\kappa)}$. The propagation regions overlap close to the maximum of the epicyclic frequency since $\Omega\approx2\kappa$ in this region (Fig.~\ref{propr}).
 
The interaction of the r~mode with the $n=2$ intermediate wave in a warped disc must be treated carefully because in this case the intermediate wave propagates between its Lindblad resonances and is absorbed at the corotation resonance. This is a radius in the disc where the potential $U(r)=-k^2$ tends to infinity and is therefore difficult to treat numerically because the wavelength tends to zero. One way of solving this problem is by including a relatively strong dissipation term in the equations for the intermediate wave. In this way the wave excited at the inner Lindblad resonance is damped before reaching corotation. This also works for the energy exchanges between the oscillations and the disc, since the $n=2$ wave has negative energy in the region where it is damped.

\subsubsection{Equations}

The following system of equations [cf. equations (\ref{motion1})--(\ref{energy1})] describes the propagation of the r~mode and $n=0$ intermediate wave, coupled by the warp, and needs to be solved for the growth rate to be determined:
\begin{equation}
\left(\frac{\partial}{\partial t}+\Omega\frac{\partial}{\partial \phi}\right)u'_{\textrm{R}r}-2\Omega u'_{\textrm{R}\phi}=-\frac{\partial h'_\textrm{R}}{\partial r}+f_{\textrm{R}r},
\label{wqpo1}
\end{equation}
\begin{equation}
\left(\frac{\partial}{\partial t}+\Omega\frac{\partial}{\partial \phi}\right)u'_{\textrm{R}\phi}+\frac{\kappa^2}{2\Omega}u'_{\textrm{R}r}=-\frac{1}{r}\frac{\partial h'_\textrm{R}}{\partial \phi}+f_{\textrm{R}\phi},
\end{equation}
\begin{equation}
\left(\frac{\partial}{\partial t}+\Omega\frac{\partial}{\partial \phi}\right)u'_{\textrm{R}z}=-\frac{\partial h'_\textrm{R}}{\partial z}+f_{\textrm{R}z},
\end{equation}
\begin{equation}
\left(\frac{\partial}{\partial t}+\Omega\frac{\partial}{\partial \phi}\right)h'_\textrm{R}-\Omega_z^2 z u'_{\textrm{R}z}=-c_\mathrm{s}^2\left[\frac{1}{r}\frac{\partial(ru'_{\textrm{R}r})}{\partial r}+\frac{1}{r}\frac{\partial u'_{\textrm{R}\phi}}{\partial\phi}+\frac{\partial u'_{\textrm{R}z}}{\partial z}\right]+f_{\textrm{R}h},
\label{wqpo2}
\end{equation}
\begin{equation}
\left(\frac{\partial}{\partial t}+\Omega\frac{\partial}{\partial \phi}\right)u'_{\textrm{I}r}-2\Omega u'_{\textrm{I}\phi}=-\frac{\partial h'_\textrm{I}}{\partial r}-\beta\Omega u'_{\textrm{I}r}+f_{\textrm{I}r},
\label{wqpo3}
\end{equation}
\begin{equation}
\left(\frac{\partial}{\partial t}+\Omega\frac{\partial}{\partial \phi}\right)u'_{\textrm{I}\phi}+\frac{\kappa^2}{2\Omega}u'_{\textrm{I}r}=-\frac{1}{r}\frac{\partial h'_\textrm{I}}{\partial\phi}-\beta\Omega u'_{\textrm{I}\phi}+f_{\textrm{I}\phi},
\end{equation}
\begin{equation}
\left(\frac{\partial}{\partial t}+\Omega\frac{\partial}{\partial \phi}\right)h'_\textrm{I}=-c_\mathrm{s}^2\left[\frac{1}{r}\frac{\partial(ru'_{\textrm{R}r})}{\partial r}+\frac{1}{r}\frac{\partial u'_{\textrm{I}\phi}}{\partial \phi}\right]-\beta\Omega h'_{\textrm{I}}+f_{\textrm{I}h},
\label{wqpo4}
\end{equation}
where
\begin{equation}
\mathbf{f}_\textrm{R}=\left(f_{\textrm{R}r},f_{\textrm{R}\phi},f_{\textrm{R}z}\right)=-\mathbf{u}'_\textrm{I}\cdot\nabla\mathbf{u}'_\textrm{W}-\mathbf{u}'_\textrm{W}\cdot\nabla\mathbf{u}'_\textrm{I},
\label{coup1}
\end{equation}
\begin{equation}
f_{\textrm{R}h}=-\mathbf{u}'_\textrm{I}\cdot\nabla h'_\textrm{W}-\mathbf{u}'_\textrm{W}\cdot\nabla h'_\textrm{I},
\label{coup2}
\end{equation}
\begin{equation}
\mathbf{f}_\textrm{I}=\left(f_{\textrm{I}r},f_{\textrm{I}\phi},f_{\textrm{I}z}\right)=-\mathbf{u}'_\textrm{R}\cdot\nabla\mathbf{u}'_\textrm{W}-\mathbf{u}'_\textrm{W}\cdot\nabla\mathbf{u}'_\textrm{R},
\label{coup3}
\end{equation}
\begin{equation}
f_{\textrm{I}h}=-\mathbf{u}'_\textrm{R}\cdot\nabla h'_\textrm{W}-\mathbf{u}'_\textrm{W}\cdot\nabla h'_\textrm{R}
\label{coup4}
\end{equation}
are the coupling terms, arising from non-linearities in the basic equations.

Since I'm interested in studying the axisymmetric r~mode, the azimuthal mode number for the intermediate wave is $1$. Also, since the simplest possible r~mode has one node in the vertical direction and the intermediate wave has $n=0$, I use the following separation of variables:

\begin{equation}
(u'_{\textrm{R}r},u'_{\textrm{R}\phi},h'_\textrm{R})=\textrm{Re}\left[\left(u_{\textrm{R}r}(r),u_{\textrm{R}\phi}(r),h_\textrm{R}(r)\right)\textrm{He}_1\left(\frac{z}{H}\right)\textrm{e}^{-\textrm{i}\omega t}\right], 
\end{equation}
\begin{equation}
u'_{\textrm{R}z}=\textrm{Re}\left[u_{\textrm{R}z}(r)\textrm{He}_0\left(\frac{z}{H}\right)\textrm{e}^{-\textrm{i}\omega t}\right],
\end{equation}
\begin{equation}
(u'_{\textrm{I}r},u'_{\textrm{I}\phi},h'_\textrm{I})=\textrm{Re}\left[\left(u_{\textrm{I}r}(r),u_{\textrm{I}\phi}(r),h_\textrm{I}(r)\right)\textrm{He}_0\left(\frac{z}{H}\right)\textrm{e}^{\textrm{i}\phi-\textrm{i}\omega t}\right],
\end{equation}
\begin{equation}
u'_{\textrm{I}z}=0 \quad \textrm{(2D wave)},
\end{equation}
\begin{equation}
(u'_{\textrm{W}r},u'_{\textrm{W}\phi},h'_\textrm{W})=\textrm{Re}\left[\left(u_{\textrm{W}r}(r),u_{\textrm{W}\phi}(r),h_\textrm{W}(r)\right) z\, \textrm{e}^{\textrm{i}\phi}\right],
\end{equation}
\begin{equation}
u'_{\textrm{W}z}=\textrm{Re}\left[u_{\textrm{W}z}(r)\, \textrm{e}^{\textrm{i}\phi}\right].
\end{equation}

This separation of variables results in having the coupling terms (\ref{coup1}) and (\ref{coup2}) proportional to $\textrm{He}_1$ only, while the coupling terms (\ref{coup3})--(\ref{coup4}) give rise to terms proportional to both $\textrm{He}_0=1$ and $\textrm{He}_2=z^2/H^2-1$. Since I'm interested in the terms that influence the mode with $n=0$, I need to project these forcing terms on to $\textrm{He}_0$. Also, the separation of variables results in coupling terms of the form
\begin{equation}
\textrm{Re}(A)\textrm{Re}(B)=\frac{1}{2}\textrm{Re}(AB+AB^*),
\end{equation}
which means that the interaction of the $m=0$ r~mode with the $m=1$ warp results in two new waves, one with $m=1$ and one with $m=-1$:
\begin{displaymath}
\textrm{r mode } (A) \propto \textrm{e}^{-\textrm{i}\omega t} \quad \textrm{and} \quad \textrm{warp } (B)\propto \textrm{e}^{\textrm{i}\phi}
\end{displaymath}
\begin{displaymath}
 \Rightarrow\quad \textrm{r mode }\times\textrm{ warp}\propto AB+AB^* \propto \textrm{e}^{\textrm{i}\phi-\textrm{i}\omega t}+\textrm{e}^{-\textrm{i}\phi-\textrm{i}\omega t}.
\end{displaymath}
Since I'm only interested in the action of this coupling on the intermediate wave with $m=1$, because the one with $m=-1$ does not have any Lindblad resonances (location where the wave should be excited) in the disc, these forcing terms are projected on to $\textrm{e}^{\textrm{i}\phi-\textrm{i}\omega t}$. Similarly, the interaction of the intermediate wave with the warp gives rise to a $m=2$ mode in addition to the axisymmetric r~mode of interest:

\begin{displaymath}
\textrm{intermediate wave } (A) \propto \textrm{e}^{\textrm{i}\phi-\textrm{i}\omega t} \quad \textrm{and} \quad \textrm{warp } (B)\propto \textrm{e}^{\textrm{i}\phi}
\end{displaymath}
\begin{displaymath}
 \Rightarrow\quad \textrm{intermediate wave }\times\textrm{ warp}\propto AB+AB^* \propto \textrm{e}^{\textrm{i}2\phi-\textrm{i}\omega t}+\textrm{e}^{-\textrm{i}\omega t}.
\end{displaymath}
Therefore, these forcing terms should be projected on to $\textrm{e}^{-\textrm{i}\omega t}$, which means that complex conjugates of warp quantities will appear in the equations.

After separating variables, and projecting the forcing terms appropriately, the equations to be solved can be written as
\begin{eqnarray}
-\textrm{i}\omega u_{\textrm{R}r}=2\Omega u_{\textrm{R}\phi}-\frac{\textrm{d}h_\textrm{R}}{\textrm{d}r}-\frac{u_{\textrm{I}r}}{2}\frac{\textrm{d}u^*_{\textrm{W}r}}{\textrm{d}r}H+\textrm{i}u_{\textrm{I}\phi}\frac{u^*_{\textrm{W}r}}{2r}H+u_{\textrm{I}\phi}\frac{u^*_{\textrm{W}\phi}}{r}H\nonumber\\-\frac{u^*_{\textrm{W}r}}{2}\frac{\textrm{d}u_{\textrm{I}r}}{\textrm{d}r}H-\textrm{i}u^*_{\textrm{W}\phi}\frac{u_{\textrm{I}r}}{2r}H,
\label{fe1}
\end{eqnarray}
\begin{equation}
-\textrm{i}\omega u_{\textrm{R}\phi}=-\frac{\kappa^2}{2\Omega} u_{\textrm{R}r}-\frac{u_{\textrm{I}r}}{2}\frac{\textrm{d}u^*_{\textrm{W}\phi}}{\textrm{d}r}H-u_{\textrm{I}\phi}\frac{u^*_{\textrm{W}r}}{2r}H-\frac{u^*_{\textrm{W}r}}{2}\frac{\textrm{d}u_{\textrm{I}\phi}}{\textrm{d}r}H-u^*_{\textrm{W}\phi}\frac{u_{\textrm{I}r}}{2r}H,
\end{equation}
\begin{equation}
-\textrm{i}\omega u_{\textrm{R}z}=-\frac{h_\textrm{R}}{H}-\frac{u_{\textrm{I}r}}{2}\frac{\textrm{d}u^*_{\textrm{W}z}}{\textrm{d}r}+\textrm{i}u_{\textrm{I}\phi}\frac{u^*_{\textrm{W}z}}{2r},
\end{equation}
\begin{equation}
-\textrm{i}\omega h_{\textrm{R}}=\Omega_z^2H u_{\textrm{R}z}-\frac{c_\mathrm{s}^2}{r}\frac{\textrm{d}(ru_{\textrm{R}r})}{\textrm{d}r}-\frac{u_{\textrm{I}r}}{2}\frac{\textrm{d}h^*_{\textrm{W}}}{\textrm{d}r}H+\textrm{i}u_{\textrm{I}\phi}\frac{h^*_{\textrm{W}}}{2r}H-\frac{u^*_{\textrm{W}r}}{2}\frac{\textrm{d}h_{\textrm{I}}}{\textrm{d}r}H-\textrm{i}u^*_{\textrm{W}\phi}\frac{h_{\textrm{I}}}{2r}H,
\end{equation}
\begin{eqnarray}
-\textrm{i}\omega u_{\textrm{I}r}=-(\textrm{i}+\beta)\Omega u_{\textrm{I}r}+2\Omega u_{\textrm{I}\phi}-\frac{\textrm{d}h_\textrm{I}}{\textrm{d}r}-\frac{u_{\textrm{R}r}}{2}\frac{\textrm{d}u_{\textrm{W}r}}{\textrm{d}r}H-\textrm{i}u_{\textrm{R}\phi}\frac{u_{\textrm{W}r}}{2r}H\nonumber\\+u_{\textrm{R}\phi}\frac{u_{\textrm{W}\phi}}{r}H-u_{\textrm{R}r}\frac{u_{\textrm{W}z}}{2H}-\frac{u_{\textrm{W}r}}{2}\frac{\textrm{d}u_{\textrm{R}r}}{\textrm{d}r}H-u_{\textrm{R}z}\frac{u_{\textrm{W}r}}{2},
\end{eqnarray}
\begin{eqnarray}
-\textrm{i}\omega u_{\textrm{I}\phi}=-(\textrm{i}+\beta)\Omega u_{\textrm{I}\phi}-\frac{\kappa^2}{2\Omega} u_{\textrm{I}r}-\frac{\textrm{i}h_\textrm{I}}{r}-\frac{u_{\textrm{R}r}}{2}\frac{\textrm{d}u_{\textrm{W}\phi}}{\textrm{d}r}H-\textrm{i}u_{\textrm{R}\phi}\frac{u_{\textrm{W}\phi}}{2r}H\nonumber\\-u_{\textrm{R}\phi}\frac{u_{\textrm{W}r}}{2r}H-\frac{u_{\textrm{W}r}}{2}\frac{\textrm{d}u_{\textrm{R}\phi}}{\textrm{d}r}H-u_{\textrm{W}\phi}\frac{u_{\textrm{R}r}}{2r}H-u_{\textrm{R}z}\frac{u_{\textrm{W}\phi}}{2}-u_{\textrm{W}z}\frac{u_{\textrm{R}\phi}}{2H},
\end{eqnarray}
\begin{eqnarray}
-\textrm{i}\omega h_{\textrm{I}}=-(\textrm{i}+\beta)\Omega h_{\textrm{I}}-\frac{c_\mathrm{s}^2}{r}\frac{\textrm{d}(ru_{\textrm{I}r})}{\textrm{d}r}-c_\mathrm{s}^2\frac{\textrm{i}u_{\textrm{I}\phi}}{r}-\frac{u_{\textrm{R}r}}{2}\frac{\textrm{d}h_{\textrm{W}}}{\textrm{d}r}H-\textrm{i}u_{\textrm{R}\phi}\frac{h_{\textrm{W}}}{2r}H\nonumber\\-\frac{u_{\textrm{W}r}}{2}\frac{\textrm{d}h_{\textrm{R}}}{\textrm{d}r}H-u_{\textrm{W}z}\frac{h_\textrm{R}}{2H}-u_{\textrm{R}z}\frac{h_\textrm{W}}{2}.
\label{fe2}
\end{eqnarray}
This system is linear in the unknowns for the r~mode and the intermediate wave. The warp, which couples these waves together, is assumed to be known (cf. section \ref{warps}).

For the interaction with the $n=2$ intermediate wave, similar equations need to be solved. After separating variables and projecting forcing terms appropriately, the equations describing this interaction read
\begin{eqnarray}
-\textrm{i}\omega u_{\textrm{R}r}=2\Omega u_{\textrm{R}\phi}-\frac{\textrm{d}h_\textrm{R}}{\textrm{d}r}-u_{\textrm{I}r}\frac{\textrm{d}u^*_{\textrm{W}r}}{\textrm{d}r}H+\textrm{i}u_{\textrm{I}\phi}\frac{u^*_{\textrm{W}r}}{r}H+2u_{\textrm{I}\phi}\frac{u^*_{\textrm{W}\phi}}{r}H \nonumber\\-u^*_{\textrm{W}r}\frac{\textrm{d}u_{\textrm{I}r}}{\textrm{d}r}H-\textrm{i}u^*_{\textrm{W}\phi}\frac{u_{\textrm{I}r}}{r}H-\frac{u^*_{\textrm{W}z}}{H}u_{\textrm{I}r}+\frac{u^*_{\textrm{W}r}}{2}u_{\textrm{I}z},
\label{int1}
\end{eqnarray}
\begin{eqnarray}
-\textrm{i}\omega u_{\textrm{R}\phi}=-\frac{\kappa^2}{2\Omega} u_{\textrm{R}r}-u_{\textrm{I}r}\frac{\textrm{d}u^*_{\textrm{W}\phi}}{\textrm{d}r}H-u_{\textrm{I}\phi}\frac{u^*_{\textrm{W}r}}{r}H-u^*_{\textrm{W}r}\frac{\textrm{d}u_{\textrm{I}\phi}}{\textrm{d}r}H-u^*_{\textrm{W}\phi}\frac{u_{\textrm{I}r}}{r}H \nonumber\\-\frac{u^*_{\textrm{W}z}}{H}u_{\textrm{I}\phi}+\frac{u^*_{\textrm{W}\phi}}{2}u_{\textrm{I}z},
\end{eqnarray}
\begin{equation}
-\textrm{i}\omega u_{\textrm{R}z}=-\frac{h_\textrm{R}}{H}-\frac{u^*_{\textrm{W}r}}{2}\frac{\textrm{d}u_{\textrm{I}z}}{\textrm{d}r}H-\textrm{i}u^*_{\textrm{W}\phi}\frac{u_{\textrm{I}z}}{2r}H-\frac{u^*_{\textrm{W}z}}{2H}u_{\textrm{I}z},
\end{equation}
\begin{eqnarray}
-\textrm{i}\omega h_{\textrm{R}}=\Omega_z^2H u_{\textrm{R}z}-\frac{c_\mathrm{s}^2}{r}\frac{\textrm{d}(ru_{\textrm{R}r})}{\textrm{d}r}-u_{\textrm{I}r}\frac{\textrm{d}h^*_{\textrm{W}}}{\textrm{d}r}H+\textrm{i}u_{\textrm{I}\phi}\frac{h^*_{\textrm{W}}}{r}H-u^*_{\textrm{W}r}\frac{\textrm{d}h_{\textrm{I}}}{\textrm{d}r}H \nonumber\\-\textrm{i}u^*_{\textrm{W}\phi}\frac{h_{\textrm{I}}}{r}H-\frac{u^*_{\textrm{W}z}}{H}h_\textrm{I}-\frac{h^*_\textrm{W}}{2}u_{\textrm{I}z},
\end{eqnarray}
\begin{eqnarray}
-\textrm{i}\omega u_{\textrm{I}r}=-(\textrm{i}+\beta)\Omega u_{\textrm{I}r}+2\Omega u_{\textrm{I}\phi}-\frac{\textrm{d}h_\textrm{I}}{\textrm{d}r}-\frac{u_{\textrm{R}r}}{2}\frac{\textrm{d}u_{\textrm{W}r}}{\textrm{d}r}H-\textrm{i}u_{\textrm{R}\phi}\frac{u_{\textrm{W}r}}{2r}H \nonumber\\+u_{\textrm{R}\phi}\frac{u_{\textrm{W}\phi}}{r}H-\frac{u_{\textrm{W}r}}{2}\frac{\textrm{d}u_{\textrm{R}r}}{\textrm{d}r}H,
\end{eqnarray}
\begin{eqnarray}
-\textrm{i}\omega u_{\textrm{I}\phi}=-(\textrm{i}+\beta)\Omega u_{\textrm{I}\phi}-\frac{\kappa^2}{2\Omega} u_{\textrm{I}r}-\frac{\textrm{i}h_\textrm{I}}{r}-\frac{u_{\textrm{R}r}}{2}\frac{\textrm{d}u_{\textrm{W}\phi}}{\textrm{d}r}H-\textrm{i}u_{\textrm{R}\phi}\frac{u_{\textrm{W}\phi}}{2r}H \nonumber\\-u_{\textrm{R}\phi}\frac{u_{\textrm{W}r}}{2r}H-\frac{u_{\textrm{W}r}}{2}\frac{\textrm{d}u_{\textrm{R}\phi}}{\textrm{d}r}H-u_{\textrm{W}\phi}\frac{u_{\textrm{R}r}}{2r}H,
\end{eqnarray}
\begin{equation}
-\textrm{i}\omega u_{\textrm{I}z}=-(\textrm{i}+\beta)\Omega u_{\textrm{I}z}-\frac{2h_\textrm{I}}{H}-\frac{u_{\textrm{R}r}}{2}\frac{\textrm{d}u_{\textrm{W}z}}{\textrm{d}r}-i\frac{u_{\textrm{W}z}}{2r}u_{2\phi}-\frac{H}{2}u_{\textrm{W}r}\frac{\textrm{d}u_{\textrm{R}z}}{\textrm{d}r},
\end{equation}
\begin{eqnarray}
-\textrm{i}\omega h_{\textrm{I}}=-(\textrm{i}+\beta)\Omega h_{\textrm{I}}-\frac{c_\mathrm{s}^2}{r}\frac{\textrm{d}(ru_{\textrm{I}r})}{\textrm{d}r}-c_\mathrm{s}^2\frac{\textrm{i}u_{\textrm{I}\phi}}{r}+\Omega_z^2Hu_{\textrm{I}z}-\frac{u_{\textrm{R}r}}{2}\frac{\textrm{d}h_{\textrm{W}}}{\textrm{d}r}H \nonumber\\-\textrm{i}u_{\textrm{R}\phi}\frac{h_{\textrm{W}}}{2r}H-\frac{u_{\textrm{W}r}}{2}\frac{\textrm{d}h_{\textrm{R}}}{\textrm{d}r}H.
\label{int2}
\end{eqnarray}
As before, the system is linear in the unknowns for the r~mode and the intermediate wave. It should be noted that although the same notation is used in the systems (\ref{fe1})--(\ref{fe2}) and (\ref{int1})--(\ref{int2}) to represent the intermediate wave quantities, they refer to two different waves: both with the same frequency and azimuthal mode number $m=1$, but with different vertical mode number ($n=0$ in the first system and $n=2$ in the second).

\subsubsection{Method and results}

To find the r~mode growth rate resulting from these interactions I solve the systems of coupled equations (\ref{fe1})--(\ref{fe2}) and (\ref{int1})--(\ref{int2}) for the interactions of
the r~mode with the $n=0$ and $n=2$ intermediate wave respectively. By
considering the warp to have a fixed amplitude and neglecting the
feedback of the r~mode and intermediate waves on the warp, a linear system of equations is still obtained, although now the r~mode and
intermediate waves are coupled through the warp.  The $n=0$
and $n=2$ intermediate waves are treated separately, although in practice both
coupling mechanisms act simultaneously and the net growth rate is the
sum of the rates due to the individual mechanisms.

These systems are solved numerically, using the Chebyshev method described in section \ref{trapping}. For the r~mode the same boundary conditions as before are used. Similar conditions are applied to the intermediate waves, i.e., $u_{\textrm{I}r}=0$ at $r_\mathrm{in}$ and $\textrm{d}u_{\textrm{I}r}/\textrm{d}r=\textrm{i}k_\textrm{I}u_{\textrm{I}r}$ at $r_\mathrm{out}$, where $k_\textrm{I}$ is given by the dispersion relation (\ref{ssdr}) at $r_\mathrm{out}$ for $m=1$ and $n=0$ or $n=2$, depending on the intermediate wave under consideration. As for the r~mode, I choose the sign of $k_\textrm{I}$ so that the outgoing or exponentially decaying wave at $r_\mathrm{out}$ is chosen. The choice of the inner boundary condition for the radial component of the velocity of the $n=2$ mode is justified by the fact that this mode is exponentially decaying there. This choice is harder to justify for the $n=0$ mode since it is oscillatory at $r_\mathrm{in}$. This means that if $u_{\textrm{I}r}=0$ there then the wave is reflected at the inner boundary. Since the conditions at the marginally stable orbit are not clear, one cannot be sure of the physical validity of this condition; it is chosen because of its simplicity. (Note that this problem is revisited in chapter \ref{reflect}.) Rigorously more boundary conditions would be needed to solve this problem, since more than 4 derivatives appear in each system. However, since the coupling terms are expected to be small, $u_{\textrm{R}\phi}$ is roughly proportional to $u_{\textrm{R}r}$ (and similarly for other quantities), thus the boundary conditions imposed for the latter will be indirectly imposed to the former.

\begin{figure}
\begin{center}
\includegraphics[width=0.485\linewidth]{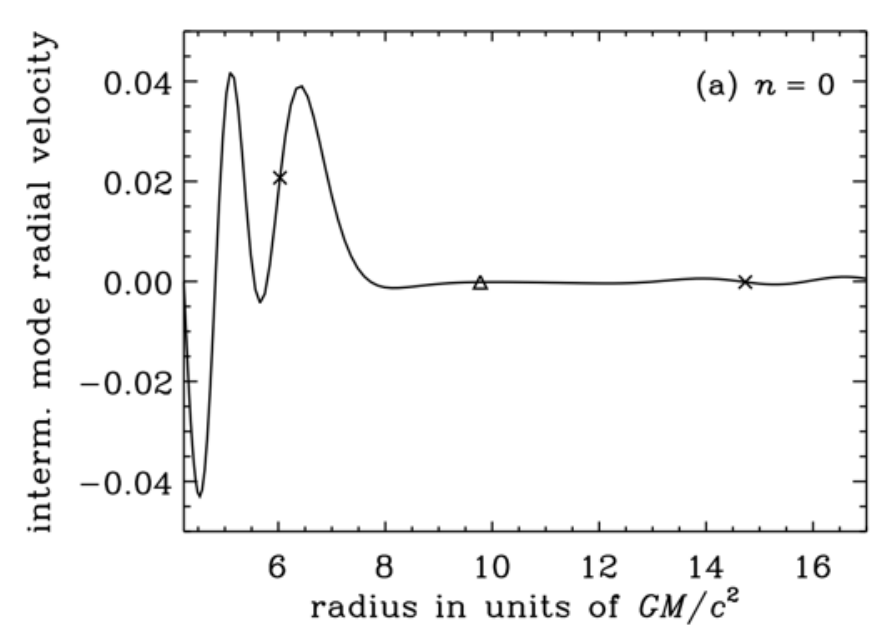} \hspace{2mm}
\includegraphics[width=0.485\linewidth]{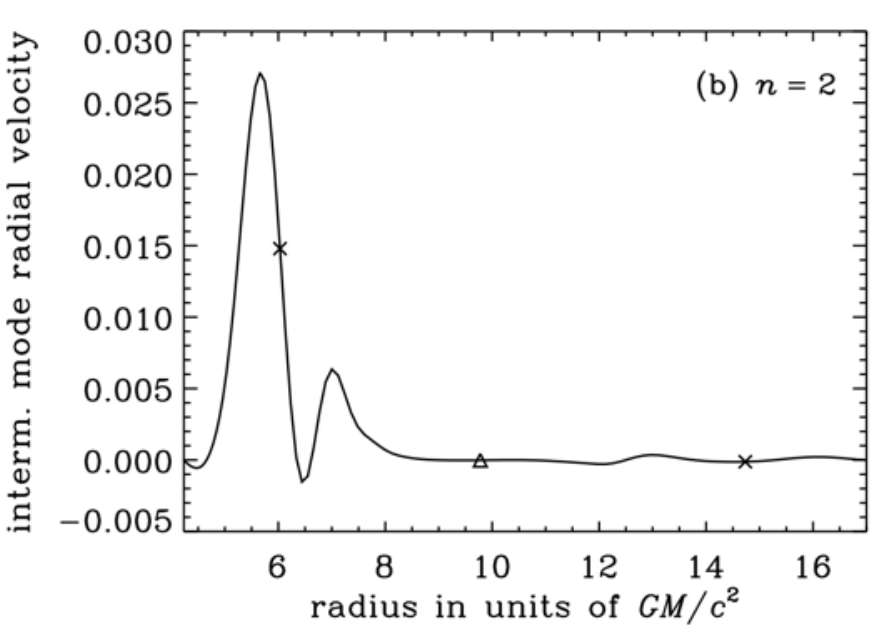}
\caption{Variation of the real part of the radial component of the (a) $m=1$, $n=0$, and (b) $m=1$, $n=2$ intermediate wave velocity with radius for $c_\mathrm{s}/c=0.01$, $a=0.5$, $W(r_\mathrm{in})=0.003$, and $\beta=0.1$. The triangle indicates the radius of the corotation resonance, and the crosses the Lindblad resonances.}
\label{intermediate}
\end{center}
\end{figure} 

\begin{figure}
\begin{center}
\includegraphics[width=0.49\linewidth]{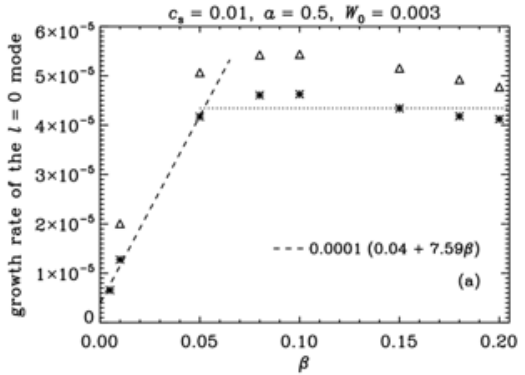} \hskip0.01\linewidth
\includegraphics[width=0.49\linewidth]{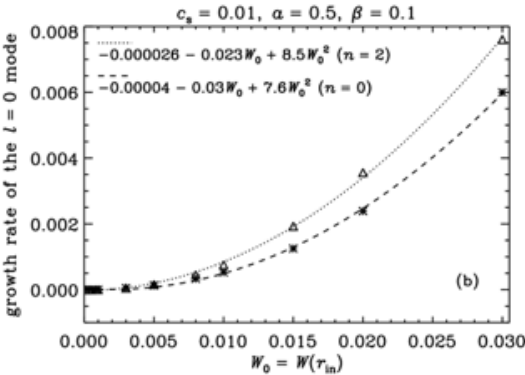} \\
\vspace{5mm}
\includegraphics[width=0.49\linewidth]{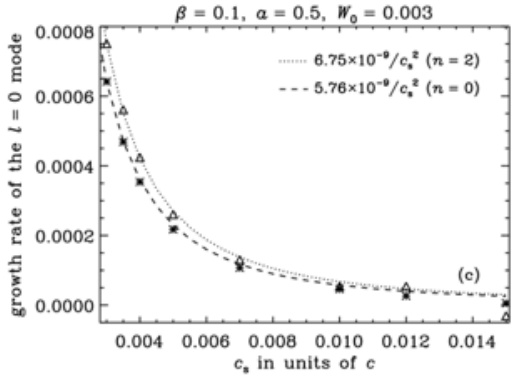}\hskip0.015\linewidth
\includegraphics[width=0.49\linewidth]{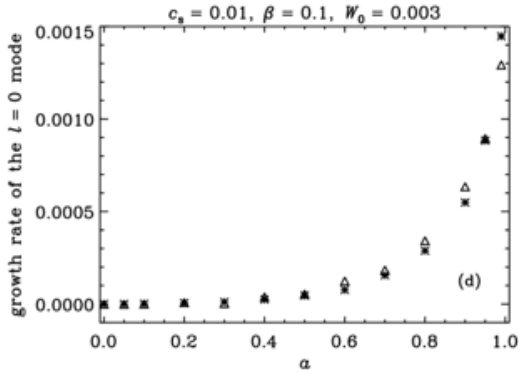}
\caption{Variation of the growth rate of the simplest trapped r~mode, $(l,m,n)=(0,0,1)$, with (a) dissipation factor $\beta$, (b) warp amplitude, $W$, at inner boundary, (c) sound speed in the disc,  and (d) spin of the black hole. The triangles show the results for the interaction with the $n=2$ mode, while the stars are due to the interaction with the $n=0$ mode. For the former, the variation of the growth rate with the dissipation factor is not shown for small values of $\beta$ because of the influence of the corotation resonance in that case.}
\label{relwarp}
\end{center}
\end{figure}

The aim is to find the frequencies $\omega$ for which the solutions corresponding to the r~mode are trapped, i.e., for which $u_{\textrm{R}r}$ resembles the parabolic cylinder functions as in Fig.~\ref{freerm} (a), which is expected if the coupling terms are small when compared to the other terms in the equations. The imaginary part of $\omega$ then gives the growth rate (or damping rate, if it's negative) of the trapped r~mode. In Fig.~\ref{intermediate} I show the $n=0$ and $n=2$ intermediate waves involved in the coupling process, when the dissipation is strong. It should be noted that when the dissipation is weak, the $n=2$ intermediate wave develops a very short wavelength as it approaches the corotation resonance. If $\beta$ is too small, the length-scale on which this wave dissipates is not resolved by the numerical method employed. The variation of the growth rate with several parameters is shown in Fig.~\ref{relwarp}. These results are discussed in section \ref{grdisc}.

\subsection{Eccentric discs}
 
Recently, \cite{kato2008} argued that one-armed global oscillations (eccentric modes), symmetric with respect to the $z=0$ plane, can also excite trapped oscillations. His conclusions are based on analytical, Lagrangian calculations and are too crude to allow for more than simple estimates for the growth rates. In this section I describe an excitation mechanism similar to the one reported previously, but where an $(m=1,n=0)$ eccentric mode has the role that previously belonged to the $(m=1,n=1)$ warp wave. I calculate the trapped r~mode growth rates using the same numerical method as before.

\begin{figure}
\begin{center}
\includegraphics[width=0.35\linewidth]{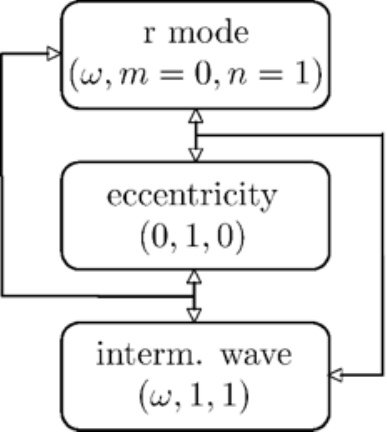}
\caption{Diagram representing the interactions involved in the coupling mechanism involving an eccentric disc.}
\label{diagrame}
\end{center}
\end{figure}

\subsubsection{Coupling mechanism}

The excitation mechanism is somewhat similar to the one discussed in the previous section: a global deformation mode, which is now the eccentricity mode, characterized by $(\omega,m,n)=(0,1,0)$, interacts with a trapped r~mode $(\omega,0,1)$ giving rise to an intermediate wave. The latter then couples with the global mode to feedback on to the trapped r~mode.

Coupling rules require the intermediate wave to have the same frequency as the trapped oscillation and, as before, $m_\textrm{I}=1$ in the case where the trapped r~mode is axisymmetric. As for the vertical dependence, since the eccentric mode has $n=0$, the intermediate wave can only have the same vertical mode number as the trapped mode, i.e., $n_\textrm{I}=1$ (Fig.~\ref{diagrame}). The propagation region for this wave is the same as for the $(\omega,1,2)$ intermediate wave, present in the interaction of the r~mode with the warp. As in that case, a relatively large damping term must be included in the equations for the intermediate wave so that it dissipates on a resolved scale before reaching the corotation resonance. In this case the energy exchanges are similar to the ones discussed for the interaction in a warped disc. \enlargethispage{\baselineskip}

\subsubsection{Equations and results}

After separating variables, and projecting the forcing terms appropriately, as done before for the interactions in a warped disc, the equations describing the coupling between the trapped r~mode, eccentric disc and $n=1$ intermediate wave read
\begin{equation}
-\textrm{i}\omega u_{\textrm{R}r}=2\Omega u_{\textrm{R}\phi}-\frac{\textrm{d}h_\textrm{R}}{\textrm{d}r}-\frac{u_{\textrm{I}r}}{2}\frac{\textrm{d}u^*_{\textrm{E}r}}{\textrm{d}r}-\frac{u^*_{\textrm{E}r}}{2}\frac{\textrm{d}u_{\textrm{I}r}}{\textrm{d}r}-\frac{\textrm{i}u^*_{\textrm{E}\phi}}{2r}u_{\textrm{I}r}+\frac{u_{\textrm{I}\phi}}{2r}(\textrm{i}u^*_{\textrm{E}r}+u^*_{\textrm{E}\phi}),
\label{inte1}
\end{equation}
\begin{equation}
-\textrm{i}\omega u_{\textrm{R}\phi}=-\frac{\kappa^2}{2\Omega} u_{\textrm{R}r}-\frac{u_{\textrm{I}r}}{2}\left(\frac{\textrm{d}u^*_{\textrm{E}\phi}}{\textrm{d}r}+\frac{u^*_{\textrm{E}\phi}}{r}\right)-\frac{u_{\textrm{I}\phi}}{2}\frac{u^*_{\textrm{E}r}}{r}-\frac{u^*_{\textrm{E}r}}{2}\frac{\textrm{d}u_{\textrm{I}r}}{\textrm{d}r},
\end{equation}
\begin{equation}
-\textrm{i}\omega u_{\textrm{R}z}=-\frac{h_\textrm{R}}{H}-\frac{\textrm{i}u_{\textrm{I}z}}{2}\frac{u^*_{\textrm{E}\phi}}{r}-\frac{u^*_{\textrm{E}r}}{2}\frac{\textrm{d}u_{\textrm{I}z}}{\textrm{d}r},
\end{equation}
\begin{equation}
-\textrm{i}\omega h_{\textrm{R}}=\Omega_z^2H u_{\textrm{R}z}-\frac{c_\mathrm{s}^2}{r}\frac{\textrm{d}(ru_{\textrm{R}r})}{\textrm{d}r}-\frac{u_{\textrm{I}r}}{2}\frac{\textrm{d}h^*_\textrm{E}}{\textrm{d}r}+\frac{\textrm{i}u_{\textrm{I}\phi}}{2r}\frac{h^*_\textrm{E}}{r}-\frac{u^*_{\textrm{E}r}}{2}\frac{\textrm{d}h_\textrm{I}}{\textrm{d}r}-\frac{\textrm{i}h_\textrm{I}}{2}\frac{u^*_{\textrm{E}\phi}}{r},
\end{equation}
\begin{equation}
-\textrm{i}\omega u_{\textrm{I}r}=-(\textrm{i}+\beta)\Omega u_{\textrm{I}r}+2\Omega u_{\textrm{I}\phi}-\frac{\textrm{d}h_\textrm{I}}{\textrm{d}r}-\frac{u_{\textrm{R}r}}{2}\frac{\textrm{d}u_{\textrm{E}r}}{\textrm{d}r}-\frac{u_{\textrm{E}r}}{2}\frac{\textrm{d}u_{\textrm{R}r}}{\textrm{d}r}-\frac{u_{\textrm{R}\phi}}{2r}(\textrm{i}u_{\textrm{E}r}-u_{\textrm{E}\phi}),
\end{equation}
\begin{eqnarray}
-\textrm{i}\omega u_{\textrm{I}\phi}=-(\textrm{i}+\beta)\Omega u_{\textrm{I}\phi}-\frac{\kappa^2}{2\Omega} u_{\textrm{I}r}-\frac{\textrm{i}h_\textrm{I}}{r}-\frac{u_{\textrm{R}r}}{2}\frac{\textrm{d}u_{\textrm{E}\phi}}{\textrm{d}r}-\frac{u_{\textrm{R}r}}{2}\frac{u_{\textrm{E}\phi}}{r}\nonumber\\-\frac{u_{\textrm{R}\phi}}{2r}(\textrm{i}u_{\textrm{E}\phi}+u_{\textrm{E}r})-\frac{u_{\textrm{E}r}}{2}\frac{\textrm{d}u_{\textrm{R}\phi}}{\textrm{d}r},
\end{eqnarray}
\begin{equation}
-\textrm{i}\omega u_{\textrm{I}z}=-(\textrm{i}+\beta)\Omega u_{\textrm{I}z}-\frac{h_\textrm{I}}{H}-\frac{u_{\textrm{E}r}}{2}\frac{\textrm{d}u_{\textrm{R}z}}{\textrm{d}r},
\end{equation}
\begin{eqnarray}
-\textrm{i}\omega h_{\textrm{I}}=-(\textrm{i}+\beta)\Omega h_{\textrm{I}}-\frac{c_\mathrm{s}^2}{r}\frac{\textrm{d}(ru_{\textrm{I}r})}{\textrm{d}r}-c_\mathrm{s}^2\frac{\textrm{i}u_{\textrm{I}\phi}}{r}+\Omega_z^2Hu_{\textrm{I}z}-\frac{u_{\textrm{R}r}}{2}\frac{\textrm{d}h_{\textrm{E}}}{\textrm{d}r}-\textrm{i}u_{\textrm{R}\phi}\frac{h_{\textrm{E}}}{2r}\nonumber\\-\frac{u_{\textrm{E}r}}{2}\frac{\textrm{d}h_{\textrm{R}}}{\textrm{d}r}.
\label{inte2}
\end{eqnarray}
To find the growth rates that result from this interaction, I solve equations (\ref{inte1})--(\ref{inte2}), using the same numerical method and boundary conditions as before. 

\begin{figure}
\begin{center}
\includegraphics[width=0.49\linewidth]{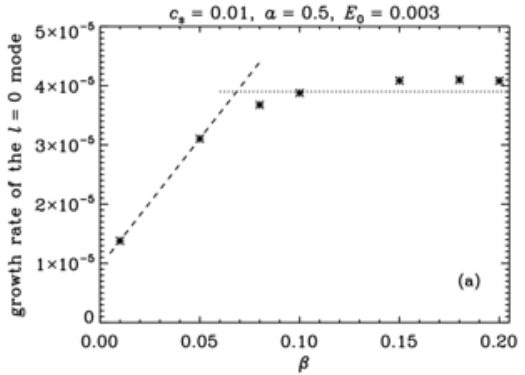} \hskip0.01\linewidth
\includegraphics[width=0.49\linewidth]{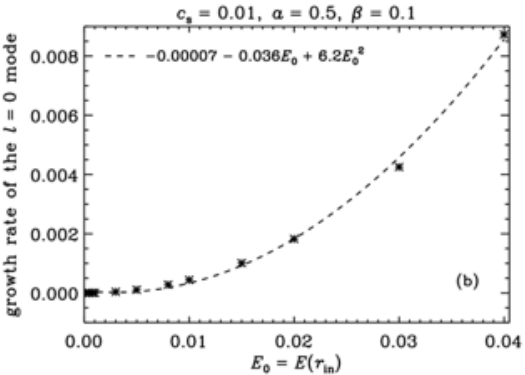}\\
\vspace{5mm}
\includegraphics[width=0.5\linewidth]{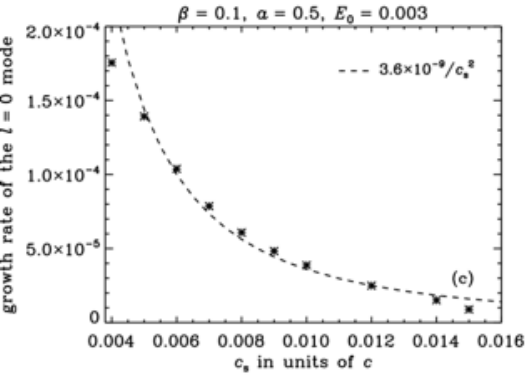}\hskip0.01\linewidth
\includegraphics[width=0.49\linewidth]{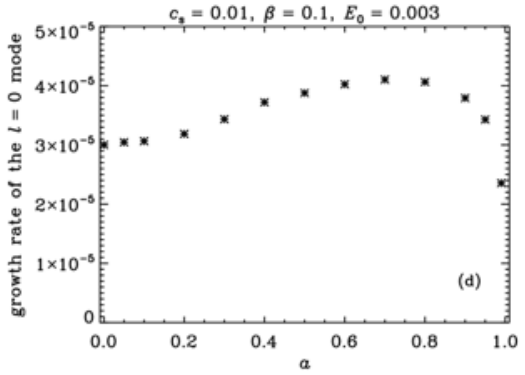}
\caption{Variation of the growth rate of the simplest trapped r~mode,
$(l,m,n)=(0,0,1)$,
 with (a) dissipation factor $\beta$, (b) eccentricity amplitude, $E$, at inner boundary, (c) sound speed of the disc,  and (d) spin of the black hole.}
\label{relecc}
\end{center}
\end{figure}

The variation of the growth rate with the inner eccentricity, sound speed, spin of black hole and dissipation factor is shown in Fig.~\ref{relecc}. The values of the growth rate achieved in the interaction of the trapped wave with the eccentric mode are, in general, similar to the values obtained in the interaction with the warp, if the inner inclination and eccentricity are similar. 

\section{Discussion}
\label{grdisc}

In this section I discuss the results shown in Figs~\ref{relwarp} and \ref{relecc}, where the dependence of the $l=0$ r~mode growth rate with several parameters is represented.

\subsection{Growth rates in warped discs}

In the variation of the growth rate with the dissipation factor, two regimes can be considered. In the weak dissipation case ($\beta \lesssim 0.05$), the variation is approximately linear, while in the strong dissipation regime ($\beta \gtrsim 0.05$), the growth rate remains approximately constant when $\beta$ varies [Fig.~\ref{relwarp} (a)]. In the former case, the $n=0$ intermediate wave is launched at its inner Lindblad resonance and propagates, being slightly attentuated, until it reaches the inner boundary where it is reflected. Owing to the attenuation, the reflected wave amplitude is smaller than the incident one, and therefore the intermediate wave does not cancel itself, leaving a small amount of energy available for the r~mode to be excited. In the strong dissipation regime, the $n=0$ mode is dissipated before reaching the marginally stable orbit. In this case, all the energy carried by this wave becomes available to excite the trapped mode. As for the $n=2$ intermediate wave, in the strong dissipation regime, the wave is completely dissipated before reaching the corotation resonance. Physically, no dissipation term is expected to be necessary in this case. An arbitrarily small amount of dissipation should lead in principle to the complete absorption of the wave at the corotation resonance. However, this cannot be verified numerically because of the difficulty in resolving the wavelength of the intermediate wave as the singularity at the corotation resonance is approached.

For small warp amplitudes, i.e., before the coupling terms start affecting the structure of the eigenfunctions, the growth rate grows with the square of the warp amplitude at the inner boundary [Fig.~\ref{relwarp} (b)]. This is expected since the coupling mechanism relies on the ``use'' of the warp twice: first on the interaction with the r~mode to give rise to the intermediate waves, and then again on the interaction with the latter to feed back on the former (Fig.~\ref{diagram}).

The excitation mechanism discussed here is similar to the well known parametric instability, in the case where one of the modes is strongly damped. The parametric instability is a type of resonant coupling between three modes satisfying $\omega_\textrm{p}\approx\omega_{\textrm{d}1}+\omega_{\textrm{d}2}$, where the subscripts p and d refer to parent and daughter modes, respectively. The parametric instability results in the transfer of energy from the former to the latter, when the daughter modes have small amplitude. The equations describing the evolution of the mode amplitudes read \citep[adapted from][]{wugoldreich2001},
\begin{equation}
\frac{\textrm{d}A_\textrm{p}}{\textrm{d}t}=+\delta_\textrm{p}A_\textrm{p}-\textrm{i}\omega_\textrm{p}A_{\textrm{p}}+\textrm{i}\omega_\textrm{p}\sigma A_{\textrm{d}1} A_{\textrm{d}2},
\end{equation}
\begin{equation}
\frac{\textrm{d}A_{\textrm{d}1}}{\textrm{d}t}=-\delta_{\textrm{d}1}A_{\textrm{d}1}-\textrm{i}\omega_{\textrm{d}1}A_{\textrm{d}1}+\textrm{i}\omega_{\textrm{d}1}\sigma A_{\textrm{p}} A^*_{\textrm{d}2},
\end{equation}
\begin{equation}
\frac{\textrm{d}A_{\textrm{d}2}}{\textrm{d}t}=-\delta_{\textrm{d}2}A_{\textrm{d}2}-\textrm{i}\omega_{\textrm{d}2}A_{\textrm{d}2}+\textrm{i}\omega_{\textrm{d}2}\sigma A^*_{\textrm{d}1} A_{\textrm{p}},
\end{equation}
where $\delta_j>0$ is the linear amplitude growth/damping rate of mode $j$ and $\sigma$ is the non-linear coupling constant. I'll consider a simplified case where the amplitude of the parent mode is approximately constant in time (because the daughter modes are of small amplitude), and $\omega_{\textrm{d}1}=-\omega_{\textrm{d}2}=\omega$, $\delta_{\textrm{p}}=\delta_{\textrm{d}1}=0$ and $\delta_{\textrm{d}2}=\delta$. In this case, the parent mode can be compared to the warp while the daughter modes 1 and 2 can be compared with the r~mode and intermediate wave, respectively. Assuming $A_{\textrm{d}1}\propto\exp{(st)}$, the growth rate is
\begin{equation}
\textrm{Re}(s)=-\frac{\delta}{2}+\left(\frac{\delta^2}{4}+|A_{\textrm{p}}|^2\sigma^2\omega^2\right)^{1/2}.
\end{equation}
If
$\delta\ll|A_\mathrm{p}|\sigma\omega$,
$\textrm{Re}(s)\approx|A_{\textrm{p}}|\sigma\omega-\frac{\delta}{2}$,
i.e., the growth rate is linearly related to the amplitude of the parent mode. On the other hand, if 
$\delta\gg|A_\mathrm{p}|\sigma\omega$,
$\textrm{Re}(s)\approx|A_\mathrm{p}|^2\sigma^2\omega^2/\delta$, i.e., the growth rate is proportional to the square of the amplitude of the parent mode. The latter case is the one similar to the excitation mechanism under discussion here. It should be noted that this parametric instability analysis gives a dependence of the growth rate in $\delta$ which is not in agreement with the numerical results (considering $\delta$ to be equivalent to $\beta$), because the dependence of the spatial structure of the intermediate wave on the dissipation is not considered in this simplistic analysis. Also, the parametric instability analysis suggests that the daughter modes gain energy from the parent mode. This is not true of the coupling mechanism under consideration since here the differential rotation of the disc, and not the warp, is the ultimate source of energy for the r~mode. Although the parametric instability analysis gives a dependence of the growth rate on the disturbance amplitude in agreement with the numerical results, it is simplistic and does not consider all the details of the coupling mechanism. A parallel between the parametric instability and the excitation mechanism is therefore not straightforward.

In the strong dissipation regime, the intermediate wave dissipates completely, and does not influence the variation of the r~mode growth rate with both the sound speed of the disc and the spin of the black hole. In this regime, the growth rate decreases with increasing $c_\mathrm{s}$, as expected. The hotter the disc is, the wider the modes get, which means that if the sound speed is high, the modes are not as well trapped. More importantly, the shape of the warp changes when the sound speed changes since its wavelength $\lambda_\textrm{W}$ is proportional to $c_\textrm{s}$. The interaction relies on the use of the warp twice, therefore it can be argued that the growth rate is proportional to $|dW/dr|^2$ ($|W|$ only represents the tilt and not the ``true'' warp). Since $|dW/dr|^2\propto W_\textrm{in}^2/\lambda_\textrm{W}^2\propto W_\textrm{in}^2/c_\textrm{s}^2$, for fixed sound speed the growth rate is proportional to the square of the inner warp amplitude [Fig.~\ref{relwarp} (b)]  and for fixed $W_\textrm{in}$, the growth rate varies with $1/c_\textrm{s}^2$ [Fig.~\ref{relwarp} (c)]. The small changes to the $1/c_\mathrm{s}^2$ law are justified by the fact that, for large sound speed, the decay rate due to the ``leakage''  at $r_\textrm{out}$ is considerable.

As for the variation of the growth rate with the spin of the black hole, an important conclusion is that, in fact, as argued in the beginning of this chapter, there is no mode excitation if the black hole is non-rotating. This was not evident in Kato's simple estimates for the growth rates. The fact that the growth rate increases with $a$ is also expected, not only because the waves are better confined for larger values of $a$, but mainly because the average warp amplitude in the trapped region is larger for larger $a$. Also, the wavelength of the warp decreases as $a$ increases because the Lense--Thirring frequency [cf. equation (\ref{lt})] increases. Therefore, for the same inner $W$, a larger $dW/dr$ is achieved. In light of these very preliminary results, it could be argued that HFQPOs would preferentially be detected in black hole candidates with large spin, if the interaction of the trapped r~mode with a warped disc is the mechanism responsible for the excitation of the former.

\subsection{Growth rates in eccentric discs}

For the interaction between the r~mode and the intermediate wave in an eccentric disc, the variation of the growth rate with the dissipation factor is similar to the same variation for the interaction in a warped disc [Figs~\ref{relecc} (a) and \ref{relwarp} (a)]. Also, the change of the growth rate with the inner eccentricity [Fig.~\ref{relecc} (b)] is very similar to the variation of the growth rate with the warp tilt at the inner radius, i.e., there is a square dependence on the eccentricity amplitude at $r_\mathrm{in}$. This is expected as the eccentric mode plays, in the wave interaction described in this section, the role of the warp in the interactions described previously. Similarly, the variation with the sound speed [Fig.~\ref{relecc} (c)] is also the expected one. 

The main difference between the interaction with the warp and the interaction with the eccentric mode is in the variation of the growth rate with the spin of the black hole [Fig.~\ref{relecc} (d)]. The warp, being a $n=1$ mode, has a variation with radius that strongly depends on the Lense--Thirring precession frequency, and therefore, that strongly depends on $a$. On the other hand, the variation of the eccentricity amplitude with radius, given by equation (\ref{eqe}), is less dependent on the spin of the black hole. Therefore, the variation of $a$ only causes variations of a factor of, at maximum, two in the growth rate obtained for the interaction in an eccentric disc. A very important difference is the fact that, in this interaction, a reasonable growth rate can be obtained in the case where $a=0$. Therefore, in slowly rotating black holes, HFQPOs might be detected if the disc is eccentric, and if this excitation mechanism is responsible for the increase in the amplitude of oscillations.

\section{Conclusions}
\label{grconc}

In this chapter I have described an excitation mechanism for trapped inertial modes, based on a non-linear coupling mechanism between these waves and global deformations in accretion discs. It was seen that the interaction of a trapped r~mode with an intermediate wave and a deformation in the disc, results in growth of the trapped mode, if there is some process capable of making the intermediate wave dissipate into the disc. Dissipation is required so that this mode can remove rotational kinetic energy from the disc, which becomes available for the r~mode to grow. Depending on the values of the sound speed, spin of the black hole and amplitude of the deformation at the inner radius, reasonable growth rates can be obtained for warps of modest amplitude. In a warped disc, since the growth rate varies significantly with the spin of the compact object, growth rates as large as $\omega/10$, where $\omega$ is the oscillation frequency, can be obtained. If $a=0$, no oscillations are excited in these discs. However, it is still possible to detect HFQPOs in non-rotating black holes, if the discs around them are eccentric.

The coupling process described here works as an excitation mechanism for trapped inertial waves, under a wide range of conditions, provided global deformations reach the inner disc region with non-negligible amplitude.  \enlargethispage{\baselineskip} The propagation of global modes in a more realistic disc model is the subject of the next chapter. 

Here I considered the excitation of trapped waves due to a non-linear coupling mechanism with global deformations in a simple disc model. While this effect is responsible for the growth of these modes, it has to compete with others that contribute to the damping of these waves. For example, since the conditions at the marginally stable orbit are unknown, it is possible for a ``leakage''  of the trapped mode (similar to the one considered in the potential barrier analogy) through $r_\textrm{ms}$ to exist. This effect is to be considered in chapter \ref{reflect}. Also viscous dissipation in the disc can cause damping of these modes. A simple estimate gives a damping rate of $\alpha\Omega$. For small enough values of $\alpha$ and large enough warp or eccentricity, net growth can occur.

Another point to be discussed is the applicability of the results obtained here to observed discs. A very simple, isothermal disc model and wave perturbations for which $\gamma=1$ was considered. In a more realistic disc, the vertical structure of the waves is changed while they propagate radially. The wave energy concentrates either near the surface of the disc \citep{lubowogilvie1998} or towards the disc mid-plane \citep{korycanskypringle1995}, which could potentially difficult the propagation of intermediate waves away from the Lindblad resonance, where they are excited, and make the coupling less efficient. However, the process of ``wave channelling'' mentioned by \cite{lubowogilvie1998} is only relevant at a distance from the resonance of $\sim r_\textrm{L}/m$, where $r_\textrm{L}$ is the radius of the Lindblad resonance. Since the intermediate waves have $m=1$, this effect is not important in the region where wave coupling occurs. The same is expected for discs where the energy concentrates towards the disc mid-plane. Therefore, I believe that the results of this chapter, obtained in a simple disc model, are still qualitatively valid in more realistic, observed discs. This matter is discussed further in chapter \ref{conc}, where the fully-relativistic, MHD simulations of tilted discs of \cite{heniseyetal2009} are commented on.

\thispagestyle{empty}
\chapter{Warp and Eccentricity}
\label{warpecc}

\section{Introduction}

\subsection{Background}

Discs around black holes are warped if the spin axis of the central object and the angular momentum of the accreting matter are misaligned. The fluid elements in tilted orbits will be subject to Lense--Thirring (gravitomagnetic) precession which tends to twist and warp the disc \citep{bardeenpetterson1975}. In several X-ray binaries, jets (which are thought to be aligned with the rotation axis of the black hole) are observed to be misaligned with respect to the binary rotation axis \citep[e.g. GRO J1655-40,][]{maccarone2002}. This is strong evidence for a warped disc around the compact object in the binary \citep{mtp2008}. As mentioned previously, warping might also be induced by radiation pressure forces \citep{pringle1996}. Despite being more likely for discs around neutron star primaries \citep{ogilviedubus2001}, radiation-driven warping might also occur around black holes provided the disc extends to large enough radius, as in the case of GRS 1915+105. \cite{rauetal2003} report the discovery of a possible, long-term periodicity in this X-ray binary which may be interpreted as the precession of a radiation-induced warp in the disc.

Eccentric accretion discs are also believed to exist in black hole binaries or, more generally, systems with mass ratio $q\lesssim0.3$ (section \ref{eccdiscs}). Eccentricity can result from an instability involving the orbiting gas and the tidal potential of the companion star. Discs may become eccentric if they are large enough to extend to the 3:1 resonance, the radius in the disc where its angular velocity equals three times that of the binary \citep{lubow1991a,lubow1991b}. The strongest observational evidence for eccentric discs is the detection, in the light curves of accreting binary systems, of long-period modulations known as superhumps, which can be explained by the action of tidal stresses on a precessing eccentric disc \citep{whitehurst1988}. Although this phenomenon is usually associated with cataclysmic variable stars, superhumps have been detected in an increasing number of black hole binaries \citep{donocharles1996,haswelletal2001,uemura2002,neiletal2007,zurita2008}. In fact, black hole binaries are likely to have mass ratios $q\lesssim0.3$ and therefore to have eccentric discs during at least some stages of their outbursts. \cite{haswelletal2001} noted that the mechanism responsible for superhump luminosity variations differs in X-ray binaries and cataclysmic variables, but the underlying dynamics is the same.

The goal of this chapter is to describe the propagation of global eccentric and warping disturbances, under a variety of conditions, from the outer parts of the disc where they originate to the inner parts. In the outer parts of the disc, the motion is Keplerian and stationary or slowly precessing global deformations are supported. Since the Lense--Thirring precession frequency increases with decreasing $r$, tilted orbits at different radii will tend to precess at different rates, which tends to twist the disc. Hydrodynamic stresses counteract this effect, resulting in wavelike warp propagation from the outer disc to the inner region. Similarly, for eccentric discs a transition has to be made between the Keplerian region where eccentric instabilities are driven, and the inner region where general relativistic effects dominate the precession of elliptical orbits. It is this connection between different regions of the disc, with different characteristics, that I attempt to describe.

In chapter \ref{oscilintro} I worked with a simplified set of hydrodynamic equations to describe the warp and eccentricity as propagating global modes. Such a simplified description was introduced in order to deal with the non-linear couplings of different wave modes of chapter \ref{excmech}. Within this framework an equation for the warp tilt $W$ at each radius \citep[similar to the one given by][]{papaloizoulin1995} was obtained assuming the warp to be a zero-frequency mode propagating in a strictly isothermal, relativistic disc. In addition to being valid only where $c_\textrm{s}$ is constant and where viscous effects are negligible, the equation derived does not hold for large radius, where the wavelength of the warp becomes comparable to $r$, since some terms were omitted in the local approximation applied. A similar method was used to derive an equation for the eccentricity at each radius $E$ valid only for an inviscid and strictly isothermal disc.

Here I consider the more general description of a stationary, wave-like warp and eccentricity given by \cite{lubowetal2002} and \cite{goodchildogilvie2006}, respectively. These theories are less general than the ones formulated by \cite{ogilvie2000,ogilvie2001} since they consider deformations to have small amplitude, and not all viscous or turbulent effects are included. Also, the \cite{goodchildogilvie2006} calculation uses a simplified, 2D disc model. However, the secular theories of \cite{lubowetal2002} and \cite{goodchildogilvie2006} are appropriate for thin discs of any given structure, and allow for viscous (turbulent) damping of deformations.

\subsection{Black hole states and high-frequency QPOs}

As seen in the introduction of this thesis, observations indicate that black hole systems can be found in different spectral states, going from the quiescent state to low, intermediate, high and very high states as the mass accretion rate increases \citep[see][and introduction of this thesis]{bhbbook}. The very high state, dominating close to the Eddington limit, is the one where high-frequency QPOs are almost exclusively detected.

The accretion flow has different characteristics in different black hole states. If the disc properties are such that the warp and eccentricity can propagate to the inner region, the interaction between relativistic effects and global deformations can give rise to interesting phenomena. In particular, the warp and eccentricity can have a fundamental role in exciting trapped inertial modes, which may explain high-frequency QPOs \citep{nowaketal1997}, as previously suggested by \cite{katowarp2004,kato2008} and by the results of chapter \ref{excmech}. 

Here I show that the inward propagation of warp and eccentricity is facilitated for high accretion rate. For fixed viscosity, global deformations reach the inner region of the disc with a modest amplitude when the accretion rate is high, and can take part in the excitation mechanism for inertial modes trapped in this region. If high-frequency QPOs can be identified with these trapped inertial waves, it can be argued that they are predominantly detected in the very high state, where the accretion rate is close to Eddington, because only in this case global deformations can reach the inner region with non-negligible amplitude and excite the trapped modes.

\subsection{Chapter outline} 

In section \ref{discmodel}, I introduce the disc model in which the propagation of global deformations is studied. I consider a disc with a polytropic structure in the vertical direction, constant opacity, and with gas and radiation contributing to the total pressure. In section \ref{propagation} I describe the propagation of global disturbances by solving equations for the warp tilt and eccentricity at each radius. To mimic several possible conditions in a disc with a fixed viscosity $\alpha$ around a black hole of a given mass and spin, I vary the accretion rate and the damping of the warp and eccentricity. The results obtained are discussed in section \ref{wediscussion}. Conclusions are presented in section \ref{wediscussion}.

\section{Disc model}
\label{discmodel}

The disc model considered in this section is similar to the innermost regions of the model introduced by \cite{ss73}. The flow is assumed to be geometrically thin and optically thick with constant opacity but the vertical structure is treated more carefully than in the standard model. The reason for this is that a rather accurate treatment of the vertical structure is necessary since, e.g., a factor 2 error in the disc thickness makes a considerable difference to the propagation of global deformations to the inner region. I use the traditional viscosity prescription, $\mu=\alpha p/\Omega_\mt{K}$, that is, as in the original model, the viscous torques responsible for angular momentum transfer are taken to be proportional to the total pressure. Even though such discs are believed to be viscously \citep{le74} and thermally \citep{ss76} unstable in the radiation pressure dominated regime, simulations \citep{hiroseetal2009} don't show signatures of such instabilities (see discussion at the end of section \ref{instability}). In any case, and for comparison, I also calculate a disc model where the stress scales with the gas pressure only.

It should be noted that the model presented in this section is Newtonian. The effect of the black hole gravity shows up only in the choice of the inner radius, which is taken to be the marginally stable orbit and is, therefore, dependent on the rotation of the compact object. The relativistic correction factors of \cite{nt73} are not of great significance here since the most important relativistic effects in the study of warp and eccentricity propagation are the apsidal and nodal precession. Therefore, throughout this section, $\Omega=\Omega_\mt{K}=\Omega_z=\sqrt{GM/R^3}$. However, relativistic expressions for the characteristic frequencies are employed in the remainder of the chapter to correctly represent the apsidal and nodal precession in the equations for warp and eccentricity.

\subsection{Vertical structure}

Independently of the viscosity prescription, the radiation-pressure dominated region might suffer from another type of instability aside from the classic thermal and viscous instabilities. In the Shakura \& Sunyaev model, the dissipation rate per unit volume is independent of $z$. If the dissipation rate per unit mass is also vertically constant, then the density is independent of $z$, vanishing abruptly at the vertical boundaries of the disc, $z=z_0(R)$ \citep{ss73}. 
A disc with these properties is subject to convective instability \citep{bb77}.

Simulations of radiation-dominated discs by \cite{turner2004} indicate that neither is the dissipation per unit mass constant nor is the density independent of $z$. In fact, the computed density profile resembles more that of a polytropic model with index $s=3$ \citep[see also][]{agoletal2001}. Using the polytropic law, $p\propto \rho^{4/3}$, in the hydrostratic equilibrium equation (\ref{hydroeq}) it is straightforward to get

\begin{equation}
\rho=\rho_0(R) \left[1-\left(\frac{z}{z_0}\right)^2\right]^3 \quad (-z_0<z<z_0).
\label{rho}
\end{equation}
This result was used by \cite{bb77}, who argued that convection in the radiation-pressure dominated region would establish a vertically isentropic structure ($T^3/\rho=$ constant in $z$).

The $s=3$ polytropic structure has the convenient property of allowing for a ratio of radiation to gas pressure,

\begin{equation}
\beta_\mt{rg}=\frac{p_\textrm{r}}{p_\textrm{g}}=\frac{4\sigma\mu_\textrm{m} m_{\textrm{p}}}{3ck_B}\frac{T^3}{\rho}, 
\end{equation}
that is independent of $z$. Using (\ref{rho}) in the hydrostratic balance equation it is simple to get $p=p_0(R)\left[1-\left(z/z_0\right)^2\right]^4$, where

\begin{equation}
p_0=\frac{\Omega_z^2z_0^2}{8}\rho_0=\frac{\Omega^2z_0^2}{8}\rho_0.
\label{he2}
\end{equation}
This is consistent with the equation of state (\ref{eqstate}) provided $T=T_0(R)\left[1-\left(z/z_0\right)^2\right]$, and independently of the variation of $\beta_\mt{rg}$ with radius. Therefore, the polytropic vertical structure with $s=3$ can be used to model not only the radiation pressure dominated regime but also the gas pressure dominated region. In fact, a full treatment of the vertical structure of a gas pressure dominated disc indicates that the resulting profiles are very similar (except perhaps $F(z)$ for large $z$) to those obtained for a polytropic model with $s\approx 2.7$ \citep{ogilvienotes}. 
This result, and the simulations in radiation-pressure dominated regions, indicate that assuming a polytropic vertical structure with $s=3$ throughout the radial extent of the disc is a good approximation, provided the opacity remains constant. This is true of the innermost regions [(a) and (b) in the Shakura \& Sunyaev model] of the disc. Since I'm most interested in the propagation of global deformations in these regions, I take the opacity to be constant throughout the disc, i.e., region (c) is neglected in the model presented. In fact, the physics of region (c) is more complicated than assumed by \cite{ss73}, because of the presence of different sources of opacity and the importance of irradiation and partial ionisation.

\subsection{Radial structure}

The assumptions of hydrostatic equilibrium and vertical isentropy (or equivalently, $s=3$ polytropic structure in $z$), together with the equation of state, are enough to determine the vertical structure of the disc. In other words, it is possible to determine the variation of $\rho$, $p$ and $T$ with $z$ without the requirement of radiative balance \citep[see discussion in section 3.1 of][]{agoletal2001}. However, to fully determine the radial structure, $\rho_0(R)$, $p_0(R)$, $T_0(R)$ and $z_0(R)$, two more equations are needed since the polytropic relation is not valid in the radial direction.

Since matter is being accreted in the disc, there are radial drift motions which transport mass and angular momentum. In steady thin discs the conservation of both mass and angular momentum can be expressed in the form (\ref{rstructure1}): $\bar{\nu}\Sigma=(\dot{M}/3\pi)(1-\sqrt{R_\textrm{in}/R})$. Here, $R_\textrm{in}$ is taken to be the marginally stable orbit and the (density-weighted) mean kinematic viscosity $\bar{\nu}(R)$ is defined as usual by

\begin{equation}
\bar{\nu}\Sigma(R)=\int^{z_0}_{-z_0}\mu(R,z) \textrm{d}z=\alpha\frac{P}{\Omega},
\end{equation}
where

\begin{equation}
\Sigma(R)=\int^{z_0}_{-z_0}\rho(R,z)\textrm{d}z=\frac{32}{35}\rho_0z_0,
\end{equation}
\begin{equation}
P(R)=\int^{z_0}_{-z_0}p(R,z) \textrm{d}z=\frac{256}{315}p_0z_0.
\end{equation}

The final equation comes from energy considerations. It is traditional to assume that the disc is in radiative equilibrium. However, convection, turbulence or other motions can also contribute to the transport of energy to the disc surface. In fact, if the disc has a polytropic vertical structure with $s=3$, the radiative diffusion law gives $F_{\textrm{rad}}$ proportional to $z$. If $F=F_{\textrm{rad}}$, i.e., if radiative diffusion carries the entire energy flux, then (\ref{eb}) gives a dissipation rate per unit volume independent of $z$. This is neither consistent with the $\alpha$ viscosity prescription assumed here since $p=p_0(R)\left[1-\left(z/z_0\right)^2\right]^4$, nor is in agreement with simulations of radiation-pressure dominated discs \citep{turner2004}. Therefore, here I use the energy balance equation in the form

\begin{equation}
\int_{0}^{z_0}\frac{\partial}{\partial z}\left(F_{\textrm{rad}}+F_{\textrm{extra}}\right)\textrm{d}z=\int_{0}^{z_0}\mu\left(R\Omega'\right)^2\textrm{d}z,
\end{equation}
where the energy transport by other motions rather than by radiative diffusion is encompassed in $F_{\textrm{extra}}$. Since radiation is supposed to carry the entire heat flux at the photosphere $z=z_0$, I assume that the extra term integrates to zero so that the energy balance can be written as

\begin{equation}
F_{\textrm{rad}}(R,z_0)=\frac{1}{2}(R\Omega')^2\left(\alpha\frac{P}{\Omega}\right),
\end{equation} 
since $F_{\textrm{rad}}(R,z=0)=0$ by symmetry.

In summary, the equations necessary to fully determine the radial structure of the disc (hydrostatic equilibrium, equation of state, energy balance, conservation law) can be written in the form

\begin{equation}
\frac{P}{H^2\Sigma}=\frac{GM}{R^3},
\label{rs1}
\end{equation}
\begin{equation}
\beta_\mt{rg}(1+\beta_\mt{rg})^3\frac{\Sigma^4}{H P^3}=\frac{729}{140}\frac{\sigma}{c}\left(\frac{\mu_\textrm{m} m_{\textrm{p}}}{k_B}\right)^4,
\label{rs2}
\end{equation}
\begin{equation}
\frac{P^4}{\Sigma^5(1+\beta_\mt{rg})^4}=\frac{35}{1458\pi}\frac{\kappa_{\textrm{T}}}{\sigma}\left(\frac{k_B}{\mu_\textrm{m} m_{\textrm{p}}}\right)^4\frac{GM}{R^3}\dot{M}f,
\label{rs3}
\end{equation}
\begin{equation}
P=\sqrt{\frac{GM}{R^3}}\frac{\dot{M}}{3\pi\alpha}f.
\label{rs4}
\end{equation}

Here I take $\mu_\textrm{m}=0.615$, $\kappa_T=0.33\textrm{ cm}^2\textrm{g}^{-1}$, and use non-dimensional parameters to represent the dependence on mass, accretion rate and radius. The radial structure of the disc can be represented in terms of $\Sigma(R)$, $P(R)$, $\beta_\mt{rg}(R)$ and $H(R)$, the (density-weighted) scale-height of the disc defined by

\begin{equation}
H^2=\frac{\int_{-z_0}^{z_0}\rho z^2\textrm{d}z}{\int_{-z_0}^{z_0}\rho\textrm{d}z}=\frac{z_0^2}{9}.
\label{dt}
\end{equation}
The equations read:
\begin{equation}
\Sigma=2.5\times10^5\, \dot{m}^{3/5}\alpha^{-4/5}m^{1/5}f^{3/5}r^{-3/5} (1+\beta_\mt{rg})^{-4/5},
\label{rsc}
\end{equation} 
\begin{equation}
P=3.0\times10^{22}\,\dot{m}\,\alpha^{-1}fr^{-3/2},
\label{rsd}
\end{equation} 
\begin{equation}
\beta_\mt{rg} (1+\beta_\mt{rg})^{-3/5}=3.6\times10^2\, \dot{m}^{4/5}\alpha^{1/10}m^{1/10}f^{4/5}r^{-21/20},
\label{rsa}
\end{equation}
\begin{equation}
H=1.7\times10^3\,  \dot{m}^{1/5}\alpha^{-1/10}m^{9/10}f^{1/5}r^{21/20} (1+\beta_\mt{rg})^{2/5},
\label{rsb}
\end{equation} 
where $m$ is the mass in units of $M_\odot$, $\dot{m}$ is the accretion rate in units of the Eddington accretion rate assuming an efficiency\footnote{Here the ``accretion efficiency'' is assumed slightly higher than assumed in chapter \ref{intro} ($0.06$) as black holes of non-zero spin are being considered.} of $0.1$, $f=1-\sqrt{r_\textrm{in}/r}$, where $r_\textrm{in}$ is the dimensionless, spin dependent, radius of the marginally stable orbit; $H$, $\Sigma$ and $P$ are in cgs units. It should be noted that in the limits $\beta_\mt{rg}\gg1$ ($P_{\textrm{r}}$ dominates) and $\beta_\mt{rg}\ll1$ ($P_{\textrm{g}}$ dominates) the same dependencies in $m$, $\dot{m}$, $\alpha$, $r$ and $f$ as in regions (a) and (b), respectively, of the Shakura--Sunyaev model are recovered [cf. equations (\ref{ss1})--(\ref{ss2})]: \newline
\newline
\textbf{(a) $\vc{\beta_\mt{rg}\gg1}$:}
\begin{equation}
\Sigma=1.9\,\alpha^{-1}\dot{m}^{-1}r^{3/2}f^{-1},
\end{equation}
\begin{equation}
\beta_\mt{rg} =2.4\times10^6\, \dot{m}^{2}\alpha^{1/4}m^{1/4}f^{2}r^{-21/8},
\end{equation}
\begin{equation}
H=6.1\times10^5\dot{m}\,m\,f.
\label{ha}
\end{equation}
\vspace{5pt}
\textbf{(b) $\vc{\beta_\mt{rg}\ll1}$:}
\begin{equation}
\Sigma=2.5\times10^5\,\alpha^{-4/5}\dot{m}^{3/5}m^{1/5}r^{-3/5}f^{3/5},
\label{sdb}
\end{equation}
\begin{equation}
\beta_\mt{rg}=3.6\times10^2\, \dot{m}^{4/5}\alpha^{1/10}m^{1/10}f^{4/5}r^{-21/20},
\end{equation}
\begin{equation}
H=1.7\times10^3\,\alpha^{-1/10}\dot{m}^{1/5}m^{9/10}r^{21/20}f^{1/5}.
\label{hb}
\end{equation}
The expression for the total pressure in both regions is the same and is given by (\ref{rsd}). The differences in the numerical factors are mainly due to the treatment of the vertical structure.

\begin{figure*}
\begin{center}
\includegraphics[width=0.49\linewidth]{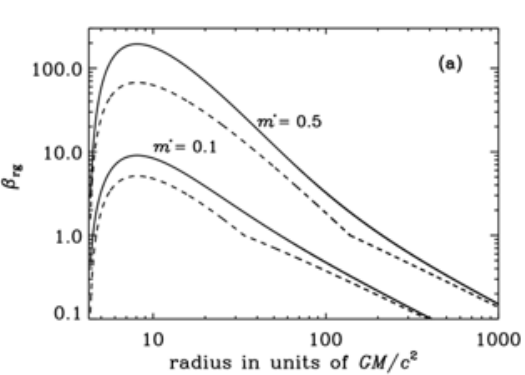}
\includegraphics[width=0.49\linewidth]{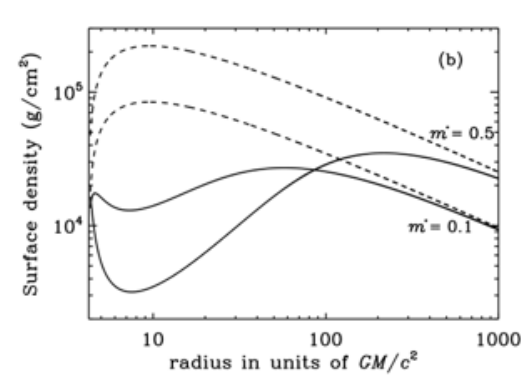} \\

\includegraphics[width=0.49\linewidth]{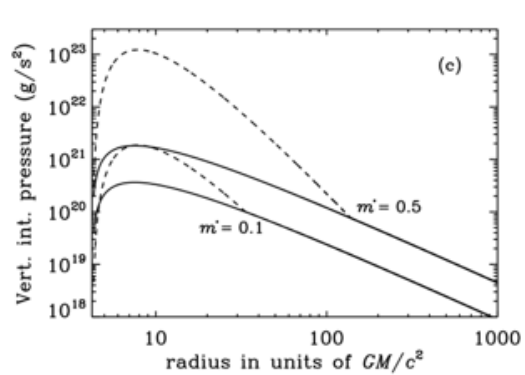}
\includegraphics[width=0.49\linewidth]{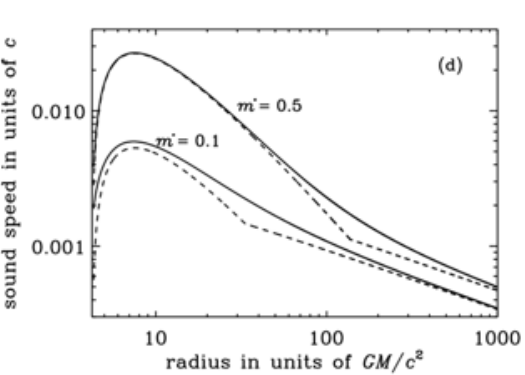}
\caption{Log-log plot of the variation of (a) $\beta_\mt{rg}=P_\textrm{r}/P_\textrm{g}$, (b) surface density, (c) vertically integrated pressure and (d) sound speed in units of $c$ with radius for $\dot{m}=0.1,\,0.5$, $\alpha=0.1$, $a=0.5$ and $m=10$. The full lines represent the disc model with a combination of gas and radiation pressure when the stress scales with total pressure, while the dashed lines correspond to solutions in regions (a) and (b) when the stress scales with the gas pressure only.}
\label{model}
\end{center}
\end{figure*}

In order to get a smooth transition between the two regions, and to allow for a combination of gas and radiation pressure throughout the disc, equation (\ref{rsa}) is solved numerically. 
The resultant function, $\beta_\mt{rg}(r)$, is then used in (\ref{rsb}) and (\ref{rsc}) in order to get expressions for $H(r)$ and $\Sigma(r)$. The radial disc structure obtained here is used in the study of the warp and eccentricity propagation, which is described in the following section.

Care is taken over the basic disc model because it is important to obtain a smooth density profile for the disc in order to calculate the propagation of the warp and eccentricity. Also, since $H^2$ appears in the equations for those quantities, the factor of $1/9$ in equation (\ref{dt}) makes a significant difference.

For comparison, I also include in this section the expressions obtained for region (a), if the stress scales with the gas pressure instead of total pressure ($\mu=\alpha p_g/\Omega$). The semi-thickness of the disc is independent of the viscosity prescription, so that (\ref{ha}) is still valid. The surface density is given by (\ref{sdb}) while $\beta_\mt{rg}$ and $P$ are given by

\begin{equation}
P=3.9\times10^{27}\, \dot{m}^{13/5}\alpha^{-4/5}m^{1/5}f^{13/5}r^{-18/5},
\end{equation}
\begin{equation}
\beta_\mt{rg} =1.3\times10^5\, \dot{m}^{8/5}\alpha^{1/5}m^{1/5}f^{8/5}r^{-21/10}.
\end{equation}

Fig.~\ref{model} shows how $\beta_\mt{rg}$, the surface density, the vertically integrated pressure and the isothermal sound speed (squared), $c^2_{\mathrm{s}}=p_0/\rho_0=\frac{9}{8}\Omega^2H^2$, vary with radius for a disc with $\alpha=0.1$, $a=0.5$ and $m=10$, for two different values of the accretion rate. The differences in the disc structure when the stress scales with the total pressure or with the gas pressure are more evident in the innermost region of the disc, and for higher accretion rate, i.e., where radiation pressure is more important. The exception is the sound speed, proportional to the thickness of the disc, since $H$ in the radiation-pressure dominated region is independent of the viscosity prescription.

\section{Stationary propagation of warp and eccentricity}
\label{propagation}

In the general case, the warp tilt and the eccentricity are not only functions of space but also of time, and their evolution has been studied by (e.g.) \cite{lubowetal2002} and \cite{ogilvie2001}, respectively. However, the characteristic time-scale for the precession of a global warp or eccentricity is small when compared to the orbital frequency in the binary system, which implies that the frequency of these global deformations is negligible compared to the characteristic frequencies in the inner part of the disc. Therefore, I choose to study the steady shape of a warped or eccentric disc around a black hole.

\subsection{Equations}

In this section I introduce the equations that are used to describe the stationary wave-like propagation of warp and eccentricity in the disc model presented in the previous section. 

The variation of the tilt of the disc, $W$, with radius can be obtained from \citep{lubowetal2002}

\begin{equation}
\frac{\textrm{d}}{\textrm{d}R}\left[\left(\frac{PR^3\Omega^2}{\Omega^2-\kappa^2+2i\alpha_\textrm{W}\Omega^2}\right)\frac{\textrm{d}W}{\textrm{d}R}\right]=\Sigma R^3(\Omega_z^2-\Omega^2)W.
\label{luboweq}
\end{equation}
This equation describes how propagating bending waves communicate the warp through the disc and is valid for small-amplitude warps; $W$ describes the amplitude and phase of the inclination of the disc. The local wavelength of these bending waves is approximately the one given by the dispersion relation (\ref{ssdr}) for $(\omega,m,n)=(0,1,1)$, if $\alpha_\textrm{W}=0$. The warp propagation is subject to viscous decay, which is described here by a dimensionless viscosity parameter designated $\alpha_\textrm{W}$. In the general case of non-isotropic viscosity, $\alpha_\textrm{W}\neq\alpha$ as the former is related to the $T_{rz}$ and $T_{\phi z}$ components of the stress tensor \textbf{T} while the latter is related to $T_{r\phi}$ \citep[see][and references therein]{lubowetal2002}.

The stationary propagation of a small eccentricity through the disc can be described, in the simplest case, by \citep[e.g.][]{goodchildogilvie2006}

\begin{equation}
\frac{\textrm{d}}{\textrm{d}R}\left[\left(\gamma-\textrm{i}\alpha_\textrm{E}\right)PR^3\frac{\textrm{d}E}{\textrm{d}R}\right]+R^2\frac{\textrm{d}P}{\textrm{d}R}E=\Sigma R^3(\kappa^2-\Omega^2)E,
\label{ecceq}
\end{equation}
where $E$ is a (possibly) complex function representing the amplitude and phase of the eccentricity at a given radius. In a strictly isothermal disc, the ratio of specific heats $\gamma=1$ and the global modes described by this equation have a local wavelength which is approximately the one given by the dispersion relation (\ref{ssdr}) for $(\omega,m,n)=(0,1,0)$, if $\alpha_\textrm{E}=0$. The equation describing the eccentricity propagation is based on, and agrees with the local dispersion relation of, a 2D disc; 3D effects are discussed by \cite{ogilvie2001,ogilvie2008}. The parameter $\alpha_\textrm{E}$ in equation (\ref{ecceq}) is essentially a bulk viscosity. Effects of shear viscosity are not included as its direct implementation may lead to growing eccentric waves \citep[viscous overstability,][]{kato1978} \citep[see also][references therein and Part \ref{ta} of this thesis]{ogilvie2001}. The process of turbulent eccentricity damping is poorly known and the simplest way of describing it is by using a dimensionless bulk viscosity parameter. 

In order to include relativistic effects in the problem, I take $\Omega^{-1}=(GM/c^3) (r^{3/2}+a)$ and use expressions (\ref{relkappa}) and (\ref{omegazfreq}) for the radial and vertical epicyclic frequencies, respectively. For the surface density $\Sigma$ and vertically integrated pressure $P$, expressions (\ref{rsc}) and (\ref{rsd}) are used.

It may be argued that a limitation of the disc model used in the calculations and introduced in section \ref{discmodel} is the fact that it does not include relativistic effects. For example, the law for angular momentum conservation in relativistic discs \citep{nt73} is slightly different from the one used here [cf. (\ref{rstructure1})]. However, relativistic corrections are small and their inclusion is not expected to significantly affect the final results. Also, since the terms on the right-hand side of equations (\ref{luboweq}) and (\ref{ecceq}) dominate the warp and eccentricity propagation, it is more important to introduce relativistic effects in these equations by using expressions (\ref{relomega})--(\ref{omegazfreq}) for the characteristic frequencies, as they correctly describe apsidal and nodal relativistic precession rates, related to $\Omega_z^2-\Omega^2$ and $\kappa^2-\Omega^2$.

\subsection{Numerical method}

Equations (\ref{luboweq}) and (\ref{ecceq}) are solved numerically, using a
4th order Runge--Kutta method. In the case of the warp, the boundary condition that corresponds to zero torque at the inner edge, $\mathrm{d}W/\mathrm{d}R(R_\mathrm{in})=0$, is employed. In order for the amplitude of the solution to be fixed, I specify the value $W_0=W(R_\mathrm{in})$ at the inner edge, which corresponds to the (small) inclination of the inner edge of the disc with respect to the equator of the black hole. For the eccentricity I use similar boundary conditions: $\mathrm{d}E/\mathrm{d}R(R_\mathrm{in})=0$ and $E_0=E(R_\mathrm{in})$. To avoid the formal singularity of the equations at the marginally stable orbit, here I take $R_\textrm{in}=R_\textrm{ms}+\delta R$, where $\delta R\ll R_\textrm{in}$ is an arbitrary value; the solutions for $W(R)$ and $E(R)$ are practically independent of the choice of $\delta R$. The solutions are linear and can be renormalized to obtain any desired $W_\textrm{out}$ and $E_\textrm{out}$, the values of disc tilt and eccentricity at the outer radius. This normalisation is particularly meaningful in the case of $W(R)$ since $W_\textrm{out}$ is the (constant) disc tilt at large radius, which can be related to observations.

To solve the warp and eccentricity equations I consider a black hole with 10 solar masses, i.e. $m=10$, and spin $a=0.5$. The ratio of specific heats is given by \citep[e.g.][]{stellar}

\begin{equation}
\gamma=\frac{5+40\beta_\mt{rg}+32\beta_\mt{rg}^2}{3+27\beta_\mt{rg}+24\beta_\mt{rg}^2}.
\end{equation}
In regions (a) and (b) this expression gives $\gamma\approx4/3$ and $\gamma\approx5/3$, respectively. The dimensionless viscosity parameter $\alpha$ is regarded as constant throughout the disc as in \cite{ss73}. Although simulations suggest values of $\alpha$ of the order $10^{-2}$, according to \cite{kingetal2007} observations indicate a typical range of $\alpha\sim0.1-0.4$. Here I choose to fix the viscosity parameter to $0.1$. Parameters $\alpha_\textrm{W}$, $\alpha_\textrm{E}$ and $\dot{m}$ can be varied. To mimic the transition between different black hole states I fix the viscous damping and vary the mass accretion rate. The results and corresponding discussion are presented in the next section.

\section{Results and Discussion}
\label{wediscussion}

\subsection{Undamped propagation}

\begin{figure*}
\begin{center}
\includegraphics[width=0.49\linewidth]{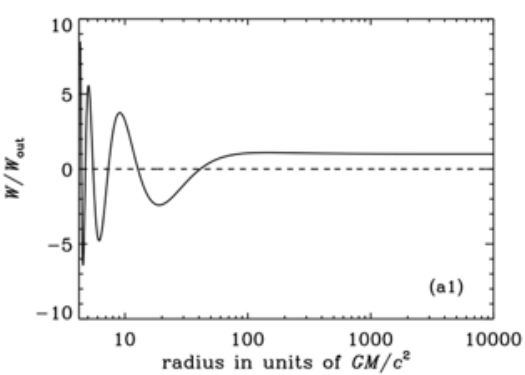}
\includegraphics[width=0.49\linewidth]{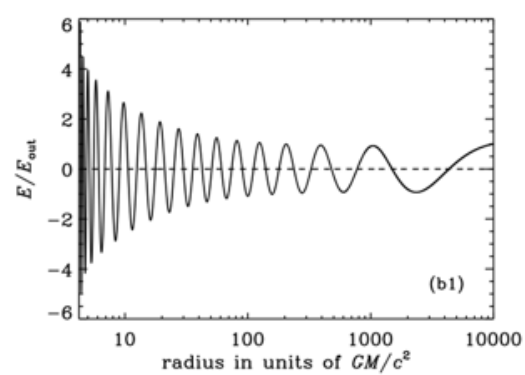}\\

\includegraphics[width=0.49\linewidth]{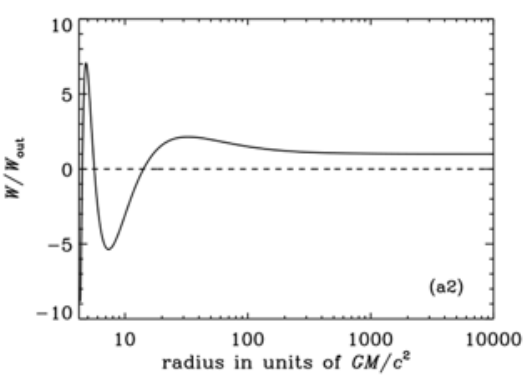}
\includegraphics[width=0.49\linewidth]{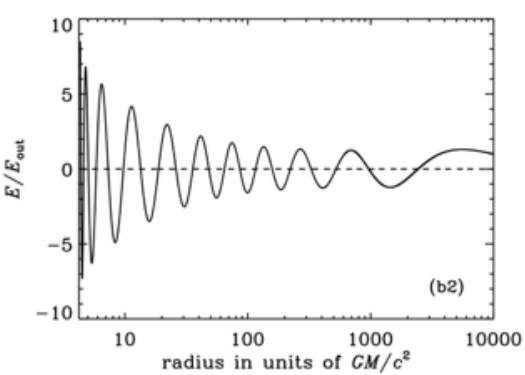}\\

\includegraphics[width=0.49\linewidth]{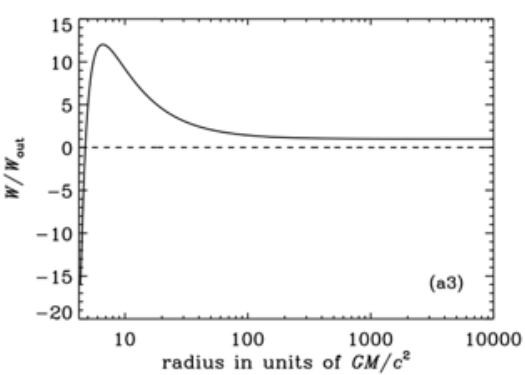}
\includegraphics[width=0.49\linewidth]{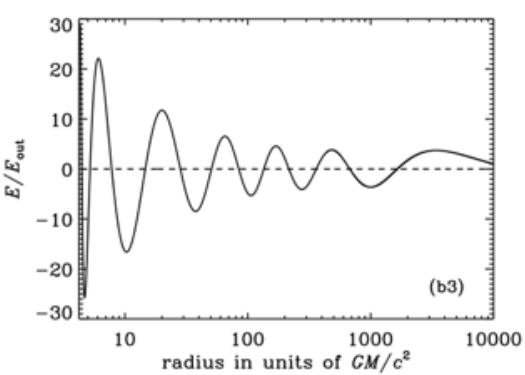}
\caption{Radial variation of (a) the warp tilt and (b) eccentricity normalised to their values at the outer radius for $\alpha_\textrm{W}=0= \alpha_\textrm{E}$ (i.e., no viscous damping) for (1) $\dot{m}=0.2$, (2) $\dot{m}=0.4$, (3) $\dot{m}=0.8$. The full line represents the real part of the disturbance while the imaginary part is represented by the dashed line. A logarithmic scale is used for the x-axis.}
\label{free}
\end{center}
\end{figure*}

\begin{figure*}
\begin{center}
\includegraphics[width=0.49\linewidth]{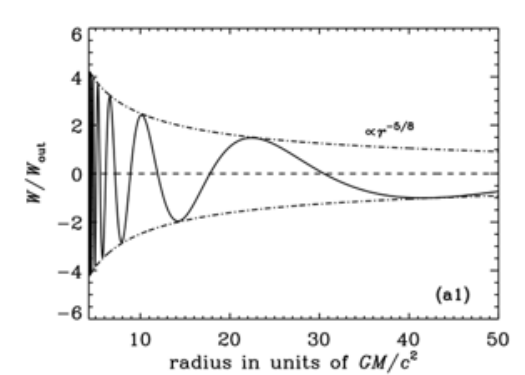}
\includegraphics[width=0.49\linewidth]{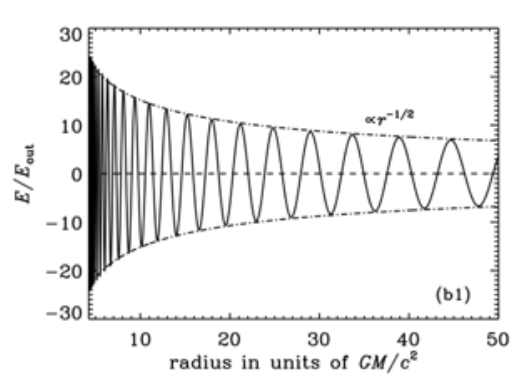} \\

\includegraphics[width=0.49\linewidth]{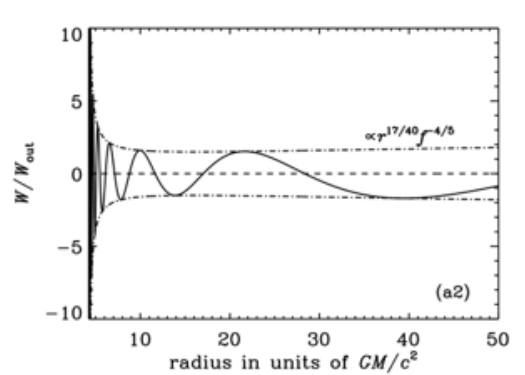}
\includegraphics[width=0.49\linewidth]{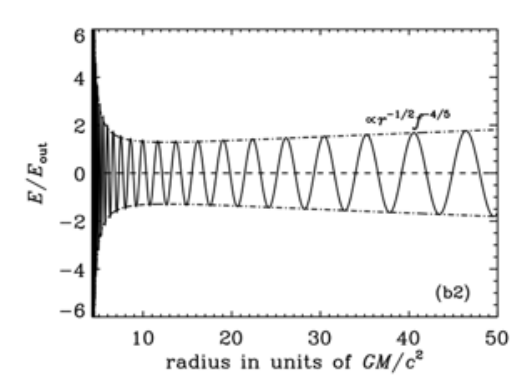}
\caption{Solutions for (a) warp and (b) eccentricity propagation in region (a) ($\beta_\mt{rg}\gg1$) for $\alpha_\textrm{W}=0= \alpha_\textrm{E}$, $m=10$, $a=0.5$, $\dot{m}=0.1$, when the stress scales with the (1) total pressure and the (2) gas pressure. The full line represents the real part of the disturbance while the imaginary part is represented by the dashed line. The dot-dashed curves show the radial variation of the amplitude of the deformations, as predicted by the WKB analysis.}
\label{wkb}
\end{center}
\end{figure*}

Fig.~\ref{free} shows the variation with radius of the warp tilt and eccentricity for increasing accretion rate when $\alpha_\textrm{W}=0= \alpha_\textrm{E}$. In this case no dissipation is present in the equations for global deformations, so they propagate everywhere with non-negligible amplitude. Since the imaginary part of the solutions is zero, the warp comprises a pure tilt, i.e., the disc is not twisted \citep{lubowetal2002}. For small accretion rate the warp has an oscillatory shape as previously described by \cite{ivanovillarionov1997}. The wavelength increases with radius and, at large $R$, $W$ tends to a constant value, the inclination of the outer disc with respect to the equator of the black hole. Note, however, that contrary to the results of \cite{bardeenpetterson1975} the inner disc is not necessarily aligned with the mid-plane of the compact object, as previously shown by \cite{lubowetal2002}. The simulations of \cite{globalsimulations} also found hints of an oscillatory structure characterised by a long wavelength with no indication of the Bardeen-Petterson effect. Eccentricity also has an oscillatory structure but its wavelength is shorter (yet much longer than the semi-thickness of the disc) and has a slower increase with radius [Fig.~\ref{eccwarp} (b)]. The shortening of the wavelengths at small radius of both the warp and eccentricity implies a sharp increase of the gradients $\textrm{d}W/\textrm{d}R$ and $\textrm{d}E/\textrm{d}R$ in the inner region of the disc. This is an important feature in terms of wave coupling in this region since the growth rate for inertial modes that are trapped below the maximum of the epicyclic frequency, and interact with global deformations, is proportional to $|\textrm{d}W/\textrm{d}R|^2$ or $|\textrm{d}E/\textrm{d}R|^2$ (chapter \ref{excmech}).

The basic behaviour of the numerical solutions presented in this section can be explained using WKB theory. In particular, it is possible to predict how the wavelength and amplitude of the several solutions scale with radius. A WKB analysis of the secular theories can be performed by assuming the deformations $W$ and $E$ to be proportional to $\exp(\int\textrm{i}\,b\,\textrm{d}R)$, where $b$ is a function of $R$ to be determined. The $\textrm{d}P/\textrm{d}R$ term in equation (\ref{ecceq}) can be ignored in the WKB analysis since relativistic precession dominates the propagation of eccentricity. Letting $D_\textrm{W}=W(R)$ and $D_\textrm{E}=E(R)$, equations (\ref{luboweq}) and (\ref{ecceq}) can be written in the form

\begin{equation}
\frac{\textrm{d}}{\textrm{d}R}\left(g_i\frac{\textrm{d}D_i}{\textrm{d}R}\right)+h_iD_i=0, \quad i=\textrm{W},\textrm{E}.
\end{equation}
The WKB solution is valid where the radial variation of $g_i$ and $h_i$ is slow and is given by
\begin{equation}
D_i\approx(g_ih_i)^{-1/4}\exp\left[\pm \int \textrm{i}\sqrt{h_i/g_i}\textrm{d}R\right].
\end{equation}
In the inviscid case ($\alpha_\textrm{W}=0=\alpha_\textrm{E}$), $g_\textrm{W}\approx PR^3 / 6 r^{-1}$, $h_\textrm{W}\approx\Sigma R^3\Omega^24ar^{-3/2}$, $g_\textrm{E}=\gamma PR^3$, and $h_\textrm{E}\approx\Sigma R^3\Omega^26r^{-1}$, keeping only the lowest-order terms in the expressions for $\Omega^2-\kappa^2$ and $\Omega^2-\Omega_z^2$. Remembering that $P=\Omega^2H^2\Sigma$ [from equation (\ref{rs1})], one can write 

\begin{equation}
W\propto r^{1/8}(\Sigma H)^{-1/2}\exp\left[\pm \int \textrm{i}\sqrt{\frac{24a}{r^{5/2}H^2}}\textrm{d}R\right],
\label{wwkb}
\end{equation}
\begin{equation}
E\propto r^{1/4}(\Sigma H)^{-1/2}\exp\left[\pm \int \textrm{i}\sqrt{\frac{6}{\gamma r H^2}}\textrm{d}R\right],
\label{ewkb}
\end{equation}
where the variation of $\gamma$ with radius was ignored. If the stress scales with the total pressure, these expressions indicate that the amplitude of the warp should be proportional to $r^{-5/8}$ in region (a) and to $r^{-1/10}f^{-2/5}$ in region (b). The amplitude of the eccentricity is proportional to $r^{-1/2}$ in the radiation-pressure dominated regime and to $r^{1/40}f^{-2/5}$ when gas pressure dominates. If the stress scales with the gas pressure only, the amplitudes in region (a) are proportional to $r^{17/40}f^{-4/5}$ and $r^{11/20}f^{-4/5}$ in the case of the warp and eccentricity, respectively. The results for region (a) can be confirmed in Fig.~\ref{wkb}. In region (b) $r\gg r_{\textrm{in}}$, $f\approx 1$, so that the radial variation of both warp and eccentricity amplitudes is slow, as seen in Fig.~\ref{free}. The eccentricity does not tend to a constant at large radius because its amplitude is slowly increasing when $r$ increases. The warp amplitude decreases with $r$ but does not tend to zero as indicated by the WKB scaling. The reason for this apparent contradiction is in the failure of the WKB approximation at large radii where the wavelength tends to infinity.

The radial variation of the wavelength of the deformations can also be understood in light of the WKB theory. From equations (\ref{wwkb}) and (\ref{ewkb}), it can be seen that the wavelengths are given by $\lambda_\textrm{W}\approx H \pi r^{5/4}/\sqrt{6 a}$, and $\lambda_\textrm{E}\approx 2H\pi r^{1/2}\sqrt{\gamma/6}$, for the warp and eccentricity respectively. These dependencies are in agreement with the results shown in Fig.~\ref{free}, where it is clear that the wavelengths of the deformations increase with radius and with the disc thickness [or equivalently with the accretion rate, cf. (\ref{ha}), (\ref{hb})]. The increase in $\lambda$ when $\dot{m}$ increases is evident: in particular, for $\dot{m}=0.8$ the wavelength of the warp becomes so large that it practically loses its oscillatory character, but still maintains the sharp increase in $\textrm{d}W/\textrm{d}R$. Even in this case, the inner disc is not aligned with the equator of the black-hole as in the \cite{bardeenpetterson1975} picture. Unfortunately the behaviour of the solutions very close to the marginally stable orbit cannot be trusted in detail because the singularity of the disc model there ought to be resolved by the transition to a supersonic inflow (see Part \ref{ta} and chapter \ref{reflect}).

\subsection{Damped propagation}

\begin{figure*}
\begin{center}
\includegraphics[width=0.49\linewidth]{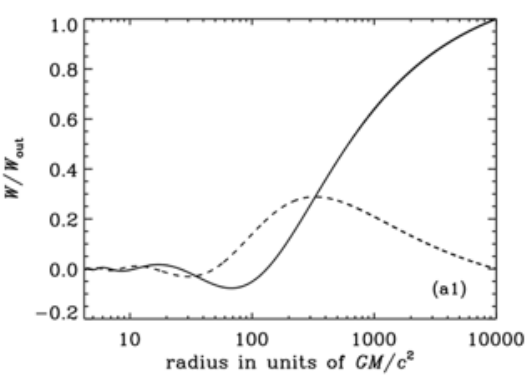}
\includegraphics[width=0.49\linewidth]{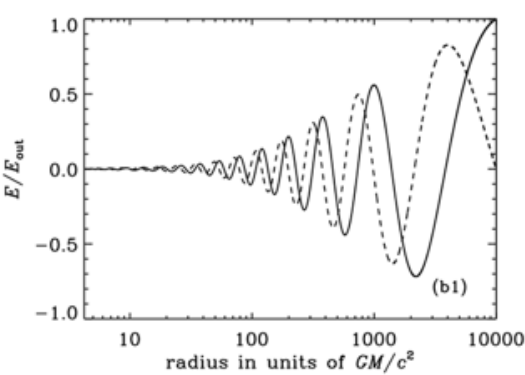} \\

\includegraphics[width=0.49\linewidth]{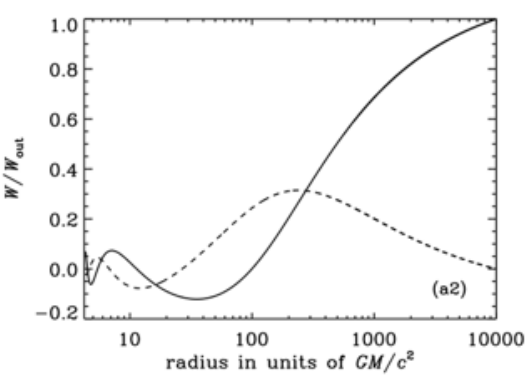}
\includegraphics[width=0.49\linewidth]{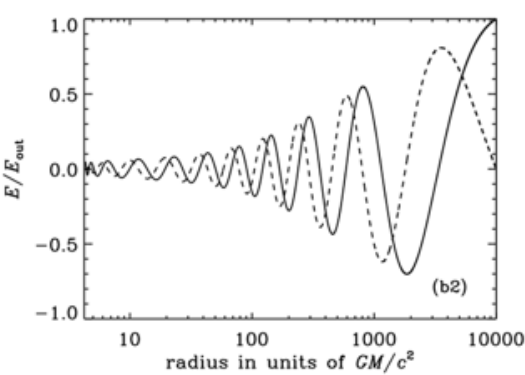} \\

\includegraphics[width=0.49\linewidth]{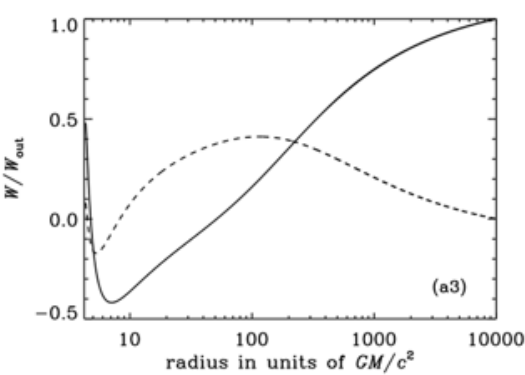}
\includegraphics[width=0.49\linewidth]{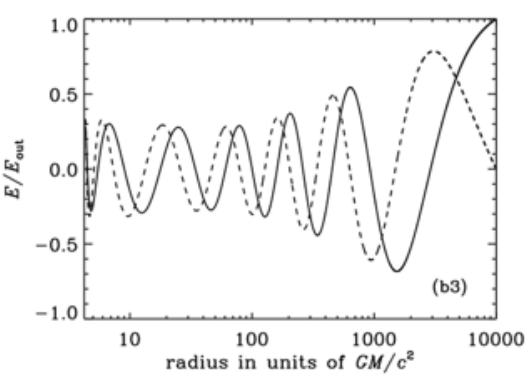}
\caption{Radial variation of (a) the warp tilt and (b) eccentricity normalised to their values at the outer radius for $\alpha_\textrm{W}=0.15$ and $\alpha_\textrm{E}=0.25$ for (1) $\dot{m}=0.2$, (2) $\dot{m}=0.4$, (3) $\dot{m}=0.8$. The full line represents the real part of the disturbance while the imaginary part is represented by the dashed line. A logarithmic scale is used for the x-axis.}
\label{damp}
\end{center}
\end{figure*}

Fig.~\ref{damp} shows the variation with radius of the warp tilt and eccentricity in the case where a significant attenuation is present. The values of $\alpha_\textrm{W}=0.15$ and $\alpha_\textrm{E}=0.25$ are both larger than the viscosity value $\alpha$ used to model the disc, and are chosen so that the warp and eccentricity have negligible amplitude at the inner region of the disc, for small accretion rate ($\dot{m}=0.2$). For dissipation values smaller than these ones, the global deformations propagate everywhere with non-negligible amplitude even for small accretion rate, although with larger amplitude in the inner region for larger $\dot{m}$. 

An important difference between the results presented here and the ones in Fig.~\ref{free} is the existence of a non-zero imaginary part in the solutions. This shows that the reflection of the global deformations from the stress-free boundary is no longer perfect. They are affected by viscous attenuation and, in consequence, their amplitude is reduced as they propagate to smaller radii; a wave with inward group velocity, as opposed to a standing wave, is set up. 

It is evident from Fig.~\ref{damp} that when the accretion rate increases the global deformations propagate to the inner region of the disc more easily, i.e., the amplitude of warp and eccentricity in the inner region increases with $\dot{m}$. This result can be explained in light of the WKB analysis introduced in 4.1. If a small viscosity is present ($\alpha_\textrm{W}, \, \alpha_\textrm{E}\neq 0$), the WKB solution for the warp and eccentricity includes an attenuation factor due to the presence of a imaginary wavenumber $k_\textrm{i}$,

\begin{equation}
k_\textrm{i}^\textrm{Warp}\approx \pm\frac{\alpha_\textrm{W}}{3H}\sqrt{\frac{6a}{r^{1/2}}}, \qquad k_\textrm{i}^\textrm{Ecc}\approx \pm\frac{\alpha_\textrm{E}}{2\gamma H}\sqrt{\frac{6}{\gamma r}}.
\end{equation}
If $H/r$ varies slowly with radius, the logarithm of the attenuation factor is

\begin{equation}
\int^{\infty}_{R_\textrm{in}} |k_\textrm{i}| \textrm{d}R\sim\frac{\alpha_i}{H/r} \quad (\alpha_i=\alpha_\textrm{W},\,\alpha_\textrm{E}),
\end{equation}
indicating that, for fixed viscosity, the attenuation is smaller if the thickness of the disc (or equivalently $\dot{m}$) is larger.

The results presented in this section indicate that global deformations can reach the inner region of accretion discs under a wide variety of conditions. In particular, even when subject to large viscous attenuation, both warp and eccentricity can reach the inner disc provided the black hole is accreting mass at a large enough rate.

Some caveats accompany the solutions obtained for $W(R)$ and $E(R)$. The equations used assume that these quantities are small enough so that non-linear effects can be neglected. They also assume that the quantities $|1-\Omega_z^2/\Omega^2|$ and $|1-\kappa^2/\Omega^2|$ are smaller than, or of the order of, $H/r$ which is not true of the inner region of relativistic discs. Another important caveat relates to the effects of viscosity in the propagation of global deformations. Here they are parametrised using viscous coefficients $\alpha_\textrm{W}$ and $\alpha_\textrm{E}$; this is the simplest way of describing the poorly understood process of turbulent damping in accretion discs. In the particular case of the eccentricity equations only a bulk viscosity is present so that the complications of viscous overstability can be avoided. However, the latter may be present resulting in growth (as opposed to decay) of eccentricity at small radii.

\subsection{Relation to global modes}

The results shown above relate to strictly stationary warp deformations such as those induced by the misalignment between the rotational axis of the black hole and that of the binary orbit. They also apply to slowly precessing global eccentric modes similar to those computed by \cite{goodchildogilvie2006} but including relativistic expressions for the characteristic frequencies. When these relativistic expressions are included, the solution for global warping and eccentric modes in the inner disc resembles the stationary solutions described above, while the outer part of the solution is practically unaffected. In addition, the precession rate (and, when viscosity is included, the decay rate) of global modes is almost the same as in a Newtonian model.

\section{Conclusion}
\label{weconclusion}

In this chapter I studied the propagation of warp and eccentricity in discs around black holes to determine the conditions under which these disturbances can propagate to the inner regions of accretion discs. High-frequency QPOs have previously been identified with inertial oscillations trapped in the inner region of discs, and are detected mainly when black holes are in the very high state where accretion rate is maximum. I find the accretion rate to have a vital role in the damped propagation of global deformations. The results suggest that the activation of the inner region, and consequent excitation of trapped oscillations by these disturbances, may be possible only when the accretion rate is close to its Eddington value, i.e., when the black hole is in the very high state.

When the propagation of global disturbances (found to have an oscillatory structure in the radial direction) is not affect by viscous damping, the increase in mass accretion rate gives rise to a lengthening in their wavelength, in agreement with the WKB analysis of the warp and eccentricity equations. The most interesting results of the calculations described here are obtained when the more realistic situation of damped propagation of global disturbances is considered. In this case, the increase of accretion rate facilitates the propagation of warp and eccentricity, i.e., their amplitudes in the inner region increase with $\dot{m}$. In particular, when the accretion rate is only a small fraction of Eddington and the viscous damping is strong enough to completely suppress the propagation of global deformations, an increase in $\dot{m}$ to the values expected in the very high state results in their amplitude in the inner region being increased to a non-negligible value. 



\thispagestyle{empty}
\chapter{Influence of background inflow on wave propagation}
\label{reflect}

\section{Introduction}

In the previous chapters of Part \ref{os} I studied linear perturbations of a disc which was assumed to terminate at the marginally stable orbit and where the background radial inflow was neglected. However, the dynamical importance of the radial velocity in the inner regions of black-hole accretion discs was highly emphasised in Part \ref{ta}. Moreover, in the transonic models of chapter \ref{timeaccretion}, the sonic point --- as opposed to the marginally stable orbit --- is regarded as the inner boundary of the subsonic disc. Inertial and inertial-acoustic waves propagating in the inner disc, considered in chapters \ref{oscilintro} and \ref{excmech}, are likely to be affected when this transonic background is taken into account. 

In the present chapter, I aim to understand how linear perturbations of a steady (i.e., with viscosity parameter $\alpha<\alpha^*$) transonic flow behave and how they are influenced by the presence of a sonic point. This is done in two ways as reflected in the two-fold structure of this chapter. In the first section I study the propagation of waves in a 1D, isothermal flow in a gravitational potential $\Phi(x)$. In this toy model, $\Phi$ is chosen to have a maximum at $x=0$ where the flow velocity $u(x)$ changes from subsonic to supersonic. Linear perturbations proportional to $\exp(-\mathrm{i}\omega t)$ are imposed on the flow and the reflection of these waves at $x=0$ (where equations describing such perturbations are singular) is calculated for different potential shapes and wave frequencies. In the second section I investigate how trapped inertial modes are affected by the background radial inflow and by a sonic point at the inner disc boundary.

The results obtained here have implications for the excitation mechanism of trapped inertial waves which has been the main focus of this part of the thesis. As detailed in chapter \ref{excmech}, this mechanism relies on a variety of assumptions. For example, the coupling between inertial waves, a global deformation and negative-energy intermediate oscillations  relies on damping of the latter to provide the necessary energy for sustained growth of the r~modes. In chapter \ref{excmech} this dissipation was assured by artificially placing a damping term in the equations for the intermediate waves. This term is supposed to describe the expected absorption of these waves as they approach their corotation resonance or the inner disc boundary. (While the $n=2$ intermediate mode can be damped at corotation, as determined by \cite{lietal2003}, the coupling with the $n=0$ oscillation relies on absorption at the inner boundary to remove the mode's negative energy.) The problem investigated in section \ref{toymodel} is expected to provide clues about the conditions under which reflection or absorption of waves at the sonic point is possible. When the transonic nature of the flow is considered, the disc isn't taken to artificially terminate at the marginally stable orbit, the surface density doesn't decrease suddenly to zero there and the inner boundary is no longer a totally reflecting wall. Depending on the variation of the density profile in the transonic region [mimicked by the shape of the potential $\Phi(x)$] and on the wave properties, the oscillations may either be absorbed or reflected there.

At this point it is important to note that, simultaneously with the research described here, the influence of a non-negligible radial inflow on inertial-acoustic modes trapped in the innermost region of subsonic discs was studied by \cite{laitsang2009}. They calculated the reflection of these oscillations at the sonic point and determined how the reflection coefficient depends on several parameters of the disc inner edge. Their results indicate that wave loss at that radius is significant unless the surface density decays rapidly at the sonic point. Even taking this study into account, the analysis done in the 1D isothermal flow in a potential is of interest to determine the relation between the flow characteristics at the sonic point, the wave properties and the reflection coefficient under more general conditions.

Returning to the implications of a non-negligible radial inflow to the excitation mechanism, perhaps most important is how the r~modes themselves are affected. For the growth of inertial waves to be effective the ``leakage'' from the epicyclic trapping barrier to both the outer and inner parts of the disc has to be relatively small. While it was shown before that the ``outer leakage'' results in a small decay rate (cf. Table \ref{tablefree}) which can easily be surpassed by the growth due to the coupling with the warp or eccentricity, the ``inner leakage'' is yet to be considered. The presence of the radial inflow and of a singular sonic point (as opposed to a rigid wall) at the inner boundary are of crucial importance to determine how much the inertial waves decay as they leak into the plunging region. The analysis of section \ref{uiw} shows that, depending on the location of the sonic point, the decay rate can have a strong dependence on the sound speed and may, in some situations, be larger than the growth attained by coupling with a global deformation. In addition, for large enough sound speeds and for the cases where the sonic point is outside the marginally stable orbit, the Gaussian structure of the simplest possible r~mode may be deeply modified or even destroyed. The implications of these results are discussed in section \ref{uiwconclusions}.

\section{Wave reflection at the sonic point --- a toy model}
\label{toymodel}

\subsection{Oscillations of a 1D flow}

Consider a one-dimensional isothermal and inviscid hydrodynamic flow under the influence of a gravitational potential $\Phi(x)$. In a steady state, the equations describing mass and momentum conservation can be expressed in the combined form
\begin{equation}
(u^2-c_\mathrm{s}^2)\frac{\mathrm{d}u}{\mathrm{d}x}=-u\frac{\mathrm{d}\Phi}{\mathrm{d}x},
\label{basic}
\end{equation}
where $u(x)<0$ is the inflow velocity and $c_\mathrm{s}$ is the isothermal sound speed. The density $\rho$ is given by $\rho=-\dot{M}/u$, where $\dot{M}>0$ is the constant mass flux. It is clear from this equation that the sonic point, where $u(x)=-c_\mathrm{s}$, is located where the derivative of $\Phi$ is zero. By differentiating (\ref{basic}) it is simple to obtain the value of the inflow derivative at the sonic point $x=x_0$,
\begin{equation}
\left(\frac{\mathrm{d}u}{\mathrm{d}x}\right)_0=\sqrt{-\frac{1}{2}\left(\frac{\mathrm{d}^2\Phi}{\mathrm{d}x^2}\right)_0}.
\label{dusp}
\end{equation}
Physical flows have $(\mathrm{d}u/\mathrm{d}x)_0$ real implying that the sonic point has to be located at a maximum of the gravitational potential.

To study the propagation of waves in this medium and to see how they behave at the sonic point, linear perturbations are imposed on the steady flow,
\begin{equation}
u(x,t)\rightarrow u(x)+\mt{Re}[u'(x)\exp(-\mathrm{i}\omega t)],
\end{equation}
\begin{equation}
q(x,t)\rightarrow q(x)+\mt{Re}[q'(x)\exp(-\mathrm{i}\omega t)],
\end{equation}
where $q=\ln\rho$ and $\omega$ is the wave frequency. Neglecting self-gravitation and non-linear terms on the perturbed quantities, the equations for the perturbations can be written as
\begin{equation}
-\mathrm{i}\omega u'+u'\frac{\mathrm{d}u}{\mathrm{d}x}+u\frac{\mathrm{d}u'}{\mathrm{d}x}=-c_\mathrm{s}^2\frac{\mathrm{d}q'}{\mathrm{d}x},
\label{w1}
\end{equation}
\begin{equation}
-\mathrm{i}\omega q'+u'\frac{\mathrm{d}q}{\mathrm{d}x}+\frac{\mathrm{d}u'}{\mathrm{d}x}=-u\frac{\mathrm{d}q'}{\mathrm{d}x},
\label{w2}
\end{equation}
where $q(x)$ can be obtained from the steady state mass conservation equation, $\mt{d}q/\mt{d}x=-(1/u)\mt{d}u/\mt{d}x$, and $u(x)$ is given by equation (\ref{basic}). 

\begin{figure*}
\begin{center}
\includegraphics[width=0.49\linewidth]{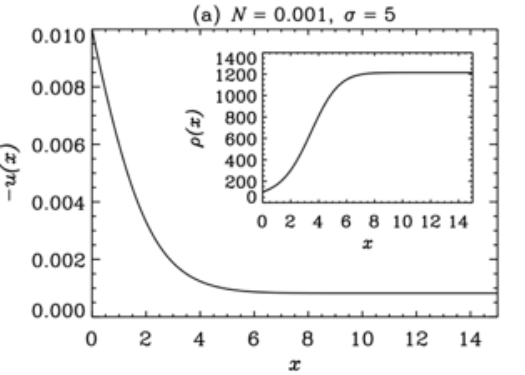}
\includegraphics[width=0.49\linewidth]{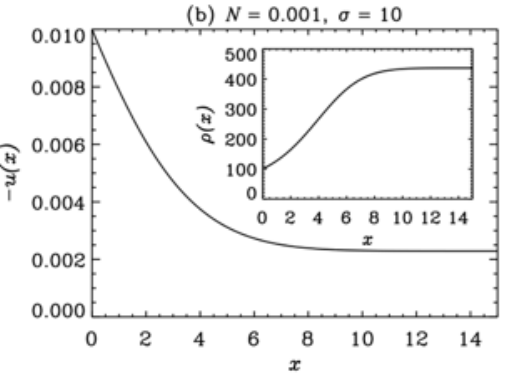} \\
\vspace{2mm}
\includegraphics[width=0.49\linewidth]{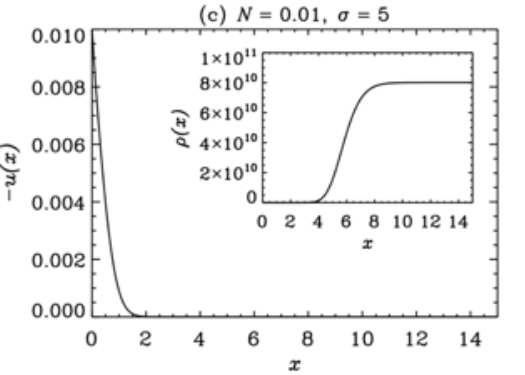}
\caption{Profile of the inflow velocity $-u(x)$ and density $\rho(x)$ in the subsonic region for (a) $N=0.001$, $\sigma=5$, (b) $N=0.001$, $\sigma=10$ and (c) $N=0.01$, $\sigma=5$. The sonic point is at $x=0$ where $u(x)=-c_\mt{s}$, which is taken to be $0.01$. The velocities are in units of $c$ while the units of $\rho$ are arbitrary (for the purpose of these plots, $\dot{M}$ was arbitrarily chosen to be 1).}
\label{ufromphi}
\end{center}
\end{figure*}

For convenience, I take the gravitational potential to have a maximum at $x=0$ and to tend to a constant at large $|x|$. For example, $\Phi(x)$ can be assumed to be a Gaussian of the form:

\begin{equation}
\Phi(x)=N\exp\left(-\frac{x^2}{2\sigma^2}\right),
\label{phi}
\end{equation}
where $N$ is the height of the potential which has units of a velocity squared and $\sigma$ is a measure of the width of the Gaussian and has the same (arbitrary) units as $x$. For the purpose of the wave reflection calculations, $N$ is to be compared with $c_\mt{s}^2$ while $\sigma$ is to be contrasted with the wavelength of the incident wave. Fig.~\ref{ufromphi} shows solutions of equation (\ref{basic}) when $\Phi$ is given by (\ref{phi}) for different values of $N$ and $\sigma$. As it can be seen, for fixed $\sigma$, increasing $N$ results in a steeper profile of $u$ and $\rho$ in the inner region. For a potential of fixed height, the background flow varies on a small/long length scale if $\sigma$ is small/large.
  
The region of positive $x$ is taken to be subsonic while $x<0$ is the supersonic side. In a realistic situation of an accretion flow surrounding a black hole, $u(x)<0$ everywhere and the region $x>0$ can be thought of as the accretion disc itself, where the radial inflow is approximately constant as mimicked in this toy model. The plunging region is located at $x<0$ and the transition occurs at $x=0$ where $u=-c_\mathrm{s}$. I am interested in studying the behaviour of waves propagating in the subsonic region and at the transition point where they can be reflected and/or transmitted. 

\begin{figure*}
\begin{center}
\includegraphics[width=0.7\linewidth]{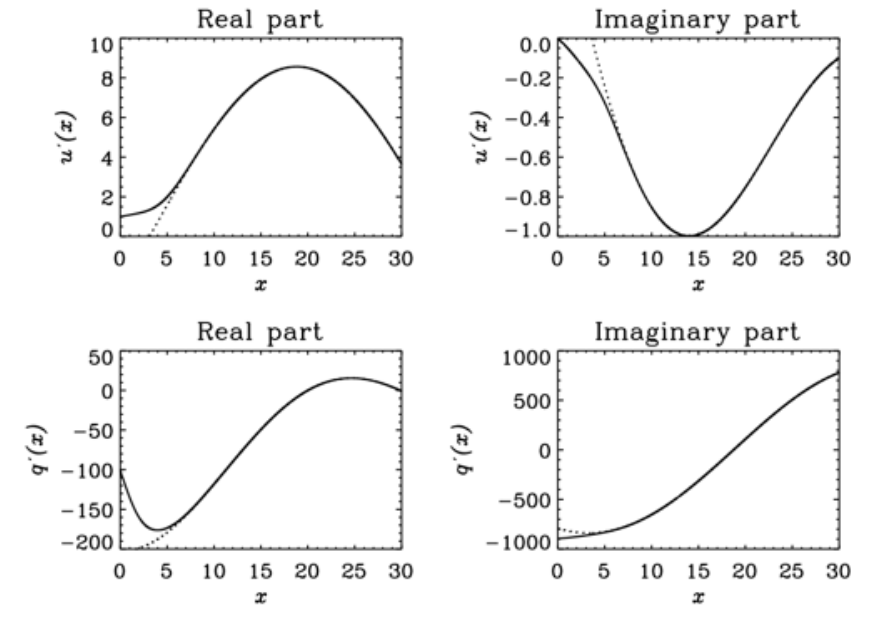} \\
\vspace{5mm}
\includegraphics[width=0.7\linewidth]{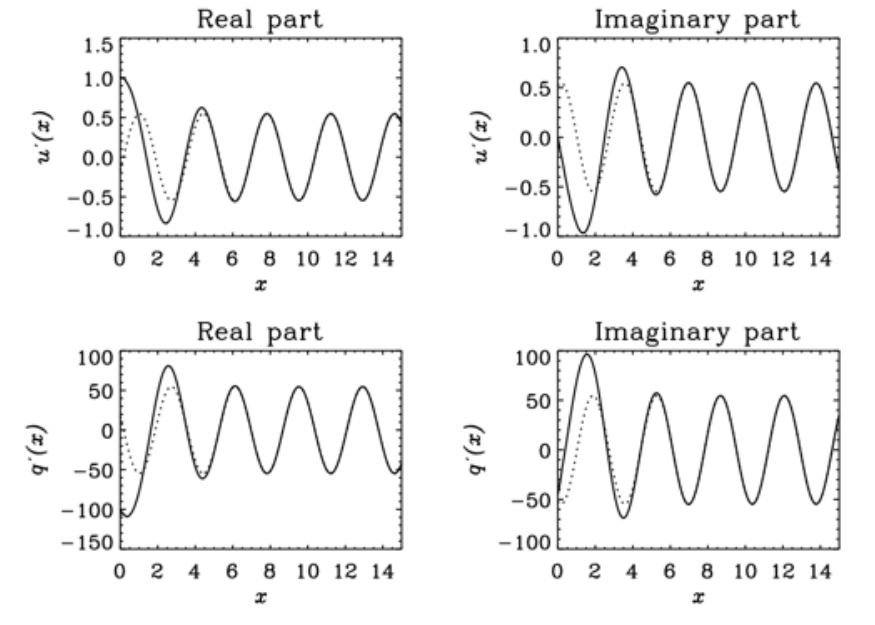}
\caption{Solutions of equations (\ref{w3}) and (\ref{w4}) for $c_\mt{s}=0.01$, $\sigma=5$, $N=10c_\mt{s}^2=0.001$ [Fig.~\ref{ufromphi} (a)], $u_0'=1.0$, and $\omega=0.5c_\mt{s}/\sigma$ (upper 4 panels, representing a wave which is reflected at the sonic point) and $\omega=10c_\mt{s}/\sigma$ (lower 4 panels, representing a wave which is absorbed at $x=0$). The dashed lines represent the analytical solution given by (\ref{wkb1}) and (\ref{wkb2}) and match the numerical solution at large $x$, as expected.}
\label{1dwaves}
\end{center}
\end{figure*}

Using the steady state mass conservation, equations for linear perturbations (\ref{w1}) and (\ref{w2}) can be written in the form

\begin{equation}
\frac{\mathrm{d}q'}{\mathrm{d}x}=\frac{1}{u^2-c_\mathrm{s}^2}\left[\mathrm{i}\omega u q'-\mathrm{i}\omega u'+2\frac{\mathrm{d}u}{\mathrm{d}x}u'\right],
\label{w3}
\end{equation}
\begin{equation}
\frac{\mathrm{d}u'}{\mathrm{d}x}=\frac{1}{u^2-c_\mathrm{s}^2}\left[-\mathrm{i}\omega c_\mathrm{s}^2 q'+\mathrm{i}\omega u u'-\left(1+\frac{c_\mathrm{s}^2}{u^2}\right)\frac{\mathrm{d}u}{\mathrm{d}x}uu'\right],
\label{w4}
\end{equation}
which shows that the ODEs are singular at the sonic point. (\ref{w3}) and (\ref{w4}) can be solved numerically everywhere except near $x=0$. The desired solution is regular at this point. To find this solution, the following analytical expansion around the sonic point is used

\begin{equation}
u(x)\approx-c_\mathrm{s}+Ax,
\end{equation}
\begin{equation}
u'(x)\approx u_0'+u_1'x, \quad u_1'=-\frac{\mathrm{i}\omega}{2c_\mathrm{s}}u_0',
\end{equation}
\begin{equation}
q'(x)\approx q_0'+q_1'x, \quad q_0'=\frac{\mathrm{i}\omega-2A}{-\mathrm{i}\omega c_\mathrm{s}}u_0',\quad q_1'=\frac{\mathrm{i}\omega-2A}{2c_\mathrm{s}^2}u_0',
\end{equation}
where $A=\left(\mathrm{d}u/\mathrm{d}x\right)_0$ [which can be found to be $\sqrt{N/2\sigma^2}$ using (\ref{dusp})] and $u_0'$ is an arbitrary constant. Given input parameters $N$, $c_\mt{s}$, $\sigma$, $\omega$ and $u_0'$, these expansions are used in a small region of length $\delta x$ around the sonic point and equations (\ref{basic}), (\ref{w3}) and (\ref{w4}) are integrated numerically for $x\ge\delta x$.  Two typical solutions are shown in Fig.~\ref{1dwaves}.

\subsection{Reflection coefficient}

At large $|x|$ where $\Phi$ is a constant (and therefore $\mathrm{d}u/\mathrm{d}x=0\Rightarrow u(x)=u$, constant), equations for the linear perturbations can be solved analytically. The solution of (\ref{w1}) and (\ref{w2}) is then

\begin{equation}
q'(x)=k_1\exp{\left(\frac{\mathrm{i}\omega}{u-c_\mathrm{s}}x\right)}+k_2\exp{\left(\frac{\mathrm{i}\omega}{u+c_\mathrm{s}}x\right)},
\label{wkb1}
\end{equation}
\begin{equation}
\frac{u'(x)}{c_\mathrm{s}}=-k_1\exp{\left(\frac{\mathrm{i}\omega}{u-c_\mathrm{s}}x\right)}+k_2\exp{\left(\frac{\mathrm{i}\omega}{u+c_\mathrm{s}}x\right)}.
\label{wkb2}
\end{equation}
In the subsonic region, $u+c_\mt{s}>0$ ($u-c_\mt{s}<0$ everywhere) and the waves are composed of outgoing and ingoing parts. The former propagate to the right and can therefore be thought of as waves reflected from the sonic point while the latter are incident as they propagate to the left. Assuming, without loss of generality, that $\omega$ is positive, the term with constant $k_1$ represents the incident wave (which has a wavelength  $\sim c_\mt{s}/\omega$) while the term with $k_2$ represents the reflected wave. The reflection coefficient can then be defined as

\begin{equation}
R=\frac{|k_2|^2}{|k_1|^2}.
\end{equation}
The constants $k_1$ and $k_2$ can be found by matching the numerical solution at some large distance $x_\mathrm{l}$ with (\ref{wkb1}) and (\ref{wkb2}):

\begin{equation}
k_1=\frac{q'(x_\mt{l})-u'(x_\mt{l})/c_\mt{s}}{2\exp\left[\mathrm{i}\omega  x_\mathrm{l}/(u-c_\mathrm{s})\right]},
\end{equation}
\begin{equation}
k_2=\frac{q'(x_\mt{l})+u'(x_\mt{l})/c_\mt{s}}{2\exp\left[\mathrm{i}\omega x _\mathrm{l}/(u+c_\mathrm{s})\right]},
\end{equation}
where $q'(x_\mt{l})$ and $u'(x_\mt{l})$ are the values of the numerical solution of (\ref{w3}) and (\ref{w4}) at $x_\mt{l}$. Note that because this solution coincides with that given by (\ref{wkb1}) and (\ref{wkb2}) at large $x$ (say for $x>x_\mathrm{match}$, as seen in Fig.~\ref{1dwaves}), the exact value of $x_\mt{l}$ is irrelevant provided it is chosen to be larger than $x_\mt{match}$. 

Note that because the reflection coefficient is calculated at large radii, it measures the gradual reflection of waves as they approach the sonic point from the subsonic region rather than the reflection at the inner boundary exactly. For example, in the case represented in Fig.~\ref{ufromphi} (c), the reflection probably occurs mainly in the region between $x=4$ and $x=8$ because that is where the density drops steeply. In any case, in a realistic disc, this region where the density drops rapidly is expected to be closer to the sonic point. The reflection coefficient calculated in the case of Fig.~\ref{ufromphi} (c) would, in a realistic disc, correspond to the reflection from the transonic region.

\begin{figure*}
\begin{center}
\includegraphics[width=110mm]{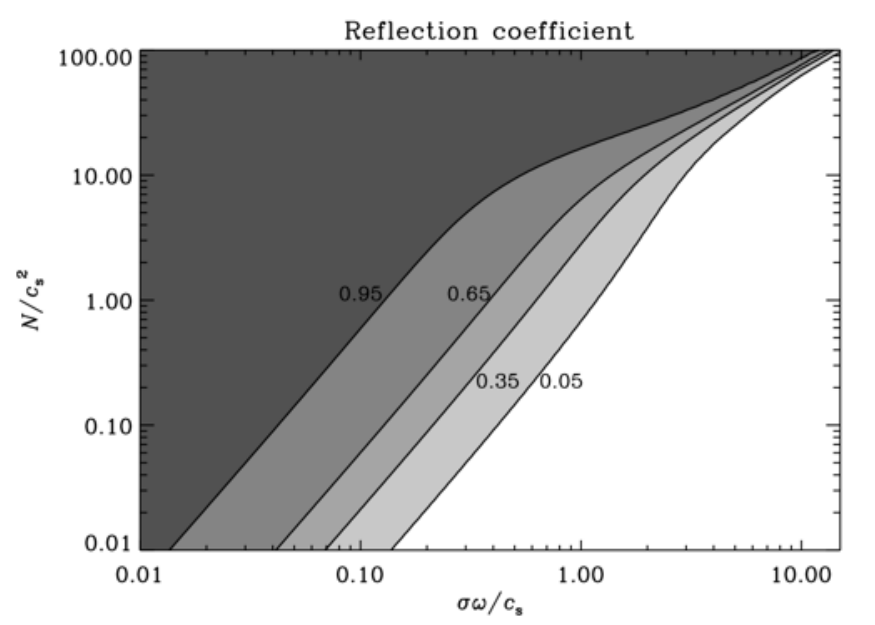}
\caption{Variation of the reflection coefficient with $N/c_\mt{s}^2$ and $\sigma/(c_\mt{s}/\omega)$. In the white region $R$ is practically null while in the dark grey area the reflection is close to total ($R>0.95$).}
\label{refcoef}
\end{center}
\end{figure*}

In Fig.~\ref{refcoef}, I show how the reflection coefficient changes for a range of values of $N/c_\mt{s}^2$ and $\sigma/(c_\mt{s}/\omega)$. For the purpose of these calculations, $c_\mt{s}=0.01$ and $\sigma=5$ were fixed while $N$ and $\omega$ were varied. 

\subsection{Discussion and conclusions}

The calculations made here provide important information about the relation between the properties of the flow at the sonic point, the wave characteristics and the reflection of oscillations as they approach the inner edge. 

Since $\sigma$ and $c_\mt{s}$ were kept constant in the determination of the reflection coefficient, changes in $N/c_\mt{s}^2$ target changes in the profile of the background flow $u$ and density $\rho$. In particular, as $N/c_\mt{s}^2$ increases, $u$ and $\rho$ become steeper in the inner region. On the other hand, the values of the x-axis of the plot of Fig.~\ref{refcoef} roughly compare the width of the potential to the wavelength of the incident wave. When $\sigma/(c_\mt{s}/\omega)$ is small, the wavelength is longer than the scale on which the background varies.

The results obtained are those intuitively expected. A tall, narrow potential results in good reflection and a low, wide one allows for transmission. More specifically, the waves are reflected effectively when the wavelength is comparable to or longer than the typical length scale for background variations. This is in line with the findings of \cite{laitsang2009}. Moreover, if the potential is sufficiently high, indicating that the background density drops abruptly as the inner edge is approached, essentially all waves are reflected.

An interesting feature of Fig.~\ref{refcoef} is the difference between the separation between the regions of $R<0.05$ and $R>0.95$ for small and large $N/c_\mt{s}^2$. When the potential is high, the waves are either completely reflected or absorbed for practically all values of $N/c_\mt{s}^2$ and $\sigma\omega/c_\mt{s}$. On the other hand, for small $N/c_\mt{s}^2$, partial reflection is also possible depending on the wavelength of the incident wave. Moreover, the slope of the reflection contours changes depending on whether the potential is high or low. These differences can be understood by remembering that when $N/c_\mt{s}^2$ is small, the density drops smoothly and the reflection occurs gradually. When the background density is very steep in the inner region it works as a totally reflecting wall there if the incident wave has a wavelength long compared to $\sigma$.

Generalising these results to the inertial-acoustic waves that work as intermediate oscillations in the coupling mechanism described in chapter \ref{excmech}, it seems that absorption is possible under some conditions. More calculations, similar to those done by \cite{laitsang2009} are necessary to determine if non-axisymmetric intermediate waves can indeed be damped or partially damped at the inner boundary under realistic conditions. Nonetheless, in the stable, steady state disc profiles calculated in chapter 3, the background radial inflow and density vary smoothly at the sonic point indicating that a realistic disc is probably in the low $N/c_\mt{s}^2$ regime where most waves are absorbed as they approach the inner boundary. These are encouraging results for the effectiveness of an excitation mechanism for trapped inertial modes that relies on damping of inertial-acoustic waves at the inner boundary.

\section{Inertial waves}
\label{uiw}

In this section I investigate how the presence of background radial flow affects trapped inertial modes. In what follows, the usual scalings apply: velocities are in units of $c$, the enthalpy $h=c_\mt{s}^2\ln\rho$ in units of $c^2$, lengths are multiples of $GM/c^2$ and frequencies of $c^3/GM$. In these units, particles moving in a $a=0$ \cite{pw1980} potential have a marginally stable orbit at $r=6$. \enlargethispage{\baselineskip}

\subsection{Equations and method}

To understand the behaviour of linear perturbations of a stable transonic flow, I start by considering the hydrodynamic equations describing an isothermal and inviscid disc, (\ref{motion1})--(\ref{energy1}). As in chapter \ref{excmech}, the disc is assumed thin and the equilibrium state is taken to be axisymmetric and time-independent. However, now I consider this equilibrium to be characterised by $\vc{u}=\left(u(r),\Omega(r) r,0\right)$ and $h=h_\mathrm{b}(r)-\Omega_z^2z^2/2$.

For the purpose of the calculations presented here, the basic state quantities $u(r)$, $h_b(r)$ and $\Omega(r)$ are obtained from equations (\ref{eqap1})--(\ref{eqap2}) of Part \ref{ta} in the case where the system reaches a steady state. From the numerical calculations made in chapter \ref{timeaccretion} and the quasi-analytical treatment of \cite{ap2003}, it is known that for low enough $\alpha$ the system reaches a steady state. In that case $u(r)$ is approximately constant and subsonic at large radii, equals the isothermal sound speed at the sonic point, which is close to the marginally stable orbit, and becomes supersonic in the plunging region. Even though viscosity is ignored for the purpose of the linear perturbation equations, its effect is indirectly present in the calculations since the value of $\alpha$ influences (although weakly for $\alpha<\alpha^*$) the background flow profile. More specifically, it slightly changes the location of the sonic point which, as will be seen later, is significant.

I assume that small perturbations of the usual form are imposed to the basic state. Neglecting self-gravitation and non-linear terms in the perturbed quantities, and separating variables as in chapters \ref{oscilintro} and \ref{excmech}, the equations describing the behaviour of these perturbations read

\begin{equation}
-\mathrm{i}\hat{\omega} u_r+u\frac{\mathrm{d}u_r}{\mathrm{d}r}+u_r\frac{\mathrm{d}u}{\mathrm{d}r}-2\Omega u_\phi=-\frac{\mathrm{d}h}{\mathrm{d}r},
\label{p1}
\end{equation}
\begin{equation}
-\mathrm{i}\hat{\omega} u_\phi+u\frac{\mathrm{d}u_\phi}{\mathrm{d}r}+\frac{\kappa^2}{2\Omega}u_r+\frac{u_\phi u}{r}=-\frac{\mathrm{i}mh}{r},
\end{equation}
\begin{equation}
-\mathrm{i}\hat{\omega} u_z+u\frac{\mathrm{d}u_z}{\mathrm{d}r}=-\frac{nh}{H},
\end{equation}
\begin{equation}
-\mathrm{i}\hat{\omega} h+u\frac{\mathrm{d}h}{\mathrm{d}r}+u_r\frac{\mathrm{d}h_\mathrm{b}}{\mathrm{d}r}-\Omega_z^2Hu_z=-c_\mathrm{s}^2\left[\frac{1}{r}\frac{\mathrm{d}}{\mathrm{d}r}(u_r r)+\frac{\mathrm{i}m u_\phi}{r}\right],
\label{p2}
\end{equation}
where the wave quantities $(u_r,u_\phi,u_z,h)$ are functions of $r$ only. As before, $H=c_\mt{s}/\Omega_z$ is the semi-thickness of the disc, $\hat{\omega}=\omega-m\Omega$ is the Doppler-shifted wave frequency and $m$ and $n$ are the azimuthal and vertical wave numbers, respectively. Note that this system of equations reduces to (\ref{free1})--(\ref{free2}) if the radial inflow and $\mt{d}h_\mt{b}/\mt{d}r$ are negligible. The vertical frequency $\Omega_z$ is assumed to be equal to $\Omega$ as in the \cite{pw1980} potential and the epicyclic frequency is calculated from the basic state $\Omega$ using the usual Newtonian formula, $\kappa=\sqrt{4\Omega^2+2r\Omega\mt{d}\Omega/\mt{d}r}$. As seen in Part \ref{ta} (Fig.~\ref{stable}), the fluid's angular velocity and epicyclic frequency are everywhere nearly equal to equivalent particle-orbit expressions except in the transonic and supersonic regions. Although with minor changes, the maximum of $\kappa$ is still located close to $r=7.5$.

As in chapter \ref{excmech}, I consider the simplest trapped inertial mode with $m=0$ and $n=1$. In this case, equations (\ref{p1}) and (\ref{p2}) can be written in the form,

\begin{equation}
\frac{\mathrm{d}h}{\mathrm{d}r}=\frac{u f-c_\mathrm{s}^2 g}{u^2-c_\mathrm{s}^2},
\label{pv1}
\end{equation}
\begin{equation}
\frac{\mathrm{d}u_r}{\mathrm{d}r}=\frac{u g-f}{u^2-c_\mathrm{s}^2},
\label{pv2}
\end{equation}
where
\begin{equation}
f=\mathrm{i}\omega h-u_r\frac{\mathrm{d}h_\mathrm{b}}{\mathrm{d}r}+\Omega_z^2Hu_z-c_\mathrm{s}^2\frac{u_r}{r},
\end{equation}
\begin{equation}
g=\mathrm{i}\omega u_r+2\Omega u_\phi-\frac{\mathrm{d}u}{\mathrm{d}r}u_r.
\end{equation}
It is evident that the equations for perturbations are singular at the sonic point, $r=r_0$, where $u(r)=-c_\mt{s}$. The regular solution satisfies $f+c_\mt{s}g=0$ at $r=r_0$ so that both numerators of (\ref{pv1}) and (\ref{pv2}) vanish there. In other words,
\begin{equation}
-\left[\left(\frac{\mathrm{d}u}{\mathrm{d}r}\right)_0 c_\mathrm{s}+\left(\frac{\mathrm{d}h_\mathrm{b}}{\mathrm{d}r}\right)_0+\frac{c_\mathrm{s}^2}{r_0}\right]u_{r,0}+c_\mathrm{s}\Omega_{z,0} u_{z,0}+2c_\mathrm{s}\Omega_0u_{\phi,0}=-\mathrm{i}\omega(h_0+c_\mathrm{s}u_{r,0}),
\label{bcinertial}
\end{equation}
where the subscript zero indicates that the quantities are evaluated at the sonic point. I solve equations (\ref{p1})--(\ref{p2}) with $(m,n)$=$(0,1)$ numerically using the Chebyshev method already applied in chapter 5. The inner boundary is now taken to be located at $r_\mt{in}=r_0+\delta r$ and (\ref{bcinertial}) is taken as an inner boundary condition (this is equivalent to using an order zero expansion for the wave quantities from $r_0$ to $\delta r\ll r_0$). Condition (\ref{bcinertial}) replaces the one employed previously, $u_{r,0}=0$. The outer boundary condition remains as before. The goal is to see how the decay rate of the simplest trapped inertial mode and its structure are affected when the radial inflow is taken into account and the inner boundary, now essentially at the sonic point, can no longer be approximated by a rigid wall.

The reader is reminded that, as in chapter \ref{excmech}, the method used to solve the system of equations is an eigenvalue problem where $-\mathrm{i}\omega$ is the complex eigenvalue. Because I want to focus my attention on the simplest possible r~mode, that with lowest radial wave number $l$, I look for the mode with a Gaussian structure similar to that of Fig.~\ref{freerm} (a). However, as it will be seen in the next section, this structure is likely to be modified by the radial inflow. Nonetheless, the real part of the frequency of this mode should still be close to the maximum of the epicyclic frequency. This value changes slightly depending on the background solution but should be close to the particle-orbit $\mt{max}(\kappa)$ which is $0.0347$ in the units used here. Since no excitation mechanism is employed here, the imaginary part of the frequency gives the decay rate of the mode. When no background flow is considered this rate is due to the boundary condition used at outer radius, which selects the outgoing wave only; these values should be equivalent to those presented in Table \ref{tablefree}.

\subsection{Results and discussion}

In Table \ref{tableinflow}, I show the dependence of the complex frequency of the simplest trapped r~mode, that with wave numbers $(l,m,n)=(0,0,1)$, on the sound speed. The three rightmost columns of the table correspond to the three different cases considered. Once again, it should be made clear that the effect of viscosity on the waves is not being calculated here; the decay rates result from the presence of the radial inflow only.
 
\begin{table}
\begin{center}
\begin{tabular}{ccccccc}
\hline
$c_\mathrm{s}/c$ & & No background flow & & $r_0<r_\mt{ms}=6$ & & $r_0>r_\mt{ms}=6$ \\
\hline
$0.003$ & & $0.0343+0.0\textrm{i}$ & & $0.0343-5.12\times 10^{-6}\textrm{i}$ & & $0.0344-6.59\times 10^{-5}\textrm{i}$ \\
$0.004$ & & $0.0342-1.60 \times 10^{-11}\textrm{i}$ & & $0.0340-1.18 \times 10^{-5}\textrm{i}$ & & $0.0345-2.87 \times 10^{-4}\textrm{i}$\\
$0.005$ & & $0.0340-5.63 \times 10^{-10}\textrm{i}$ & & $0.0338-1.74 \times 10^{-5}\textrm{i}$ & & $0.0345-5.93 \times 10^{-4}\textrm{i}$\\
$0.006$ & & $0.0339-6.14 \times 10^{-9}\textrm{i}$ & & $0.0336-2.66 \times 10^{-5}\textrm{i}$ & & $0.0345-9.38 \times 10^{-4}\textrm{i}$\\
$0.007$ & & $0.0338-3.64 \times 10^{-8}\textrm{i}$ & & $0.0334-3.55 \times 10^{-5}\textrm{i}$ & & $0.0343-1.29 \times 10^{-3}\textrm{i}$\\
$0.008$ & & $0.0337-1.25 \times 10^{-7}\textrm{i}$ & & $0.0332-4.53 \times 10^{-5}\textrm{i}$ & & $0.0342-1.69 \times 10^{-3}\textrm{i}$\\
$0.01$ & & $0.0334-7.70 \times 10^{-7}\textrm{i}$ & & $0.0327-6.59 \times 10^{-5}\textrm{i}$ & & N/A \\
$0.02$ & & $0.0321-4.34 \times 10^{-5}\textrm{i}$ & & $0.0310-3.61 \times 10^{-4}\textrm{i}$ & & N/A \\
\hline
\end{tabular} 
\caption{Values of the (complex) frequency of the simplest trapped inertial mode for different values of sound speed and background flow appropriate for a \emph{non-rotating} black hole. The second column corresponds to the case where no radial inflow is considered and the modes are damped simply due to the outgoing boundary condition applied at $r_\mt{out}$. The third column corresponds to the case where the radial inflow was calculated for a small enough value of viscosity and where the sonic point is located inwards from the marginally stable orbit. In the fourth column the radial inflow was determined for slightly larger values of $\alpha$ in which case the sonic point is located at $r>6$. The N/A indicates that no mode with a Gaussian structure could be identified.}
\label{tableinflow}
\end{center}
\end{table} 

\begin{figure*}
\begin{center}
\includegraphics[width=0.95\linewidth]{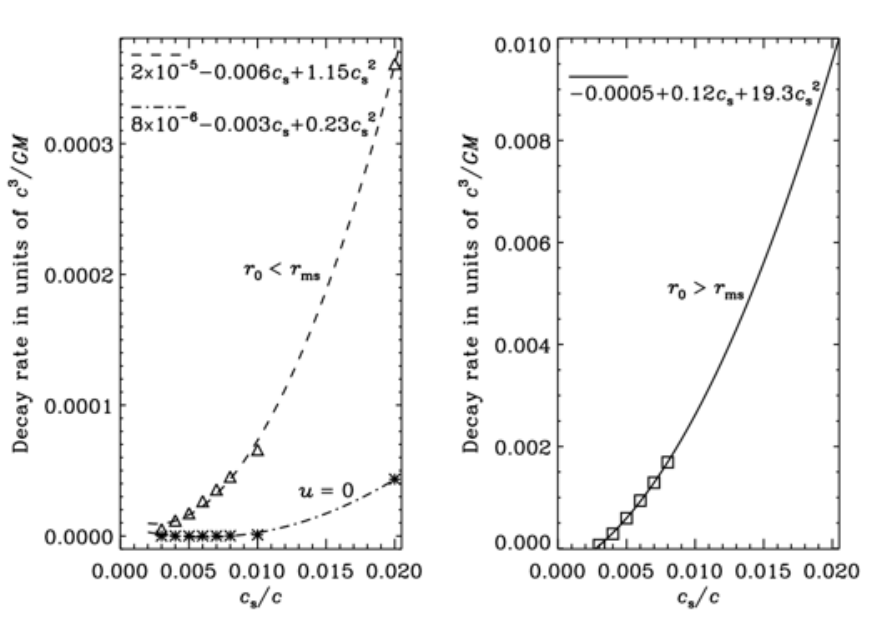}
\caption{Decay rate as a function of sound speed for the 3 different cases presented in Table \ref{tableinflow}. The stars, triangles and squares correspond to the values of decay rate of columns 2, 3 and 4, respectively. In all cases, the variation can be reasonably well fitted by a polynomial of degree 2 as indicated by the lines represented. The extrapolation of the variation of decay rate with $c_\mt{s}$ up to 0.02 is included in the plot on the right even though this corresponds to the N/A values of the table which cannot be found numerically.}
\label{decayfit}
\end{center}
\end{figure*}

The first of these columns shows the frequency of the r~mode when no background flow is considered and where the waves decay solely due to the use of an outgoing wave outer boundary condition. The imaginary parts of the frequencies are, as expected, equivalent to those of Table \ref{tablefree}. The real parts are larger in the present case because the maximum of the epicyclic frequency used here is higher than that of chapter \ref{excmech}, where a different expression for $\kappa$ was used [cf. (\ref{kappapw}), (\ref{relkappa})]. Even so, the general behaviour is the same as that indicated in Table \ref{tablefree}: the smaller the sound speed, the closer the mode is trapped to the maximum of the epicyclic frequency which results in a smaller trapping region and less significant outer ``leakage''.

The values of the $r_0<r_\mt{ms}$ and $r_0>r_\mt{ms}$ columns, were obtained by considering perturbations to a transonic flow which has a value of viscosity such that the sonic point is located inside and outside the marginally stable orbit, respectively. As seen in Part \ref{ta} of the thesis, the profile of the transonic radial inflow is dependent on the values of viscosity $\alpha$ and isothermal sound speed $c_\mt{s}$, even in those cases where a stable flow is considered. The introduction of chapter \ref{steadyintro} and references therein indicate that two scenarios are possible. For fixed sound speed, a flow with a low value of $\alpha$ goes through a saddle-type critical point which is located at a radius smaller than $r_\mt{ms}$. As $\alpha$ is increased, the characteristics of the flow change and it goes through a nodal sonic point located outside the marginally stable orbit. As seen in \cite{ap2003} and in chapter \ref{timeaccretion}, the passage can still be stable provided $\alpha<\alpha^*=0.14 (100 c_\mt{s})^{1/3}$. According to \cite{ap2003}, the limiting value of $\alpha$ separating the cases of saddle and nodal passages is $\alpha_\mt{sn}=0.08 (100 c_\mt{s})^{1/3}$. In summary, the values presented in the third column of Table \ref{tableinflow} were calculated using a background inflow with $\alpha<\alpha_\mt{sn}$ while those in the last column were determined using $u(r)$ corresponding to a viscosity $\alpha>\alpha_\mt{sn}$ (but still smaller than $\alpha^*$ so that the transonic background is stable)\footnote{The reader should recall that the values of $\alpha$ indicated in this paragraph correspond to a stress tensor such that $\tau_{R\phi}\approx-2\alpha P\left(\Omega/\Omega_\mt{K}\right)$.}.

A comparison between the values of the last two columns reveals that the location of the sonic point has a significant influence on the decay rates of the r~mode. When the background flow is such that the sonic point is located inside the marginally stable orbit, the frequency of the simplest inertial mode has a variation with sound speed similar to that of the case where the basic flow isn't included. As the sound speed increases, the real part of the frequency decreases indicating that the mode is trapped in a wider region and the decay rate increases. The difference is that the decay rate is higher and increases faster in the case where the background inflow is taken into account, as seen in the left panel of Fig.~\ref{decayfit}. When $u(r)$ is such that the sonic point is at $r_0>r_\mt{ms}$, the increase of the decay rate with sound speed is extremely significant. Even for moderate values of $c_\mt{s}$, the imaginary part of the frequency can be as high as 10 times the real part. In addition, the extrapolated curve presented in the the right panel of Fig.~\ref{decayfit} reveals that for $c_\mt{s}=0.02$, the inertial oscillation, if present, would have a decay rate comparable to its frequency. This indicates that the inertial wave is deeply affected, even at moderate sound speeds, when the sonic point is at $r_0>r_\mt{ms}$: even if present, the inertial structure can no longer be referred to as a ``mode'' since it is damped before completing a single oscillation.

\begin{figure*}
\begin{center}
\includegraphics[width=0.7\linewidth]{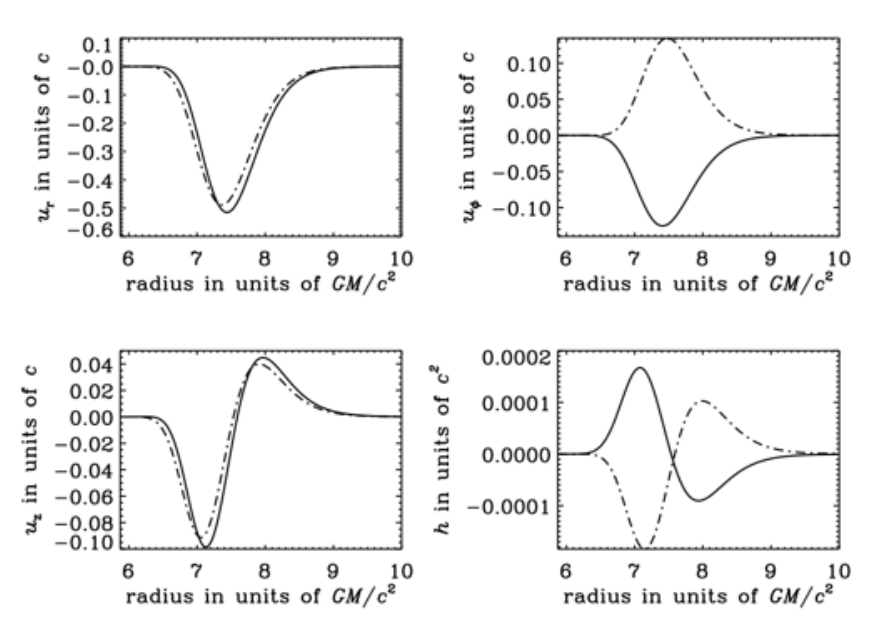} \\
\vspace{5mm}
\includegraphics[width=0.7\linewidth]{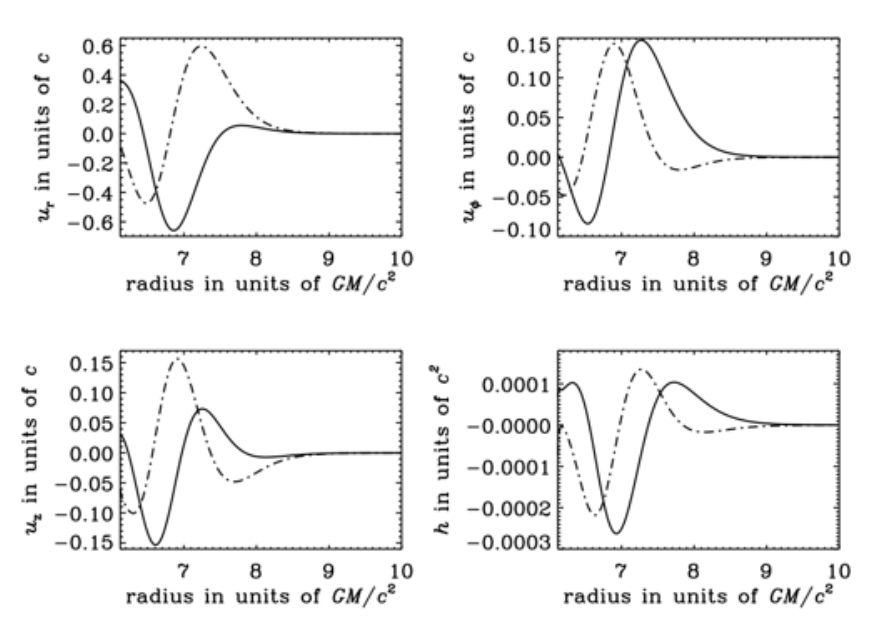}
\caption{Profile of the wave quantities $(u_r,u_\phi,u_z,h)$ for $c_\mt{s}=0.004$ corresponding to the simplest trapped inertial mode. The upper 4 panels represent a perturbation of a stable transonic flow with a sonic point inside the marginally stable orbit while the lower 4 panels represent the case $r_0>r_\mt{ms}$. In all plots, $r_0\lesssim r<10$ and the full line represents the real part of the wave while the dash-dotted curve represents the imaginary part.}
\label{figinflow}
\end{center}
\end{figure*}

The differences between the cases $r_0<r_\mt{ms}$ and $r_0>r_\mt{ms}$ can be better understood by looking at the eigenfunctions of the trapped modes. This is represented in Fig.~\ref{figinflow}. As it can be seen by looking at the 4 upper panels, when the background inflow locates the sonic point inside $r_\mt{ms}$, the structure of the r~mode is unaffected [cf. Fig.~\ref{freerm} (a)]. As in the case without inflow, $u_r$ still has a simple Gaussian structure centred at the maximum of $\kappa$. This is, however, not the case when $r_0>r_\mt{ms}$. Although the Gaussian structure is still an identifiable feature (at least for the small value of $c_\mt{s}$ considered in the plots), it is no longer centred at $\mt{max}(\kappa)$. Unlike the previous cases, the real and imaginary parts of $u_r$ and $u_\phi$ no longer peak at the same point: the real part of $u_r$ and the imaginary part of $u_\phi$ seem to be pulled inwards by the radial inflow. Other wave quantities are also slightly changed in the inner region in comparison to the case where $r_0<r_\mt{ms}$. \enlargethispage{-\baselineskip}These changes become more evident as the sound speed is increased until, eventually, the Gaussian structure is completely destroyed.  

It should be noted that, in the $r_0>r_\mt{ms}$ case, the destruction of the Gaussian profile at large $c_\mt{s}$ is unrelated to the proximity of the sonic point to the maximum of $\kappa$. Indeed, the background profile used in the case represented in the figure had a viscosity such that $r_0$ was located at about $6.07$ while for $c_\mt{s}=0.02$ (in which case the Gaussian structure couldn't be identified), $\alpha$ was such that $r_0=6.01$. The simple profile is more affected at large $c_s$ because the background inflow is more significant when the sound speed is larger.

Nonetheless, when the cases $r_0<r_\mt{ms}$ and $r_0>r_\mt{ms}$ are compared, it is clear that the location of the sonic point, inside or outside the marginally stable orbit, is important. In the former case, the transition from subsonic to supersonic occurs far away enough from the trapping region and, as a result, the structure of the r~mode isn't affected. In the latter case, the maximum of $\kappa$ is dangerously close to the transonic region indicating that the radial inflow is more significant where the mode is trapped than in the previous case. As a consequence, the r mode is deeply modified or even destroyed and has a significant decay rate.

\subsection{Conclusions and implications}
\label{uiwconclusions}

The results obtained reveal that inertial modes can be significantly affected when a background radial inflow is considered. If the conditions are such that the sonic point is located inside the marginally stable orbit (low viscosity), the structure of the trapped r~modes isn't modified but their decay rate is higher than in the case where no background inflow is taken into account. Presumably this is due to the ``leakage'' of the wave to the inner disc and plunging region. The inertial modes are more affected by a radial flow which has the sonic point located outside $r_\mt{ms}$ (high viscosity) and therefore closer to the trapping region. In this case, the decay rates are more significant and, for moderate to high sound speeds, the structure of the inertial modes is deeply altered and possibly destroyed.  

These results have implications to the excitation of trapped inertial modes in realistic discs. As explained previously, the coupling mechanism described in chapter \ref{excmech} relies on the protective trapping region created by the non-monotonic variation of the epicyclic frequency. In an idealised thin disc with no background inflow, the r~modes are trapped between their two Lindblad resonances, far enough away from the uncertain conditions of the marginally stable orbit, and the wave ``leakage'' out of the trapping region isn't significant. Global deformations can then couple with these trapped modes and they can grow as a result.

But a real disc is far from ideal. Here, I will refrain from mentioning the complications associated with the presence of magnetic fields and MRI turbulence and will instead focus on the scope of this chapter: the influence of the radial inflow. The results of this section are not exactly in favour of the coupling mechanism of chapter \ref{excmech}, but they are also far from being devastating. Observations indicate that black-hole accretion discs might be thick in some accretion states and have a relatively high $\alpha$ (see chapter 1 and references therein). Under these conditions, the trapped inertial modes have, at the very least, a significant decay rate, possibly comparable with the growth achieved by coupling with a global deformation. In the worst case scenario, realistic conditions are such that the flow is relatively thick and has a sonic point outside $r_\mt{ms}$ and the inertial modes are destroyed. In any case, the actual properties of real discs are uncertain and one can only speculate about the typical values of $c_\mt{s}$ and $\alpha$ in black-hole accretion flows. If the conditions are such that the radial inflow only influences the decay rate of the r~mode and/or only modifies its structure slightly, the coupling in a warped or eccentric disc may still be effective depending, in particular, on the amplitude of the deformation.

The fact that high-frequency QPOs are only detected in certain states of black-hole accretion indicates  that the mechanism in their origin is fragile. The results obtained here, together with the model that identifies HFQPOs with trapped inertial modes excited in a warped or eccentric disc, are in agreement with this statement. HFQPOs may only be visible in states where the accretion flow has specific properties, namely when the viscosity and sound speed are such that the sonic point is inside $r_\mt{ms}$ and global deformations are able to propagate to the inner disc with reasonable amplitudes.


\thispagestyle{empty}
\part{Conclusion}
\label{c}

\thispagestyle{empty}
\chapter{Conclusion}
\label{conc}

\section{Summary and conclusions}

This dissertation has addressed the problem of variability in black-hole accretion discs by means of various theoretical studies which focused on the dynamical processes occurring in these objects. Particular attention was given to those processes that result in quasi-periodic variability of high frequency which are specific to matter orbiting a relativistic compact object. Two classes of problems were investigated. One concerned the transonic nature of accretion around black holes and the influence of viscosity in its steadiness and stability. The other focused on the propagation and excitation of oscillations in these objects and how they are influenced by strong gravitational field effects. \newline

The work done on the stability of transonic black-hole accretion was described in chapters \ref{steadyintro} and \ref{timeaccretion}. In the former a comprehensive review of the topic was made in order to properly introduce the research described in the latter. This introduction revealed that the radial structure of transonic accretion discs is not as simple as that of Shakura--Sunyaev discs. If the radial component of the fluid velocity is included in the equations, they have a critical point $R_0$ where the flow goes from subsonic ($R>R_0$) to supersonic ($R<R_0$). Physically acceptable steady solutions are equivalent to Shakura and Sunyaev discs at large radii, go through the sonic point regularly for each set of flow parameters, and become supersonic close to the black hole. However, at least in isothermal discs with $\alpha P$ stress, such solutions cannot be found for $\alpha>\alpha^*\propto c_\mt{s}^{1/3}$, not because transonic accretion is no longer possible but because it proceeds unsteadily. 

The time-dependent calculations of chapter \ref{timeaccretion} indicate that persistent inertial-acoustic waves with frequency close to the maximum of the epicyclic frequency are found in the disc for $\alpha>\alpha^*$. These waves grow due to viscous overstability which is most likely to act effectively where these waves slow down, close to the point where $\kappa$ peaks. Simulations show that waves propagate both inwards and outwards from this point. Steady accretion is possible for smaller values of $\alpha$, even though the disc is locally unstable to viscous overstability, because growing disturbances propagate out of the domain of interest. 

The review of chapter \ref{steadyintro} showed that inertial-acoustic waves propagating in viscous discs have long been known to be locally viscously overstable for all values of $\alpha$. Nonetheless, the conditions under which the instability becomes global --- that is, it is revealed in a system with open boundaries via which waves are likely to escape before reaching considerable amplitudes --- are more difficult to quantify. Numerical calculations made in the 90s show that viscous overstability is global when $\alpha>\alpha_\mt{vo}$. This limiting viscosity depends on the mass accretion rate or, equivalently, the sound speed in the disc; $\alpha_\mt{vo}$ is lower for lower $\dot{M}$ or $c_\mt{s}$. The numerical calculations presented in chapter \ref{timeaccretion} showed that in isothermal discs with $\alpha P$ stress $\alpha_\mt{vo}=\alpha^*\propto c_\mt{s}^{1/3}$, that is, accretion is globally viscously overstable because it can't proceed steadily in a physically acceptable way. 

This relation may, however, be exclusive to isothermal, $\alpha P$ discs --- the very existence of a limit $\alpha^*$ is probably related to the simplicity of such model, as speculated by \cite{ap2003}. Notwithstanding the possible lack of generality of the relation $\alpha_\mt{vo}=\alpha^*$, the study presented in Part \ref{ta} of the thesis was useful to obtain the exact relation $\alpha_\mt{vo}(c_\mt{s})$ in the case of the disc model considered and to understand the conditions under which such discs may become viscously overstable. Future work is important to look for a relation $\alpha_\mt{vo}(c_\mt{s}\;\mathrm{or}\;\dot{M})$ applicable to more general flows. The ideas discussed in Part \ref{ta} are also useful to interpret future simulations of turbulent discs. \newline

The work on oscillations in accretion discs around black holes was described in the various chapters that constitute Part \ref{os} of this thesis. The topic is introduced in chapter \ref{oscilintro} where I describe the mathematical theory of oscillations in discs and mention possible wave-trapping regions, present due to the non-monotonic variation of the epicyclic frequency with radii in relativistic discs. Throughout most of Part \ref{os}, the background radial inflow is neglected for simplicity although it is considered in chapter \ref{reflect}.

An excitation mechanism for the simplest of the trapped inertial modes is described in chapter \ref{excmech}. I find that these waves can achieve reasonable growth rates when they couple with a global deformation in the disc. (Deformations are simply described as very low-frequency modes for the purpose of solving the complicated non-linear coupling equations.) The energy necessary for the growth of the r~mode is drawn from the disc by dissipation of a negative-energy intermediate wave involved in the coupling. I conclude that inertial modes can be excited in warped or eccentric discs for a variety of sound speeds and black-hole spins. Although the results of this chapter were obtained in a simple disc model, they are expected to hold in more realistic flows and therefore be observationally relevant, as it will be discussed in section \ref{obrel}.

There is a fundamental requirement for the excitation mechanism described to be efficient: the global deformations have to reach the inner disc region, where the r~modes are trapped, so that the coupling can occur. With this condition in mind, I found it necessary to study the propagation of such modes from the outer disc where they originate to the inner parts. This is done in chapter \ref{warpecc} where a more general description of a stationary wave-like warp and eccentricity, allowing for viscous damping, is made. A model similar to \cite{ss73} discs but with a more accurate description of the vertical structure is introduced and wave propagation is considered within it. I deduce from the results obtained that global deformations can reach the inner parts of the disc and effectively interact with the r~modes when the mass accretion rate is high (i.e., the disc is thick) and the viscous damping is low. 

Finally, in chapter \ref{reflect}, I study the influence of a background radial inflow and the presence of a sonic point in the propagation of diskoseismic modes. I start by analysing the reflection properties at the critical point in a simple toy model. The intermediate waves of the excitation mechanism of chapter \ref{excmech} need, for some of the coupling processes considered, to be absorbed at the inner disc boundary in order for energy to be drawn from the disc. In a more realistic flow description, where the radial inflow is considered, this boundary is at the sonic point. Calculations of the reflection coefficient within the toy model introduced in chapter \ref{reflect} show that absorption is indeed possible under some conditions, being favoured when the background density isn't too steep in the inner region. 

More interesting is the influence of the background inflow on the trapped inertial modes. Results show that these modes leak into the plunging region and decay as a result. Even though this damping rate can be considerable, the excitation mechanism can still work if, e.g., the warp or eccentricity amplitude is significant. Less favourable to growth of inertial modes are flows which have the sonic point located outside the marginally stable orbit. In these cases, the radial velocity in the trapping region is higher and, as a result, the structure of inertial modes is altered and possibly destroyed. This result has serious implications for the detection of r~modes in discs with considerable radial inflow. Nonetheless, detection of these waves is still possible if radial advection is less significant.

\section{Observational relevance}
\label{obrel}

This thesis intended to be a contribution to the theoretical knowledge of variability in black-hole accretion discs. However, no study aiming to play a part in the understanding of astrophysical phenomena can exist isolated from observational data. Doing research in theoretical astrophysics should be more than doing interesting mathematical calculations; it should always endeavour to explain observational events.

In the core of this dissertation is rapid X-ray variability, in particular the phenomenon of high-frequency quasi-periodic oscillations. Fundamental characteristics of these peaks seen in the X-ray spectra of some black-hole binaries are their frequency, $\mathcal{O}(100\;\mt{Hz})$ and stability against luminosity variations, which connects these oscillations to the very inner, strong gravitational field region. Any model claiming to explain HFQPOs has to, at the very least, be intrinsically relativistic and explain the range and stability of frequencies. At the heart of the search for theoretical models for these quasi-periodic oscillations is precisely their connection with the strong gravitational field of the compact object, making them promising probes of this region and precise tools to measure black-hole spin.

In Part \ref{ta}, I investigate the possibility that HFQPOs can be identified with inertial-acoustic waves excited due to viscous overstability in non-steady accretion flows. On the other hand, in Part \ref{os} I associate high-frequency variability with trapped inertial waves excited in a warped or eccentric disc. The two models presented are intrinsically relativistic as they both rely on the non-monotonic variation of the epicyclic frequency with radii. Viscous overstability seems to only attain a global character because inertial-acoustic waves slow down in the region around the maximum of $\kappa$. On the other hand, inertial waves are trapped precisely because of the existence of this maximum. In both models the frequency of the QPO can roughly be identified with $\mt{max}(\kappa)$ which is in the expected range and depends only on the mass and spin of the black hole. The frequency stability is corroborated by the fact that changes in sound speed (within the thin-disc range) don't significantly alter this scenario.

The models based on both inertial-acoustic and inertial waves not only explain the basic properties of HFQPOs but provide a straightforward correspondence between their frequency and black-hole spin. Unfortunately, that's not the end of the story. As seen in chapter \ref{intro}, the current observational status of X-ray variability demands for far more complete QPO models. These oscillations are known to be detected almost exclusively when the black hole is in the very-high state of emission which dominates at X-ray luminosities close to the Eddington limit (high mass accretion rate). In this state, the spectrum has a blackbody-like component at low energies and a steep power law tail to higher energies. Although the physical origin of this state is unknown \citep{remimcclin2006}, a possible model explains the spectral characteristics by assuming that the cool, optically thick disc (responsible for the blackbody-like component) is surrounded by a hot, optically thin corona (responsible for the power law tail) in the inner region \citep{donereview}. HFQPOs may originate in the disc but they seem to only be revealed when the cool flow interacts with the hot corona.

The trapped inertial waves model may still fit these observational constraints: these oscillations are only excited in the very high state because only then does the warp or eccentricity reach the inner region with reasonable amplitude. This is justified by the study presented in chapter \ref{warpecc} where it was seen that the propagation of these global deformations is, at fixed viscous damping, favoured when $\dot{M}$ is high. The model considered is, however, too crude to allow for a proper description of the steep power law state and to consider the influence of a hot corona on the inertial oscillations.

In contrast with the previous model, a drawback of the inertial-acoustic wave theory is precisely the fact that viscous overstability is favoured at low accretion rates. This doesn't necessarily mean that these oscillations couldn't be detected in the very high state since at high enough viscosity they can still exist for reasonable $\dot{M}$. But it also doesn't explain why viscously-overstable inertial-acoustic oscillations would only be detected in the high luminosity state. But it should be mentioned that the physical origin of this state is far from being known for certain and its observational characteristics vary slightly from object to object. It is therefore complicated to try to relate a QPO model with the properties of the flow in this state.

A possible problem with the trapped inertial oscillations model regards the influence of radial inflow on these waves. While the transonic nature of the flow is at the heart of the viscous overstability model presented in Part \ref{ta}, the consideration of a background radial velocity may hinder the theory of chapter \ref{excmech} of Part \ref{os}. As seen in chapter \ref{reflect}, if the sonic point is located outside the marginally stable orbit, the simplest trapped inertial mode suffers considerable damping and its structure may be deeply altered by the radial inflow. Despite that, it is not known how strong this inflow may be in the very high state or how the mode would be affected in more realistic deformed discs in which accretion seems to occur through two plunging streams that start at high latitudes with respect to the midplane of the disc \citep{globalsimulations}.

To terminate the reference to observational characteristics, I should mention the infamous 3:2 ratio which is not explained within any of the models considered. As seen in the introduction, a considerable number of the sources that exhibit HFQPOs displays pairs of these oscillations with frequencies in a 3:2 ratio. Curiously enough, these pairs are most often not detected simultaneously and in some of the sources the lowest frequency QPO appears when the power law flux is very strong while the highest frequency appears when the flux is weaker \citep{remimcclin2006}. Perhaps this is indicative of different, yet correlated, mechanisms in the origin of each component of the pair. While the models presented in this thesis don't explain the frequency commensurability they may still explain the origin of one of these HFQPOs.

\section{Theoretical difficulties and numerical simulations}

More than aiming at explaining observational characteristics, a good model of HFQPOs has to surpass theoretical barriers. In other words, the mechanism that gives rise to quasi-periodic variability cannot be an artefact of the approximations used to model the accretion flow, i.e., it has to be shown to operate in more realistic discs. \enlargethispage{\baselineskip}

Throughout this thesis, the process of accretion was characterised somewhat unrealistically: magnetic fields were ignored and angular momentum transport was described simplistically or, at times, completely neglected. Moreover, heating and cooling mechanisms were ignored altogether as isothermal flows were studied in both Part \ref{ta} and Part \ref{os}, and relativistic effects were considered within a pseudo-relativistic approximation. These simplifications reflect the theoretical difficulties involved in modelling any phenomenon related to accretion discs. In an analytical or quasi-analytical framework, problems are only tractable within certain approximations. Still, this type of work provides valuable insight into understanding the physics of accretion flows and should always accompany, if not precede, the less-simplistic numerical simulations.

Sophisticated numerical calculations which use fewer approximations than the ones considered here can be used to see how the models studied are altered when more realistic discs are considered. To date, the most realistic simulation looking into viscous overstable modes in black-hole accretion discs is perhaps that of \cite{oneill2009}. While they show, in agreement with previous results, that waves may be excited by viscous overstability for large $\alpha$ and $\dot{M}$, they still don't consider MRI turbulence or full relativistic effects. Indeed, MHD simulations indicate that HFQPOs may be damped or hidden by MRI turbulence since their observational amplitude falls below the level of turbulent noise \citep{rm2009}. Viscous overstable modes may still be visible in MHD simulations if they feature larger effective viscosities and are run for longer to improve the signal-to-noise ratio \citep{oneill2009}. Simpler numerical work as that described in chapter \ref{timeaccretion} 
could provide some insight into the exact value of viscosity (and its dependence on $\dot{M}$ or $c_\mt{s}$) above which viscous overstability can be efficient.

A considerably more realistic simulation has looked into the excitation of trapped inertial modes in tilted discs, providing a numerical test of the model described in chapter \ref{excmech}. The fully relativistic, MHD work of \cite{heniseyetal2009}, already mentioned in the introduction, is the only combining state-of-the-art MRI simulations with tilted discs. It provides preliminary confirmation of the inertial-waves model since simulations of tilted discs show an excess of inertial power when compared to equivalent untilted simulations. My contribution to this paper was precisely that of helping interpret the origin of this excess inertial power in light of the work on the excitation of trapped inertial modes in deformed discs. Indeed, simulations show a particularly interesting trapped feature with a frequency close to $\max{(\kappa)}$ which is found in all fluid variables analysed. Unfortunately, it is not straightforward to identify this feature with a trapped r~mode since its characteristics are not all of those expected from the analytic theory of thin-disc oscillations. The exact cause of the excess inertial power seen in the simulations is difficult to assess due to the complicated background characteristic of a tilted, thick accretion flow. It may be that it is simply due to the differences in the accretion process in tilted and untilted flows. Alternatively, the excess inertial power may be due to trapped inertial waves excited by coupling with the warp which have different characteristics from those predicted in simple analytic calculations precisely due to the complicated background. This would should that inertial modes can resist both MRI turbulence and radial inflow in a deformed disc. Future work, both numerical and analytic, is required to explain all the features of the simulations, to understand how different processes influence inertial modes in tilted flows, and if HFQPOs may indeed be explained in terms of these oscillations.\newline

With the increasingly sophisticated numerical simulations and the flow of observational data from modern X-ray observatories, it is an exciting time for the theoretical modelling of black-hole accretion discs. High-frequency variability models are of particular interest because of the connection between HFQPOs and the strong gravitational field region. Scientists in the field aim not only to comprehend the physics of accretion onto compact objects, but also to contribute to the understanding of black holes themselves and to play a part in testing or disproving the theory of general relativity. With the advancement of numerical simulations, the further understanding of observational data, the valuable insight of analytic theoreticians and the constant collaboration between the three parts involved, this goal will soon be achieved.


\clearpage

\bibliographystyle{mn} 
\renewcommand{\bibname}{References} 
\bibliography{mn-jour,references} 

\begin{thebibliography}{188}
\expandafter\ifx\csname natexlab\endcsname\relax\def\natexlab#1{#1}\fi

\bibitem[{Abramowicz {et~al.}(1996)Abramowicz, Brandenburg, \&
  Lasota}]{abl1996}
Abramowicz M.~A., Brandenburg A., Lasota J.-P., 1996, \emph{MNRAS}, 281, L21

\bibitem[{Abramowicz {et~al.}(1980)Abramowicz, Calvani, \&
  Nobili}]{abramowiczetal1980}
Abramowicz M.~A., Calvani M., Nobili L., 1980, \emph{ApJ}, 242, 772

\bibitem[{Abramowicz \& Kato(1989)}]{ak1989}
Abramowicz M.~A., Kato S., 1989, \emph{ApJ}, 336, 304

\bibitem[{Abramowicz \& Klu\'zniak(2001)}]{abramowiczkluzniak2001}
Abramowicz M.~A., Klu\'zniak W., 2001, \emph{A\&A}, 374, L19

\bibitem[{Abramowicz {et~al.}(2004)Abramowicz, Klu\'zniak, Stuchl\'ik, \&
  T\"or\"ok}]{aetal2004}
Abramowicz M.~A., Klu\'zniak W., Stuchl\'ik, T\"or\"ok G., 2004,
  \emph{astro-ph/0401464v1}

\bibitem[{Abramowitz \& Stegun(1972)}]{abramowitzstegun}
Abramowitz M., Stegun I.~A., 1972, \emph{Handbook of Mathematical Functions}.
  Dover Press, New York

\bibitem[{Afshordi \& Paczy\'nski(2003)}]{ap2003}
Afshordi N., Paczy\'nski B., 2003, \emph{ApJ}, 592, 354

\bibitem[{Agol {et~al.}(2001)Agol, Krolik, Turner, \& Stone}]{agoletal2001}
Agol E., Krolik J., Turner N.~J., Stone J.~M., 2001, \emph{ApJ}, 558, 543

\bibitem[{Arras {et~al.}(2006)Arras, Blaes, \& Turner}]{arrasetal2006}
Arras P., Blaes O., Turner N.~J., 2006, \emph{ApJ}, 645, L65

\bibitem[{Balbus \& Hawley(1991)}]{mri}
Balbus S.~A., Hawley J.~F., 1991, \emph{ApJ}, 374, 214

\bibitem[{Balbus \& Hawley(1998)}]{reviewmri}
---, 1998, \emph{Rev. Mod. Phys.}, 70, 1

\bibitem[{Bardeen \& Petterson(1975)}]{bardeenpetterson1975}
Bardeen J.~M., Petterson J.~A., 1975, \emph{ApJ}, 195, L65

\bibitem[{Belloni {et~al.}(2006)Belloni, Soleri, P.~Casella, M\'endez, \&
  Migliari}]{bellonietal2006}
Belloni T., Soleri P., P.~Casella P., M\'endez M., Migliari S., 2006,
  \emph{MNRAS}, 369, 305

\bibitem[{Bisnovatyi-Kogan \& Blinnikov(1977)}]{bb77}
Bisnovatyi-Kogan G.~S., Blinnikov S.~I., 1977, \emph{A\&A}, 59, 111

\bibitem[{Blaes(1987)}]{blaes1987}
Blaes O.~M., 1987, \emph{MNRAS}, 227, 975

\bibitem[{Blaes(2003)}]{blaesnotes}
---, 2003, in Beskin V., Henri G., Menard F., Dalibard J., eds, \emph{Accretion
  Disks, Jets, and High Energy Phenomena in Astrophysics}, EDP Science, Paris,
  p. 137

\bibitem[{Blumenthal {et~al.}(1984)Blumenthal, Yang, \& Lin}]{blumenthal1984}
Blumenthal G.~R., Yang L.~T., Lin D. N.~C., 1984, \emph{ApJ}, 287, 774

\bibitem[{Bondi(1952)}]{bondi}
Bondi H., 1952, \emph{MNRAS}, 112, 195

\bibitem[{Boyd(2001)}]{boydbook}
Boyd J.~P., 2001, \emph{Chebyshev and Fourier Spectral Methods}. Dover Press,
  New York

\bibitem[{Chandrasekhar(1960)}]{chandrasekhar1960}
Chandrasekhar S., 1960, \emph{Proc. Natl. Acad. Sci.}, 46, 253

\bibitem[{Chen \& Taam(1992)}]{chentaam1992}
Chen X., Taam R.~E., 1992, \emph{MNRAS}, 255, 51

\bibitem[{Chen \& Taam(1993)}]{chentaam1993}
---, 1993, \emph{ApJ}, 412, 254

\bibitem[{Chen \& Taam(1995)}]{chentaam1995}
---, 1995, \emph{ApJ}, 441, 354

\bibitem[{Christensen-Dalsgaard(2002)}]{cd2002}
Christensen-Dalsgaard J., 2002, \emph{Rev. Mod. Phys.}, 74, 1073

\bibitem[{Christensen-Dalsgaard(2003)}]{dalsgaard}
---, 2003, Lecture notes on stellar oscillations,
  \emph{http://astro.phys.au.dk/$\sim$jdc/oscilnotes}

\bibitem[{Connors {et~al.}(1980)Connors, Stark, \& Piran}]{csp1980}
Connors P.~A., Stark R.~F., Piran T., 1980, \emph{ApJ}, 235, 224

\bibitem[{Cox \& Everson(1983)}]{coxeverson1983}
Cox J.~P., Everson B.~L., 1983, \emph{ApJS}, 52, 451

\bibitem[{Done {et~al.}(2007)Done, Gierli\'nski, \& Kubota}]{donereview}
Done C., Gierli\'nski M., Kubota A., 2007, \emph{A\&AR}, 15, 1

\bibitem[{Dov\v{c}iak {et~al.}(2004)Dov\v{c}iak, Karas, \& Yaqoob}]{dky2004}
Dov\v{c}iak M., Karas V., Yaqoob T., 2004, \emph{ApJS}, 153, 205

\bibitem[{Esin {et~al.}(1997)Esin, McClintock, \& Narayan}]{esinetal1997}
Esin A.~A., McClintock J., Narayan R., 1997, \emph{ApJ}, 489, 865

\bibitem[{Ferrari {et~al.}(1985)Ferrari, Trussoni, Rosner, \&
  Tsinganos}]{ferrarietal1985}
Ferrari A., Trussoni E., Rosner R., Tsinganos K., 1985, ApJ, 294, 397

\bibitem[{Fragile(2005)}]{fragile2005}
Fragile P.~C., 2005, in Chen P., Bloom E., Madejski G. and Patrosian V., eds,
  \emph{Proceedings of the 22nd Texas Symposium on Relativistic Astrophysics at
  Stanford}, p. 270

\bibitem[{Fragile {et~al.}(2007)Fragile, Blaes, Anninos, \&
  Salmonson}]{globalsimulations}
Fragile P.~C., Blaes O.~M., Anninos O.~M., Salmonson J.~D., 2007, \emph{ApJ},
  668, 417

\bibitem[{Frank {et~al.}(2002)Frank, King, \& Raine}]{accretionpower}
Frank J., King A., Raine D., 2002, \emph{Accretion Power in Astrophysics}.
  Cambridge Univ. Press, Cambridge

\bibitem[{Fu \& Lai(2009)}]{fulai2009}
Fu W., Lai D., 2009, \emph{ApJ}, 690, 1386

\bibitem[{Gallimore {et~al.}(1997)Gallimore, Baum, \& O'Dea}]{agn}
Gallimore J.~F., Baum S.~A., O'Dea C.~P., 1997, \emph{Nature}, 388, 852

\bibitem[{Gammie(1999)}]{gammie99}
Gammie C.~F., 1999, \emph{ApJ}, 522, L57

\bibitem[{Gammie {et~al.}(2000)Gammie, Goodman, \& Ogilvie}]{gammieetal2000}
Gammie C.~F., Goodman J., Ogilvie G.~I., 2000, \emph{MNRAS}, 318, 1005

\bibitem[{Gerend \& Boynton(1976)}]{gerendboynton1976}
Gerend D., Boynton P.~E., 1976, \emph{ApJ}, 209, 562

\bibitem[{Gierli\'nski \& Done(2004)}]{gd2004}
Gierli\'nski M., Done C., 2004, \emph{MNRAS}, 347, 885

\bibitem[{Gierli\'nski {et~al.}(2008)Gierli\'nski, Middleton, Ward, \&
  Done}]{gierlinskietal2008}
Gierli\'nski M., Middleton M., Ward M., Done C., 2008, \emph{Nature}, 455, 369

\bibitem[{Goldreich {et~al.}(1986)Goldreich, Goodman, \&
  Narayan}]{goldreichetal1986}
Goldreich P., Goodman J., Narayan R., 1986, \emph{MNRAS}, 221, 339

\bibitem[{Goldreich \& Tremaine(1979)}]{gt1979}
Goldreich P., Tremaine S., 1979, \emph{ApJ}, 233, 857

\bibitem[{Goodchild \& Ogilvie(2006)}]{goodchildogilvie2006}
Goodchild S., Ogilvie G.~I., 2006, \emph{MNRAS}, 368, 1123

\bibitem[{Goodman(1993)}]{goodman1993}
Goodman J., 1993, \emph{ApJ}, 406, 596

\bibitem[{Haswell {et~al.}(2001)Haswell, King, J.R., \&
  Charles}]{haswelletal2001}
Haswell C.~A., King A.~R., J.R. J. R.~M., Charles P.~A., 2001, \emph{MNRAS},
  321, 475

\bibitem[{Henisey {et~al.}(2009)Henisey, Blaes, Fragile, \&
  Ferreira}]{heniseyetal2009}
Henisey K., Blaes O.~M., Fragile P.~C., Ferreira B.~T., 2009, \emph{ApJ}, in
  press

\bibitem[{Hirose {et~al.}(2009)Hirose, Krolik, \& Blaes}]{hiroseetal2009}
Hirose S., Krolik J.~H., Blaes O., 2009, \emph{ApJ}, 691, 16

\bibitem[{Honma {et~al.}(1992)Honma, Matsumoto, \& Kato}]{hmk1992}
Honma F., Matsumoto R., Kato S., 1992, \emph{PSAJ}, 529, 529

\bibitem[{Huerre \& Monkewitz(1990)}]{absconvec}
Huerre P., Monkewitz P.~A., 1990, \emph{ARFM}, 22, 473

\bibitem[{Ichimaru(1977)}]{ichimaru1977}
Ichimaru S., 1977, \emph{ApJ}, 214, 840

\bibitem[{Ivanov \& Illarionov(1997)}]{ivanovillarionov1997}
Ivanov P.~B., Illarionov A.~F., 1997, \emph{MNRAS}, 285, 394

\bibitem[{Kaaret {et~al.}(1997)Kaaret, Ford, \& Chen}]{kaaretetal1997}
Kaaret P., Ford E.~C., Chen K., 1997, \emph{ApJ}, 480, L27

\bibitem[{Kato(1978)}]{kato1978}
Kato S., 1978, \emph{MNRAS}, 185, 629

\bibitem[{Kato(1983)}]{kato1983}
---, 1983, \emph{PASJ}, 35, 249

\bibitem[{Kato(1989)}]{kato1989}
---, 1989, \emph{PASJ}, 41, 745

\bibitem[{Kato(1990)}]{kato1990}
---, 1990, \emph{PASJ}, 42, 99

\bibitem[{Kato(1994)}]{kato1994}
---, 1994, \emph{PASJ}, 46, 415

\bibitem[{Kato(2001)}]{rkato2001}
---, 2001, \emph{PASJ}, 53, 1

\bibitem[{Kato(2004)}]{katowarp2004}
---, 2004, \emph{PASJ}, 56, 905

\bibitem[{Kato(2007)}]{kato2007}
---, 2007, \emph{PASJ}, 59, 451

\bibitem[{Kato(2008)}]{kato2008}
---, 2008, \emph{PASJ}, 60, 111

\bibitem[{Kato \& Fukue(1980)}]{katofukue1980}
Kato S., Fukue J., 1980, \emph{PASJ}, 32, 377

\bibitem[{Kato \& Fukue(2006)}]{katofukue2006}
---, 2006, \emph{PASJ}, 58, 909

\bibitem[{Kato {et~al.}(1988{\natexlab{a}})Kato, Honma, \& Matsumoto}]{khm1988}
Kato S., Honma F., Matsumoto R., 1988{\natexlab{a}}, \emph{MNRAS}, 231, 37

\bibitem[{Kato {et~al.}(1988{\natexlab{b}})Kato, Honma, \&
  Matsumoto}]{khm1988b}
---, 1988{\natexlab{b}}, \emph{PASJ}, 40, 709

\bibitem[{Kato {et~al.}(1993)Kato, Wu, Yang, \& Yang}]{kxlz1993}
Kato S., Wu X.-B., Yang L.-T., Yang Z.-L., 1993, \emph{MNRAS}, 260, 317

\bibitem[{Katz(1973)}]{katz1973}
Katz J.~I., 1973, \emph{Nature Phys. Sci.}, 246, 87

\bibitem[{King {et~al.}(2007)King, Pringle, \& Livio}]{kingetal2007}
King A.~R., Pringle J.~E., Livio M., 2007, \emph{MNRAS}, 376, 1740

\bibitem[{King {et~al.}(2004)King, Pringle, West, \& Livio}]{kingetal2004}
King A.~R., Pringle J.~E., West R.~G., Livio M., 2004, \emph{MNRAS}, 348, 111

\bibitem[{Kippenhahn \& Weigert(1996)}]{stellar}
Kippenhahn R., Weigert A., 1996, \emph{Stellar Structure and Evolution}.
  Springer-Verlag, Berlin

\bibitem[{Klein-Wolf {et~al.}(2004)Klein-Wolf, Homan, \& van~der
  Klis}]{kleinwolfetal2004}
Klein-Wolf M., Homan J., van~der Klis M., 2004, Nuc. Phys. B PS, 132, 381

\bibitem[{Kley {et~al.}(1993)Kley, Papaloizou, \& Lin}]{kleyetal1993}
Kley W., Papaloizou J. C.~B., Lin D. N.~C., 1993, \emph{ApJ}, 409, 739

\bibitem[{Klu\'zniak \& Abramowicz(2001)}]{kluzniakabramowicz2001}
Klu\'zniak W., Abramowicz M.~A., 2001, \emph{astro-ph/0105057}

\bibitem[{Klu\'zniak {et~al.}(1990)Klu\'zniak, Michelson, \& Wagoner}]{kmw1990}
Klu\'zniak W., Michelson P., Wagoner R.~V., 1990, \emph{ApJ}, 358, 538

\bibitem[{Klu\'zniak \& Wagoner(1985)}]{kw1985}
Klu\'zniak W., Wagoner R.~V., 1985, \emph{ApJ}, 297, 548

\bibitem[{Korycansky \& Pringle(1995)}]{korycanskypringle1995}
Korycansky D.~G., Pringle J.~E., 1995, \emph{MNRAS}, 272, 618

\bibitem[{Krolik(1999)}]{krolik99}
Krolik J.~H., 1999, \emph{ApJ}, 515, L73

\bibitem[{Lai(1999)}]{lai1999}
Lai D., 1999, \emph{ApJ}, 524, 1030

\bibitem[{Lai \& Tsang(2009)}]{laitsang2009}
Lai D., Tsang D., 2009, \emph{MNRAS}, 393, 979

\bibitem[{Laney(1998)}]{gasdynamics}
Laney C.~B., 1998, \emph{Computational Gasdynamics}. Cambridge Univ. Press,
  Cambridge

\bibitem[{Latter \& Ogilvie(2006)}]{latterogilvie2006}
Latter H.~N., Ogilvie G.~I., 2006, \emph{MNRAS}, 372, 1829

\bibitem[{Li {et~al.}(2003)Li, Goodman, \& Narayan}]{lietal2003}
Li L.-X., Goodman J., Narayan R., 2003, \emph{ApJ}, 593, 980

\bibitem[{Liang \& Thompson(1980)}]{lt1980}
Liang E. P.~T., Thompson K.~A., 1980, \emph{ApJ}, 240, 271

\bibitem[{Lightman \& Eardley(1974)}]{le74}
Lightman A.~P., Eardley D.~M., 1974, \emph{ApJ}, 187, L1

\bibitem[{Lin \& Shu(1964)}]{linshu}
Lin C.~C., Shu F.~H., 1964, \emph{ApJ}, 140, 646L

\bibitem[{Lubow(1991{\natexlab{a}})}]{lubow1991a}
Lubow S.~H., 1991{\natexlab{a}}, \emph{ApJ}, 381, 259

\bibitem[{Lubow(1991{\natexlab{b}})}]{lubow1991b}
---, 1991{\natexlab{b}}, \emph{ApJ}, 381, 268

\bibitem[{Lubow \& Ogilvie(1998)}]{lubowogilvie1998}
Lubow S.~H., Ogilvie G.~I., 1998, \emph{ApJ}, 504, 983

\bibitem[{Lubow {et~al.}(2002)Lubow, Ogilvie, \& Pringle}]{lubowetal2002}
Lubow S.~H., Ogilvie G.~I., Pringle J.~E., 2002, \emph{ApJ}, 337, 706

\bibitem[{Lubow \& Pringle(1993)}]{lubowpringle1993}
Lubow S.~H., Pringle J.~E., 1993, \emph{ApJ}, 409, 360

\bibitem[{Lynden-Bell(1969)}]{lyndenbell1969}
Lynden-Bell D., 1969, \emph{Nature}, 223, 690

\bibitem[{Lynden-Bell(1978)}]{lyndenbell1978}
---, 1978, \emph{Phys. Scrip.}, 17, 185

\bibitem[{Maccarone(2002)}]{maccarone2002}
Maccarone T.~J., 2002, \emph{MNRAS}, 336, 1371

\bibitem[{Martin {et~al.}(2008)Martin, Tout, \& Pringle}]{mtp2008}
Martin R.~G., Tout C.~A., Pringle J.~E., 2008, \emph{MNRAS}, 387, 188

\bibitem[{Matsumoto {et~al.}(1984)Matsumoto, Kato, Fukue, \&
  Okazaki}]{mkfo1984}
Matsumoto R., Kato S., Fukue J., Okazaki A.~T., 1984, \emph{PASJ}, 36, 71

\bibitem[{Matsumoto {et~al.}(1988)Matsumoto, Kato, \& Honma}]{mkh1988}
Matsumoto R., Kato S., Honma F., 1988, in Tanaka Y., ed, \emph{Physics of
  Neutron Stars and Black Holes}, Tokyo: Universal Academy, Tokyo, p. 155

\bibitem[{McClintock \& Remillard(2006)}]{bhbbook}
McClintock J.~E., Remillard R.~A., 2006, in Lewin W. H. G., van der Klis M.,
  eds, \emph{Compact Stellar X-ray Sources}, Cambridge Univ. Press, Cambridge,
  p. 157

\bibitem[{McHardy {et~al.}(2006)McHardy, Knigge, Uttley, \&
  Fender}]{mchardyetal2006}
McHardy I.~M., Knigge E. K.~C., Uttley P., Fender R.~P., 2006, \emph{Nature},
  444, 730

\bibitem[{Merloni \& Fabian(2002)}]{mf2002}
Merloni A., Fabian A.~C., 2002, \emph{MNRAS}, 332, 165

\bibitem[{Middleton {et~al.}(2006)Middleton, Done, Gierli\'nski, \&
  Davis}]{mdgd2006}
Middleton M., Done C., Gierli\'nski M., Davis S.~W., 2006, \emph{MNRAS}, 373,
  1004

\bibitem[{Miller {et~al.}(1996)Miller, Lamb, \& Psaltis}]{mlp1996}
Miller M.~C., Lamb F.~K., Psaltis D., 1996, \emph{AAS}, 189, 1329

\bibitem[{Milsom \& Taam(1996)}]{mt1996}
Milsom J.~A., Taam R.~E., 1996, \emph{MNRAS}, 283, 919

\bibitem[{Milsom \& Taam(1997)}]{mt1997}
---, 1997, \emph{MNRAS}, 286, 358

\bibitem[{Muchotrzeb(1983)}]{m1983}
Muchotrzeb B., 1983, \emph{AcA}, 33, 79

\bibitem[{Muchotrzeb \& Paczy\'nski(1982)}]{mp1982}
Muchotrzeb B., Paczy\'nski B., 1982, \emph{AcA}, 32, 1

\bibitem[{Muchotrzeb-Czerny(1986)}]{m1986}
Muchotrzeb-Czerny B., 1986, \emph{AcA}, 36, 1

\bibitem[{Narayan(1992)}]{nara1992}
Narayan R., 1992, \emph{ApJ}, 394, 261

\bibitem[{Narayan {et~al.}(1998)Narayan, Mahadevan, \& Quataert}]{nmqbook}
Narayan R., Mahadevan R., Quataert E., 1998, in Abramowicz M. A., Bj\"ornsson
  G., Pringle J. E., eds, \emph{Theory of Black Hole Accretion Discs},
  Cambridge Univ. Press, Cambridge, p. 148

\bibitem[{Narayan {et~al.}(1996)Narayan, McClintock, \& Yi}]{nmy1996}
Narayan R., McClintock J., Yi I., 1996, \emph{ApJ}, 457, 821

\bibitem[{Narayan \& Yi(1994)}]{adaf}
Narayan R., Yi I., 1994, \emph{ApJ}, 428, L13

\bibitem[{Neil {et~al.}(2007)Neil, Bailyn, \& Cobb}]{neiletal2007}
Neil E.~T., Bailyn C.~D., Cobb B.~E., 2007, \emph{ApJ}, 657, 409

\bibitem[{Novikov \& Thorne(1973)}]{nt73}
Novikov I.~D., Thorne K.~S., 1973, in DeWitt C., DeWitt B. S., eds, \emph{Black
  Holes, Les Astres Occlus}, Gordon and Breach Science Publishers, New York, p.
  343

\bibitem[{Nowak \& Lehr(1998)}]{nowaklehrchapter}
Nowak M.~A., Lehr D.~E., 1998, in Abramowicz M. A., Bj\"ornsson G., Pringle J.
  E., eds, \emph{Theory of Black Hole Accretion Discs}, Cambridge Univ. Press,
  Cambridge, p. 233

\bibitem[{Nowak \& Wagoner(1991)}]{nowakwagoner1991}
Nowak M.~A., Wagoner R.~V., 1991, \emph{ApJ}, 378, 656

\bibitem[{Nowak \& Wagoner(1992)}]{nowakwagoner1992}
---, 1992, \emph{ApJ}, 393, 697

\bibitem[{Nowak {et~al.}(1997)Nowak, Wagoner, Begelman, \&
  Lehr}]{nowaketal1997}
Nowak M.~A., Wagoner R.~V., Begelman M.~C., Lehr D.~E., 1997, \emph{ApJ}, 477,
  L91

\bibitem[{O'Donoghue \& Charles(1996)}]{donocharles1996}
O'Donoghue D., Charles P.~A., 1996, \emph{MNRAS}, 282, 191

\bibitem[{Ogilvie(1998)}]{ogilvie1998}
Ogilvie G.~I., 1998, \emph{MNRAS}, 297, 291

\bibitem[{Ogilvie(2000)}]{ogilvie2000}
---, 2000, \emph{MNRAS}, 317, 607

\bibitem[{Ogilvie(2001)}]{ogilvie2001}
---, 2001, \emph{MNRAS}, 325, 231

\bibitem[{Ogilvie(2005)}]{ogilvienotes}
---, 2005, Lecture notes on accretion discs, \emph{Part III of the Mathematical
  Tripos, University of Cambridge,
  http://www.damtp.cam.ac.uk/user/gio10/accretion.html}

\bibitem[{Ogilvie(2008)}]{ogilvie2008}
---, 2008, \emph{MNRAS}, 388, 1372

\bibitem[{Ogilvie \& Dubus(2001)}]{ogilviedubus2001}
Ogilvie G.~I., Dubus G., 2001, \emph{MNRAS}, 320, 485

\bibitem[{Okazaki {et~al.}(1987)Okazaki, Kato, \& Fukue}]{okazakietal1987}
Okazaki A.~T., Kato S., Fukue J., 1987, \emph{PASJ}, 39, 457

\bibitem[{Okuda \& Mineshige(1991)}]{okudamineshige1991}
Okuda T., Mineshige S., 1991, \emph{MNRAS}, 249, 684

\bibitem[{O'Neill {et~al.}(2009)O'Neill, Reynolds, \& Miller}]{oneill2009}
O'Neill S.~M., Reynolds C.~S., Miller M.~C., 2009, \emph{ApJ}, 693, 1100

\bibitem[{Ortega-Rodr\'iguez \& Wagoner(2000)}]{orw2000}
Ortega-Rodr\'iguez M., Wagoner R.~V., 2000, \emph{ApJ}, 537, 922

\bibitem[{Paczy\'nski(1987)}]{nature}
Paczy\'nski B., 1987, \emph{Nature}, 327, 303

\bibitem[{Paczy\'nski(2000)}]{notorque}
---, 2000, \emph{astro-ph/0004129v1}

\bibitem[{Paczy\'nski \& Bisnovatyi-Kogan(1981)}]{pbk1981}
Paczy\'nski B., Bisnovatyi-Kogan G., 1981, \emph{AcA}, 31, 283

\bibitem[{Paczy\'nski \& Wiita(1980)}]{pw1980}
Paczy\'nski B., Wiita P.~J., 1980, \emph{A\&A}, 88, 23

\bibitem[{Papaloizou \& Szuszkiewicz(1994)}]{ps1994}
Papaloizou J., Szuszkiewicz E., 1994, \emph{MNRAS}, 268, 29

\bibitem[{Papaloizou \& Lin(1995)}]{papaloizoulin1995}
Papaloizou J. C.~B., Lin D. N.~C., 1995, \emph{ApJ}, 438, 841

\bibitem[{Papaloizou \& Pringle(1984)}]{ppi1}
Papaloizou J. C.~B., Pringle J.~E., 1984, \emph{MNRAS}, 208, 721

\bibitem[{Papaloizou \& Pringle(1985)}]{ppi2}
---, 1985, \emph{MNRAS}, 213, 799

\bibitem[{Papaloizou \& Stanley(1986)}]{papaloizoustanley1986}
Papaloizou J. C.~B., Stanley G. Q.~G., 1986, \emph{MNRAS}, 220, 593

\bibitem[{Papaloizou \& Terquem(1995)}]{papterquem1995}
Papaloizou J. C.~B., Terquem C., 1995, \emph{MNRAS}, 274, 987

\bibitem[{Patterson {et~al.}(2005)Patterson, Kemp, Harvey, Fried, Rea, Monard,
  Cook, Skillman, Vanmunster, Bolt, Armstrong, McCormick, Krajci, Jensen, Gunn,
  Butterworth, Foote, Bos, Masi, \& Warhurst}]{pattersonetal2005}
Patterson J., Kemp J., Harvey D.~A., Fried R.~E., Rea R., Monard B., Cook
  L.~M., Skillman D.~R., Vanmunster T., Bolt G., Armstrong E., McCormick J.,
  Krajci T., Jensen L., Gunn J., Butterworth N., Foote J., Bos M., Masi G.,
  Warhurst P., 2005, \emph{PASP}, 117, 1204

\bibitem[{Perez {et~al.}(1997)Perez, Silbergleit, Wagoner, \&
  Lehr}]{perezetal1997}
Perez C.~A., Silbergleit A.~S., Wagoner R.~V., Lehr D.~E., 1997, \emph{ApJ},
  476, 589

\bibitem[{Prendergast \& Burbridge(1968)}]{pb1968}
Prendergast K.~H., Burbridge G.~R., 1968, \emph{ApJ}, 151, L83

\bibitem[{Pringle(1981)}]{pringle1981}
Pringle J.~E., 1981, \emph{ARA\&A}, 19, 137

\bibitem[{Pringle(1996)}]{pringle1996}
---, 1996, \emph{MNRAS}, 281, 357

\bibitem[{Pringle \& Rees(1972)}]{pr1972}
Pringle J.~E., Rees M.~J., 1972, \emph{A\&A}, 21, 1

\bibitem[{Psaltis {et~al.}(1999)Psaltis, Belloni, \& van~der Klis}]{pbk1999}
Psaltis D., Belloni T., van~der Klis M., 1999, \emph{ApJ}, 520, 262

\bibitem[{Raine \& Thomas(2005)}]{bh}
Raine D., Thomas E., 2005, \emph{Black Holes, An Introduction}. Imperial
  College Press, London

\bibitem[{Rau {et~al.}(2003)Rau, Greiner, \& McCollough}]{rauetal2003}
Rau A., Greiner J., McCollough M.~L., 2003, \emph{ApJ}, 590, L37

\bibitem[{Rebusco(2008)}]{rebusco2008}
Rebusco P., 2008, \emph{NewAR}, 51, 855

\bibitem[{Rees {et~al.}(1982)Rees, Begelman, Blandford, \& Phinney}]{rbbp1982}
Rees M.~J., Begelman M.~C., Blandford R.~D., Phinney E.~S., 1982,
  \emph{Nature}, 295, 17

\bibitem[{Reis {et~al.}(2009)Reis, Fabian, Ross, \& Miller}]{reisetal2009}
Reis R.~C., Fabian A.~C., Ross R.~R., Miller J.~M., 2009, \emph{MNRAS}, 395,
  1257

\bibitem[{Remillard \& McClintock(2006)}]{remimcclin2006}
Remillard R.~A., McClintock J.~E., 2006, \emph{ARA\&A}, 44, 49

\bibitem[{Remillard {et~al.}(2002)Remillard, Muno, McClintock, \&
  Orosz}]{remillardmicroquasar2002}
Remillard R.~A., Muno M.~P., McClintock J.~E., Orosz J.~A., 2002, \emph{ApJ},
  580, 1030

\bibitem[{Reynolds \& Miller(2009)}]{rm2009}
Reynolds C.~S., Miller M.~C., 2009, \emph{ApJ}, 692, 869

\bibitem[{Reynolds \& Nowak(2003)}]{rn2003}
Reynolds C.~S., Nowak M.~A., 2003, \emph{Phys. Rep.}, 377, 389

\bibitem[{Rezzolla {et~al.}(2003)Rezzolla, Yoshida, Maccarone, \&
  Zanotti}]{rezzollaetal2003}
Rezzolla L., Yoshida S., Maccarone T.~J., Zanotti O., 2003, \emph{MNRAS}, 344,
  L37

\bibitem[{Robinson \& Nather(1979)}]{robinsonnather1979}
Robinson E.~L., Nather R.~E., 1979, \emph{ApJS}, 39, 461

\bibitem[{Rodriguez {et~al.}(2002)Rodriguez, Varni\`ere, Tagger, \&
  Durouchoux}]{vrtd2002}
Rodriguez J., Varni\`ere P., Tagger M., Durouchoux P., 2002, \emph{A\&A}, 387,
  487

\bibitem[{Rothschild {et~al.}(1974)Rothschild, Boldt, Holt, \&
  Serlemitsos}]{rothschildetal1974}
Rothschild R.~E., Boldt E.~A., Holt S.~S., Serlemitsos P.~J., 1974, \emph{ApJ},
  189, L13

\bibitem[{Sakimoto \& Coroniti(1981)}]{sc81}
Sakimoto P.~J., Coroniti F.~V., 1981, \emph{ApJ}, 247, 19

\bibitem[{Schnittman(2005)}]{s2005}
Schnittman J.~D., 2005, \emph{ApJ}, 621, 940

\bibitem[{Schnittman \& Bertschinger(2004)}]{sb2004}
Schnittman J.~D., Bertschinger E., 2004, \emph{ApJ}, 606, 1098

\bibitem[{Shakura \& Sunyaev(1973)}]{ss73}
Shakura N.~I., Sunyaev R.~A., 1973, \emph{A\&A}, 24, 337

\bibitem[{Shakura \& Sunyaev(1976)}]{ss76}
---, 1976, \emph{MNRAS}, 175, 613

\bibitem[{Shklovsky(1967)}]{shklovsky1967}
Shklovsky I.~S., 1967, \emph{ApJ}, 148, L1

\bibitem[{Stella \& Vietri(1998)}]{sv1998}
Stella L., Vietri M., 1998, \emph{ApJ}, 492, L59

\bibitem[{Stella {et~al.}(1999)Stella, Vietri, \& Morsink}]{svm1999}
Stella L., Vietri M., Morsink S.~M., 1999, \emph{ApJ}, 524, L63

\bibitem[{Taam \& Lin(1984)}]{tl1984}
Taam R.~E., Lin D. N.~C., 1984, \emph{ApJ}, 287, 761

\bibitem[{Tagger \& Varni\`ere(2006)}]{tv2006}
Tagger M., Varni\`ere P., 2006, \emph{ApJ}, 652, 1457

\bibitem[{Tanaka {et~al.}(2002)Tanaka, Takeuchi, \& Ward}]{tanakaetal2002}
Tanaka H., Takeuchi T., Ward W.~R., 2002, \emph{ApJ}, 565, 1257

\bibitem[{Thorne(1974)}]{thorne1974}
Thorne K.~S., 1974, \emph{ApJ}, 191, 507

\bibitem[{Toomre(1977)}]{toomre1977}
Toomre A., 1977, \emph{ARAA}, 15, 437

\bibitem[{Turner(2004)}]{turner2004}
Turner N.~J., 2004, \emph{ApJ}, 605, L45

\bibitem[{Uemura {et~al.}(2002)Uemura, Kato, Ishioka, Yamaoka, Schmeer, Krajci,
  Starkey, Torii, Kawai, Urata, Kohama, Yoshida, Ayani, Kawabata, Tanabe,
  Matsumoto, Kiyota, Pietz, Vanmunster, Oksanen, \& Giambersion}]{uemura2002}
Uemura M., Kato T., Ishioka R., Yamaoka H., Schmeer P., Krajci T., Starkey
  D.~R., Torii K., Kawai N., Urata Y., Kohama M., Yoshida A., Ayani K.,
  Kawabata T., Tanabe K., Matsumoto K., Kiyota S., Pietz J., Vanmunster T.,
  Oksanen A., Giambersion A., 2002, \emph{PASJ}, 54, 599

\bibitem[{van~der Klis(2006)}]{klisbook}
van~der Klis M., 2006, in Lewin W. H. G., van der Klis M., eds, \emph{Compact
  Stellar X-ray Sources}, Cambridge Univ. Press, Cambridge, p.~39

\bibitem[{{Van Horn} {et~al.}(1980){Van Horn}, Wesemael, \& Winget}]{hww1980}
{Van Horn} H.~M., Wesemael F., Winget D.~E., 1980, \emph{ApJ}, 235, L143

\bibitem[{Varni\`ere {et~al.}(2002)Varni\`ere, Rodriguez, \& Tagger}]{rvt2002}
Varni\`ere P., Rodriguez J., Tagger M., 2002, \emph{A\&A}, 387, 497

\bibitem[{Velikhov(1959)}]{velikhov1959}
Velikhov E.~T., 1959, \emph{Sov. Phys. JETP}, 36, 995

\bibitem[{von Weizs\"acker(1948)}]{vonw1948}
von Weizs\"acker C.~F., 1948, \emph{Z. Naturforsch}, 3a, 524

\bibitem[{Wagoner(1999)}]{reldisko}
Wagoner R.~V., 1999, \emph{Phys. Rep.}, 311, 259

\bibitem[{Wallinder(1990)}]{wallinder1990}
Wallinder F.~H., 1990, \emph{A\&A}, 237, 270

\bibitem[{Wallinder(1995)}]{wallinder1995}
---, 1995, \emph{MNRAS}, 273, 273

\bibitem[{Warner(2004)}]{warner2004}
Warner B., 2004, \emph{PASJ}, 116, 115

\bibitem[{Whitehurst(1988)}]{whitehurst1988}
Whitehurst R., 1988, \emph{MNRAS}, 232, 35

\bibitem[{Wijers \& Pringle(1999)}]{wijerspringle1999}
Wijers R. A. M.~J., Pringle J.~E., 1999, \emph{MNRAS}, 308, 207

\bibitem[{Wood {et~al.}(2002)Wood, Wolff, Bjorkman, \& Whitney}]{hh30}
Wood K., Wolff M.~J., Bjorkman J.~E., Whitney B., 2002, \emph{ApJ}, 564, 887

\bibitem[{Wu \& Goldreich(2001)}]{wugoldreich2001}
Wu Y., Goldreich P., 2001, \emph{ApJ}, 546, 469

\bibitem[{Zhang {et~al.}(1997)Zhang, Cui, \& Chen}]{zcc1997}
Zhang S.~N., Cui W., Chen W., 1997, \emph{ApJ}, 482, L155

\bibitem[{Zurita {et~al.}(2008)Zurita, Durant, Torres, Shahbaz, Casares, \&
  Steeghs}]{zurita2008}
Zurita C., Durant M., Torres M. A.~P., Shahbaz T., Casares J., Steeghs D.,
  2008, \emph{ApJ}, 681, 1458

\end{thebibliography}
\addcontentsline{toc}{chapter}{References} 

\end{document}